%% file: PH2-enrichment.tex
\documentclass[tradiabstract]{aa}  

\usepackage{graphicx}
\usepackage[dvipsnames]{xcolor}

\usepackage{txfonts}
\usepackage{appendix}
\begin{document}

%User shortcuts:
\newcommand{\Ho}{\mbox{$H_0$}}
\newcommand{\ang}{\mbox{{\rm \AA}}}
\newcommand{\abs}[1]{\left| #1 \right|} % for absolute value
\newcommand{\kms}{\ensuremath{{\rm km\,s^{-1}}}}
\newcommand{\zabs}{\ensuremath{z_{\rm abs}}}
\newcommand{\zem}{\ensuremath{z_{\rm QSO}}}
\newcommand{\cmsq}{\ensuremath{{\rm cm}^{-2}}}
\newcommand{\ergs}{\ensuremath{{\rm erg\,s^{-1}}}}
\newcommand{\ergsa}{\ensuremath{{\rm erg\,s^{-1}\,{\AA}^{-1}}}}
\newcommand{\ergscm}{\ensuremath{{\rm erg\,s^{-1}\,cm^{-2}}}}
\newcommand{\ergscma}{\ensuremath{{\rm erg\,s^{-1}\,cm^{-2}\,{\AA}^{-1}}}}
\newcommand{\msyr}{\ensuremath{{\rm M_{\rm \odot}\,yr^{-1}}}}
\newcommand{\nhi}{n_{\rm HI}}
\newcommand{\fhi}{\ensuremath{f_{\rm HI}(N,\chi)}}
\newcommand{\refs}{{\bf (refs!)}}
\newcommand{\Av}{\ensuremath{A_V}}
\newcommand{\lya}{Ly-$\alpha$}
\newcommand{\Hb}{H-$\beta$}
\newcommand{\OIII}{\ion{O}{iii}}
\newcommand{\OII}{\ion{O}{ii}}
\newcommand{\OI}{\ion{O}{i}}
\newcommand{\HI}{\ion{H}{i}}
\newcommand{\HeII}{\ion{He}{ii}}
\newcommand{\HH}{\ensuremath{{\rm H}_2}}
\newcommand{\SII}{\ion{S}{ii}}
\newcommand{\SiIII}{\ion{Si}{iii}}
\newcommand{\SiIV}{\ion{Si}{iv}}
\newcommand{\SiII}{\ion{Si}{ii}}
\newcommand{\AlIII}{\ion{Al}{iii}}
\newcommand{\AlII}{\ion{Al}{ii}}
\newcommand{\ArI}{\ion{Ar}{i}}
\newcommand{\FeII}{\ion{Fe}{ii}}
\newcommand{\ZnII}{\ion{Zn}{ii}}
\newcommand{\CrII}{\ion{Cr}{ii}}
\newcommand{\MnII}{\ion{Mn}{ii}}
\newcommand{\MgII}{\ion{Mg}{ii}}
\newcommand{\MgI}{\ion{Mg}{i}}
\newcommand{\NiII}{\ion{Ni}{ii}}
\newcommand{\NV}{\ion{N}{v}}
\newcommand{\CIV}{\ion{C}{iv}}
\newcommand{\CIII}{\ion{C}{iii}}
\newcommand{\CII}{\ion{C}{ii}}
\newcommand{\CI}{\ion{C}{i}}
\newcommand{\CaII}{\ion{Ca}{ii}}
\newcommand{\TiII}{\ion{Ti}{ii}}
\newcommand{\Jzzuc}{J0015$+$1842}
\newcommand{\Jzzucl}{SDSS\,J001514.82$+$184212.3}
\newcommand{\Jzzun}{J0019$-$0137}
\newcommand{\Jzzunl}{SDSS\,J001930.55$-$013708.4}
\newcommand{\Jzzcn}{J0059$+$1124}
\newcommand{\Jzzcnl}{SDSS\,J005917.64$+$112407.7}
\newcommand{\Jzudc}{J0125$-$0129}
\newcommand{\Jzudcl}{SDSS\,J012555.11$-$012925.0}
\newcommand{\Jzuts}{J0136$+$0440}
\newcommand{\Jzutsl}{SDSS\,J013644.02$+$044039.1}
\newcommand{\Judts}{J1236$+$0010}
\newcommand{\Judtsl}{SDSS\,J123602.11$+$001024.5}
\newcommand{\Judqh}{J1248$+$0639}
\newcommand{\Judqhl}{SDSS\,J124829.51$+$063925.6}
\newcommand{\Judcn}{J1259$+$0309}
\newcommand{\Judcnl}{SDSS\,J125917.31$+$030922.5}
\newcommand{\Juttu}{J1331$+$0206}
\newcommand{\Juttul}{SDSS\,J133111.41$+$020609.0}
\newcommand{\Jutch}{J1358$+$1410}
\newcommand{\Jutchl}{SDSS\,J135808.94$+$141053.2}
\newcommand{\Jdddh}{J2228$-$0221}
\newcommand{\Jdddhl}{SDSS\,J222807.36$-$022117.1}
\newcommand{\Jdtdc}{J2325$+$1539}
\newcommand{\Jdtdcl}{SDSS\,J232506.62$+$153929.3}
\newcommand{\Jzhch}{J0858+1749}
\newcommand{\Jzhchl}{SDSS\,J085859.67+174925.1}

%Institutes:
\newcommand{\fcla}{French-Chilean Laboratory for Astronomy, IRL 3386, CNRS and U. de Chile, Casilla 36-D, Santiago, Chile \label{fcla}}
\newcommand{\iap}{Institut d'Astrophysique de Paris, CNRS-SU, UMR\,7095, 98bis bd Arago, 75014 Paris, France \label{iap}}
\newcommand{\ioffe}{Ioffe Institute, {Polyteknicheskaya 26}, 194021 Saint-Petersburg, Russia \label{ioffe}}
\newcommand{\cral}{Centre de Recherche Astrophysique de Lyon, UMR5574, U. Lyon 1, ENS de Lyon, CNRS, 69230 Saint-Genis-Laval, France \label{cral}}

\newcommand{\obspm}{Observatoire de Paris, LERMA, Coll\`ege de France, CNRS, PSL University, Sorbonne University, 75014, Paris, France \label{obspm}}
\newcommand{\iucaa}{Inter-University Centre for Astronomy and Astrophysics, Pune University Campus, Ganeshkhind, Pune 411007, India \label{iucaa}}
\newcommand{\dawn}{Cosmic Dawn Center (DAWN) and Niels Bohr Institute, University of Copenhagen, Jagtvej 128, DK-2200, Copenhagen N, Denmark \label{dawn}}
\newcommand{\uchile}{Departamento de Astronom\'ia, Universidad de Chile, Casilla 36-D, Santiago, Chile \label{uchile}}
\newcommand{\unige}{Department of Astronomy, University of Geneva, Chemin Pegasi 51, 1290 Versoix, Switzerland \label{unige}}
\newcommand{\eso}{European Southern Observatory, Alonso de C\'ordova 3107, Vitacura, Casilla 19001, Santiago, Chile \label{eso}}
%---------------------------------------------------------
% Commands to enable all co-authors to make comments choose your color !
\definecolor{green}{rgb}{0,0.4,0}
\newcommand{\PN}[1]{{\color{orange} PN:~ #1}}
\newcommand{\SB}[1]{{\color{olive} SB:~ #1}}
\newcommand{\JK}[1]{{\color{violet} JK:~ #1}}
\newcommand{\RC}[1]{{\color{blue} RC:~ #1}}
\newcommand{\FC}[1]{{\color{green} FC:~ #1}}
\newcommand{\CL}[1]{{\color{red} CL:~ #1}}
\newcommand{\ADC}[1]{{\color{teal} AdC:~ #1}}
\newcommand{\RS}[1]{{\color{teal} RS:~ #1}}
\newcommand{\NRJ}[1]{{\color{teal} NG~ #1}}
\newcommand{\SL}[1]{{\color{teal} SL:~ #1}}
\newcommand{\test}[1]{{\color{teal} test:~ #1}}

%--------------------------------------------------

% Changes to text
\newcommand{\old}[2]{{\color[rgb]{0,0,0}\sout{#1}}{\color[rgb]{0.7,0,0.0}{\bf #2}}}
\newcommand{\new}[1]{{\bf #1}}
\newcommand{\assign}[2]{\noindent{\color{blue}#1 $\to$ #2}}

   \title{Proximate molecular quasar absorbers}

   \subtitle{Chemical enrichment and kinematics of the neutral gas
   \thanks{Based on observations collected at the European Organisation for Astronomical Research in the Southern Hemisphere under ESO programmes 103.B-0260(A) and 105.203L.001.}}

 \author{
   P.~Noterdaeme\inst{\ref{fcla},\ref{iap}}
  \and
   S.~Balashev\inst{\ref{ioffe}}
   \and
   R.~Cuellar\inst{\ref{fcla}, \ref{uchile}}
   \and
   J.-K.~Krogager\inst{\ref{cral}}
   \and
   F.~Combes\inst{\ref{obspm}}
   \and\\
      % Authors order may be changed -- Alpha for now beyond that line.
   A.~De~Cia\inst{\ref{unige}}
   \and
   N.~Gupta\inst{\ref{iucaa}}
   \and
   C.~Ledoux\inst{\ref{eso}}
   \and
   S.~L{\'opez}\inst{\ref{uchile}}
   \and
   R.~Srianand\inst{\ref{iucaa}}
          }
          
   \institute{\fcla \and \iap \and \ioffe \and \uchile \and \cral \and \obspm \and \unige \and \iucaa \and \eso}
             
   \date{\today.}

 \abstract{
 Proximate molecular quasar absorbers (P\HH) are an intriguing population of absorption systems recently uncovered through strong \HH\ absorption at small velocity separation from the background quasars. 
We performed a multi-wavelength spectroscopic follow-up of thirteen such systems with VLT/X-Shooter. Here, we present the observations and study the overall chemical enrichment measured from the \HI, H$_2$ and metal lines. We combine this with an investigation of the neutral gas kinematics with respect to the quasar host.

 We find {gas-phase} metallicities in the range 2\% to 40\% of the Solar value, i.e. in the upper-half range of \HI-selected proximate damped Lyman-$\alpha$ systems, 
 but similar to what is seen in intervening \HH-bearing systems. This is likely driven by similar selection effects that play against the detection of most metal and molecular rich systems in absorption. Differences are however seen {in the abundance of dust (from [Zn/Fe]) and} its depletion pattern, when compared to intervening systems, possibly indicating different dust production or destruction close to the AGN. {We also note the almost-ubiquitous} presence of a high-ionisation phase {traced by \NV} in proximate systems. In spite of the hard UV field from the quasars, we found no strong overall deficit of neutral argon, at least when compared to intervening DLAs. This likely results from argon being mostly neutral in the \HH\ phase, which actually accounts for a large fraction of the total amount of metals. 
 
 We measure the quasar systemic redshifts through emission lines from both ionised gas and CO(3-2) emission, the latter being detected in all 6 cases for which we obtained 3-mm data from complementary NOEMA observations. For the first time, we observe a trend 
 between line-of-sight velocity {with respect to systemic redshift} and metallicity of the absorbing gas. This suggests that high-metallicity neutral and molecular gas is more likely to be located in outflows while low-metallicity gas could be more clustered in velocity space around the quasar host, possibly with an infalling component. 
 }%{\newline}{\newline}{\newline}{} 
% 5 {} token are mandatory

   \keywords{galaxies: active -- galaxies: evolution -- quasars: general -- quasars: absorption lines -- quasars: emission lines}

   \maketitle
%
%-------------------------------------------------------------------

\section{Introduction}
\defcitealias{Noterdaeme2019}{Paper I} 	 
\defcitealias{Noterdaeme2021}{Paper II}
\defcitealias{Noterdaeme2021b}{Paper III} 	 
\defcitealias{Balashev202x}{Paper IV}

The co-evolution of active galactic nuclei (AGNs) and their host galaxies 
is regulated by galaxy processes, including secular evolution, inflow rate of gas (from the intergalactic medium to the interstellar medium and ultimately to the supermassive black hole (SMBH)) as well as merging history \citep[see][]{Hopkins2008}. 
The apparent properties of AGNs may further depend on the orientation and the evolutionary stage at which the observations are taken \citep[e.g.,][]{Urry1995}. This constitutes an entire field of modern astrophysics, motivating large efforts from the community, in both theoretical and observational grounds. Because of the large diversity of observed properties, it is common to select objects based on a few among their many characteristics (e.g., redshift, brightness, colour, radio-loudness, variability, etc.) and study the relationships between other marginalised properties. Each of these studies then bring a different and very valuable lightening angle to the overall picture.

As cold gas constitutes the fuel for both star-formation and the growth of the supermassive black holes, many works have focused on detecting the dense molecular component of the brightest AGNs --quasars-- through CO emission \citep[e.g.,][]{Omont1996, Weiss2007, Wang2016}, in order to explore the links between star-formation, total molecular mass in the host and growth and activity of the SMBH \citep[e.g.,][]{Bischetti2017}.

On the other hand, absorption studies enable detailed investigations of the gas properties along the line-of-sight to the central engine, over a wide range of densities. For example, ionised winds powered by the accretion disc can be observed as broad absorption line systems (BALs) in roughly 15\% of quasars, and have velocities from a few thousands \kms\ up to $\sim$0.3\,$c$ \citep{Hamann2018}. 
Damped Lyman-$\alpha$ systems (DLAs) are absorption systems with high enough \HI\ column density ($N(\HI) > 2\times10^{20}$~\cmsq) to be self-shielded from ionising radiation and are frequently seen along quasar lines of sight \citep[e.g., ][]{Prochaska2005,Noterdaeme2009b}. While most DLAs are unrelated to the quasar environment and only intercepted by chance (called {\sl intervening} DLAs), some DLAs are found at roughly the same redshift as the background quasar  and hence dubbed {\sl proximate} DLAs (PDLAs) where "proximate" refers to proximity in velocity space (typically $<$ a few 10$^{3}$~\kms). Because of the much smaller redshift path in which to identify PDLAs compared to that available for intervening DLAs, the former are much rarer than the latter. Nonetheless, \citet{Prochaska2008} showed the incidence of PDLAs (per unit redshift) is higher than expected from the intervening statistics meaning that they are likely associated to a clustered environment around the quasar. 

\citet{Ellison2010} performed the first systematic study of PDLAs at high-resolution and concluded that at least a fraction of them exhibit different characteristics relative to the intervening population. They however suggested that PDLAs are unlikely related to the quasar host, but sample preferably more massive galaxies in the highly clustered region around the quasar. Interestingly, \lya\ emission has been relatively frequently found in the core of several PDLAs \citep[e.g.,][]{Moller1993, Leibundgut1999, Ellison2002}, when this is more rarely seen among the several 10$^4$ known intervening DLAs (but see \citealt{Fynbo2010,Noterdaeme2012b,Ranjan2018} for some example of direct detection and \citealt{Rahmani2010,Noterdaeme2014,Joshi2017} for statistical detection through stacking). \citet{Finley2013} and \citet{Fathivavsari2018} further identified a population of PDLAs with \lya\ emission spanning a range of strengths, up to the point where the \HI\ \lya\ absorption becomes unseen, but the presence of neutral gas is confirmed by strong low-ionisation metal lines. Such emissions are unlikely to arise from star-formation but rather from \lya\ from the AGN that is not fully covered by the absorber. Because \lya\ photons scatter easily and up to large distances around quasars \citep[e.g.,][]{Battaia2018,Borisova2016}, concluding about the location, scale and geometry of the absorbing gas is not straightforward. However, signatures of high excitation of the gas suggest an origin at rather small distance from the AGN, at least for the most extreme cases.  

\citet[][hereafter \citetalias{Noterdaeme2019}]{Noterdaeme2019} presented a new approach focusing on the presence of strong molecular hydrogen absorption directly detected at the quasar redshift in Sloan Digital Sky Survey (SDSS) spectra, without prior on the presence of \HI\ or metal lines. Not only H$_2$ is sensitive to the physical and chemical conditions in the cold gas, but such an approach also differs from studies based on, e.g., CO molecular emission in the sense that no prior is made on the overall mass of dense molecular gas in the galaxy host, but rather on the interception of more diffuse cold gas by the quasar line of sight. The strong excess of proximate H$_2$ absorbers  with respect to intervening statistics shows that most of these systems must be associated with the quasar environment, although it is not clear whether these arise from the quasar host{, or} if the excess is due to strong clustering of galaxies around the quasar.

To shed light on this issue, we embarked into a follow-up study with the multi-wavelength, medium resolution spectrograph X-shooter on the Very Large Telescope (VLT). A detailed study of a first system revealed 
a likely origin in outflowing material intercepted by the line of sight to the central engine \citep[][\citetalias{Noterdaeme2021}]{Noterdaeme2021}. Remarkably strong and wide CO(3-2) emission as well as 3~mm continuum emission was detected in the same, optically bright (moderately reddened) system through NOEMA observations \citep[][\citetalias{Noterdaeme2021b}]{Noterdaeme2021b}. All this suggests that the nucleus is seen through a channel of lower density but where some of the molecular content is carried out by outflows, when the host still contains large amounts of dust and molecular gas with wide kinematics, possibly indicative of a post-merger system.

In this work, we focus on the overall chemical enrichment and kinematics of the gas of our follow-up sample. In Sect.~\ref{s:obs} we present the sample, observations, data reduction and post-processing. In Sect.~\ref{s:analysis} we present the column density and dust extinction measurements. In Sect.~\ref{s:kine}, we present the emission redshifts and absorption line kinematics. We discuss our findings and compare our results with other PDLAs as well as strong intervening H$_2$ systems in  Sect.~\ref{s:results}. We conclude in Sect.~\ref{s:conclusion}.

%--------------------------------------------------------------------
\section{Sample, observations and data reduction \label{s:obs}}

\subsection{Sample and observations}
The sample studied here is entirely drawn from \citetalias{Noterdaeme2019}. Among the statistical sample of 50 quasars  
with high-confidence proximate molecular absorbers, 22 have 
$\delta < 20^{\circ}$, that is, reachable from the Southern hemisphere. We obtained VLT/X-shooter data for {thirteen} of them: eleven based on their observability during the allocated periods of this specific programme
(P103 and P105\footnote{Because of delays related to the observatory shutdown during the Covid-19 pandemic, the observations were completed in P109.}) plus the quasars \Jzutsl\ {and \Jzhchl} for which we use data collected in P94 as part of another programme \citep{Balashev2019}\footnote{We note that \Judtsl\ had already been observed by \citet{Balashev2019}. This quasar was then re-observed under better conditions and from various position angles thanks to the availability of the X-Shooter atmospheric dispersion corrector. We use here the new data.}.
We note that since the quasars in the present sample were selected from the statistical sample upon their observability only (and not any other property) they are likely a fairly good representation of the latter.

X-shooter \citep{Vernet2011} simultaneously covers the whole optical/NIR wavelength range from 0.3 to 2.5~$\mu$m, splitting the incoming light beam into three spectrographs ('arms'): UVB ranging from 0.3 to 0.6~$\mu$m; VIS ranging from 0.6 to 1.0~$\mu$m; and NIR ranging from 1.0 to 2.5~$\mu$m. All observing blocks (OBs) consisted of 2980\,s, 3010\,s and 5$\times$600\,s exposures for the UVB, VIS and NIR arms, respectively obtained in 'stare' mode. We used 1.0 and 0.9$\arcsec$-wide slits in the UVB and VIS, with slow read-out mode and 1$\times$2 pixel binning. The NIR observations were performed with 1.2$\arcsec$-slits for the two quasars observed in P103 (\Jzzucl\ and \Jdtdcl). The remaining observations were performed with a narrower NIR slit (0.9~arcsec) and a K-band blocking filter to avoid stray light in the instrument, effectively cutting out all wavelengths longer than 2.08$~\mu$m for all remaining quasars but \Jzutsl\ and \Jzhchl\ (in P94) for which the K-band filter was not yet available.
We executed two to five OBs for each quasar at different position angles.
Most of the observations were obtained under good seeing, somewhat sharper than the slit widths at the corresponding wavelengths, resulting in higher than nominal spectral resolution.
The observing log is given in Table~\ref{t:log}. {For convenience, we will refer to these objects using a short name version in the following: SDSS\,Jhhmmss.ss+ddmmss.s will be abbreviated as Jhhmm+ddmm (e.g. \Jzzuc\ instead of \Jzzucl).}

\addtolength{\tabcolsep}{-4pt}
\begin{table*}
\caption{Quasar sample and log of X-shooter observations \label{t:log}}
\begin{tabular}{c c l l l c}
\hline \hline
{\large \strut}Quasar (SDSS)\tablefootmark{a}    & $z$  & Observing dates       & Seeing\tablefootmark{b} (arcsec) & airmass & Ref.\\ 
\hline
J001514.82+184212.3 & 2.63 & 29/07, 31/08, 31/08, 28/09 2019           &  1.67, 0.75, 0.65, 0.53       & 1.48, 1.39, 1.41, 1.42    & (1)      \\
J001930.55-013708.4 & 2.53 & 06/11, 07/11, 08/11 2021                     &  0.91, 0.72, 0.71             & 1.30, 1.26, 1.16                        & (2)     \\
J005917.64+112407.7 & 3.03 & 09/11, 09/11, 09/11 2021                     &  0.77, 0.80, 1.14             & 1.24, 1.30, 1.47                        & (2)      \\
J012555.11-012925.0 & 2.66 & 21/07, 01/09, 01/09, 11/10, 25/11 2021 &  1.14, 1.00, 0.98, 1.22, 0.92 & 1.11, 1.10, 1.17, 1.10, 1.17 & (3)      \\
J013644.02+044039.1 & 2.78 & 27/11, 28/11 2014                   & 1.15, 1.15 & 1.16, 1.26                                         & (4)   \\
J085859.67+174925.1 & 2.65 & 24/12 2014, 15/02 2015              & 0.73, 1.38 & 1.36, 1.37 & (4) \\
J123602.11+001024.5 & 3.03 & 07/02, 07/02, 07/02 2022                     &  0.73, 0.83,0.87              & 1.20,1.12,1.11                          & (2)\\
J124829.51+063925.6 & 2.53 & 08/02, 07/03, 07/03 2022                     & 1.21, 0.61,0.65               & 1.19, 1.18, 1.19                        & (2)   \\
J125917.31+030922.5 & 3.23 & 06/02, 06/02 2022                            &  0.79, 0.93      & 1.33, 1.43                            &    (2) \\
J133111.41+020609.0 & 2.92 & 08/03, 29/05, 29/05 2022           &  0.80, 0.84, 0.98      & 1.25,1.17,1.30                &    (2) \\
J135808.94+141053.2 & 2.89 & 25/02, 01/03, 07/03 2022            &  1.01, 0.79 , 0.68     &   1.29, 1.29, 1.30            &  (2)           \\
J222807.36-022117.1 & 2.77 & 31/08, 15/09, 12/10, 26/10 2021            &  0.87, 0.71, 0.84, 0.85      &  1.10, 1.12, 1.09, 1.09                 &   (2)  \\
J232506.62+153929.3 & 2.62 &     29/08, 31/08, 30/09 2019        & 0.84, 0.77, 0.67       &  1.33, 1.33, 1.34                     &  (2) \\
\hline
\end{tabular}
\tablefoot{
\tablefoottext{a}{{The actual object designation following the rules from the International Astronomy Union, includes the "SDSS" preceding the J2000 coordinates sequence provided in this column.}}
\tablefoottext{b}{Delivered seeing corrected for airmass{, as measured at the telescope by the Shack–Hartmann wavefront sensor (fits keyword \texttt{HIERARCH ESO TEL IA FWHM}).}}
}
 \tablebib{(1)~\citet{Noterdaeme2021}; (2)~this work; (3)~Balashev et al. (in prep); (4)~\citet{Balashev2019}}
\end{table*}
\addtolength{\tabcolsep}{+4pt}

\subsection{Data reduction and telluric corrections}
All X-shooter spectra were processed using the official esorex pipeline version 3.5.3 \citep{Modigliani2010} for `stare' mode reduction maintaining the default parameters and including the correction for instrument flexure (recipe `xsh\_flexcomp'). The raw images were pre-processed using the Python package astroscrappy (a reimplementation of the L.A.Cosmic algorithm by \citealt{vanDokkum2001}) to interpolate over pixels contaminated by cosmic ray hits {using the neighboring pixels}. Each arm of each OB was reduced separately using the standard star observed closest in time in order to flux calibrate the given spectrum. The 1D spectra were then extracted using optimal extraction \citep{Horne1986}.

Next, we used the ESO tool molecfit v.4.2.2a.1 \citep{Smette2015} to correct each individual 1D VIS and NIR spectra for the absorption lines produced by the Earth's atmosphere. In the VIS, the main telluric absorption features are due to O$_2$ and H$_2$O, while the NIR is strongly affected by H$_2$O and CO$_2$ bands. Following the recommendation from the molecfit manual, the correction was done by choosing a few spectral regions with clear but not strongly saturated atmospheric lines for each quasar -typically 0.68-0.69, 0.71-0.73, 0.76-0.77, 0.81-0.82 and 0.91-0.91 $\mu$m in the VIS. In the NIR, we had to adjust more the selected spectral regions for each quasar, depending on the strength of absorption bands and location of quasar emission lines. During this process, we also masked deviant pixels (e.g., cosmic remnants) as well as astrophysical lines. 

The 
best-fit model was used to calculate the overall atmospheric transmission and correct the individual VIS and NIR 1D exposures accordingly. 
The telluric-corrected spectra were then shifted to vacuum-barycentric wavelengths, rebinned to a common wavelength grid and combined using an inverse-variance weighting. The same vacuum-baryocentric correction and combination of 1D spectra was applied to UV data though without telluric correction.

We note that molecfit also provides the actual spectral resolution, that is a free parameter of the transmission model. As expected from a seeing sharper than the slit widths for most of the observations, the resolving power is typically 25\% higher than the nominal values in the documentation ($R_{\rm nom}=5400$ (UV) and 8900 (VIS)). {The average spectral resolution obtained {from individual VIS spectra} is then adopted when fitting the lines. In the UV, the spectral resolution is first estimated by scaling the nominal UV value by the observed-to-nominal VIS resolution ratio.{ We cross-checked the values are in broad agreement with those expected for a slit width that would correspond to the observed seeing.} It was then adjusted during the fitting process if necessary. While the exact values have little influence on the metal column densities, we provide the adopted values of the resolutions for the combined spectra along with the best-fit model parameters in Appendix~\ref{s:metfigs} to ease reproducibility.}

\subsection{NOEMA observations}

While the present work focuses on X-shooter data, 
we also obtained observations with the NOrthern Extended Millimeter Array (NOEMA) and the PolyFIX correlator in the 3 mm band between May 2020 and May 2021 for 6 out of the {thirteen} objects presented here. The central frequency was chosen in order to cover the expected position of the CO(3-2) emission line, which we detected in all 6 cases. 
A description of data handling and analysis of NOEMA observations {(including other quasars from the same parent sample but not observed with X-shooter)}
{will be presented in a}
future paper but we already use the redshifts from the detected CO(3-2) emission lines in the discussion (Sect.~\ref{s:results}). {The Gaussian fits to the CO(3-2) emission lines are presented in Appendix~\ref{a:zCO}.}

\section{Column-density measurements \label{s:analysis}}

We analysed the absorption lines through standard, simultaneous, multi-component Voigt-profile fitting using vpfit v.12.3 \citep{Carswell2014} to measure the column densities, Doppler parameters and redshifts. The quasar continuum was evaluated using spline functions, with nodes constrained by the unabsorbed regions and adjusted over strong lines in an iterative way when necessary.
In order to best take into account blending between metal, H$_2$ and \HI\ lines, the final absorption models were obtained by fitting all lines simultaneously. However, for the sake of presentation, we describe here the analysis separately for \HI, H$_2$ and metals. The results from the fits are presented in the Appendix, with the main derived properties summarised in Table.~\ref{t:sum}.

\subsection{Total HI and H$_2$ column densities}
\label{sect:HIH2}
The total \HI\ column densities were mostly constrained from the saturated core of the Ly-$\alpha$ line, which is little sensitive to the exact placement of the continuum on top of the Ly-$\alpha$ emission line of the quasars. Notwithstanding, thanks to the wide wavelength coverage of X-shooter and the relatively high redshifts of quasars (implied by the selection upon \HH\ in SDSS), all lines from the \HI\ Lyman series are covered by our spectra, allowing to constrain the \HI\ column density further. While the formal statistical errors from fitting the lines are very small (typically 0.01-0.02~dex), we note that the continuum placement as well as choice of fitting regions are necessarily somehow subjective, possibly introducing some systematic errors. Allowing for continuum variation using Chebychev polynomials indicate uncertainties less than 0.1~dex for similar DLA observations with X-shooter \citep[e.g.,][]{Balashev2019,Ranjan2020}. We hence added here a conservative 0.1~dex uncertainty into the error budget.

In all the systems, we confirm the presence of strong H$_2$ absorption lines from the Lyman and Werner bands. These are typically detected from rotational level J=0 to J$\sim$6-7 in our X-shooter spectra. 
The high column densities imply damped lines at least in the first few rotational levels and hence accurate column densities, since lines in this regime do not depend on the Doppler parameter $b$. We do also take into account and fit the detected high rotational levels, assuming the same Doppler parameter and redshift for all H$_2$ lines. 
However, the derived column densities for the high-$J$ levels are likely to be more sensitive to the assumptions on the Doppler parameters. In fact, $b$-values for H$_2$ lines have been found to present a dependence on the rotational level in several cases, see, e.g., \citet{Noterdaeme2007b, Balashev2009}. A more detailed study and exploration of systematics is hence needed in order to use these lines to constrain the excitation of the gas and the prevailing physical conditions. This is postponed to a future work{, while} we focus here on the total column densities, which are dominated by the low-rotational levels. The fits to the H$_2$ absorption bands is presented in Appendix~\ref{s:H2figs}.

In Fig.~\ref{f:H2_sdss_xs}, we compare the total H$_2$ and \HI\ column densities 
obtained with X-shooter, with those derived from SDSS. In the case of \HI\ the agreement is very good, and always better than 0.3~dex. For H$_2$, for which the involved columns are much lower, the agreement remains remarkable (apart for the lowest column density system with a difference of 0.85~dex), in particular given the fact that the SDSS values were obtained by visually scaling and matching a template (i.e., assuming single component with fixed Doppler parameter and excitation temperature) to the low-resolution, low S/N SDSS spectra (\citetalias[see][]{Noterdaeme2019}). This agreement is similar to what was found by \citet{Balashev2019} for intervening \HH\ systems.

\begin{figure}
    \centering
    \includegraphics[width=\hsize]{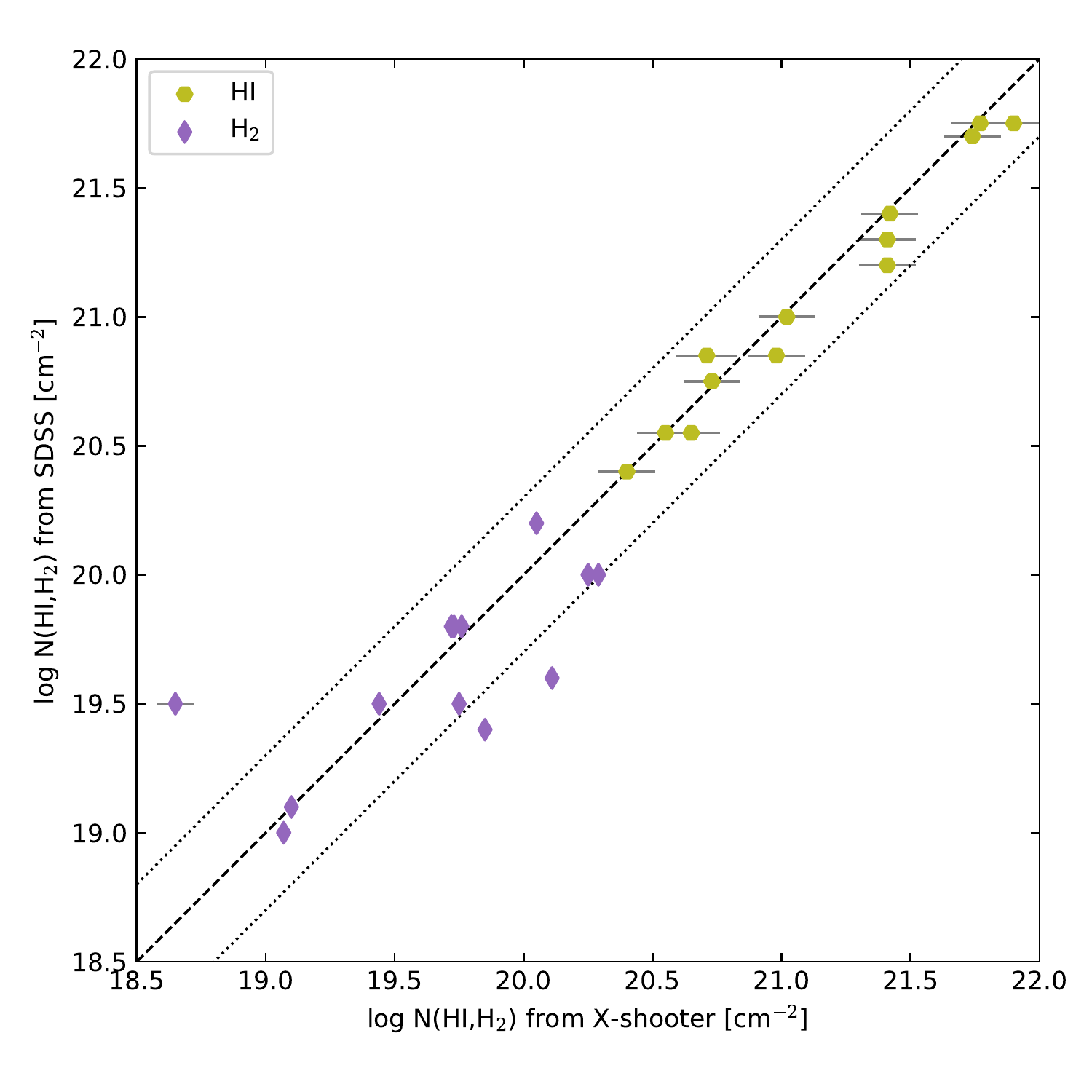}
    \caption{Comparison of H$_2$ (purple diamonds) and \HI\ (green hexagons) column densities obtained using X-shooter and SDSS spectra. 
    The dashed line depicts the one-to-one relation, and the dotted lines show $\pm0.3$~dex around it.
    }
    \label{f:H2_sdss_xs}
\end{figure}

\subsection{Low-ionisation metal species}

{In this section, we focus on the main low-ionisation metal lines that provide reliable column densities, that is, lines that have multiplets, are not strongly saturated and not in a region strongly affected by telluric lines. For example, we did not attempt to fit the only singly-ionised and very strong carbon line \CII$\lambda1334$.} 
We fitted the low-ionisation metal lines assuming a common velocity structure with turbulent broadening (i.e., tied redshifts and Doppler parameters). We first identified the velocity components using the strong \FeII\ and \SiII\ lines in the visual --that also has higher spectral resolution than UVB-- and obtained a first fit, including weaker lines. We then added other metal species 
and adjusted the fitting regions and the number of components whenever necessary to obtain the final fit. 

During this process, we also fitted simultaneously some other intervening systems that posses lines blended with that of the proximate systems, or masked the corresponding blended regions. Note also that $\ZnII\lambda$2026 is systematically blended with $\MgI\lambda2026$, which we did take into account {assuming the same components}, although \MgI\ is not a species discussed in this work. {When ever possible, we also used \MgI$\lambda$1827 and/or \MgI$\lambda$2852 to further constrain the (small) contribution of \MgI\ to the 2026 profile.}
{We also took into account partial blending of $\ZnII$ with $\CrII$ at 2062~{\AA}. }
Finally, the analysis of neutral carbon (incl. fine-structure lines), neutral chlorine and excited fine-structure lines (from \CII, \SiII\ and \OI) is post-poned to a future work since we focus here on chemical enrichment where their contribution is negligible. {That being said, we already note the clear detection of \SiII* towards \Jzzuc, \Jzzun, \Jzzcn, \Jzudc, \Judcn, \Jutch, \Jdddh\ and \Jdtdc.}
The fit to the different metal lines is presented in Appendix~\ref{s:metfigs}.

We note that the metal column densities obtained at medium spectral resolution should always be considered with some caution since 
the modelling of the velocity profile mostly implied fewer but broader ($b$ frequently as high as 20~\kms) components than typically observed in high-resolution spectra, where $b <10~\kms$. Therefore, in spite of our best efforts to model the profiles and use lines with a range of strengths, unresolved line saturation may still affect the column densities estimates for some of the species more than the formal uncertainty suggests. Anyway, metallicities are derived using weak \ZnII\ lines in all but one case, where the total column densities are less sensitive to exact velocity structure decomposition.

\subsection{Dust}
We quantify the amount of dust in the absorbers by measuring the extinction of the quasar, following a template matching procedure as done in many works \citep[e.g.,][]{Srianand2008b,%Noterdaeme2009,
Ma2015,Krogager2016,Zhang2015}.
In order to perform the dust reddening measurement using the full wavelength range provided by X-shooter, we 
combined the spectra from all three arms to obtain a joint spectra for each quasar.
The combination was performed using an inverse-variance weighting in the overlapping regions.
During the combination, all three arms were interpolated onto a common wavelength grid using
a fixed sampling of 0.2 $\AA$. The joint spectra were lastly corrected for Galactic extinction
using the map by \citet{Schlafly2011}.

The individual joint spectra were then fitted using the quasar template by \citet{Selsing2016}
assuming a fixed extinction law at the redshift of the quasar\footnote{This implicitly assumes that no other intervening absorber is causing the reddening. Such an assumption is reasonable since the proximate systems likely contain most of the dust present along the line of sight as indicated by the presence of \HH\ and the low probability to intercept by chance another rare dust-rich absorber. Indeed, we do not find signs for additional metal-rich intervening absorbers in any of the quasar spectra.}.
Since we observe no evidence of 2175~\AA\ extinction features, we used the extinction law inferred for the Small Magellanic Cloud \citep{Gordon2003}.
In order not to bias the fit we exclude regions of the spectra around the strong, broad emission lines.
We furthermore exclude regions that are affected by strong telluric absorption. 

For each target, we derive the dust extinction as the average of the $A_V$ values
for the individual observations. The statistical uncertainty is determined as the
standard deviation of the individual measurements. This uncertainty includes any uncertainty
on the flux calibration of the different spectra.
Lastly, from previous work we assign a systematic uncertainty of 0.07~mag \citep{Noterdaeme2017}
due to any possible intrinsic mismatch of the template and the given target {(i.e. accounting for scatter in the intrinsic spectral energy distribution of quasars)}.
The results are summarised in the before-last column of Table~\ref{t:sum} and the fits shown in Appendix~\ref{a:dust}.

\section{Absorption line kinematics and systemic redshifts\label{s:kine}}

\subsection{Absorption line kinematics}
We quantify the kinematics of the neutral absorbing gas using the velocity width as defined by \citet{Prochaska1997}: $\Delta v_{90} = c[\lambda_{95}-\lambda_5]/\lambda_{50}$, where $\lambda_5$, $\lambda_{50}$ and $\lambda_{95}$ are the wavelengths corresponding to, respectively, the five per cent, fifty and ninety-five percentiles of the apparent optical depth distribution. For each system, we followed the criteria from \citet{Ledoux2006} and selected a few low-ionisation absorption lines that are not too saturated --it would otherwise be impossible to measure the apparent optical depth-- nor too weak --in which case part of the gas would remain untraced--. We then checked for consistency between results. The 
$\Delta v_{90}$ for each PDLA is shown in Appendix~\ref{a:dv90} and provided in the last column of Table~\ref{t:sum}.

We note that while the smearing of the lines at medium spectral resolution can in principle result in a small overestimation of $\Delta v_{90}$ \citep{Prochaska2008,Arabsalmani2015}, this is mostly true for low velocity widths and becomes only an issue when comparing statistics of samples taken at different spectral resolutions. Moreover, the spectral resolution and $\Delta_v$ are large enough for the corrections to be here negligible, although, strictly speaking, the smallest values (for \Juttu\ and \Judcn) could be considered as upper-limits.

\subsection{Quasar redshifts \label{s:zem}}
A precise measurement of the quasar's redshift is then required to study the motion of the absorbing gas with respect to the central engine along the line of sight, and more generally, to investigate dynamical processes in the circum-nuclear region and/or in the host galaxy. Velocity shifts between the different emission lines are 
seen in quasar spectra, reflecting the diverse kinematics of the emitting regions  \citep[e.g.,][]{Gaskell1982,Tytler1992,VandenBerk2001,Shen2016}. Overall, it has been found that high-ionisation lines tend to be strongly blue-shifted with respect to the systemic redshift ($z_{\rm sys}$), while low-ionisation lines provide a better mean to measure $z_{\rm sys}$, the most reliable of which being the oxygen lines. 

We hence first focused on the [\OII] and [\OIII] lines, which are redshifted in the NIR. Unfortunately, telluric absorption bands frequently affect these lines, and in spite of our best efforts to correct for the atmospheric absorption, it is not always possible to recover the unabsorbed emission line properly.
In the cases where the lines were reasonably seen, we modelled the region covering H$\beta$ and [\OIII] using a simple linear continuum on top of which we added Gaussian lines (with more than one component whenever necessary), but tied the parameters of the two lines of the [\OIII] doublet together: same velocity, same width, and a flux ratio of 3. [\OII] was fitted independently. 
In order to remove spikes, skyline or cosmic-ray residuals as well as absorption lines, we iteratively fitted the data rejecting deviant pixels {(at more than 3\,$\sigma$) at each iteration}  
until convergence {(i.e. until no more pixels are rejected)}. We then visually compared the derived model with the rebinned data, obtained from median averaging in bins of typically 25 pixels. Unfortunately, for two [\OII] lines, sky line residuals are close to the peak and hence the [\OII] redshifts may not be as reliable as expected. This effect is mitigated for [\OIII], which is generally stronger and has two lines.

In principle, \MgII\ is the following choice at shorter wavelengths \citep{Shen2016}. We note that owing to our selection on the presence of a proximate absorber, the line is affected by strong \MgII\ absorption, which we masked during the fitting process. We then also considered the strong \CIII] and the weaker \HeII\ lines at 1908 and 1640~{\AA} rest-frame, respectively, which are free from strong absorption.  
We followed the prescription from \citet{Shen2016} and measure the lines redshifts from the peak of the emission, obtained from a model that reproduces the data. Multiple Gaussians provided convenient models thanks to their flexibility and the relatively small numbers of free parameters. 

The redshift obtained this way are given in Table~\ref{t:zem} using the same rest-frame wavelengths as in \citet{Shen2016}. In the case of [\OII], [\OIII] and \MgII, the authors found small systematic shifts ($\sim 8$, $-45$ and $-57$~\kms, respectively) compared to the systemic value, {which they determined from \CaII\ absorption.}
This is less than our measurement uncertainties so that we do not attempt any correction here. In the case of the \CIII] complex (which includes \SiIII] and \AlIII), they found a luminosity-independent systematic shift of $-$229~\kms and in the case of \HeII\, the systematic shift depends on luminosity. {Here as well, we followed the prescription from \citet{Shen2016} and applied the correction accordingly (see their Eq.~1 and best-fit parameters in their Table~2).}
We note that these empirical shifts include any possible physical shifts of the emitting regions, plus the fact that several species actually contribute to the complex.
We provide also the corrected redshifts in Table~\ref{t:zem}, in an attempt to remove the average trends seen in quasars. As the average shifts from \citet{Shen2016} are computed based on low redshift quasars observed in the SDSS, we finally did a last test by matching the quasar composite from \citet{Selsing2016} with our data in a region of the visual spectrum covering the range 1620-2450~{\AA} rest-frame, i.e., covering the features from \HeII\ to [Ne\,{\sc iv}]. 
This range is slightly reduced in a few cases to avoid bad quality regions, generally at long wavelengths. Absorption lines, bad pixels and sky-line residuals are masked during the fitting process. The only four free parameters are a scaling value, an additional continuum (power-law+constant) 
and the redshift. 

The different measurements (see Appendix \ref{a:zem}) then allow us to cross-check the emission redshift and decide on the best estimate based on ionised lines (which we call $z_{\rm syst}^i$), as given in the before-last column of Table~\ref{t:zem}. {In general, we prioritised the lines following the prescription from \citet{Shen2016} although in some cases, we preferred a second-choice line with high S/N over a line with more uncertain measurement. This is the case for \Jzuts\ and \Jzhch, where we preferred the corrected \CIII] over \MgII, and \Juttu\ where we preferred \MgII\ over [\OII]. At the end, \HeII\ and template measurement were not used, but provided for comparison and to give more confidence to the measurement.}
Finally, in the last column of this table, we indicate the redshift of the CO(3-2) emission line derived from fitting the 3-mm NOEMA data with single Gaussian profiles. This is expected to be an even better indicator of the quasar host systemic redshift, since it traces the bulk of the molecular gas \citep{Wang2016}. The corresponding fits are shown in Appendix~\ref{a:zCO}. {Interestingly, the CO-based redshifts appear to be systematically higher by about 250~\kms (average based on the 6 systems where CO(3-2) measurements are available) with respect to the systemic redshifts based on rest frame UV/optical transitions. This is discussed in Sect.~\ref{s:resultskine}.}

\input zem.tex

\section{Results and discussions \label{s:results}}

\subsection{Chemical enrichment}

In the following, we will discuss the chemical properties of our sample based on the measured column densities and further compare them with other samples of absorption systems: (1) Proximate \HI-selected DLAs observed at high spectral resolution by \citet{Ellison2010}; (2) a compilation of intervening DLAs 
\citep{DeCia2016}
with robust abundance measurements and (3) a compilation of \HH\ systems from \citet{Balashev2019} and confirmed from follow-up observations. This compilation contains H$_2$-bearing systems identified with a number of different selections: in regular DLA systems \citep[][]{Noterdaeme2008,Balashev2014,Balashev2017}, extreme \HI\ absorbers \citep{Noterdaeme2015,Ranjan2020} and \CI-selected systems \citep{Ledoux2015,Noterdaeme2018}. A discussion on the selection function is presented in \citet{Balashev2019}, so we here only consider the sample as a whole, but keeping in mind that its distribution in parameter space is a sum of several selections. We removed from the intervening samples some systems that have $\zabs\approx\zem$ and could actually be considered as proximate. In order to avoid possible evolutionary effect in our comparison, we also restrict these samples to $z>2$. {Finally, we removed known \HH-systems from the sample by \citet{DeCia2016} to avoid duplicates with \citet{Balashev2019}.
} 

We use the standard notation of {gas-phase abundances} of the different species X expressed relative to the Solar values as [X/H] = $\log(N({\rm X})/N({\rm H_{tot}})) - \log({\rm X/H})_{\odot}$, and where we take the Solar abundances from \citet{Asplund2021} 
following the recommendations of \citet{Lodders2003} on whether to take photospheric, meteoritic or average values \citep[see, e.g., table 1 of][]{Konstantopoulou2022}.  We do the usual assumption that ionisation corrections are negligible for DLAs thanks the self-shielding from $>13.6$~eV photons, i.e., that the considered species are in the main ionisation stage in the neutral gas (e.g., $N(\ZnII$) = N(Zn)). We do however include \HH\ in the total hydrogen column density, i.e., $N({\rm H}) = N(\HI) + 2N(\HH)$. 
{We remind that we consider the gas-phase abundances only when quoting abundances as [X/H]. In other words, we do not attempt any correction for dust depletion. We will further investigate this effect by considering volatile and refractory elements in the following sections.} A summary of the total abundances of the main species of interest is provided in Table~\ref{t:sum}. 

\input PH2sum

\subsubsection{Overall metallicity from volatile species: zinc and sulphur}
\begin{figure}[t]
    \centering
        \includegraphics[width=\hsize]{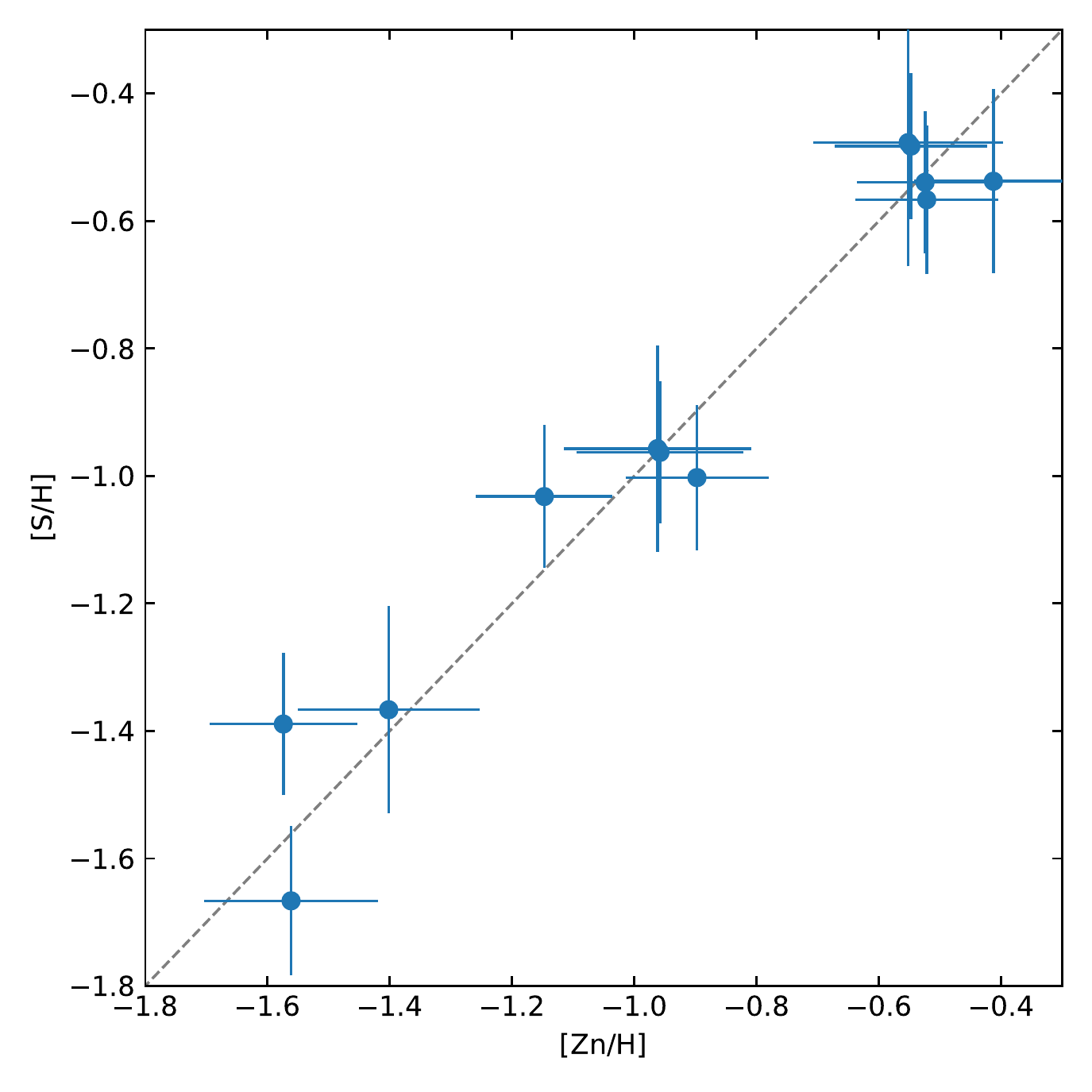} 
    \caption{Observed abundance of sulphur against that of zinc in proximate H$_2$ systems.}
    \label{f:SH_ZnH}
\end{figure}

Zinc and sulphur are generally considered as the best reference elements for the intrinsic metallicity (i.e., $\log Z/Z_{\rm \odot} \approx [$Zn/H$] \approx [$S/H$]$) since they are volatile and hence little depleted onto dust grains \citep[see, e.g.,][among many other works]{Welty1999b,DeCia2016}.  We confirm that for all PDLAs where both zinc and sulphur abundances can be measured, there is good agreement between both values, with an average measured sulphur-to-zinc abundance ratio $[$S/Zn$]\approx 0$ and dispersion 0.1~dex (see Fig.~\ref{f:SH_ZnH}). {We note that, in principle, this does not exclude depletion of both species to similar amounts. 
However, we could expect any depletion to start differing at higher [Zn/H], which is not apparent. We do not see any trend of [S/Zn] with [Zn/Fe] either. In any case, for comparison with other measurements in the literature, we will continue assuming $\rm \log Z/Z_{\rm \odot} \equiv [Zn/H]$ in the following.} 
For the only system without Zn measurement (\Jzuts, where the corresponding lines are too heavily blended with atmospheric lines), we hence use Sulphur instead. Overall, we found metallicities in proximate H$_2$ systems to span a range from $\log Z/Z_{\rm \odot} \approx -1.6$ to -0.4. This corresponds to the upper half range of \HI-selected proximate DLAs, as investigated by \citet{Ellison2010}, but similar to the metallicity of intervening \HH-bearing DLA systems \citep[e.g.][]{Petitjean2006,Balashev2019}.

\subsubsection{Dust depletion from relative abundances}
\begin{figure}[t]
    \centering
    \includegraphics[width=\hsize]{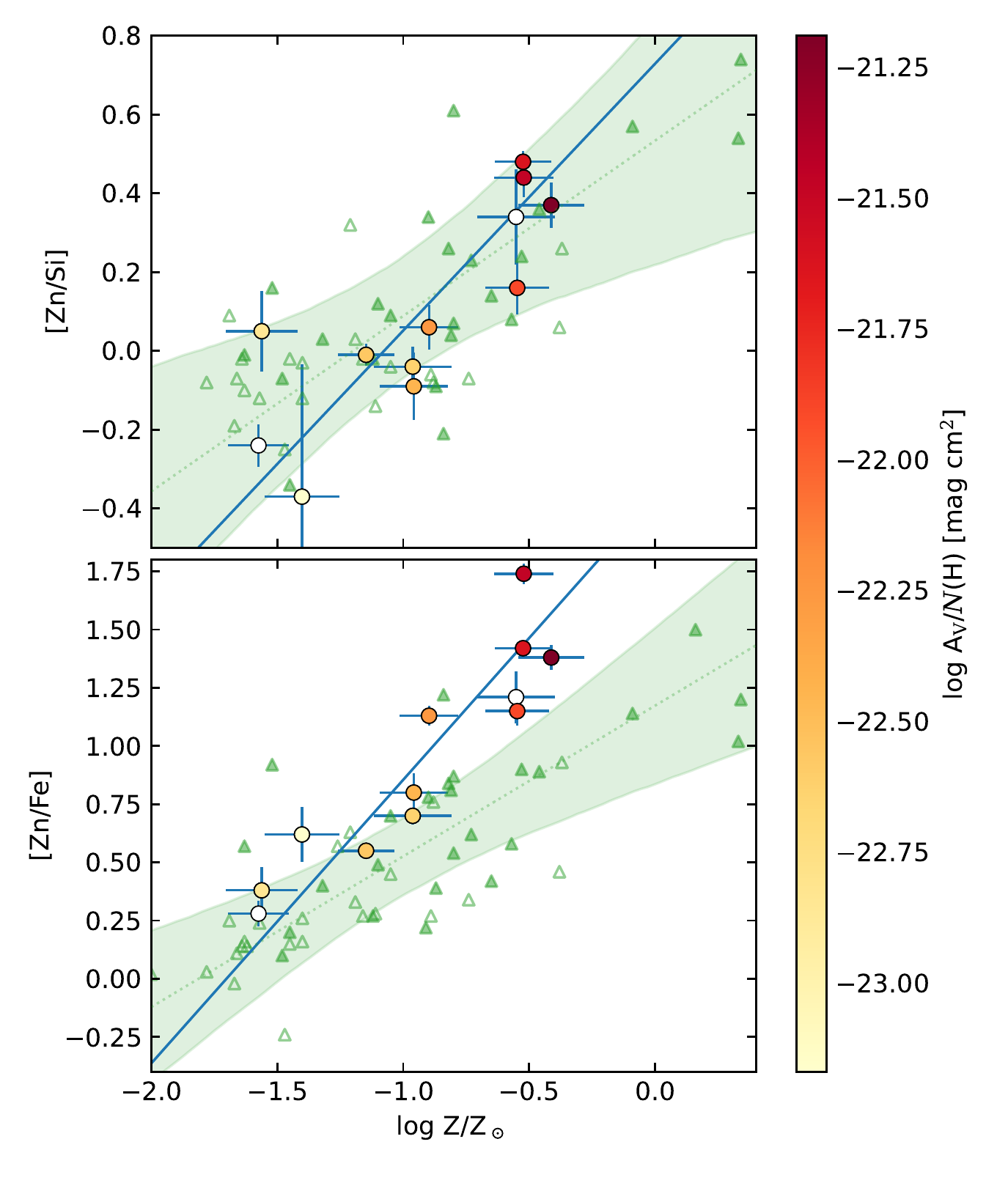}
    \caption{Zinc-to-iron {(bottom)} and zinc-to-silicon {(top) }
    for our sample of proximate H$_2$ systems (circles with error bars). The colour-scale {provide the value of A$_{\rm V}/N(\rm H)$, which is proxy of the} dust-to-gas ratio. For two systems the derived $A_{\rm V}$ values are actually negative (hence non-physical) and the corresponding points are unfilled. 
    {Green} triangles represent measurements in intervening DLAs {\citep{DeCia2016}} {with filled ones corresponding to known \HH\ intervening systems \citep{Balashev2019}}. The {solid blue} lines show the best linear fit to the value for the proximate \HH\ systems, taking into account uncertainties on both axis.{ The green lines and shadow region show the best linear fit and associated 1\,$\sigma$ uncertainty for the overall sample of intervening DLAs.}
    }
    \label{f:ZnFe_Z}
\end{figure}

Silicon and iron are refractory elements known to be respectively mildly and strongly depleted onto dust grains and their gas-phase abundances are typically lower than those of zinc and sulphur. In Fig.~\ref{f:ZnFe_Z}, we show the abundances ratios [Zn/Fe] and [Zn/Si] as a function of metallicity. There is clearly an increase of these ratios with increasing metallicity --as has been observed in a number of works on intervening DLAs (e.g., \citealt{Ledoux2003,Noterdaeme2008,DeCia2016, Balashev2019}) 
-- which can be attributed to the increasing abundance of dust with increasing metallicity. 

Iron appears to be significantly depleted even down to the lowest metallicities while [Si/Zn] becomes positive only for metallicities higher than 1/10$^{th}$ Solar. Interestingly, [Zn/Fe] in proximate \HH\ systems are clearly higher for a given metallicity than the values seen in regular intervening DLAs suggesting a higher depletion of iron in proximate systems. This remains true when comparing only to intervening \HH\ absorbers, even if the latter {tend to have high depletion among the overall (mostly non-\HH\ bearing) intervening population at a given metallicity.} It is also interesting to see a lower dispersion for proximate systems around the main trend (dotted blue line), {likely} suggesting a more homogeneous population.

On the other hand, the difference between proximate H$_2$ and intervening systems (regardless of their H$_2$ content) is less clear in the case of [Zn/Si]. Since silicon {is} an $\alpha$-element predominantly produced in massive stars, it is possible that a stronger intrinsic depletion (with respect to intervening systems) is compensated by recent star-formation, or that the depletion sequence for Si and Fe also differs.

In short, the high depletion (from [Zn/Fe]) at a given metallicity suggests a mechanism for build-up and/or destruction of dust grains that is similar across the population of proximate \HH\ systems but different from that seen in intervening DLAs. {Any correction to account for some depletion of zinc and recover the intrinsic metallicity could then also be different for proximate systems.} {We note that assuming the same depletion sequences as in intervening systems, the corrected metallicities, following \citet{DeCia2016} would be a factor two higher, on average.}

\begin{figure}[!ht]
    \centering
    \includegraphics[width=\hsize]{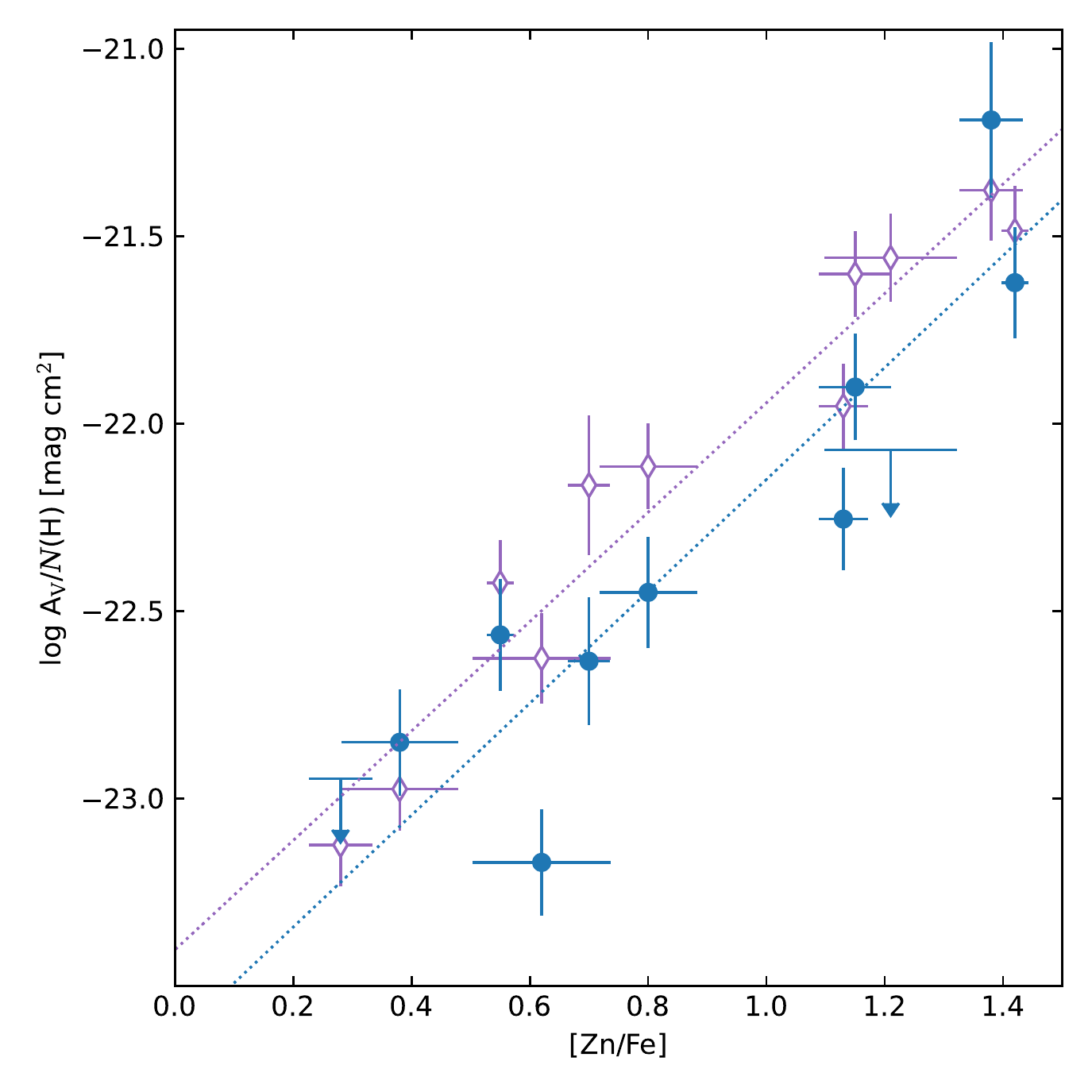}
    \caption{{Dust-to-gas ratio measured from the continuum spectrum (filled circles) and derived from the metal column densities (open diamonds, following \citealt{DeCia2016}), as a function of depletion factor. The blue (resp. purple) dotted line corresponds to a linear fit to ([Zn/Fe], $\log A_{\rm V}/N({\rm H})$) data points, where $A_v$ is based on continuum fitting (resp. metal column density).}}
    \label{f:Av_deple}
\end{figure}

The abundance of dust can be independently quantified using the extinction per hydrogen atom, $A_{\rm V}/N({\rm H})$. While the 0.07~mag uncertainty on A$_V$ measurements remains quite large compared to their values, there is a clear tendency for higher $A_{\rm V}/N({\rm H})$ at higher [Zn/Fe], [Zn/Si] and metallicities, strengthening a dust-depletion origin for the observed underabundance of iron and silicon.  
{
The trend between depletion and value of A$_{\rm V}/N(\rm H)$
turns out to be a strong correlation with Pearson coefficient $r=0.9$ and probability of chance coincidence 0.07\%, see Fig.~\ref{f:Av_deple}. The low dispersion around the approximate linear relation 
($\log A_V/N($H$) \approx 1.3[$Zn/Fe$]-23.5$) 
indicates that the quasar template matching provides reliable estimates, at least relatively to each other. Interestingly, it is also possible to indirectly derive the extinction from the gas phase abundances, assuming some scaling with respect to the Galactic values, which we also represent in Fig.~\ref{f:Av_deple}, based on calculations following \citet{DeCia2016}. These values appear to 
be in relatively good agreement with the reddening-based measurements, albeit the former being on average higher by 0.25~mag than the latter. This may be another indication for a different type of dust close to the quasar. However, a comparison of extinction derived from gas-phase abundances with that directly obtained from the spectral slope should be performed first in intervening systems to validate the prescription by \citet{DeCia2016} at such precision level.}

\subsubsection{The sulphur-to-silicon ratio in proximate absorption systems}
\begin{figure}[t]
    \centering
        \includegraphics[width=\hsize]{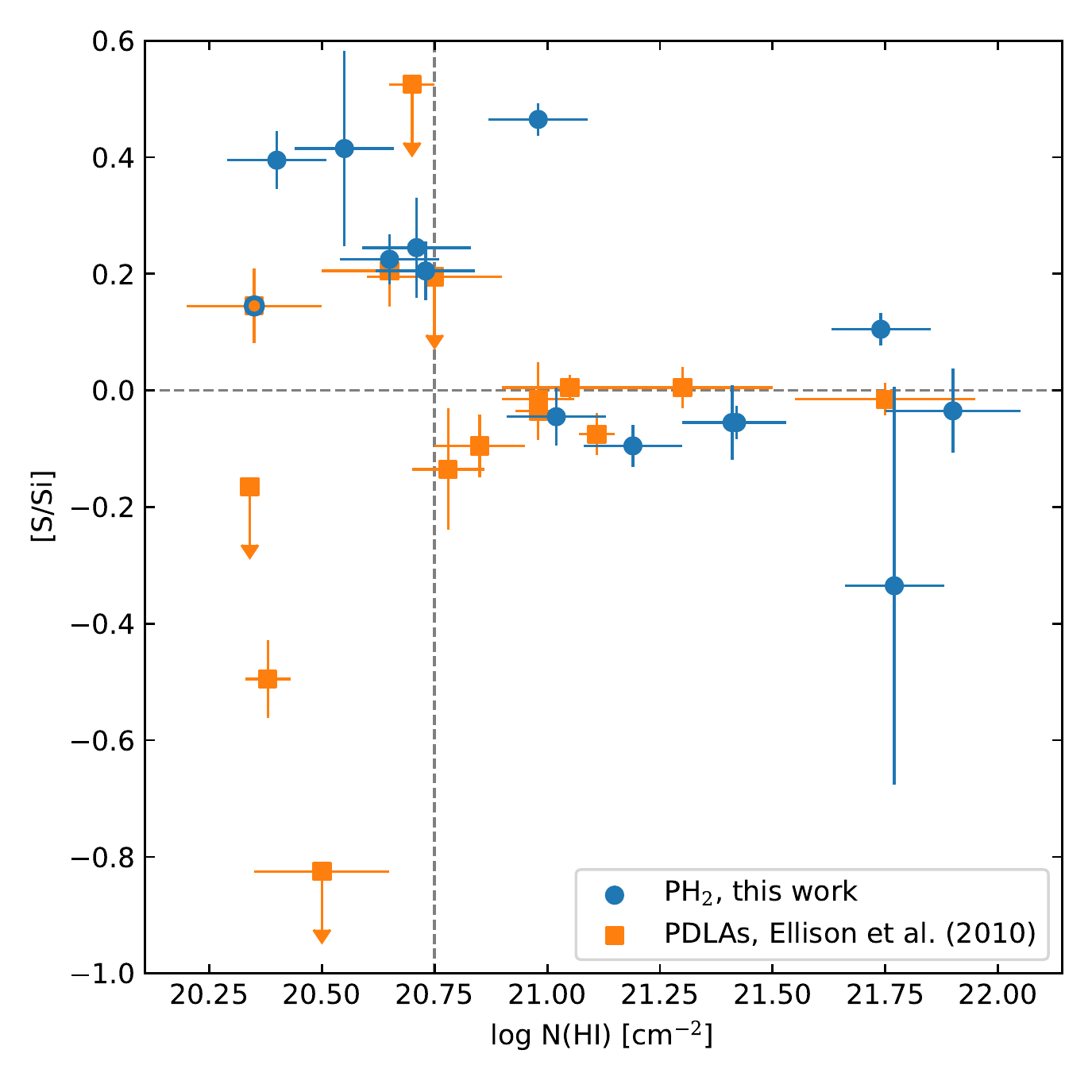} 

    \caption{Sulphur to Silicon abundance ratio versus \HI\ column density. The horizontal dashed line marks the Solar ratio while the vertical line shows the $N({\rm \HI})$ limit below which \citet{Ellison2010} observed under-Solar abundances in PDLAs (see text). Note that, while not in our sample of proximate H$_2$, one PDLA from \citet{Ellison2010}, namely J2340-00, also has strong H$_2$ and is marked with a blue circle.
    }
    \label{f:SSi_H}
\end{figure}

The [S/Si] ratio is also an interesting quantity to look at since both species are $\alpha$ elements and their intrinsic abundances are expected to follow each other. According to \citet{Ellison2010}, {based on modelling by \citet{Rix2007}}, the observed negative [S/Si] values at low \HI\ column densities in their sample of PDLAs is due to ionisation by a hard spectrum. 
We compare this ratio in our sample of proximate systems with those from \citeauthor{Ellison2010} as a function of the \HI\ column density in Fig.~\ref{f:SSi_H}. Contrary to what these authors observed in PDLAs, we do not see any drop in [S/Si] below $\log N(\HI)=20.75$ in our strong H$_2$-bearing PDLAs. While we do not probe \HI\ columns as small as these authors do, it is interesting to note that both PDLAs and PH$_2$ have close to Solar ratios at high column densities $\log N(\HI) >21$. 
In contrast with the PDLAs from \citeauthor{Ellison2010}, we see a tendency for increasing [S/Si] with decreasing \HI-column, which is opposite to what is expected if {such} ionisation corrections are at play but rather reflects an increasing depletion with decreasing \HI. This can be due to a bias against systems with high column densities of dust that would be either excluded from the parent quasar sample \citep{Ledoux2009, Augustin2018} or result in low S/N ratio of the corresponding spectra. It can also be due to a higher abundance of metals/dust required for the presence of H$_2$ at low \HI\ column densities \citep[e.g.,][]{Bialy2017}, thereby increasing the dust shielding, H$_2$ formation rate and gas cooling through atomic lines. In the low $N(\HI)$ regime, we would qualify the observed [S/Si] in (non-\HH) PDLAs as highly dispersed and possibly interpret this as a wide range of ionisation.

\begin{figure}
    \centering
    \includegraphics[width=\hsize]{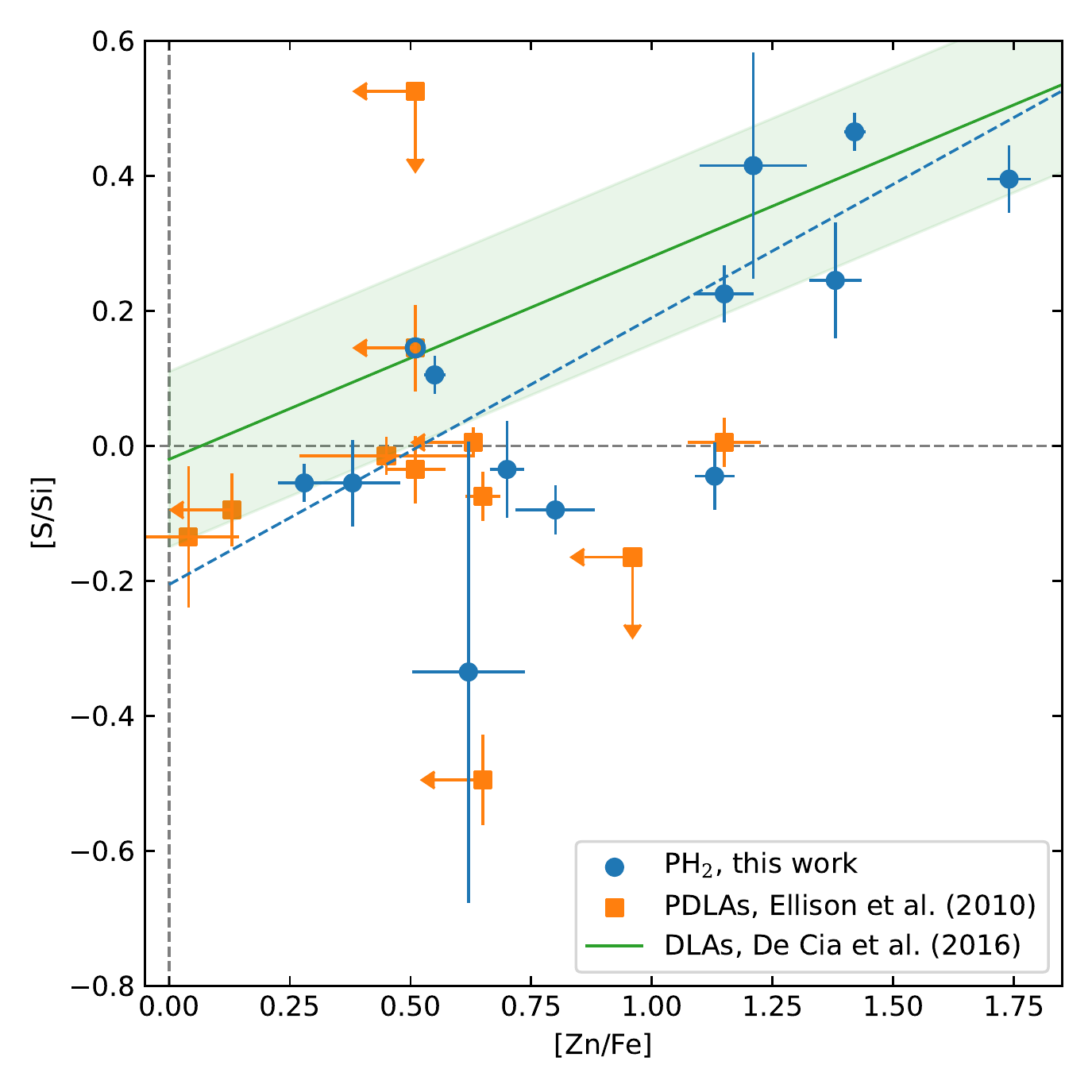}
    \caption{Sulphur-to-silicon vs zinc-to-iron abundance ratios in proximate H$_2$ and DLA systems. The green line and shaded area represent the best-fit for intervening DLAs by \citet{DeCia2016} and the 1\,$\sigma$ scatter around this relation. The dashed blue line is a linear fit to the proximate H$_2$ data. Dashed vertical and horizontal lines mark Solar values. Symbols and colours for data points are as in Fig.~\ref{f:SSi_H}.}
    \label{f:SSi_ZnFe}
\end{figure}

To investigate the effect of differential dust depletion further, we plot  [S/Si] as a function of [Zn/Fe] in Fig.~\ref{f:SSi_ZnFe}. Here, we see an apparent correlation between the two abundance ratio {in the case of proximate \HH\ systems} (Pearson correlation coefficient 
{0.76 with 0.4\% } probability to be due to chance coincidence). 
This favours depletion of silicon as a more likely origin for the high [S/Si] values. 
Fitting a linear relationship to the proximate H$_2$ data, that takes into account uncertainties on both axes, we obtain 
{${\rm [S/Si]\approx 0.39(\pm0.08)\times[Zn/Fe]-0.21(\pm 0.09)}$. 
A correlation between the two abundances ratio has also been observed in intervening DLAs (Pearson correlation $r=0.7$), for which we show the best fit linear relation obtained by \citet{DeCia2016} together with the associated intrinsic dispersion\footnote{The differences in adopted Solar values between this work and \citet{DeCia2016} only affect the [Zn/Fe] ratio by 0.005~dex and hence can safely be ignored here.}.  
However, proximate systems differ from intervening DLAs in the sense that 
they {have [S/Si]$\approx 0$ for a wider range of [Zn/Fe] values.}
This is observed for PDLAs regardless of whether they contain H$_2$ or not,  
}
when the intercept for intervening DLAs is almost zero. 

{In summary, the PH$_2$ systems with high [S/Si] values appear to behave like intervening DLAs. The five systems with highest values also have low \HI\ column densities, which 
is naturally explained if high-$\HI$ high-depletion systems are missing due to dust bias. At low [Zn/Fe] (and high $N(\HI)$), the [S/Si] values tend to differ between intervening and proximate systems, the latter having typically lower observed [S/Si] --from \SII\ and \SiII-- closer to solar. }
We conclude that the main driver of the [S/Si] ratio is depletion onto dust
--with possibly a different depletion sequence than that seen for intervening DLAs, again suggesting different dust type proximate to the quasar. However, ionisation effects are still possibly at play, meaning that part of these low-ionisation metals could actually arise from ionised gas. {Larger samples of proximate DLAs (with and without \HH) are necessary to investigate this further.}

\subsubsection{Argon as probe of hard UV ionising field}
Argon can provide an additional diagnostic on the ionising field. Owing to its high first-ionisation energy (15.76 eV), argon is expected to be mostly neutral in DLAs, which are self-shielded against \HI-ionising photons. While argon is predominantly ionised by cosmic rays in local ISM, high-energy photons significantly above 13.6~eV can still penetrate the cloud and ionise argon, that further has a low recombination to ionisation rate in diffuse atomic medium  \citep[see e.g.][]{Sofia1998}. \citet{Zafar2014b} found a deficiency of argon by about 0.4~dex (with standard deviation about 0.2~dex) with respect to other $\alpha$-elements in intervening DLAs, with a small dependency on \HI\ column density. They concluded that, since argon is extremely volatile \citep{Lodders2003}, the argon deficit is likely dominated by ionisation from the quasar metagalactic radiation, with local modulation by \HI\ column inside the absorbing galaxies. 

We constrain the \ArI\ column densities in our sample of proximate H$_2$ absorbers using the two available absorption lines (at 1048 and 1066~{\AA}). Because these fall into the Lyman-$\alpha$ forest and can easily be blended with H$_2$/HD lines, we used a slightly different procedure: we fitted the \ArI\ lines by fixing the redshifts and Doppler parameters to that previously obtained for other metal metal lines and fitting for the column densities. We included the H$_2$ absorption profile {constrained by independent fit (see Section~\ref{sect:HIH2}) and avoided blends with \lya\ forest as best as possible. In a few cases, however, we could get no meaningful constraint. The fit to argon lines are shown in Fig.~\ref{f:ArIfits}. 

The derived [Ar/S] (obtained from \ArI\ and \SII) values are shown as a function of the total \HI\ column densities and overall metallicities in Fig.~\ref{f:ArS}. 
First, we do not see any dependence of [Ar/S] on $N(\HI)$, in agreement with the expected lack of shielding by neutral hydrogen. We do not see any dependence with overall metallicity either, contrary to the possible negative (but not easily explained) trend seen by \citet{Zafar2014b}. The lack of dependence on overall metallicity is in line with the expected non-depletion of argon onto dust.
Second, [Ar/S] is found to be very similar (-0.35~dex on average) to that seen in intervening DLAs, which could be surprising given the proximity of the quasar. However, our measurements can still be affected by \lya\ interlopers so that the \ArI\ columns should in principle be considered as upper-limits --in a conservative approach\footnote{This should be less an issue at high spectral resolution as in the work by \citet{Zafar2014b} based on UVES data.}-- but more importantly, the metal column density is generally dominated by the \HH-bearing component, where we observe lower deficit of argon 
{(-0.2~dex)} 
with respect to components without detectable \HH\ (-0.7~dex). 
In Fig.~\ref{f:ArIfits}, we also show the expected \ArI\ profiles obtained assuming Solar [Ar/S] ratio. This over-predicts the absorption in most cases, but we also see that such over-prediction seems less severe at the velocity of H$_2$ components.

The relatively low deficit of argon in \HH\ components 
most likely is a result from an efficient conversion of Ar$^+$ into neutral argon through chemical reactions with \HH\ (${\rm Ar^+ + \HH \to ArH^+ + H}$ followed by $\rm ArH^+ + \HH \to Ar + H_2^+$, see \citealt{Duley1980,Schilke2014,Neufeld2016}). The situation could be similar to that with neutral chlorine \citep{Balashev2015} and all argon being in neutral form in the presence of \HH.  

{In turn, in the regions where hydrogen is predominantly atomic, the ionisation fraction of argon should be very dependent on the ionising field.
Indeed, at higher $I_{\rm UV}$ the photo-ionisation rate increases, while the \HH\ abundance decreases, impeding the molecular-ion recombination route. Hence the ionisation fraction of argon should rise at least proportionally to $I_{\rm UV}$.}
{Therefore, the systematically lower abundance in non-H$_2$-bearing components confirms that the medium is likely exposed to significantly enhanced (and hard) radiation field and/or cosmic ray ionisation.} {While quantitative statements would require a full modelling of the observed column densities and a better-assessed association between \HH\ and metal components (i.e. at high spectral resolution) --out of the scope of this paper-- we can estimate that the {$-$0.2~dex} deficit of argon means that about half of the metal column seen at the velocity of \HH\ components could come from their (\HH-free) \HI-envelope where argon would be mostly in ionised form.}

\begin{figure}
    \centering
    \begin{tabular}{c}
             \includegraphics[trim={0.5cm 0.1cm 1cm 1.2cm},clip=,width=\hsize]{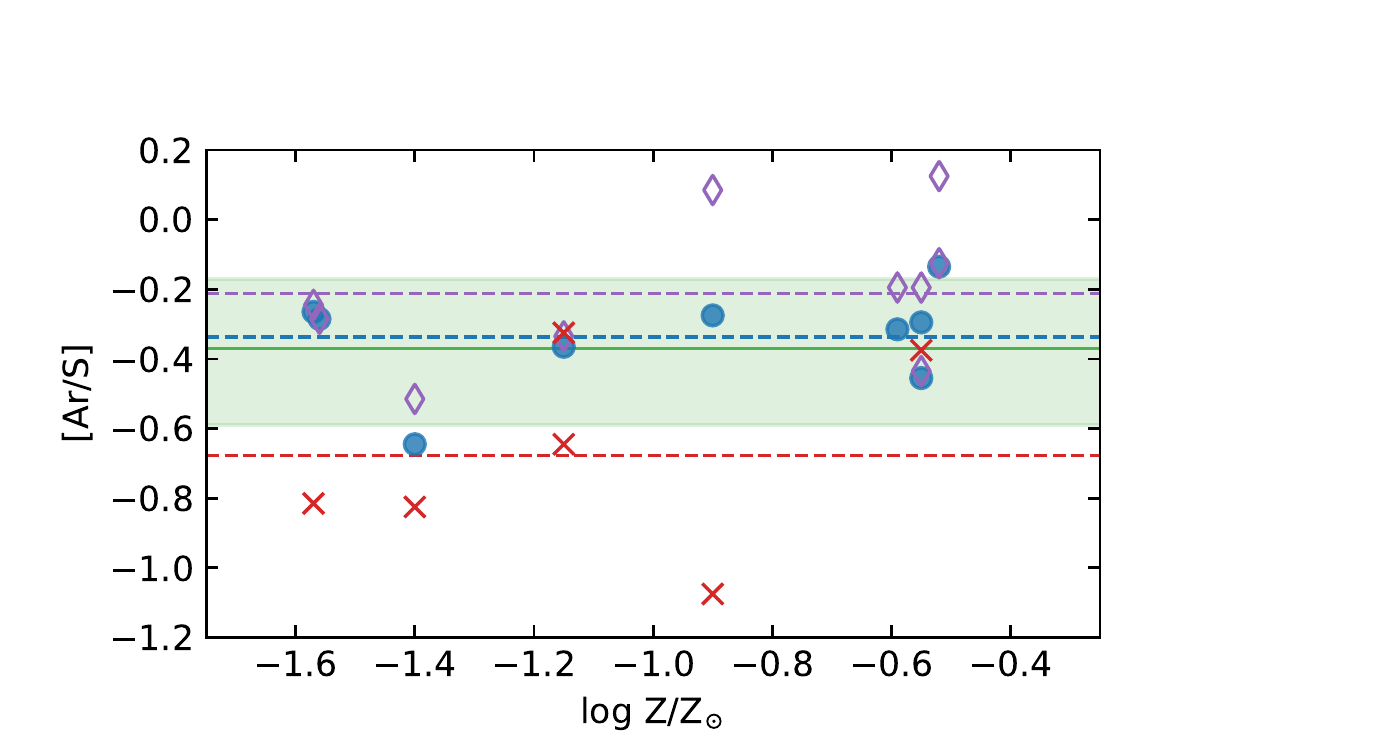} \\ 
    \includegraphics[trim={0.5cm 0cm 1cm 1.3cm},clip=,width=\hsize]{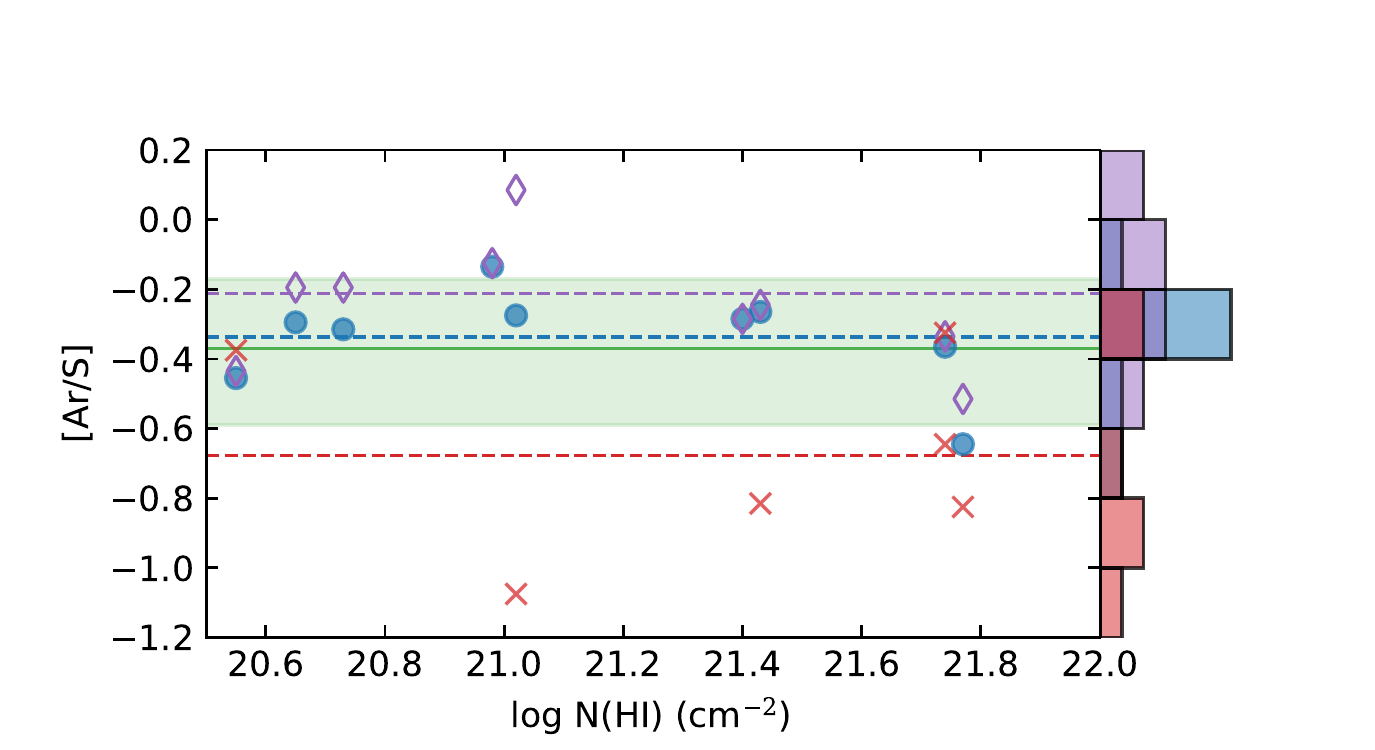}  
       \end{tabular}
    \caption{Observed [Ar/S] ratio (based on \ArI\ and \SII) as a function of overall metallicity (top) and \HI\ column density (bottom) in our sample of proximate systems. Blue circles correspond to the overall value in each system, while purple diamonds and red crosses to the \HH\ and non-\HH\ components, respectively. The dashed lines correspond to the corresponding mean values, with the distribution shown on the side panel. The green line corresponds to the mean [Ar/S] value (corrected to same Solar reference) and dispersion for intervening DLAs \citep{Zafar2014}.
    In spite of our best efforts to take into account for blends, all points should \textit{in principle} be considered as upper-limits. 
    }
 
    \label{f:ArS}
\end{figure}

\subsection{\HH\ in proximate versus intervening systems}

We compare proximate and intervening systems according to their H$_2$ and total hydrogen column densities in Fig.~\ref{f:NH_NH2}. 
The absence of proximate systems at low \HH\ columns is naturally due to our selection based on the presence of strong H$_2$ Lyman-Werner bands \citepalias{Noterdaeme2019}, and in fact, all our systems have $\log N(\HH)>18.5$. In turn, proximate 
H$_2$ were selected independently of $N(\HI)$, but as can be seen in this figure, they are limited to $\log N({\rm H})>20.5$, when 
intervening systems can have lower $N(\rm H)$ even at similar \HH\ column densities (see distributions on top of this figure). However, this is driven by a few very metal-rich intervening absorbers (as can be seen from the colour coding), two of them following a \CI-based selection, skewed towards very high metallicities. Actually metallicities tend to increase with decreasing $N$(H) and increasing $N(\HH)$. This could be naturally explained by the $1/Z$ dependence of the critical column density for \HH\ to form \citep[e.g.,][]{Krumholz2008}, but also to a bias against high-metallicity systems at high-$N(\HI)$ due to stronger dust extinction.

This is better seen in Fig.~\ref{f:NH_Z}, where the location H$_2$ systems is shown in the $N$(H)-$Z$ plane. The different lines show the ratio of UV intensity to number density as a function of the \HI\ column density in the envelope of a single H$_2$ cloud illuminated on one side\footnote{Following Eq.~2 of \citetalias{Noterdaeme2019}, based on theoretical calculations by \citet{Bialy2017} and where we used $Z/Z_{\rm \odot}$ as a proxy for $\tilde{\sigma_g}$, the grain Lyman-Werner photon absorption cross section per hydrogen nucleon normalised to the fiducial Galactic value. 
Note that such calculations do not provide the column density of \HH\ which just continues building up as more gas would be added to a given cloud. In practice, the total column is dominated by \HI, so that the plot against $N(\HI)$-only is almost identical.}. 
Three quarters (9/12) of the proximate H$_2$ systems are located between the $I_{\rm UV}/n = 10^{-2}$ and $10^{-1}$\,cm$^{3}$ lines, suggesting similar physical conditions. We caution however that the measured \HI\ column correspond to that integrated over the profiles while the calculations provide the \HI-column that participate in the shielding of \HH\ clouds. 
Notwithstanding, any atomic gas --with some amount of dust-- located upstream along the line of sight will still participate in shielding H$_2$ from the quasar field.

Determining the physical conditions from population levels is thus necessary to constrain 
the distances of the cloud from the AGNs.
In turn, owing to the different selections, intervening H$_2$ systems tend to have a wider spread of values. We also note that the lines of constant $I_{\rm UV}/n$ closely follow constant metal column density (for volatile species), and hence almost constant extinction \citep[e.g.,][]{Vladilo2006}. In other words, the conditions that facilitate the presence of H$_2$ are also those that can extinguish the background sources to the point that may be excluded from the optically-selected samples. However, since most of the intervening systems were also detected towards quasars from the SDSS, it is likely that more reddened quasars with more metal-rich proximate H$_2$ absorbers are actually already present in the SDSS database --in particular since such systems could afford lower densities to survive-- but the S/N achieved in the blue prevented their detection directly from \HH\ lines.

We also show in Fig.~\ref{f:NH_Z} the location of proximate and intervening DLAs without strong \HH\ as open symbols. In the case of intervening non-\HH\ systems, we used here the sample from \citet{Noterdaeme2008} where the presence of \HH\ has been systematically assessed.
The fact that no \HH\ is detected, even at high metallicity and \HI\ column density means that the corresponding system do not fulfil the requirements on $I_{\rm UV}/n$. For example, for a UV field about 1/10$^{th}$ of the Draine field, such systems would have low {number} densities $n<1$\,cm$^{-3}$ {(which is typical for the warm neutral phase)}, too low to build up molecular hydrogen. Alternatively, their total \HI\ column density is spread over many different clouds. We finally note that the above implicitly assumes static equilibrium, which may not necessarily be the case, in particular if \HH\ is found in outflowing gas. Indeed, for $n=100\rm\,cm^{-3}$ the characteristic timescale for H$_2$ formation will be $\sim 10^7$\,yrs, which is then comparable to the dynamical timescales.

\begin{figure}
    \centering
        \includegraphics[trim={0cm 0cm 0.5cm 1cm},clip=, width=\hsize]{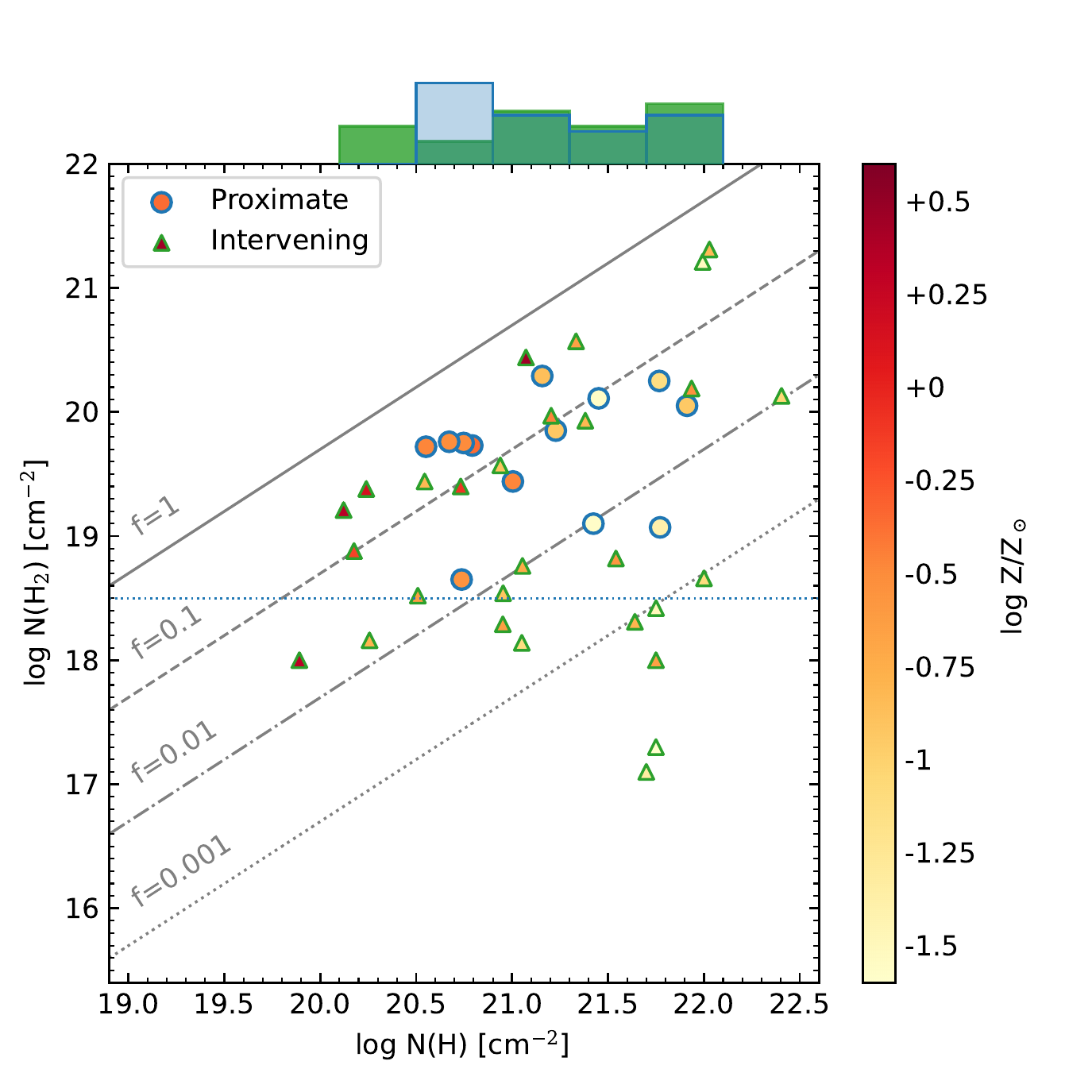} 
    \caption{Column density of molecular hydrogen versus total neutral hydrogen column density $N({\rm H})=N(\HI)+2N(\HH)$ for our sample of proximate H$_2$ absorbers and $z>2$ intervening strong H$_2$-bearing DLAs from \citet[][]{Balashev2019}. For a fair comparison, we removed systems that may potentially be proximate, keeping only systems with a conservative cut of $c\,(\zabs-\zem)/(1+\zem) < -10\,000$~\kms.
    Our selection of proximate \HH\ systems based on strong \HH\ lines directly seen in the low-resolution SDSS spectra naturally skews our sample to high \HH\ columns. 
    The blue horizontal line shows the approximate and empirical lower-limit of at $\log N(\HH)>18.5$ for the present sample. 
    {The top histograms show the $N($H)-distributions for the proximate (blue) and intervening (green) systems with $N(\HH)>18.5$.} 
    Finally, the different lines correspond to different averaged molecular fractions $f_{\rm H2} \equiv 2{N(\HH)}/(2N(\HH)+N(\HI))$.
    }
    \label{f:NH_NH2}
\end{figure}

\begin{figure}
    \centering
        \includegraphics[width=\hsize]{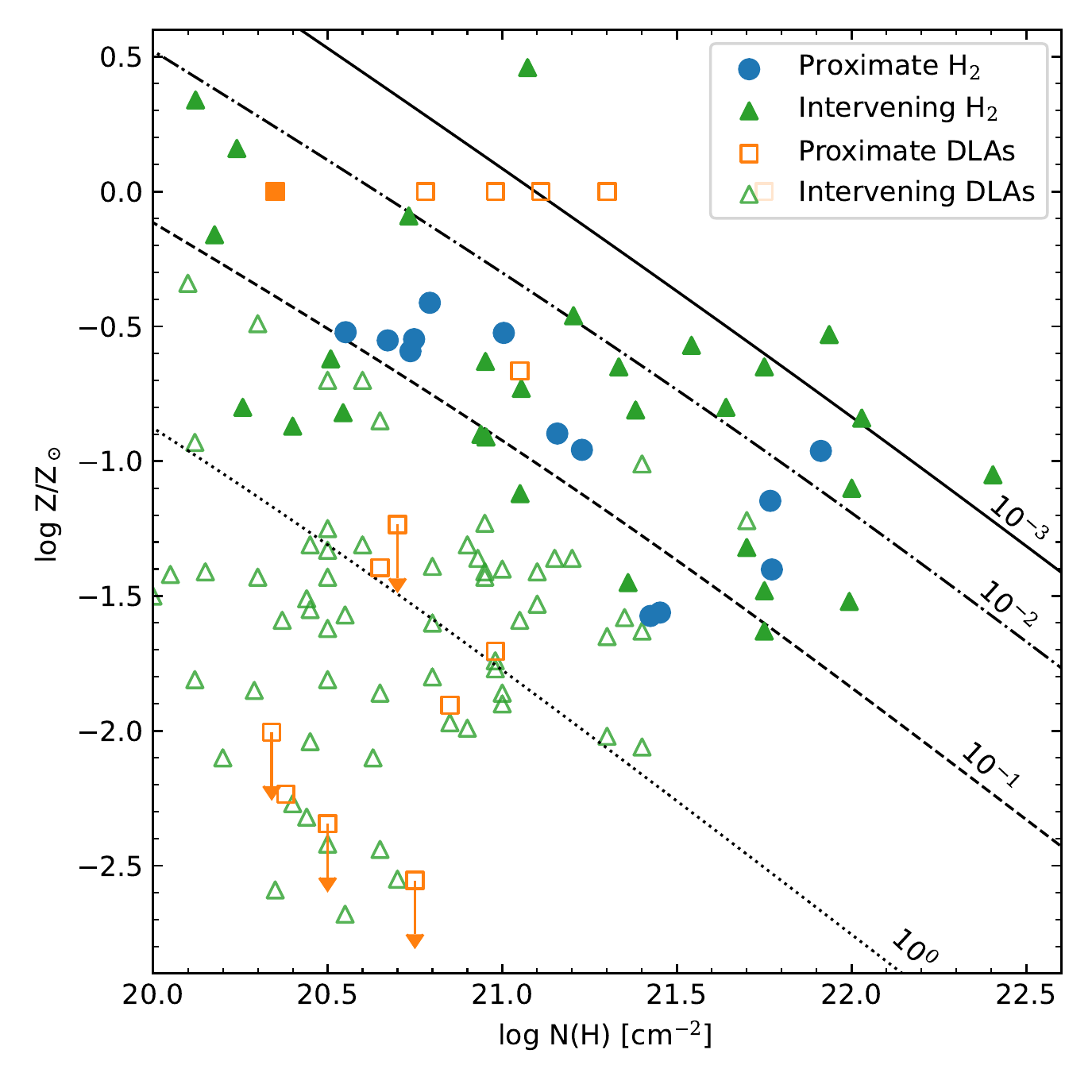} 
    \caption{Metallicity vs total hydrogen column density. The different lines correspond to the hydrogen column density of a single \HH\ cloud illuminated by a UV field for different UV to density ratios $I_{\rm UV}/n$ as indicated above each line, with $I_{\rm UV}$ in units of the Draine field and $n$ in cm$^{-3}$. Filled symbols correspond to \HH-bearing systems. {Proximate \HH\ systems are from this work, proximate DLAs from \citet{Ellison2010} and intervening systems from \citet{Noterdaeme2008} and \citet{Balashev2019}.}
    }
    \label{f:NH_Z}
\end{figure}

\subsection{Kinematics}
\label{s:resultskine}
%.....
The velocity extent of low-ionisation metal lines, $\Delta v_{90}$ is considered as a good way to quantify the kinematics of the neutral gas (since this information is not available from the \HI\ lines). In the case of intervening DLAs, \citet{Ledoux2006} have first observed a correlation between this velocity width and the gas metallicity, suggesting that this reflects an underlying relation between mass and metallicity, since the motion of low-ionisation lines is expected to be governed by gravity. This has been discussed in a number of subsequent works \citep{Moller2013, Neeleman2013,Christensen2019}, with attempts to connect this to the mass-metallicity relation seen in emission \citep[e.g.,][]{Moller2020}. While the assumption that $\Delta v_{90}$ depends primarily on the galaxy mass seem to be consolidated at least in a statistical sense, there are evidences that outflows, tidal-streams and environment in general can affect the velocity spread in a more complex way \citep[e.g.,][]{Zou2018,Nielsen2022}. In the case of proximate systems, because of the presence of the quasar, one could expect the situation to be even more complex. For example, the case of \Jzzuc\ is strongly suggestive of outflowing material (\citetalias{Noterdaeme2021}). Surprisingly enough, the distribution of $\Delta v_{90}$ in proximate \HH-systems follows the trend and dispersion seen in intervening DLAs, although limited to high-values, as expected from the presence of  \HH\ \citep{Noterdaeme2008}. Overall, the velocity spread of the neutral gas in proximate-\HH\ systems suggests an origin in massive galaxies, but without clear indication of perturbed kinematics, or at least, not more perturbed than similar high-metallicity systems in the intervening population. 

\begin{figure}
    \centering
        \includegraphics[width=\hsize]{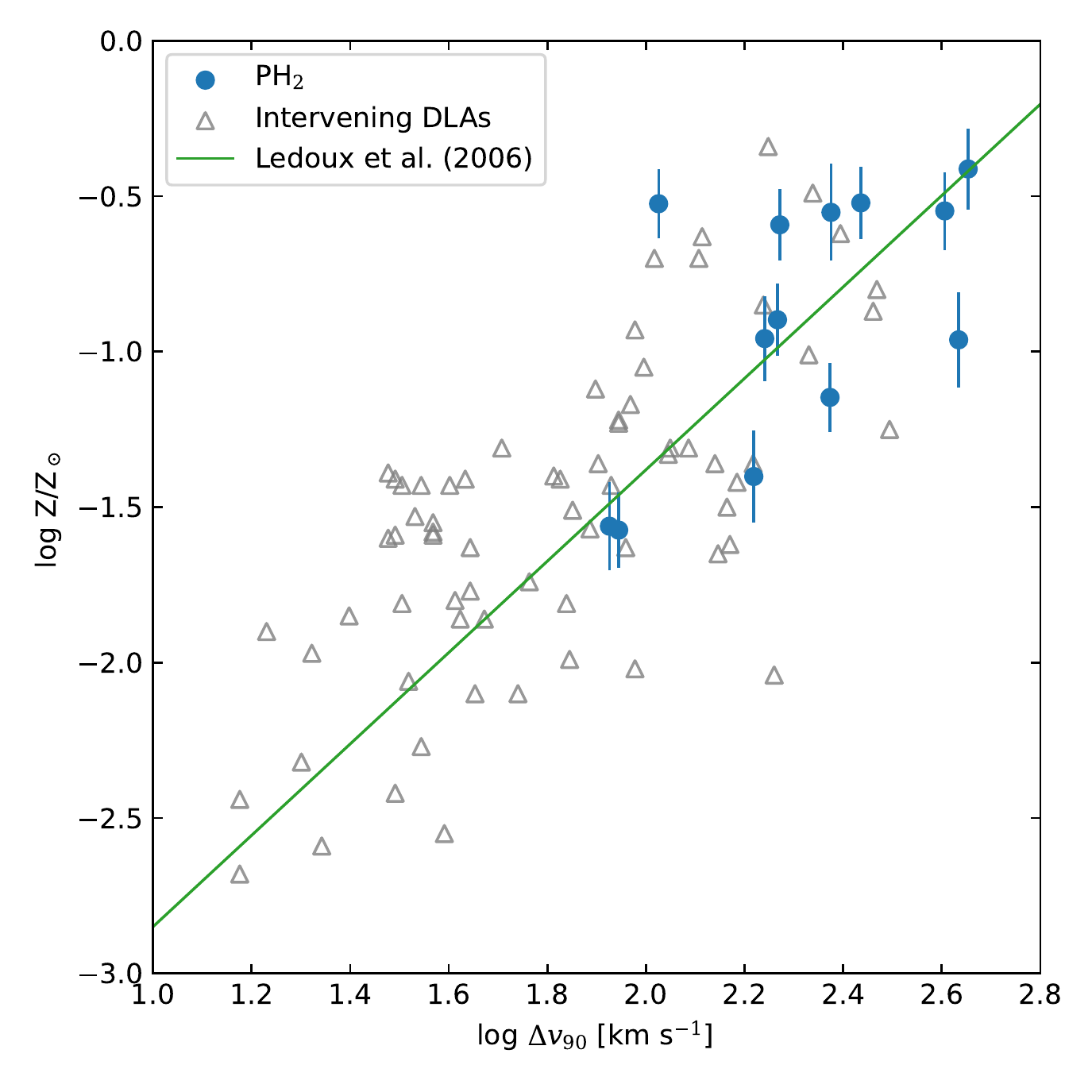} 
    \caption{Metallicity versus velocity extent of the low-ionisation metal lines for our sample proximate \HH-systems and a sample of intervening DLAs (mostly from \citealt{Ledoux2006} with some additions by \citealt{Noterdaeme2008}). The green line shows the best-fit relation for high-$z$ intervening DLAs by \citet{Ledoux2006}. 
    }
    \label{f:dv90_Z}
\end{figure}

In order to investigate now the kinematics of the gas with respect to the quasar, we represent in Fig.~\ref{f:kine} the kinematics of the neutral gas as a function of its metallicity. There is an apparent trend for high-metallicity systems to be around the quasar systemic redshift as obtained from the ionised lines, while low-metallicity systems tend to be redshifted. The Pearson correlation coefficient between $v_{50} = c(z_{\rm QSO}-z_{50})/z_{\rm QSO}$ (i.e., the velocity of the centroid of the absorption profile ) and the gas metallicity is 
{$r=-0.50$ with $p$-value $0.08$.
} 
In other words, there is only about 8\% probability of this anti-correlation to be due to chance alone. {One system, towards \Jzhch, is significantly more blueshifted than the rest of the sample. This does not drive the correlation however as removing it from the correlation test only changes the Pearson values to $r=-0.49$, $p=0.10$. In fact, the $\sim 2000$~\kms\ blueshift of this system makes it intermediate between what could be considered as clearly intervening or proximate. For example, at this velocity, the incidence of strong \HH\ absorbers already reaches that of intervening ones \citep[see Fig. 3][]{Noterdaeme2019}. In fact, the derived UV flux in this system \citep{Balashev2019} suggests that this system is located further from the quasar, at a distance of at least several hundreds kpc.} %
To account for the statistical uncertainties on metallicity and on the quasar systemic redshift, we calculated the Pearson's ($r$,$p$) values for 10\,000 random samples where the metallicity of each system is taken using a normal distribution around the best value with the corresponding uncertainties. 
Similarly, the velocity is chosen using a normal distribution around the measured value with $\sigma$ set to the intrinsic velocity dispersion of $z_{\rm QSO}$ provided by \citet{Shen2016} (i.e., 46, 56, 205 and 233~\kms for $z_{\rm QSO}$ based on [\OII], [\OIII], \MgII\ and \CIII], respectively). 
From this exercise, we find an approximate normal distribution of $r$ values, giving {$r=-0.47\pm0.09$ ($r=-0.44\pm0.11$ ignoring \Jzhch)}. The $p$-values have a median of 0.11 (0.14) but with some tail towards larger values. In short, this means that the trend remains even taking into account statistical uncertainties on metallicity and emission redshift measurements.  

In principle, the projected line-of-sight velocity towards the quasar for low-metallicity systems could be interpreted as neutral gas infalling onto the quasar host galaxy. However, it is unlikely that gas would feed directly the quasar and we may instead expect any feeding gas to orbit the quasar host galaxy without significant line-of-sight velocity component. It is possible however, that gas is orbiting the galaxy, in non-circular orbits, after a galaxy companion interaction, where outer neutral gas is coming back as a fountain after having been dragged out in a tidal tail.

Alternatively, the systemic redshift could be 
underestimated if the blueshifts of the emission lines are larger than observed by \citet{Shen2016}. 
In their work, the authors used \CaII\ absorption as reference to compute the (non)-shifts of [\OII] and [\OIII] and then the shift from other lines (\MgII, \CIII], etc.) are computed with respect to the latter. The use of SDSS data (covering wavelengths shorter than 10\,400~{\AA}) implies that the computed shifts correspond to low to intermediate redshift quasars (and hence likely not particularly luminuous). At high redshifts ($z\sim 5-7$), \citet{Schindler2020} and \citet{Eilers2021} found larger \MgII\ blueshifts in luminuous quasars, using precise systemic redshifts from [\CII]$\mu$m158 and/or CO emission line measurements from ALMA and NOEMA as reference. These lines arise respectively from star-formation and molecular gas, and hence should better represent the systemic redshift of the quasar host, which is also very likely to dominate over other, less massive galaxies in the quasar group. The sample presented here has redshifts $z\sim 3$, so we may naively expect line shifts in-between those obtained by \citet{Shen2016} and those obtained by \citet{Schindler2020} or \citet{Eilers2021}. For half of our sample, we do have CO(3-2) redshifts from our NOEMA observations, which we represent as ellipses on Fig.~\ref{f:kine}. On average, the CO redshifts are about 250~\kms\ higher than those obtained from the (corrected) ionised emission lines. If $z_{\rm CO}$ does represent better estimates of the host galaxy redshift, then the absorbing gas now appears to have no particular line-of-sight velocity at low-metallicities, while the outwards velocities at high metallicities is reinforced. This maintains the anti-correlation to $r\approx-0.6$, $p\approx0.2$ (now computed on 6 quasars). In that case, high-metallicity gas (above $\sim$25\% Solar) {is more likely to} represent gas enriched by intense star-formation activity and being expelled through associated winds, as clearly seen  in the case of \Jzzuc\  (\citetalias{Noterdaeme2021}), which has the highest metallicity among our sample. 
This would be consistent with chemically-enriched outflowing gas coming from nuclear regions, contrary to outer gas, due to the radial $Z$-gradient. 

Determining the distances between the absorbing gas and the active nucleus by modelling the excitation of molecular and atomic species should help understanding the origin of the gas with different metallicities further. Such a study using the present sample is devoted to a future paper.

\begin{figure}
    \centering
        \includegraphics[trim={0cm 0.5cm 0cm 0cm},clip=,width=\hsize]{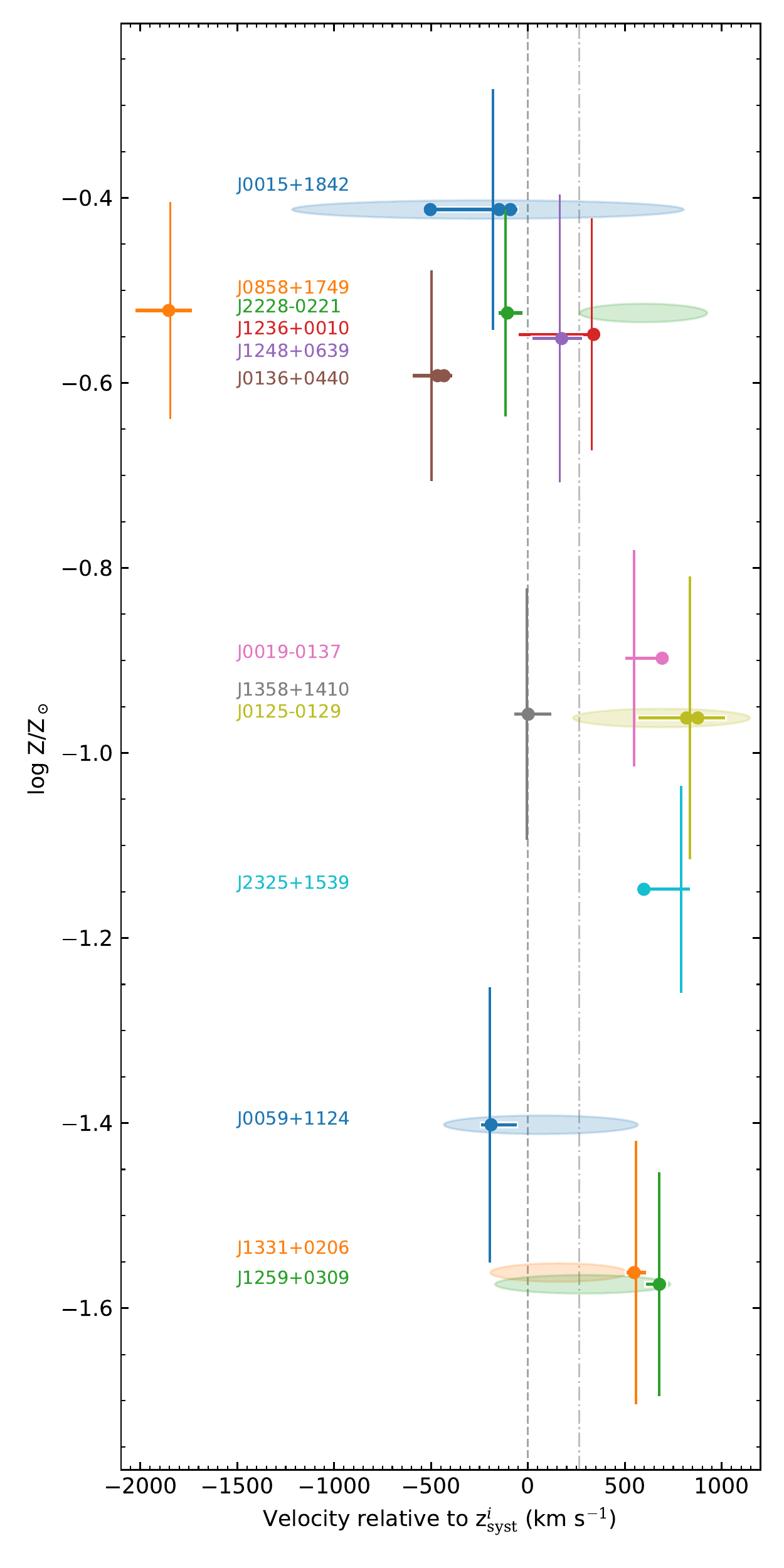}

    \caption{Summary of absorbing gas metallicity and kinematics. The different coloured segments represent the $\Delta v_{90}$ regions, with dots indicating the location of H$_2$ components. The zero of the velocity scale is set to the best estimate of the systemic redshift $z_{\rm sys}$ from ionised lines (Sect.~\ref{s:zem}). Positive velocities mean here $z>z_{\rm sys}^i$. 
    {Ellipses are shown at the location of the CO(3-2) emission whenever observed, with major axis equal to the line's FWHM. $z_{\rm CO}$ is shifted with respect to $z_{\rm syst}^i$ by about 250~\kms on average (dash-dotted line).}
    }
    \label{f:kine}
\end{figure}

\subsection{Note on the AGN proximity from the presence of \NV}

While we focus on the neutral phase (and low ionization metal species) in this paper, we briefly discuss here the presence of \NV\ absorption at the quasar redshift. Photo-ionisation of nitrogen to that stage requires an energy of 77.5\,eV, meaning a hard ionisation spectrum since stellar light falls rapidly after 54\,eV where He\,{\sc ii} ionisation occurs. This makes \NV\ one of the best diagnostic for the physical proximity of the AGN and distinguish between intervening and associated system \citep[e.g.,][]{Perrotta2016}. \citet{Fox2009} found that \NV\ is detected in only 13\% of intervening DLAs but they also found a surprisingly similar fraction in PDLAs out to 5000~\kms, suggesting that proximity in velocity does not necessarily mean physical proximity. Indeed, 5000~\kms\ in the Hubble flow corresponds already to about 16 Mpc proper distance at $z\sim3$. \citet{Ellison2010} found a tentative excess of \NV\ detection rate (2/7) at similar velocities in their PDLA sample, but note that this fraction would increase if they limited the statistics to smaller velocity separations from the quasar. Finally, using composite spectra of about a thousand \CIV\ absorbers and interpreting their results with the help of ionisation modelling, \citet{Perrotta2018} confirmed that \NV\ is only present in proximate systems and conclude it is a good tool to identify gas physically related to the quasar host. \NV\ absorption is clearly seen at the quasar redshift for nine out of our {thirteen} proximate \HH-absorbers (see Fig.~\ref{f:HI}). Note that the non-detection of \NV\ in the two systems with lowest metallicity in our sample may just be due to column density effect rather than lower ionisation. Overall, the very high incidence of \NV\ in our sample supports an origin in the close environment of the quasars. Careful ionisation modelling, together with comparative kinematic (\NV\ is typically present over wider velocity ranges than low-ionisation species) should help understanding the ionised component of \HH-selected proximate systems.

\section{Conclusion \label{s:conclusion}}

The chemical enrichment and velocity dispersion of the neutral gas in strong proximate H$_2$ systems appears not to be significantly different from what is seen in intervening systems. While the metallicities in proximate H$_2$ systems are higher than the general population of intervening DLAs, and in the upper-half when compared to the sample of proximate DLAs from \citet{Ellison2010}, they are consistent with those intervening systems that also have high \HH\ content \citep[see][]{Balashev2019}. 
The relative abundances of sulphur, silicon, iron and zinc follow the expected trends due to increasing depletion of refractory elements with increasing metallicity \citep[e.g.,][]{DeCia2016}, indicating an increased abundance of dust with increasing chemical enrichment. This is also confirmed independently by the increasing quasar extinction per hydrogen atom with depletion and metallicity. 
{We however note higher depletion of iron in proximate systems compared to intervening ones at given metallicity, suggesting different dust production and/or destruction close to the AGN.  
The ubiquitous presence of \NV\ absorption indicates indeed a strong and hard UV field, which also explains the deficit of neutral argon in non-\HH-bearing components. In turn, argon remains predominantly in neutral form in the the presence of \HH, aided by ion-molecule reactions.} 

The lack of {proximate and intervening} systems with {both} high-metallicity/depletion and hydrogen column density is %in fact 
likely due to a selection effect against highly reddened systems which would be more severe than just exclusion from {optically-selected quasar samples}. Actually, such systems may already be present in the SDSS quasar database, but not selected through \HH-absorption because of their {low} S/N in the blue. Those systems would also be interesting to recover since they should be able to survive with relaxed densities for a given intensity of UV field (i.e., distance to the central engine). 
{Upcoming surveys should also provide less biased samples of quasars and absorbers. For example, 4MOST \citep{deJong2019} will collect optical spectra of quasars selected on astrometry (The 4MOST--Gaia Purely Astrometric Quasar Sample -- 4G-PAQS; PI: Krogager), infrared or X-ray properties \citep{Merloni2019}. The higher spectral resolution and larger collecting area should push deeper into the dust-obscured population of absorbers. Similarly, the MeerKAT Absorption Line Survey \citep[MALS;][]{Gupta2016} also uncovers quasars in a dust-unbiased manner, owing to radio+IR selection of the targets \citep[see][]{Krogager2018b,Gupta2022b}. A complementary view of proximate cold gas absorbers can then be done through associated 21-cm absorption. Further, 21-cm absorption provides different (and multiple) lines of sight owing to radio-continuum emission regions distinct from that producing the optical continuum \citep{Gupta2022}.}

The significant excess of proximate H$_2$ systems with respect to the expected 
incidence from the intervening statistics showed {that most of} the former are very likely to be associated to the quasar environment (quasar host or galaxies in the quasar group). On the other hand, the similar column densities of neutral and molecular hydrogen at similar metallicities when compared to intervening H$_2$-bearing DLAs means that proximate H$_2$ should have similar UV-to-density ratios. This means that proximate \HH-systems should have significantly higher density than their intervening analogues if located within the dominating influence of the quasar UV field (i.e., {within} the host galaxy, or possibly several {hundreds of kpc away}). The higher densities if located close to the quasars would explain the higher fraction of systems with excited atomic fine-structure lines. {Indeed, \SiII* is clearly detected in eight out of the thirteen systems presented here.}  Determining the physical conditions will be necessary in that respect too.

While our parent sample was selected by having $\zabs\approx\zem$, the systems could potentially have been found within a wide velocity range around the quasar owing to the large uncertainties on the quasar emission redshifts in SDSS spectra, together with the low S/N and resolution allowing \HH-detection over a relatively wide range. It is {then} almost surprising to {find that} all systems {but one} ended up having velocities significantly {smaller} than 1000~\kms\ from the quasar, after securing precise absorption and emission redshifts. In the absence of peculiar motions (i.e., pure Hubble flow), this would correspond to at most a few Mpc. If peculiar motions contribute significantly to the observed velocities, then this becomes an upper-limit to the {actual} distance. In fact, velocities of $\zabs>\zem$ systems gives a hint on the dispersion due to peculiar velocities since this obviously cannot be due to the Hubble flow. The kinematics are then all consistent with no Hubble-flow component at all (except for \Jzhch), suggesting that most systems are indeed associated to the quasar host or its group environment.

We observe a trend for different line of sight velocities of the absorbing gas with respect to the quasar as a function of metallicity. Assuming systemic redshifts from ionised lines, this would suggest that low-metallicity gas tends to move towards the quasar. However, it is more likely that the low-metallicity gas is actually more clustered (in velocity space) around the quasar, as corroborated 
by the use of CO-based redshifts, and that the high-metallicity gas is actually more likely to arise from outflowing winds. While this remains speculative for now, if confirmed in a large sample of systems, including non-\HH\ ones, the relation between motion of neutral gas with respect to the quasar host galaxy and/or the central engine and the chemical enrichment of this gas would open a new window to understand quasar evolution.

Selection effects seem however to be complex when it comes to proximate systems. For example, the classical selection of PDLAs based on saturated cores biases the samples against systems with leaking Lyman-$\alpha$ emission, that is actually seen in about half of our proximate H$_2$ systems. If, as expected, the amount of gas and its distribution around the quasar affects the ability of Lyman-$\alpha$ photons to scatter at large distances, such a bias will also affect any conclusion on the nature of proximate systems.

We conclude that proximate systems have a strong potential to investigate many processes driving the evolution of quasars. Determining the physical conditions and distances of gas clouds from the central engine will be the next natural step on this matter.

\begin{acknowledgements}
{We thank the anonymous referee for useful comments and suggestions.
}     The research leading to these results has received support from the French \textit{Agence Nationale de la Recherche} under ANR grant 17-CE31-0011-01/project ``HIH2''. RC gratefully acknowledges support from the French-Chilean Laboratory for Astronomy (IRL\,3386). SL acknowledges support by FONDECYT grant 1191232.
\end{acknowledgements}

\bibliographystyle{aa}
\bibliography{mybib.bib}

%%%%%%%%%%%%%%%%%%%%%% APPENDIX %%%%%%%%%%%%%%%%%%%%%%
\clearpage

\begin{appendix}
%HI & H2
\section{Fit to \HI\ \lya\ and H$_2$ Lyman-Werner absorption bands \label{s:H2figs}}

\addtolength{\tabcolsep}{-3pt}
\begin{figure*}
    \centering
    \begin{tabular}{ccc}
\includegraphics[trim={0cm 0cm 0.5cm 0cm},width=0.32\hsize]{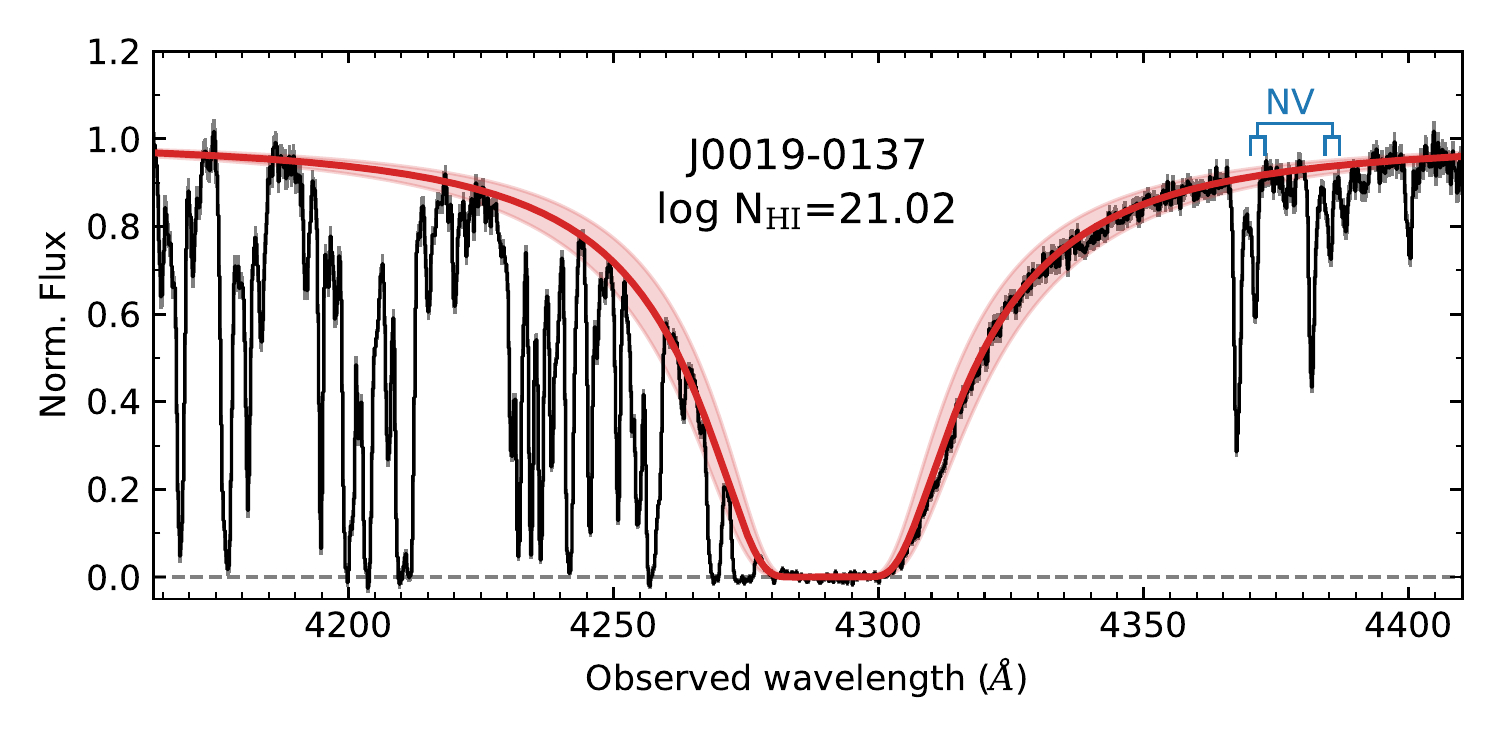} &
\includegraphics[trim={0cm 0cm 0.5cm 0cm},width=0.32\hsize]{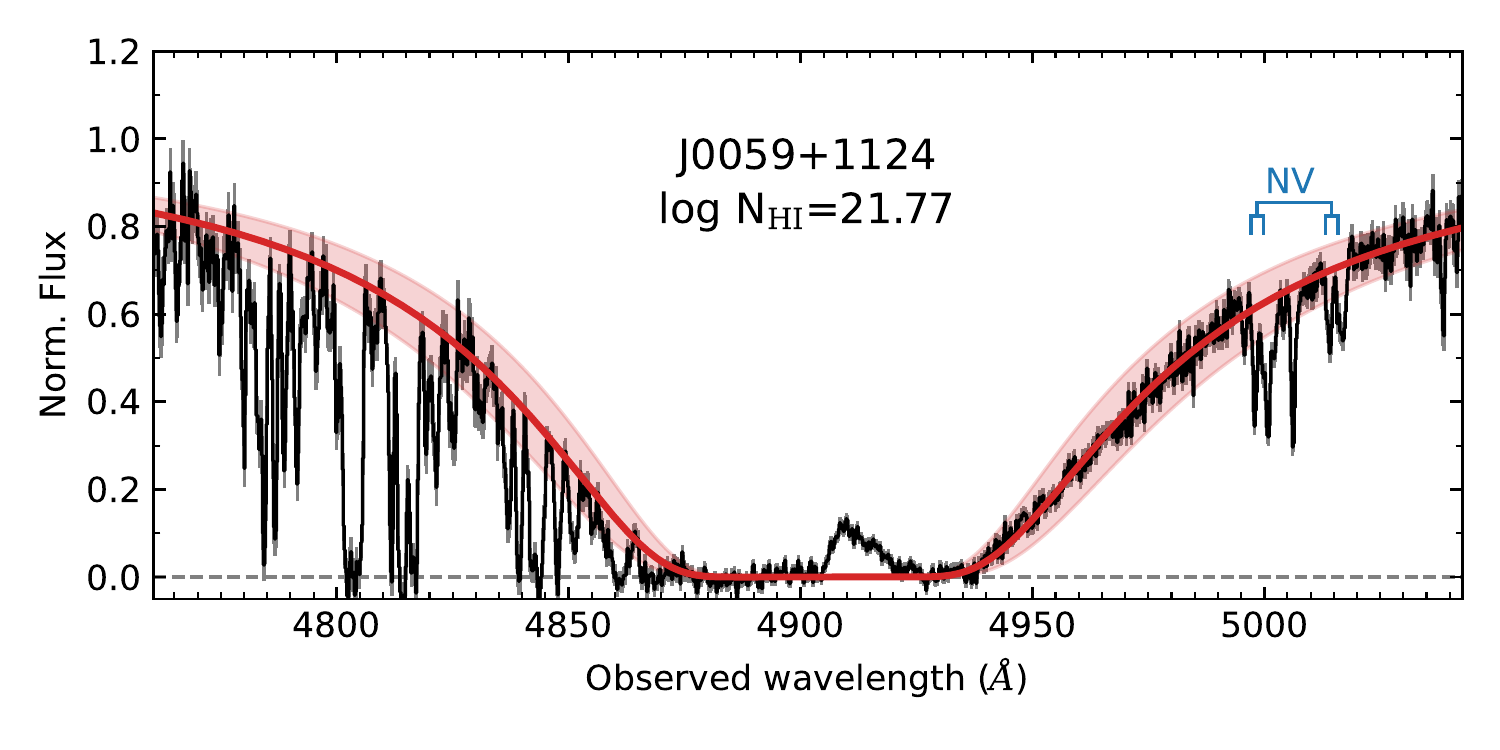} &
\includegraphics[trim={0cm 0cm 0.5cm 0cm},width=0.32\hsize]{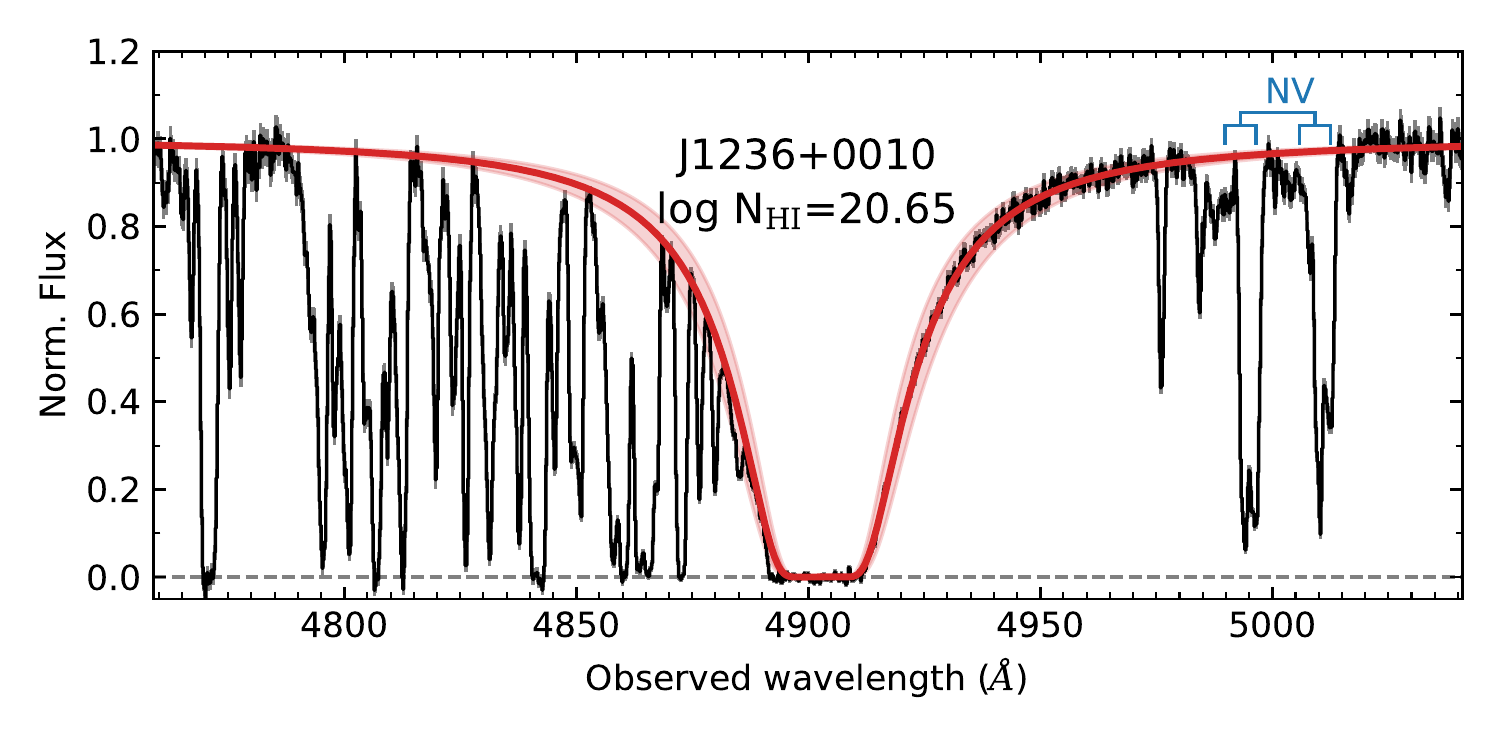}\\
\includegraphics[trim={0cm 0cm 0.5cm 0cm},width=0.32\hsize]{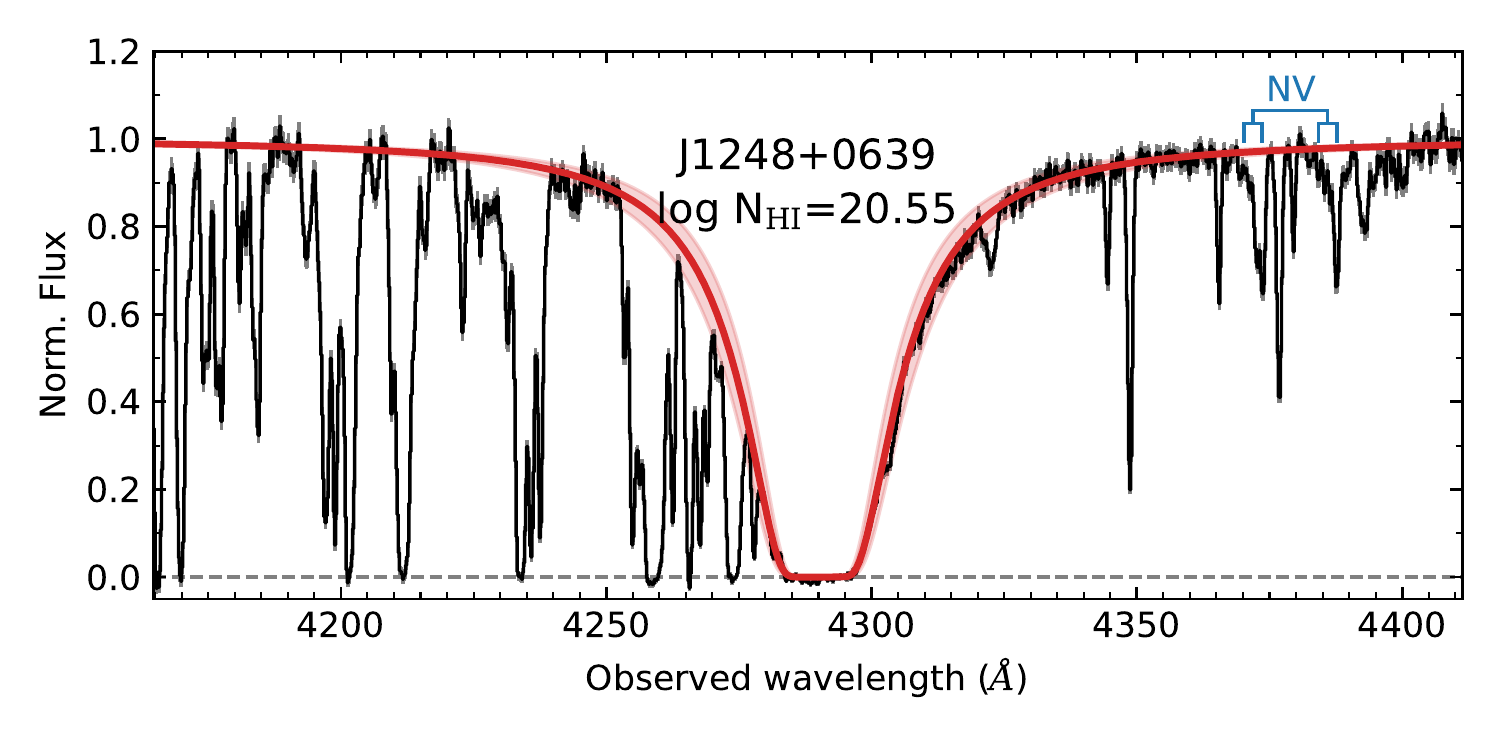} &
\includegraphics[trim={0cm 0cm 0.5cm 0cm},width=0.32\hsize]{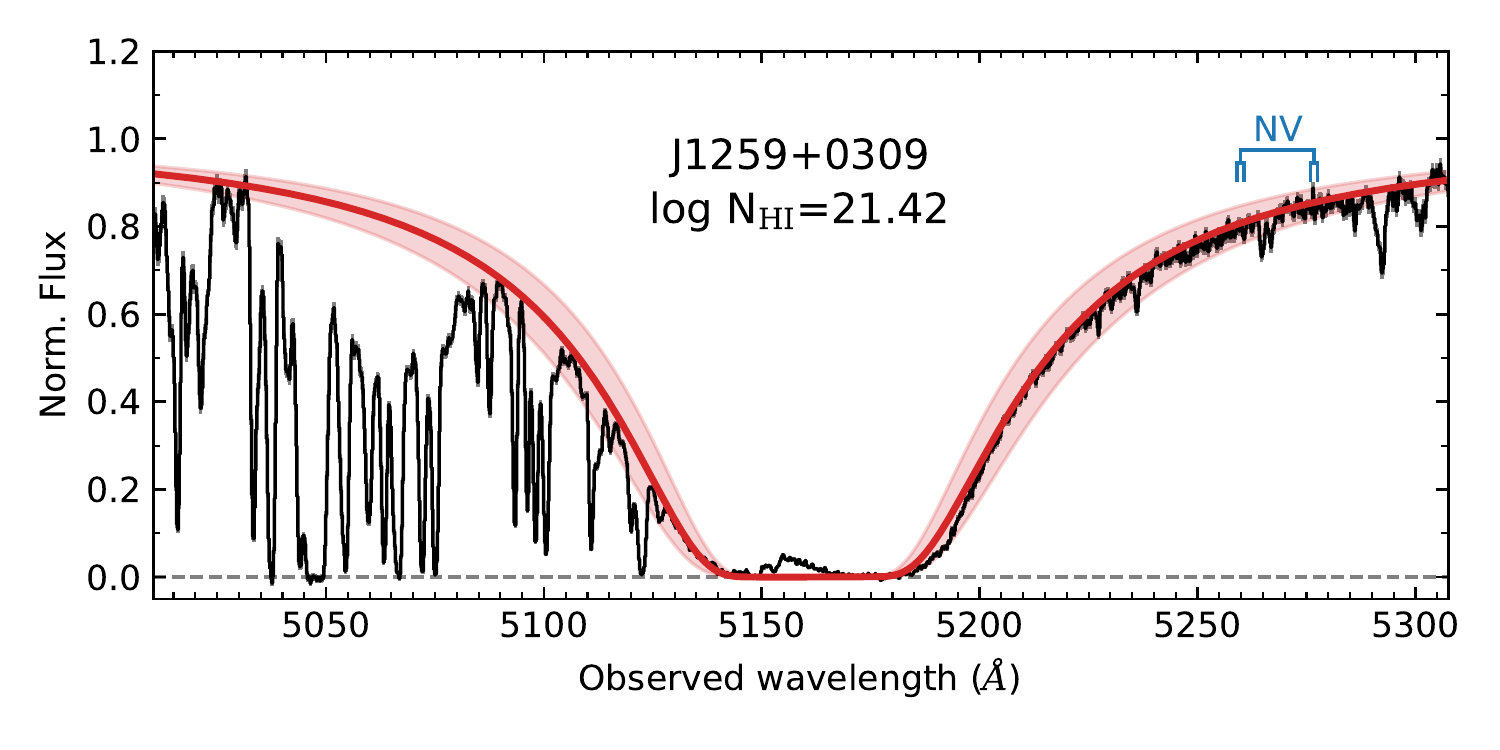} &
\includegraphics[trim={0cm 0cm 0.5cm 0cm},width=0.32\hsize]{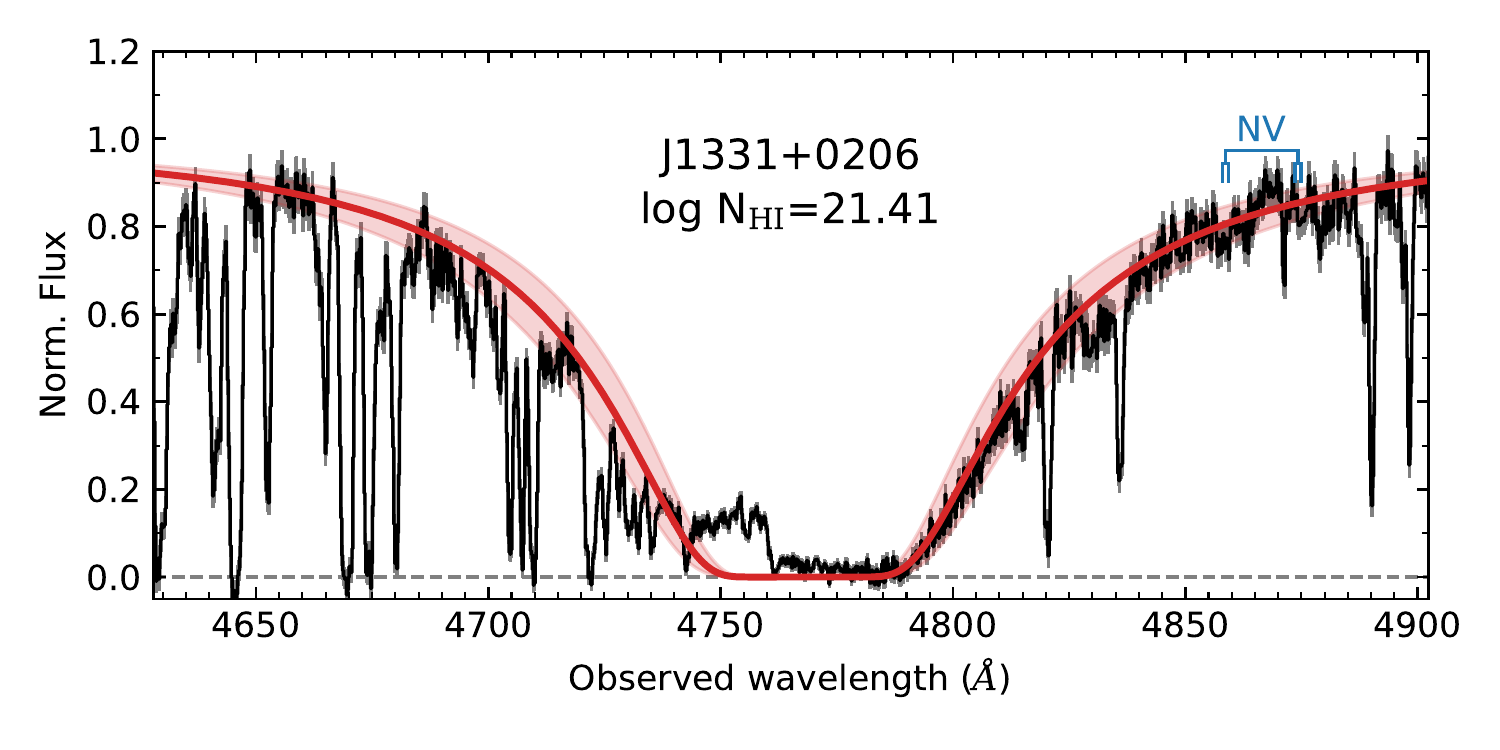} \\
\includegraphics[trim={0cm 0cm 0.5cm 0cm},width=0.32\hsize]{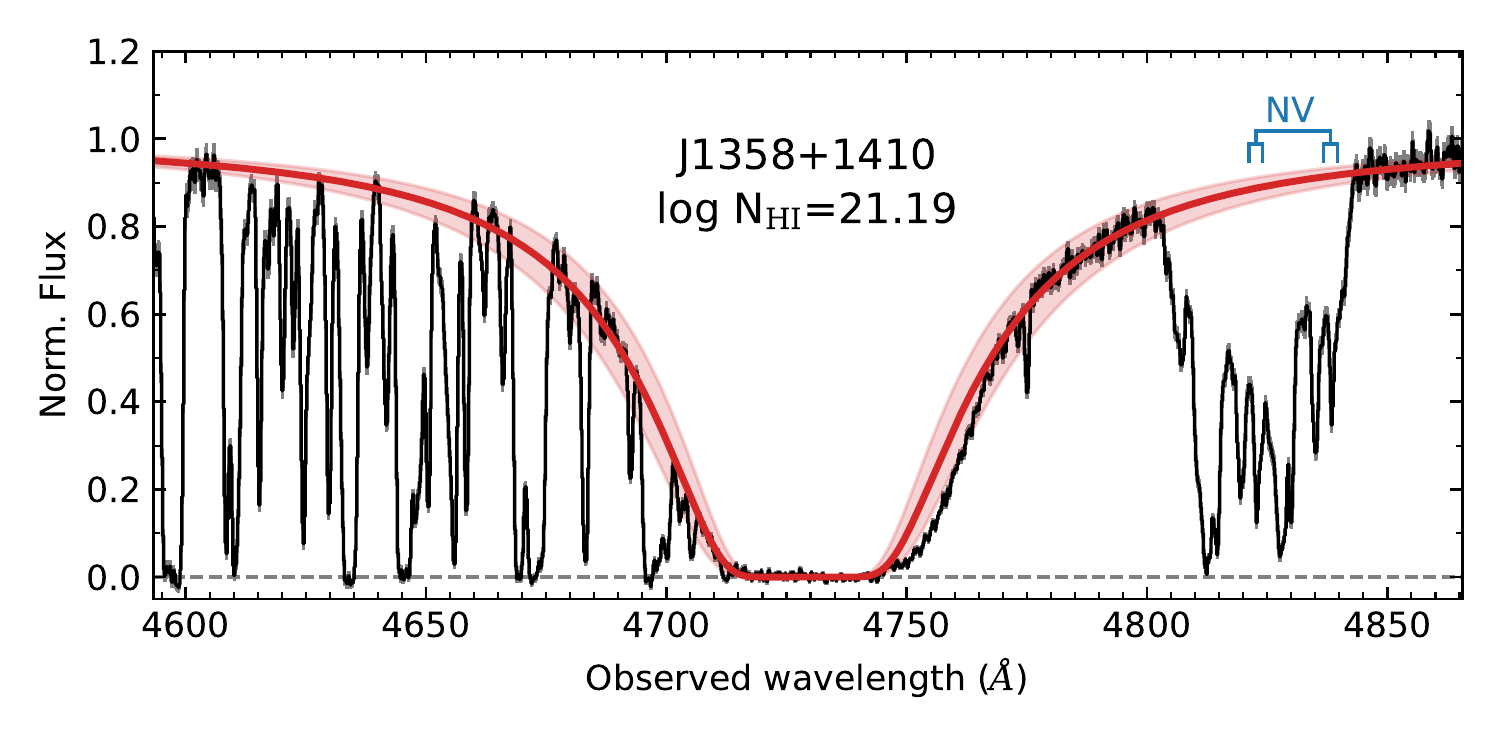} &
\includegraphics[trim={0cm 0cm 0.5cm 0cm},width=0.32\hsize]{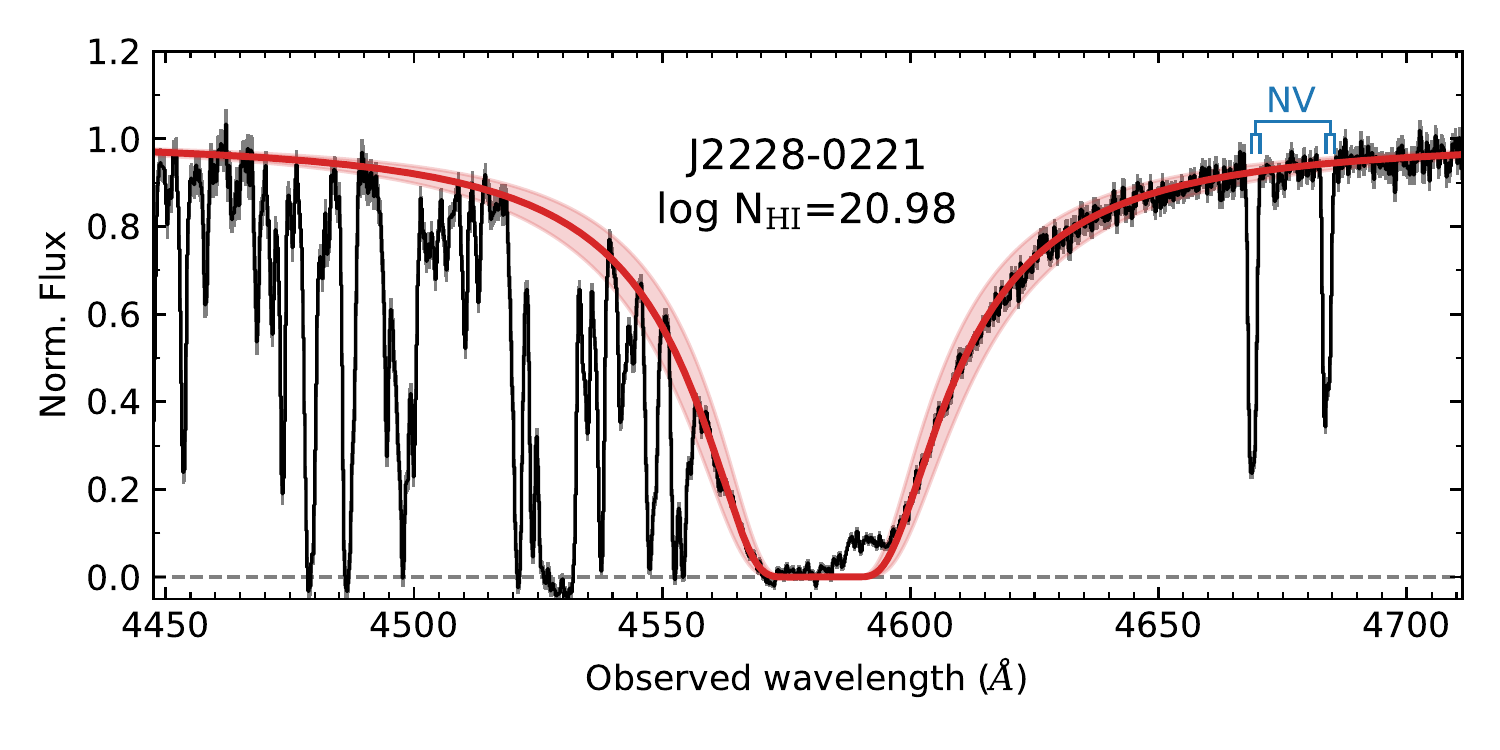} &
\includegraphics[trim={0cm 0cm 0.5cm 0cm},width=0.32\hsize]{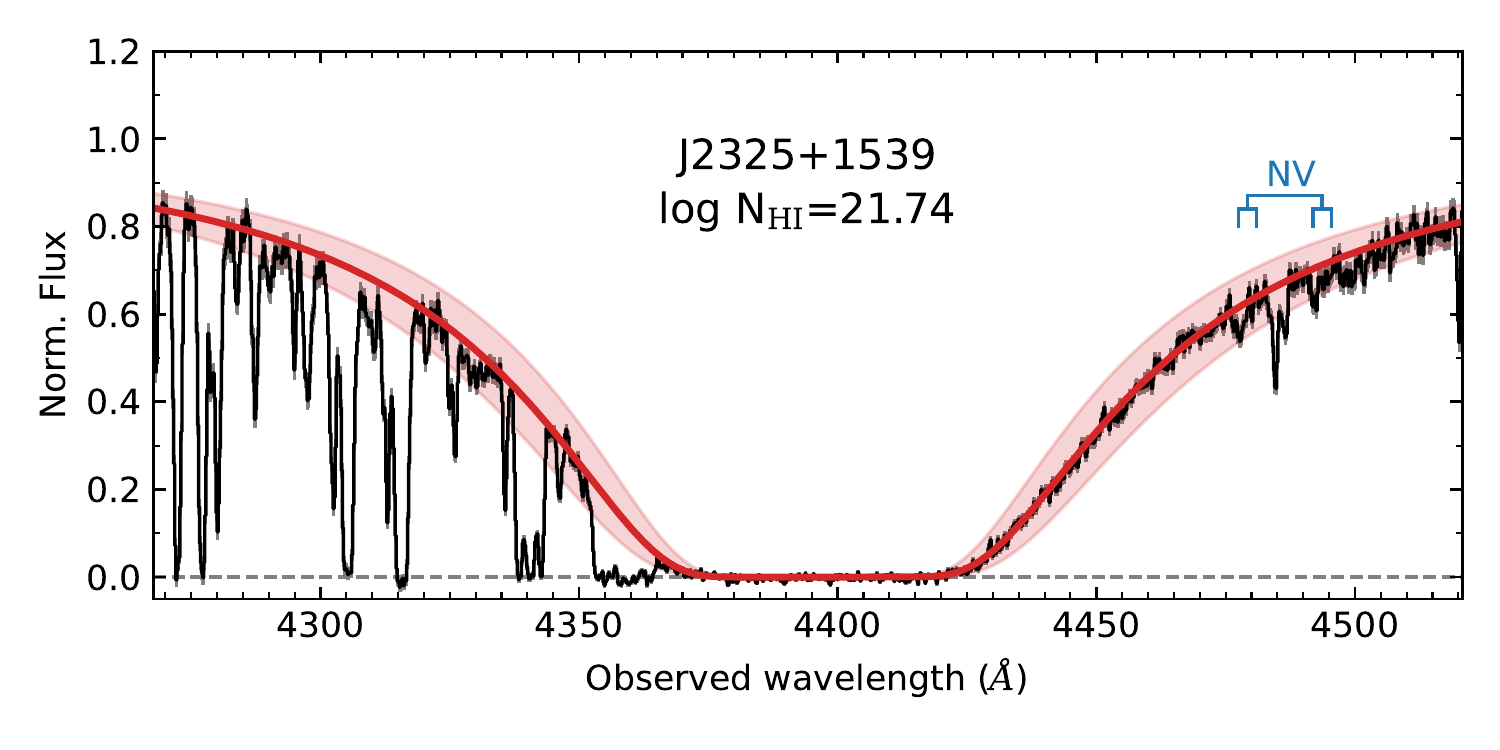} \\
    \end{tabular}
    \caption{Damped \HI-\lya\ lines for the proximate \HH\ systems analysed in this paper. The X-shooter data is shown in black with the profile corresponding to best fit shown in red. The red shadows represents the conservative range of $N(\HI)$, including continuum placement uncertainties. Note the residual \lya\ emission in the core of several DLAs. We also mark the expected position of the \NV$\lambda\lambda1238,1242$ doublet in blue, with the tick marks representing the range seen in low-ionisation species (i.e., $z_5$ and $z_{95}$). It is here detected in all cases but \Judcn\ and \Juttu.}
    \label{f:HI}
\end{figure*}
    \addtolength{\tabcolsep}{+3pt}

\begin{figure*}
    \centering
    \includegraphics[width=0.95\hsize,trim={1cm 2.0cm 0cm 2.0cm}]{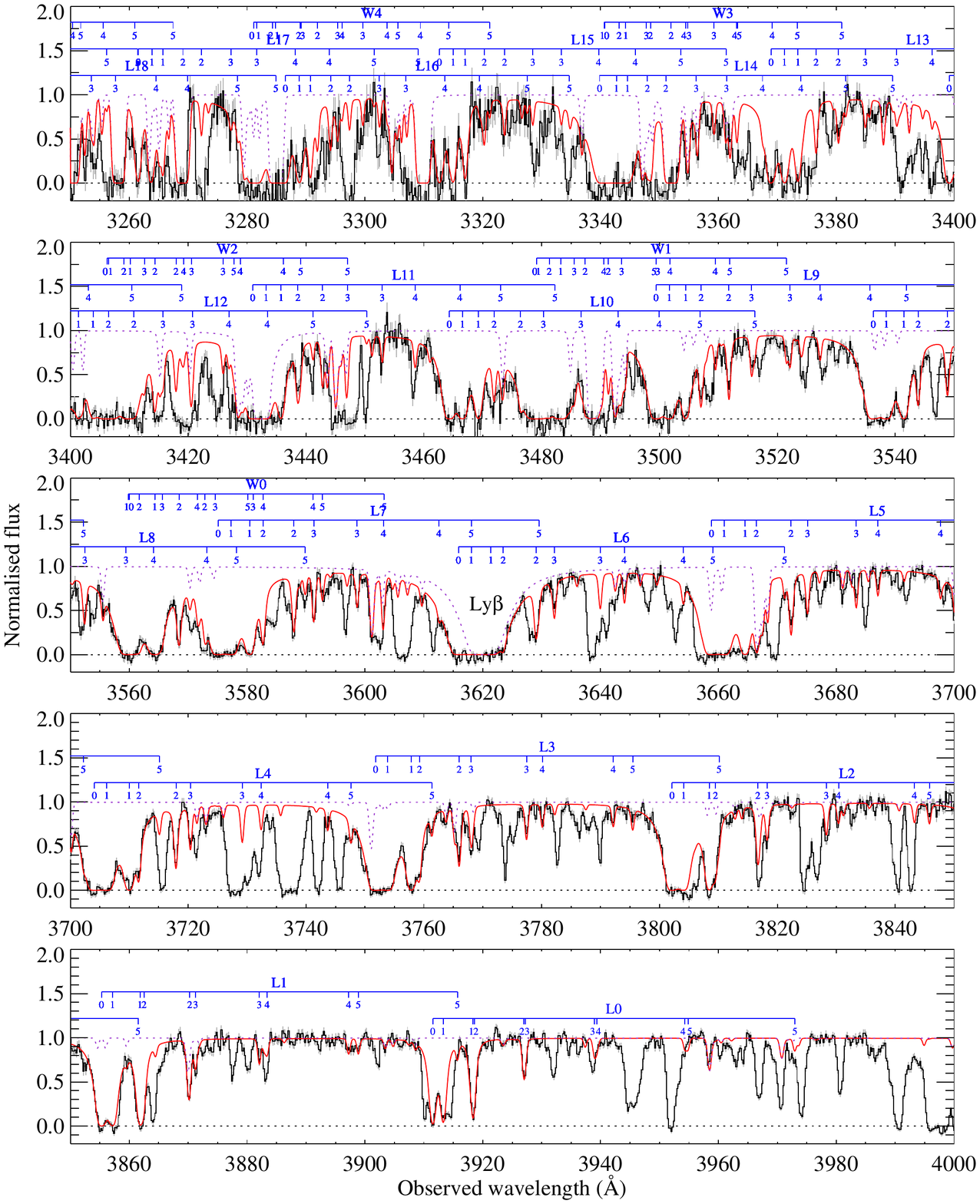}
    \caption{Portion of X-shooter spectrum of \Jzzun\ featuring H$_2$ absorption bands. The solid red line shows the overall absorption model, while contribution from non-H$_2$ lines (\HI, metals and lines from intervening systems) is shown as dashed profile. The blue segments connect lines from different rotational levels (shown here for convenience from J=0 to J=5) of a given band (indicated by the label above, 'L' for Lyman, 'W' for Werner and the following number being the upper vibrational level.}
    \label{f:J0019H2}
\end{figure*}

\begin{figure*}
    \centering
    \includegraphics[width=0.95\hsize,trim={1cm 2.0cm 0cm 2.0cm}]{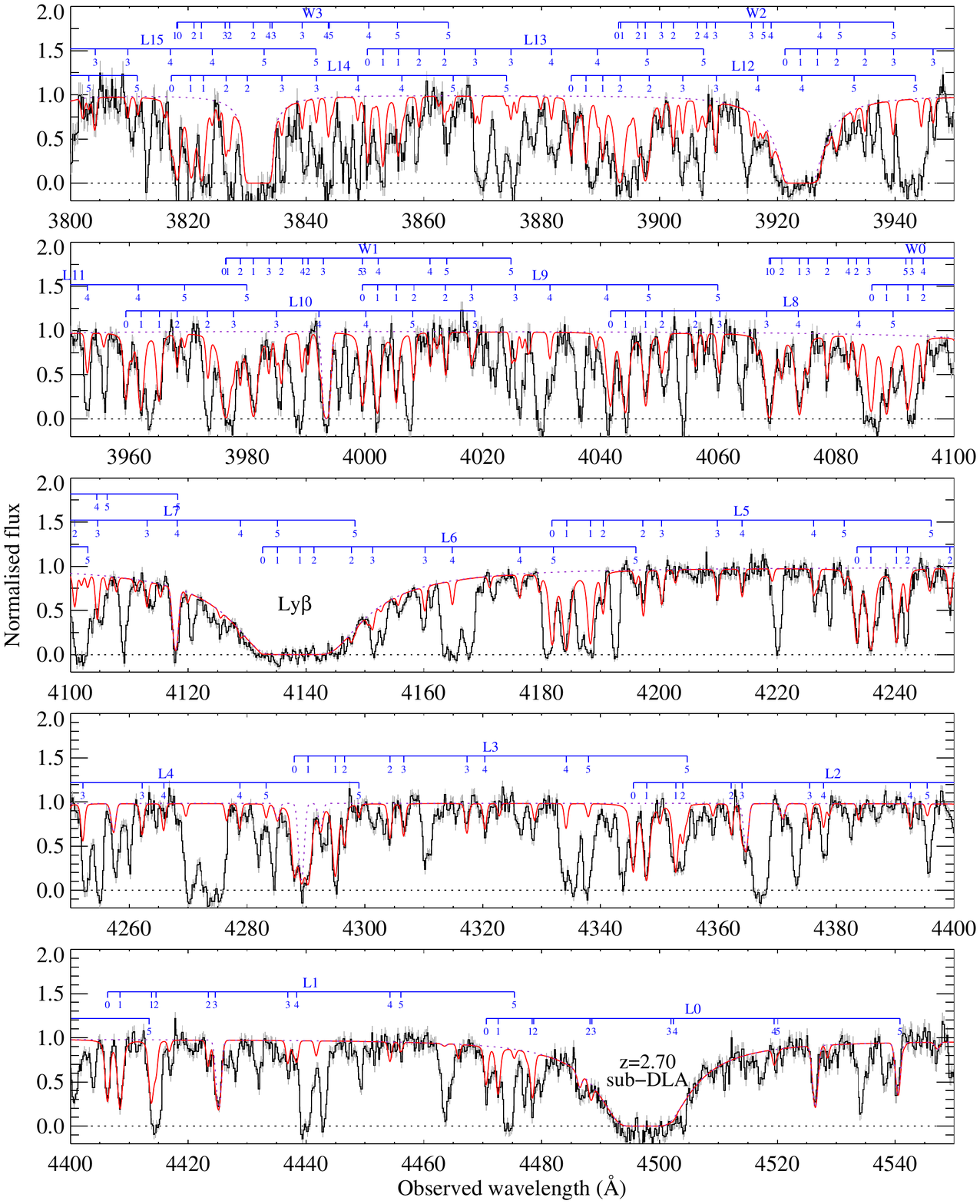}
    \caption{Same as Fig.~\ref{f:J0019H2} for the quasar \Jzzcn.}
    \label{f:J0059H2}
\end{figure*}

\begin{figure*}
    \centering
    \includegraphics[width=0.95\hsize,trim={1cm 2.0cm 0cm 2.0cm}]{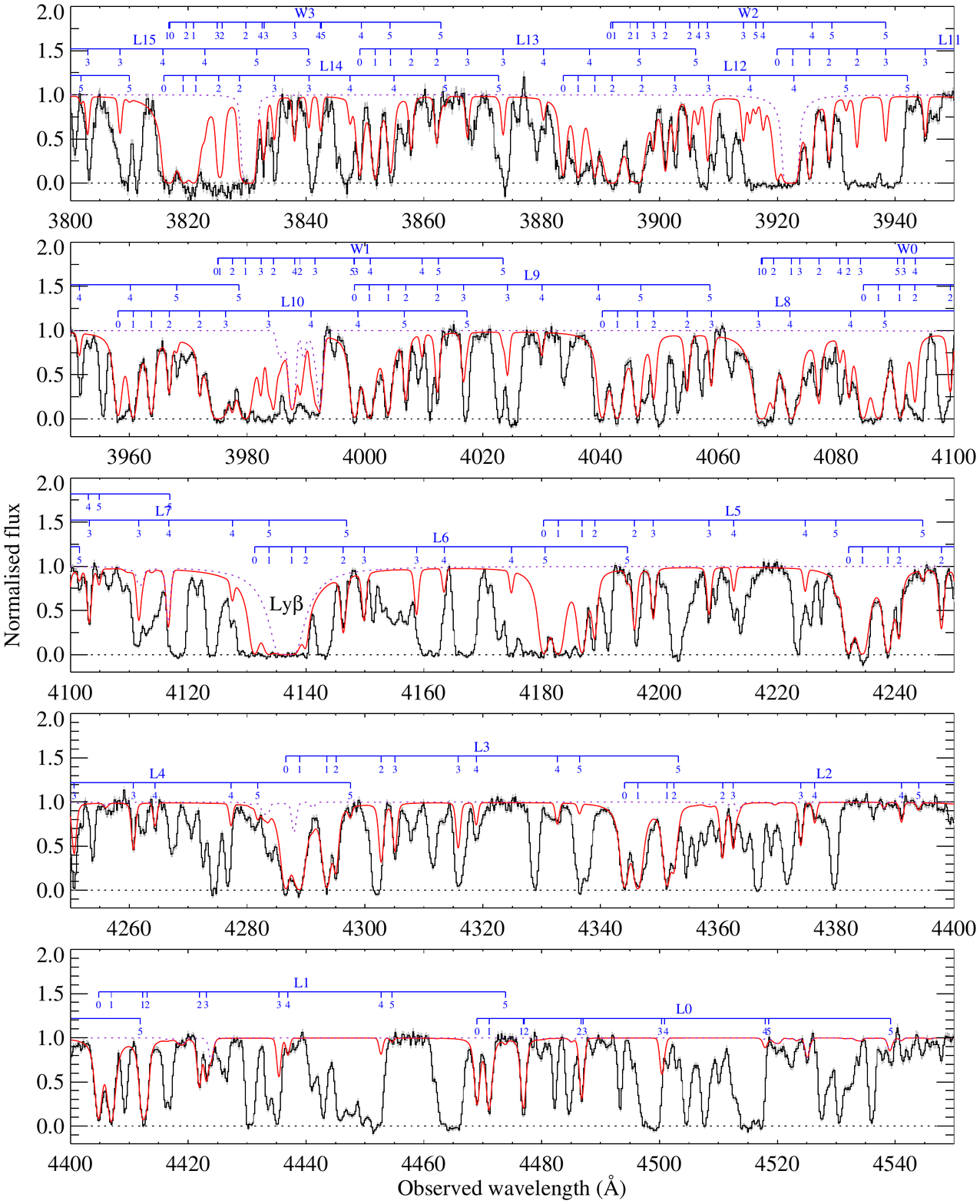}
    \caption{Same as Fig.~\ref{f:J0019H2} for the quasar \Judts.}
    \label{f:J1236H2}
\end{figure*}

\begin{figure*}
    \centering
    \includegraphics[width=0.95\hsize,trim={1cm 2.0cm 0cm 2.0cm}]{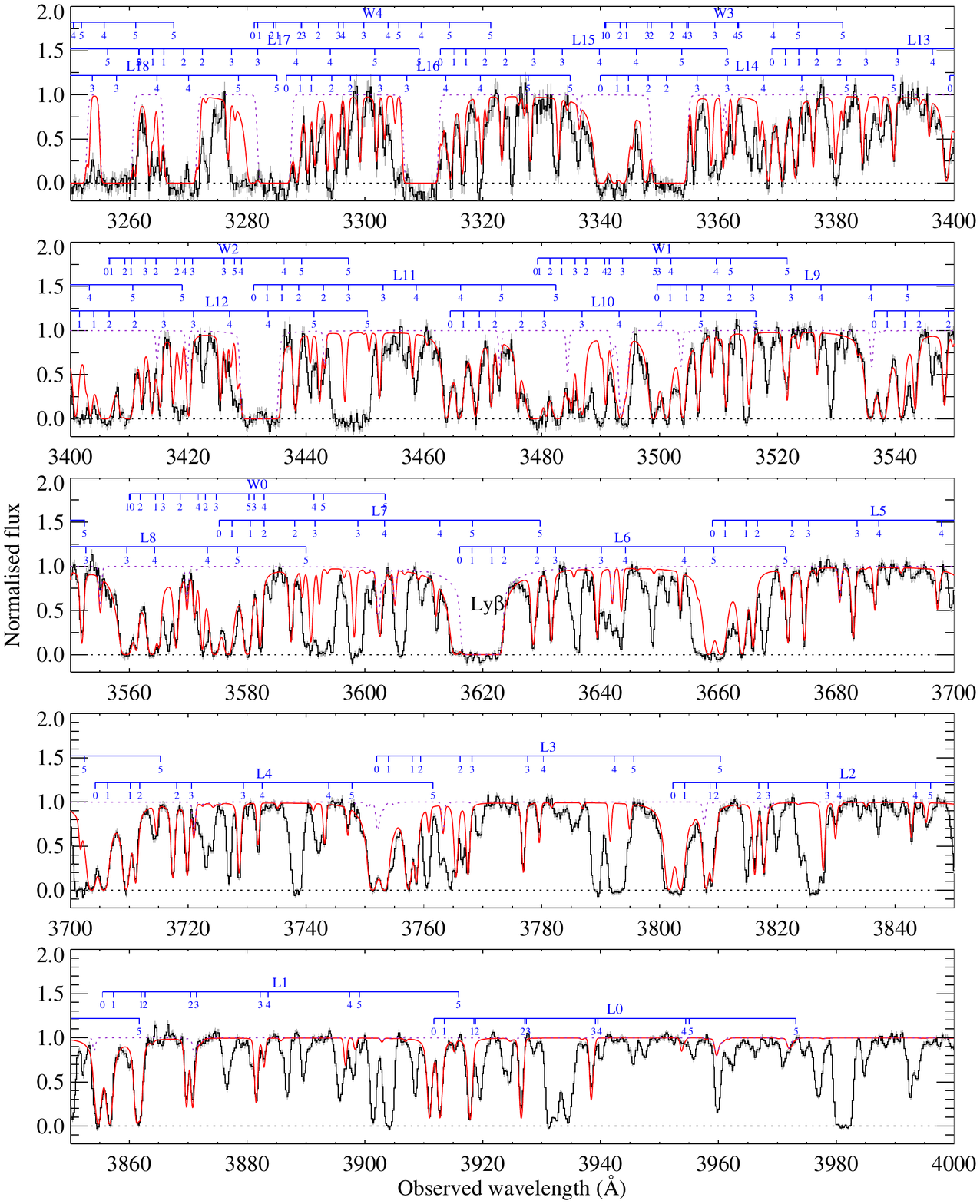}
    \caption{Same as Fig.~\ref{f:J0019H2} for the quasar \Judqh.}
    \label{f:J1248H2}
\end{figure*}

\begin{figure*}
    \centering
    \includegraphics[width=0.95\hsize,trim={1cm 2.0cm 0cm 2.0cm}]{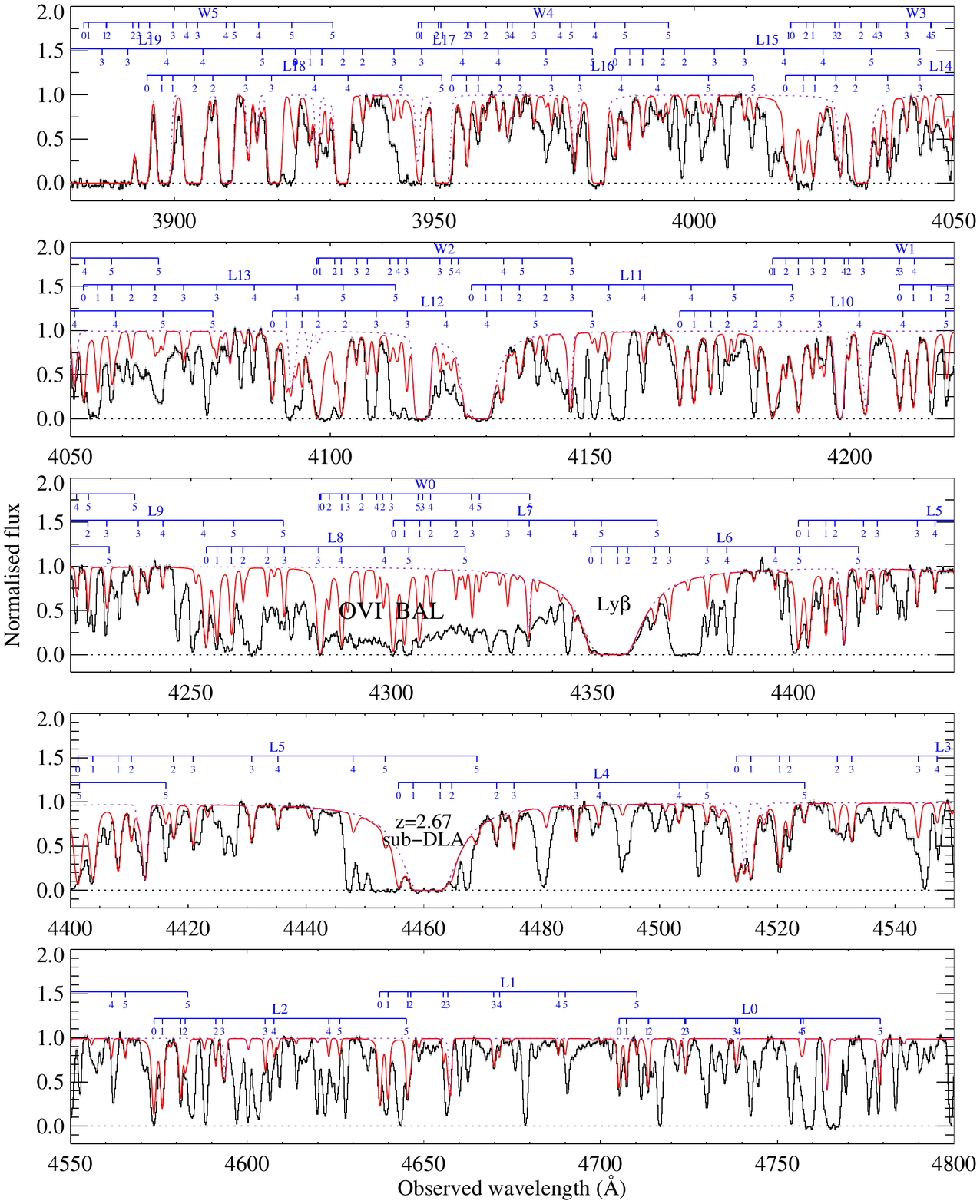}
    \caption{Same as Fig.~\ref{f:J0019H2} for the quasar \Judcn. Note the presence of a high-velocity broad absorption line (BAL) system in this quasar.}
    \label{f:J1259H2}
\end{figure*}

\begin{figure*}
    \centering
    \includegraphics[width=0.95\hsize,trim={1cm 2.0cm 0cm 2.0cm}]{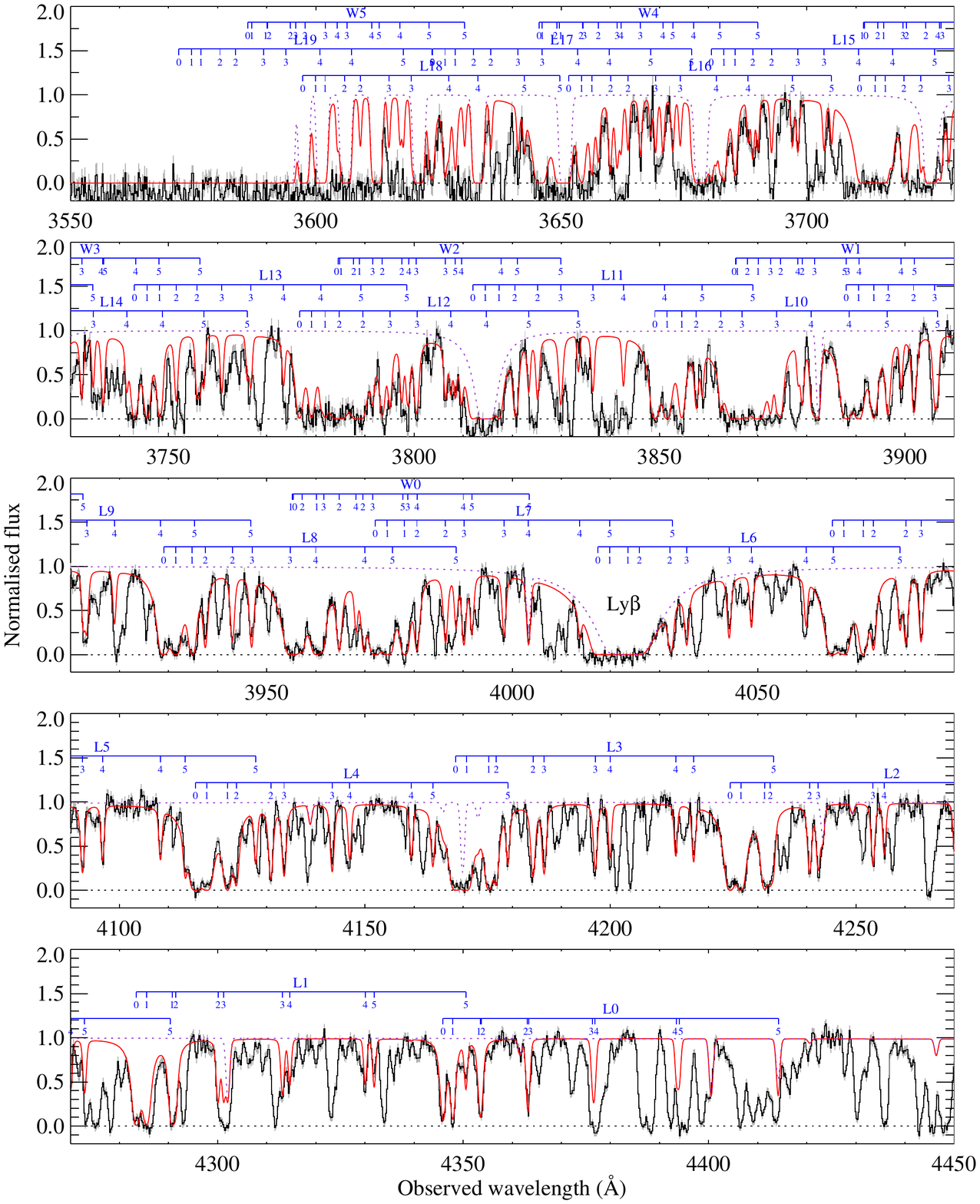}
    \caption{Same as Fig.~\ref{f:J0019H2} for the quasar \Juttu.}
    \label{f:J1331H2}
\end{figure*}

\begin{figure*}
    \centering
    \includegraphics[width=0.95\hsize,trim={1cm 2.0cm 0cm 2.0cm}]{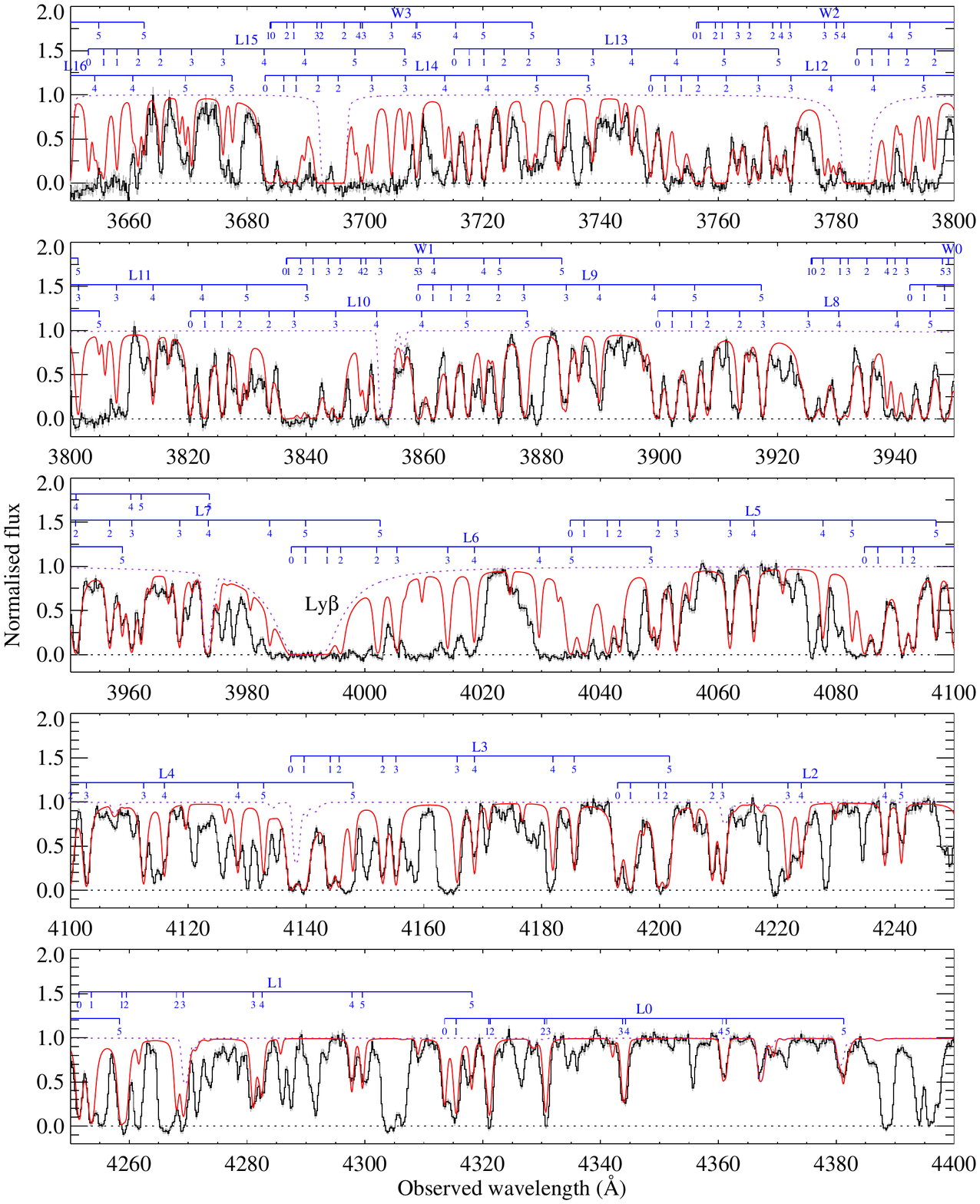}
    \caption{Same as Fig.~\ref{f:J0019H2} for the quasar \Jutch.}
    \label{f:J1358H2}
\end{figure*}

\begin{figure*}
    \centering
    \includegraphics[width=0.95\hsize,trim={1cm 2.0cm 0cm 2.0cm}]{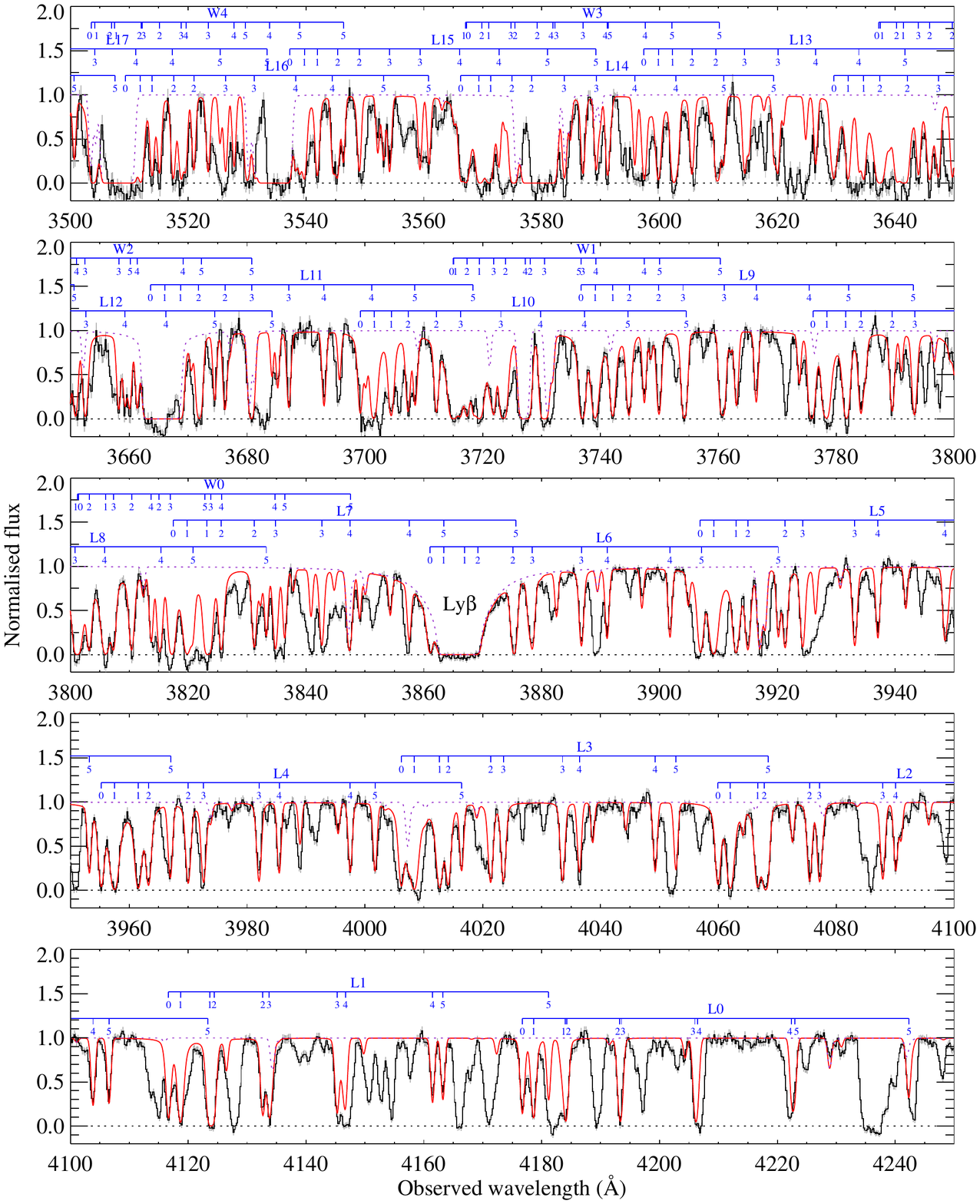}
    \caption{Same as Fig.~\ref{f:J0019H2} for the quasar \Jdddh.}
    \label{f:J2228H2}
\end{figure*}

\begin{figure*}
    \centering
    \includegraphics[width=0.95\hsize,trim={1cm 2.0cm 0cm 2.0cm}]{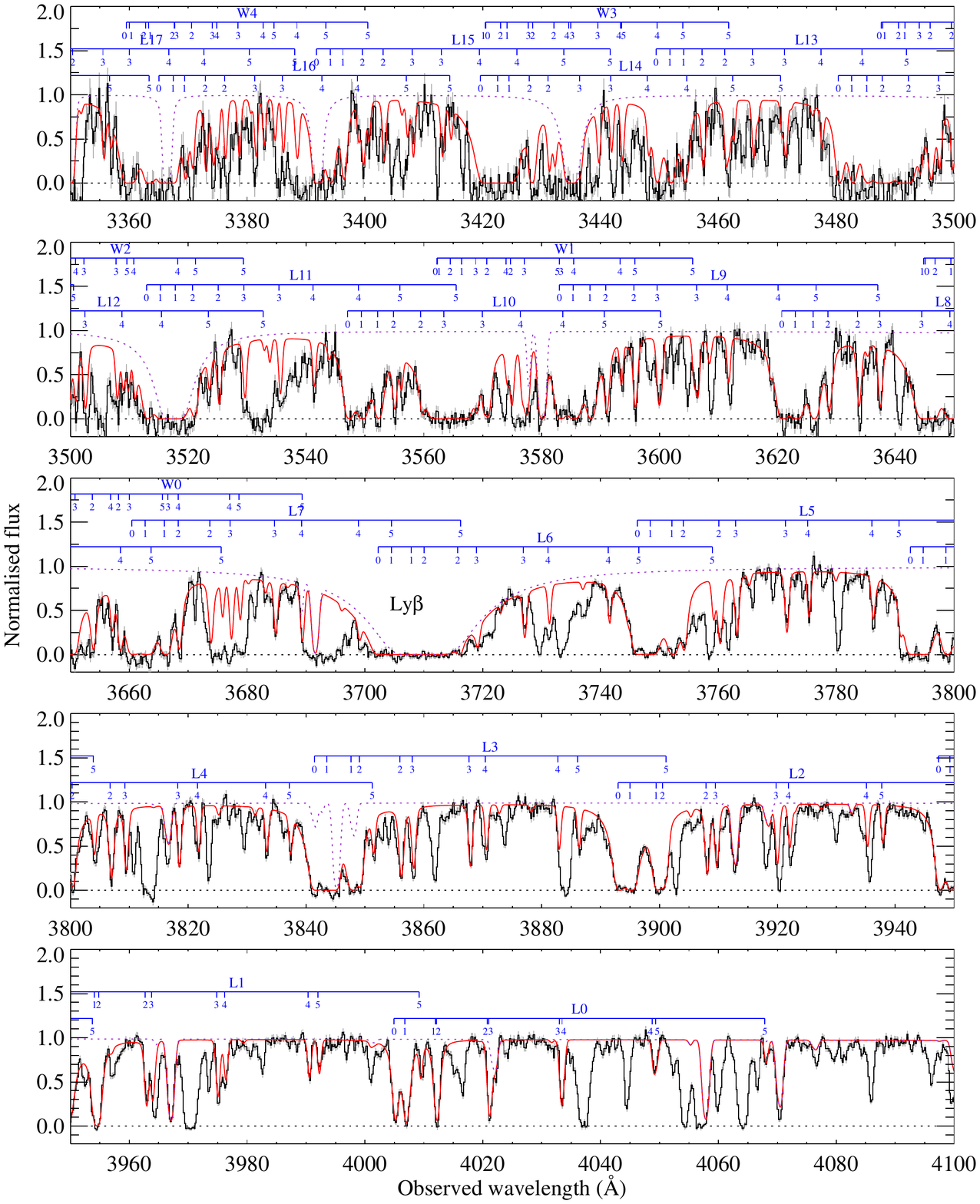}
    \caption{Same as Fig.~\ref{f:J0019H2} for the quasar \Jdtdc.}
    \label{f:J2325H2}
\end{figure*}

\clearpage

\section{Low-ionisation metal absorption lines \label{s:metfigs}}

\begin{figure}[!ht]
    \centering
    \includegraphics[width=\hsize,trim={0.5cm 7.5cm 6.5cm 0}]{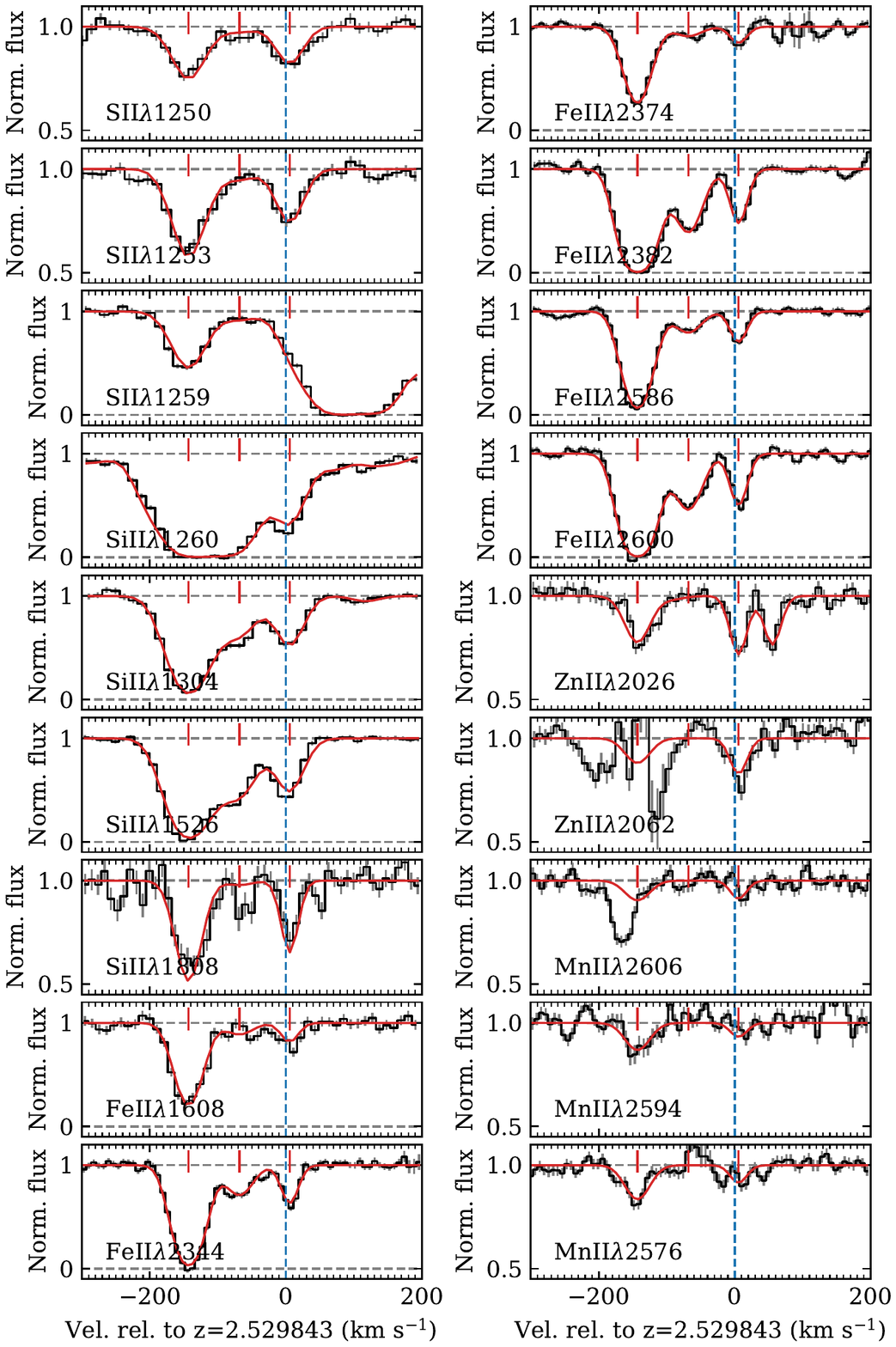}
    \caption{Main singly ionised metal lines in the proximate system towards \Jzzun\ in their ground state. The zero of the velocity scale is set to the redshift of the H$_2$ component (blue dashed line). Short red ticks mark the position of the different components used in the model (overplotted in red).}
    \label{f:J0019met}
\end{figure}

\begin{figure}[!ht]
    \centering
    \includegraphics[width=\hsize,trim={0.5cm 6.0cm 6.5cm 0}]{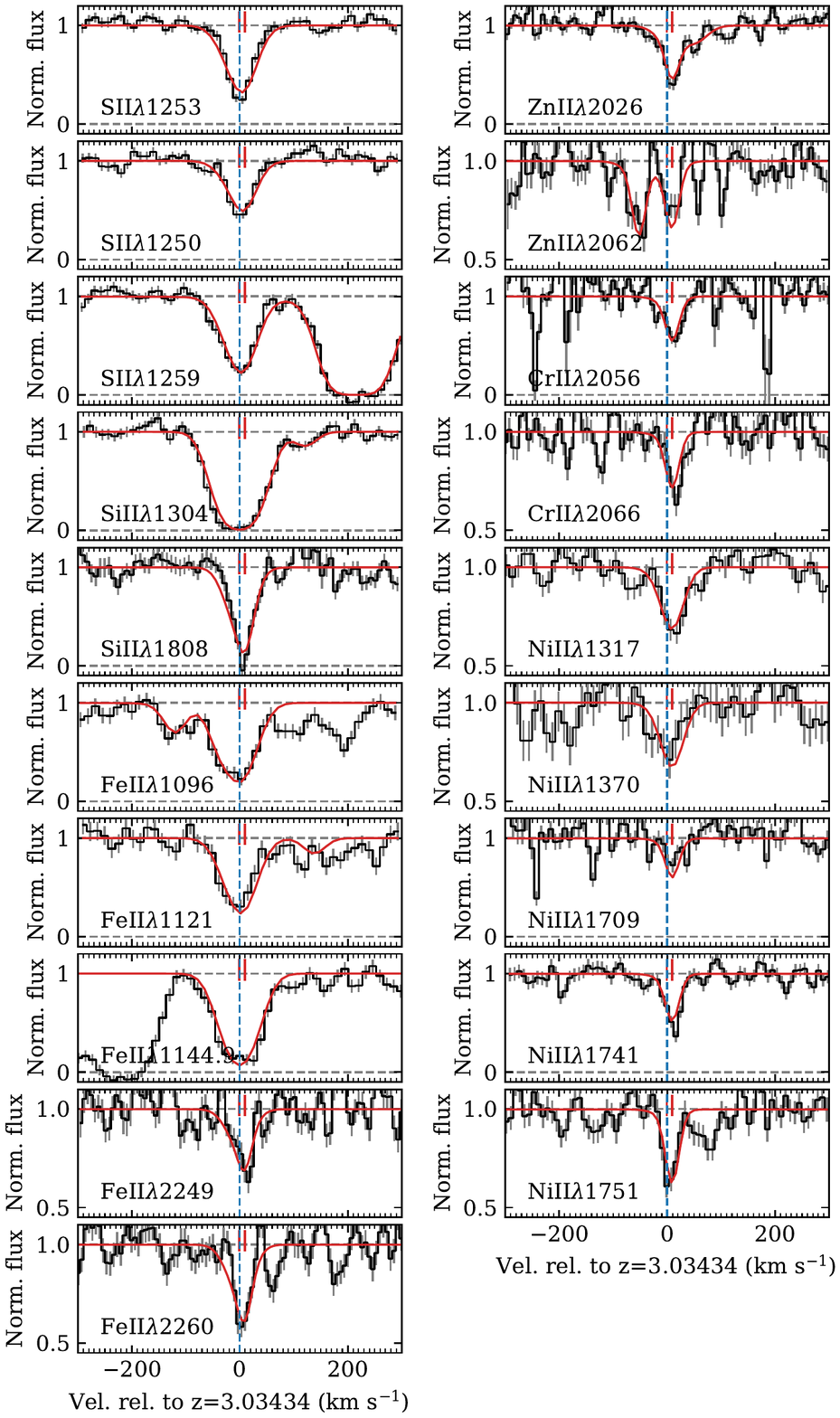}
    \caption{Singly ionised metal lines in the proximate system towards \Jzzcn.}
    \label{f:J10059met}
\end{figure}

\begin{figure}[!ht]
    \centering
    \includegraphics[width=\hsize,trim={0.5cm 10.0cm 6.5cm 0}]{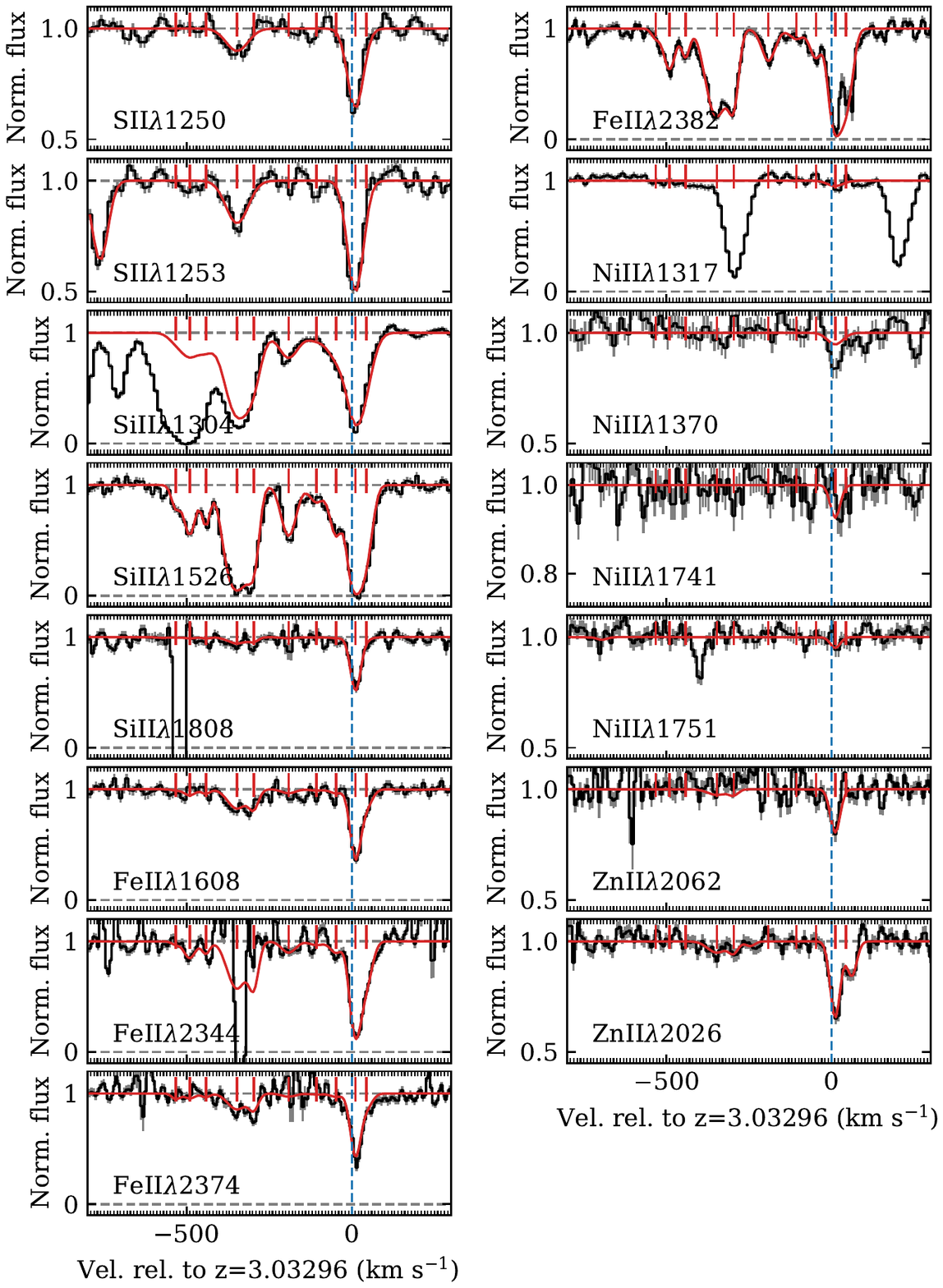}
    \caption{Singly ionised metal lines in the proximate system towards \Judts. Spike at v=0 in \FeII$\lambda$2382 is a sky line residual.}
    \label{f:J1236met}
\end{figure}

\begin{figure}[!ht]
    \centering
    \includegraphics[width=\hsize,trim={0.5cm 12.5cm 6.5cm 0}]{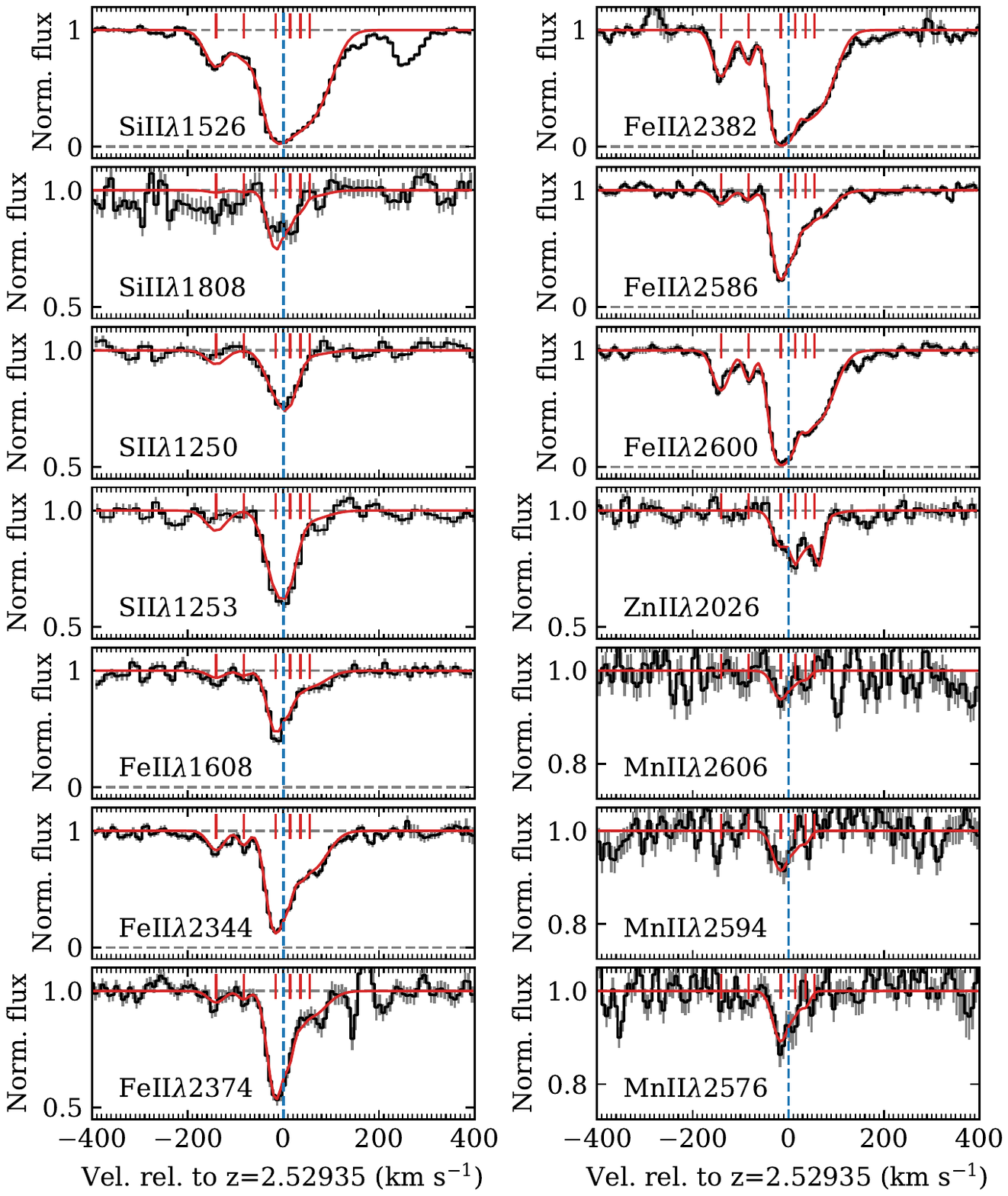}
    \caption{Singly ionised metal lines in the proximate system towards \Judqh. %\PN{Some issues being resolved...work in progress}
    }
    \label{f:J1248met}
\end{figure}

\begin{figure}[!ht]
    \centering
    \includegraphics[width=\hsize,trim={0.5cm 12.5cm 6.5cm 0}]{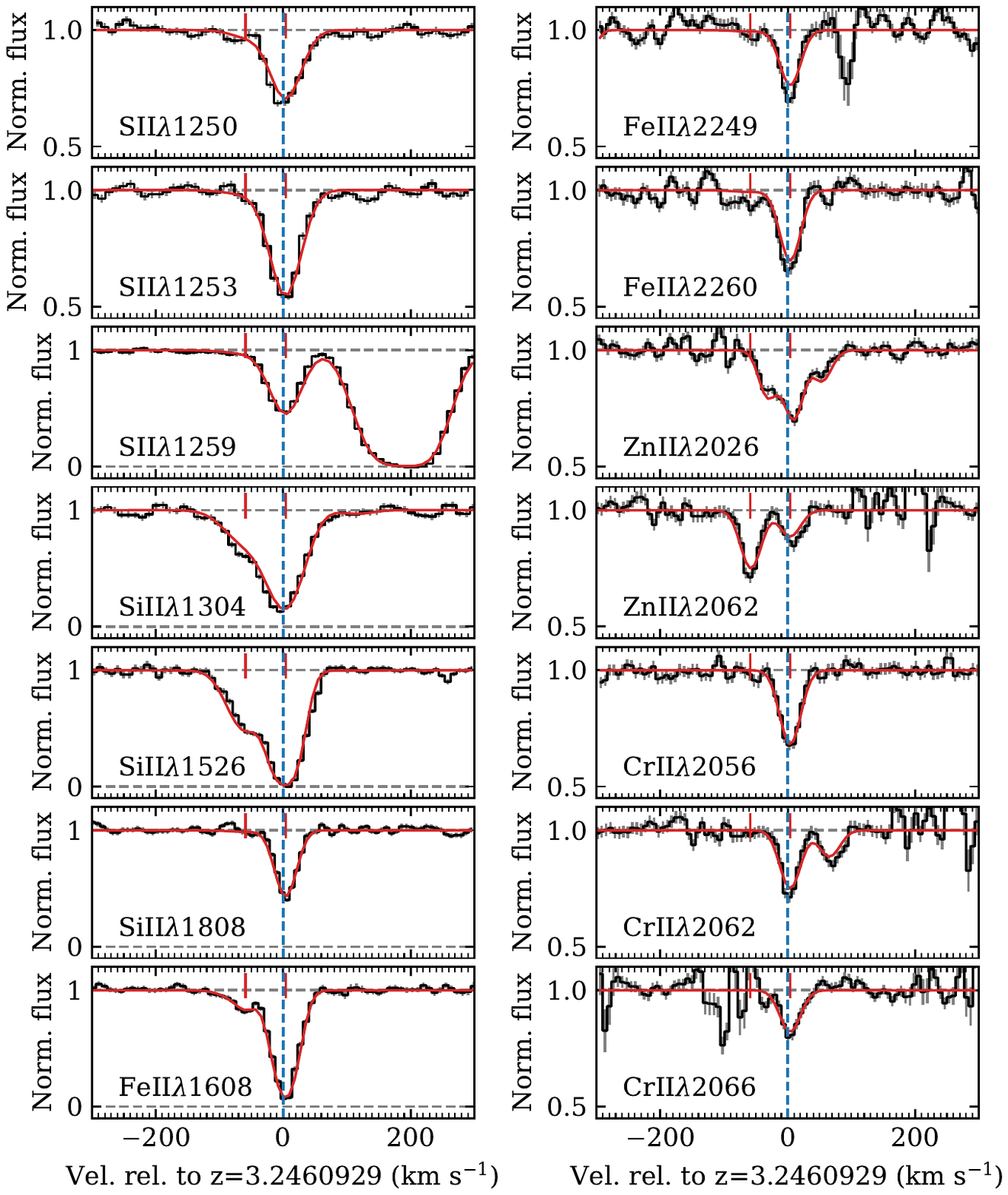}
    \caption{Singly ionised metal lines in the proximate system towards \Judcn. }
    \label{f:J1259met}
\end{figure}

\begin{figure}[!ht]
    \centering
    \includegraphics[width=\hsize,trim={0.5cm 7.5cm 6.5cm 0}]{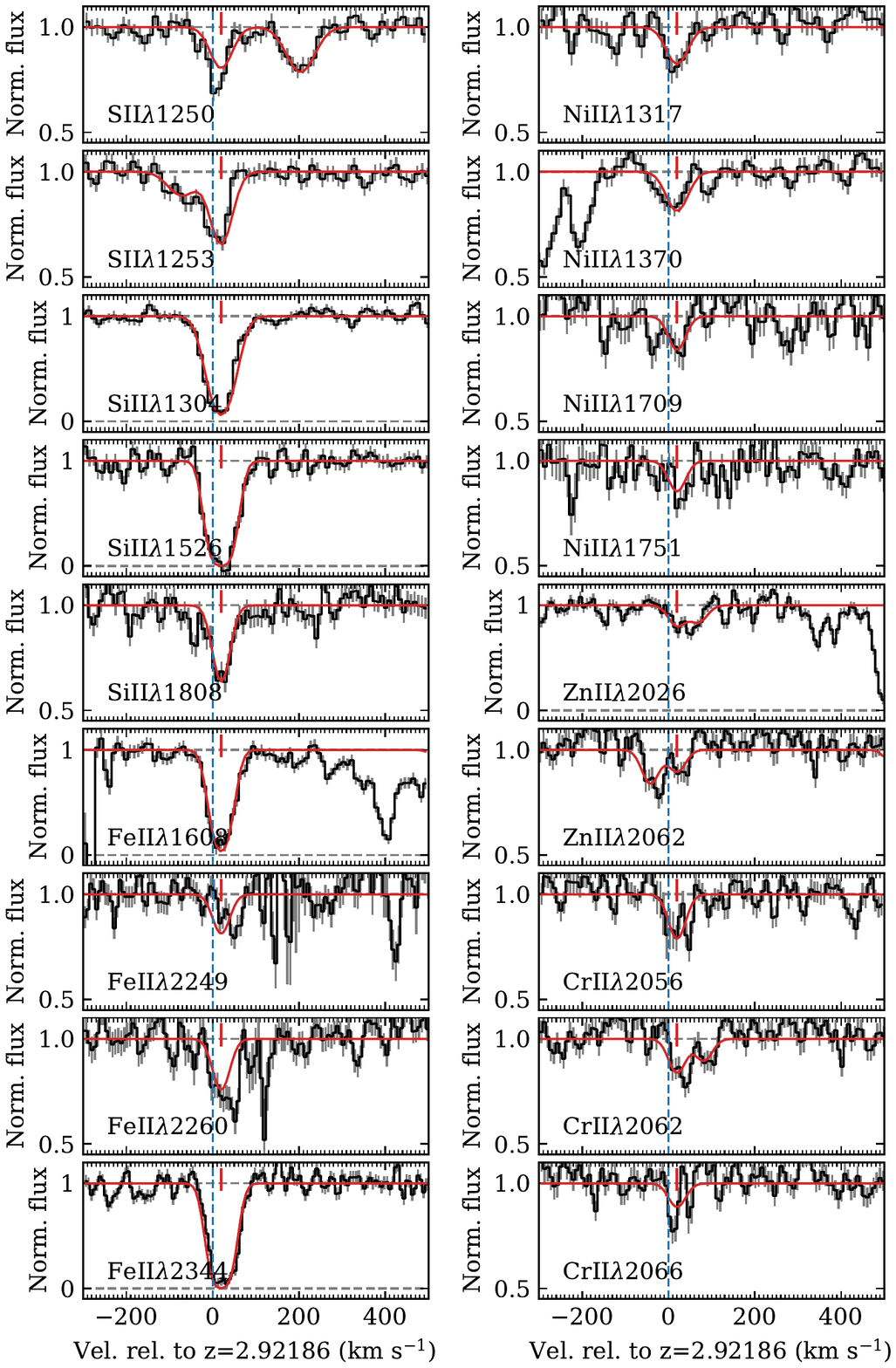}
    \caption{Singly ionised metal lines in the proximate system towards \Juttu.}
    \label{f:J1331met}
\end{figure}

\begin{figure}[!ht]
    \centering
    \includegraphics[width=\hsize,trim={0.5cm 6.0cm 6.5cm 0}]{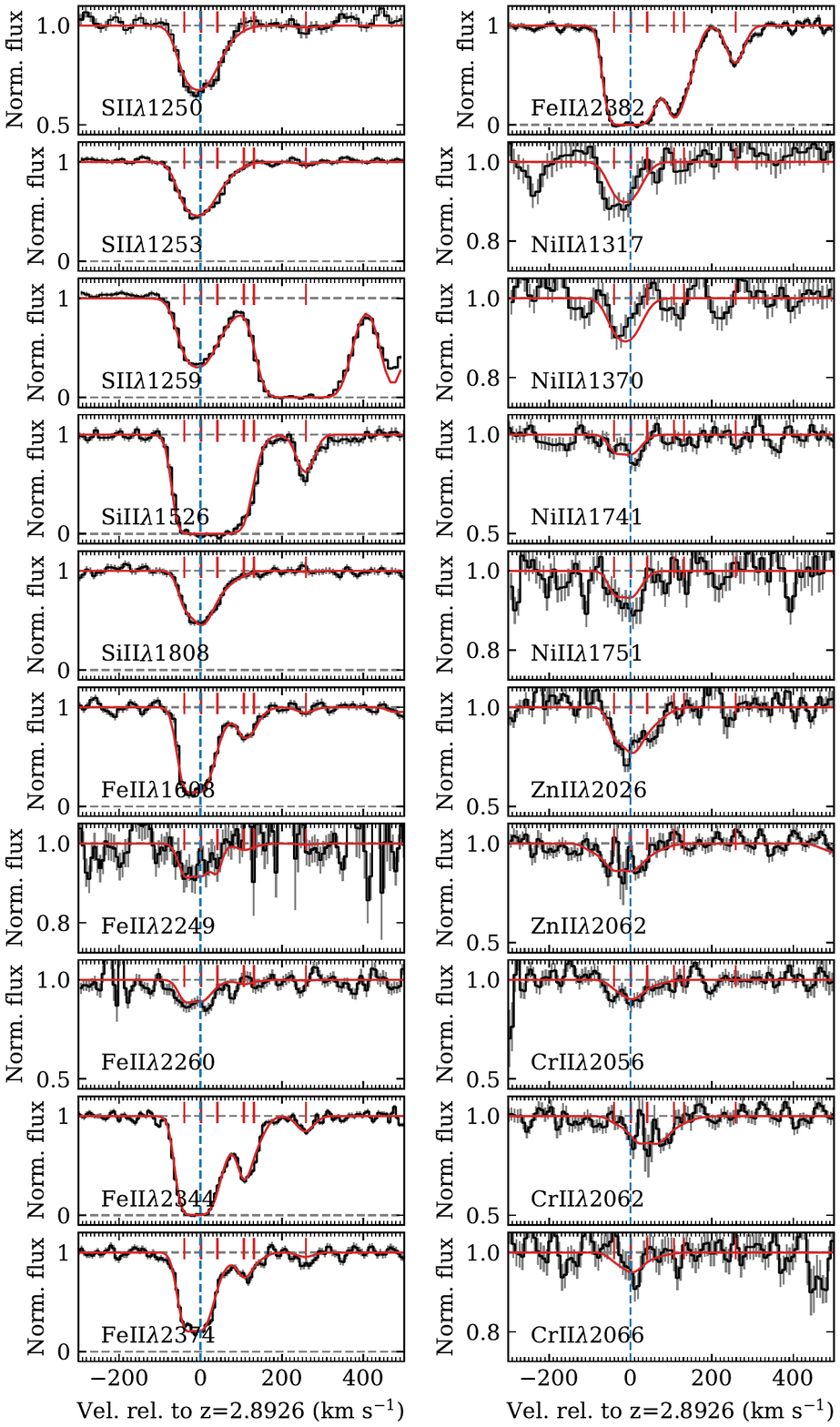}
    \caption{Singly ionised metal lines in the proximate system towards \Jutch. }
    \label{f:J1358met}
\end{figure}

\begin{figure}[!ht]
    \centering
    \includegraphics[width=\hsize,trim={0.5cm 7.5cm 6.5cm 0}]{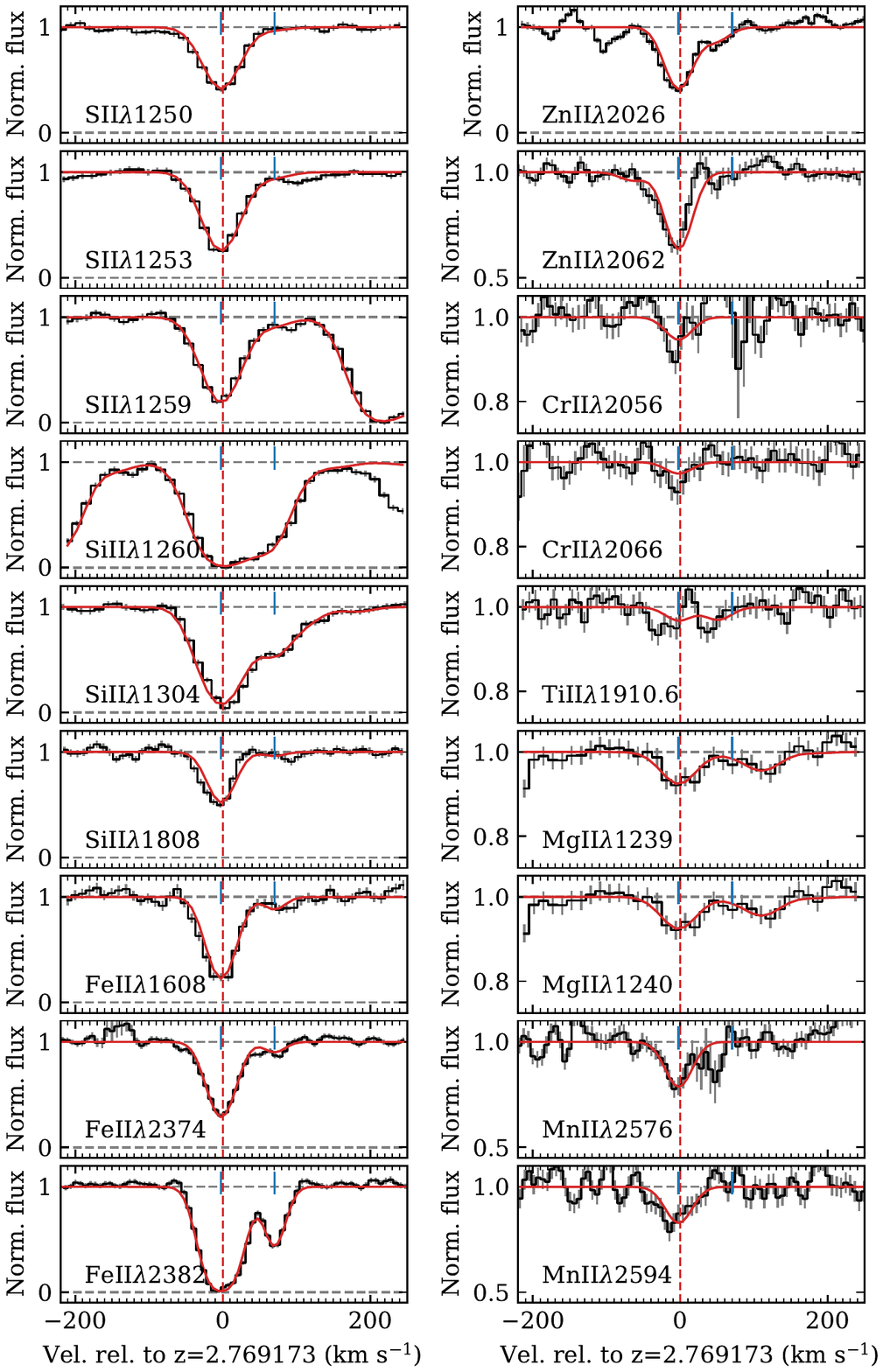}
    \caption{Singly ionised metal lines in the proximate system towards \Jdddh.}
    \label{f:J2228met}
\end{figure}

\begin{figure}[!ht]
    \centering
    \includegraphics[width=\hsize,trim={0.5cm 1.5cm 6.5cm 0}]{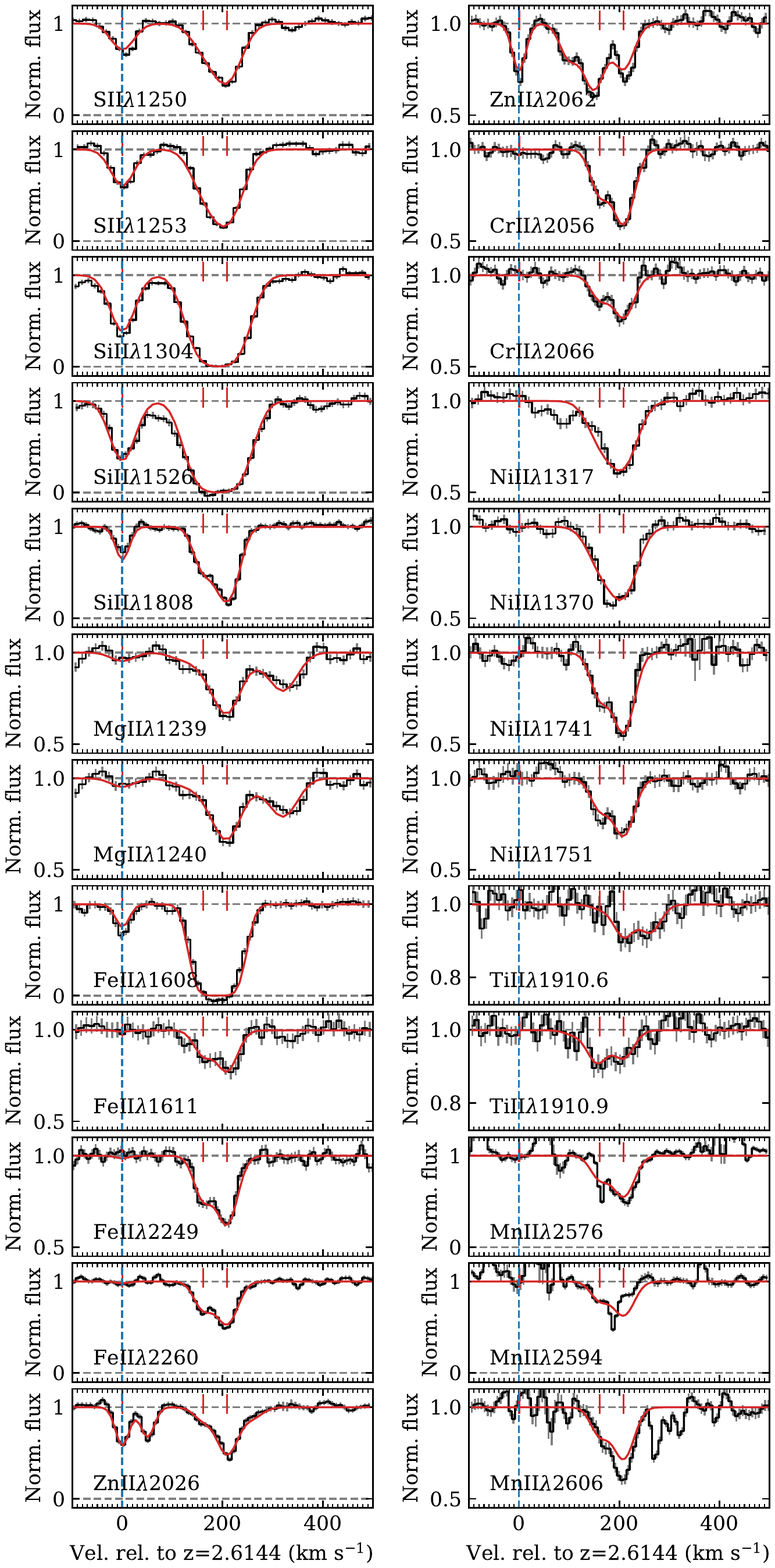}
    \caption{Singly ionised metal lines in the proximate system towards \Jdtdc. \MnII\ close to sky residuals.}
    \label{f:J2325met}
\end{figure}

%tables
\addtolength{\tabcolsep}{-3pt}
\input J0019met
\input J0059met
\input J1236met
\input J1248met
\input J1259met
\input J1331met
\input J1358met
\input J2228met
\input J2325met
\addtolength{\tabcolsep}{3pt}

\begin{figure*}
    \centering
    \addtolength{\tabcolsep}{-8pt}
    \begin{tabular}{ccccc}
\includegraphics[trim={0.0cm 18.8cm 10.5cm 0.cm}, clip=,width=0.21\hsize]{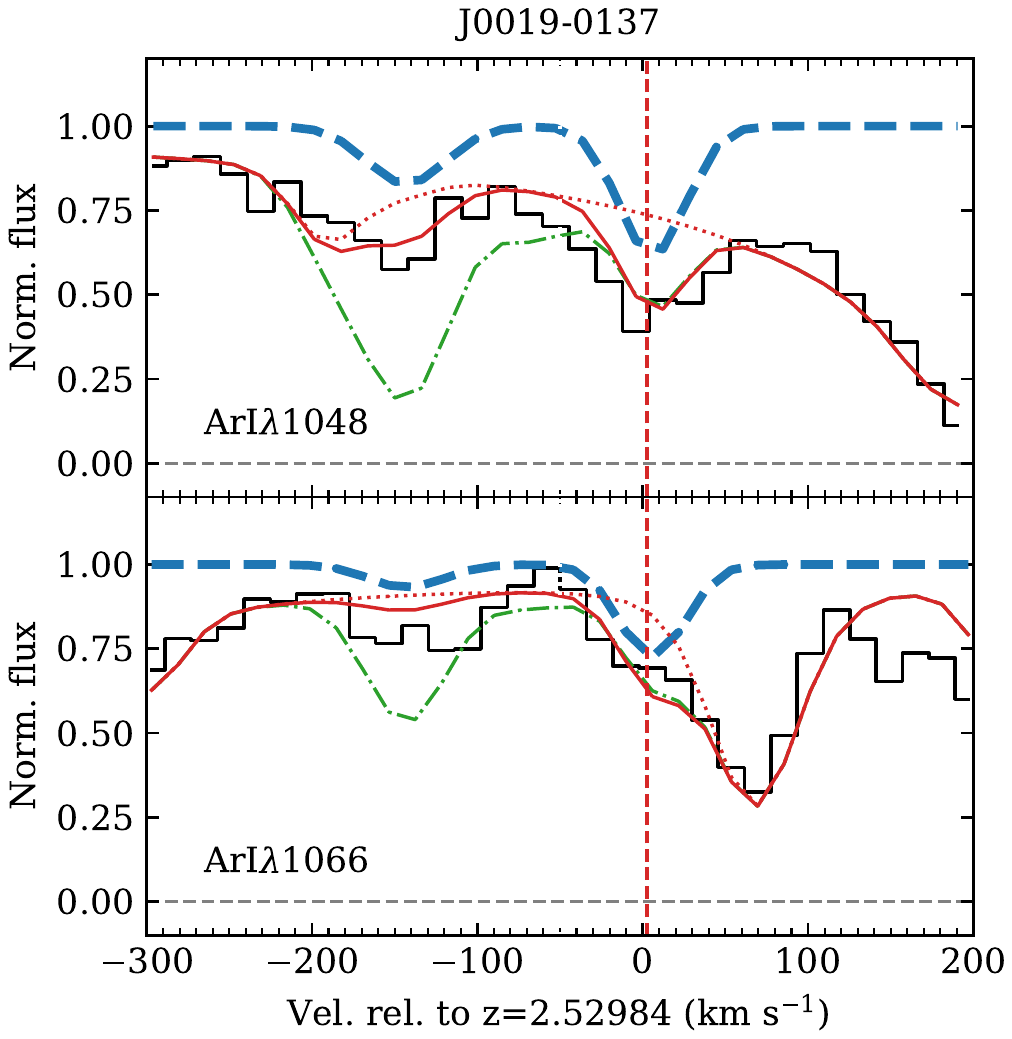}&
\includegraphics[trim={0.0cm 18.8cm 10.5cm 0.cm}, clip=,width=0.21\hsize]{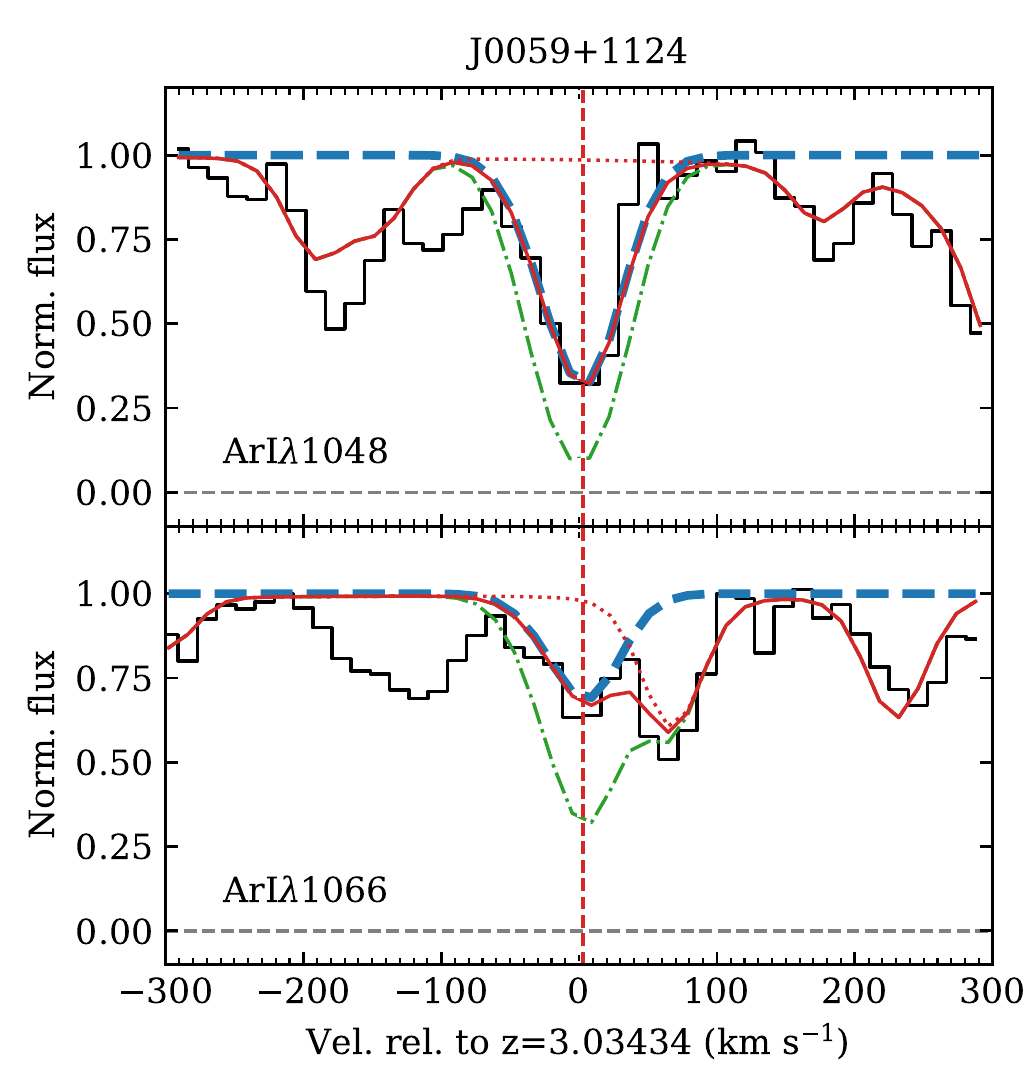}&
\includegraphics[trim={0.0cm 18.8cm 10.5cm 0.cm}, clip=,width=0.21\hsize]{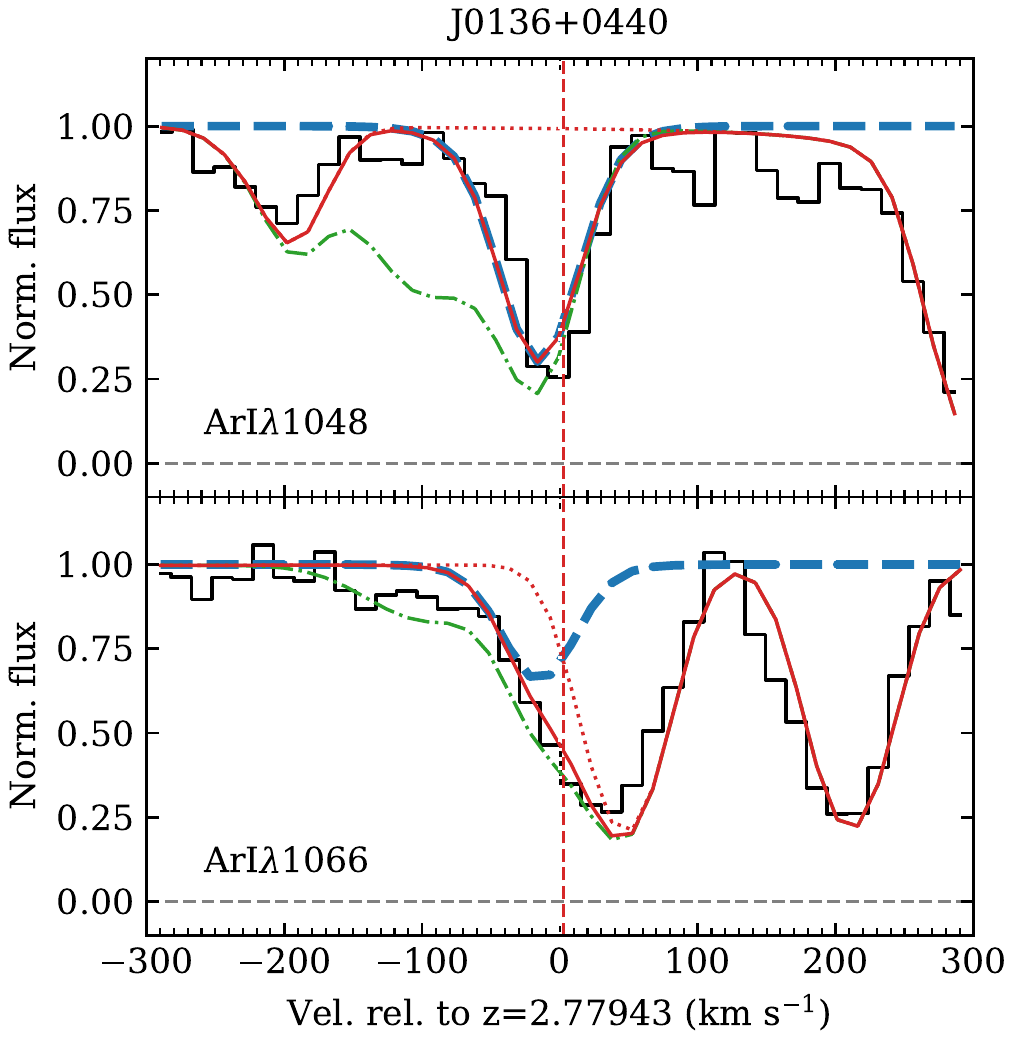}&
\includegraphics[trim={0.0cm 18.8cm 10.5cm 0.cm}, clip=,width=0.21\hsize]{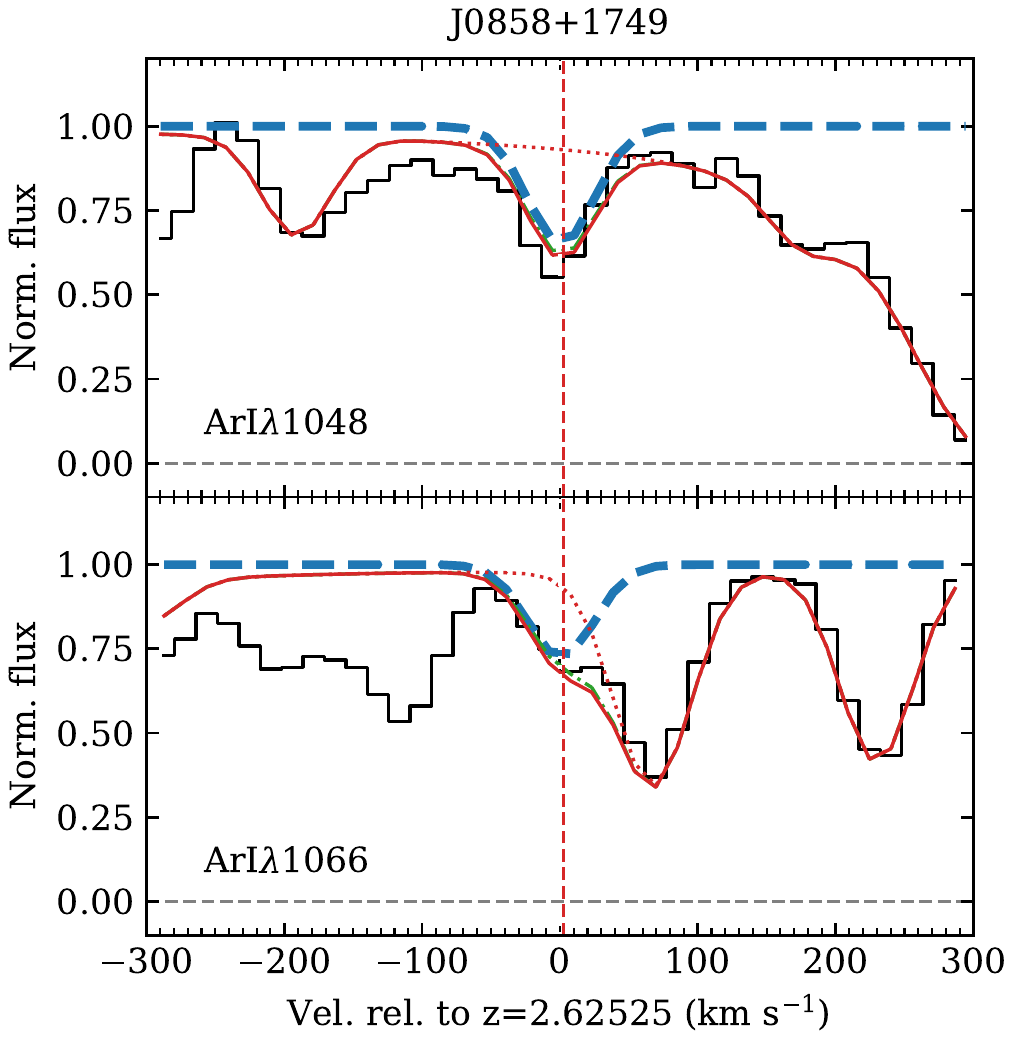}&
\includegraphics[trim={0.0cm 18.8cm 10.5cm 0.cm}, clip=,width=0.21\hsize]{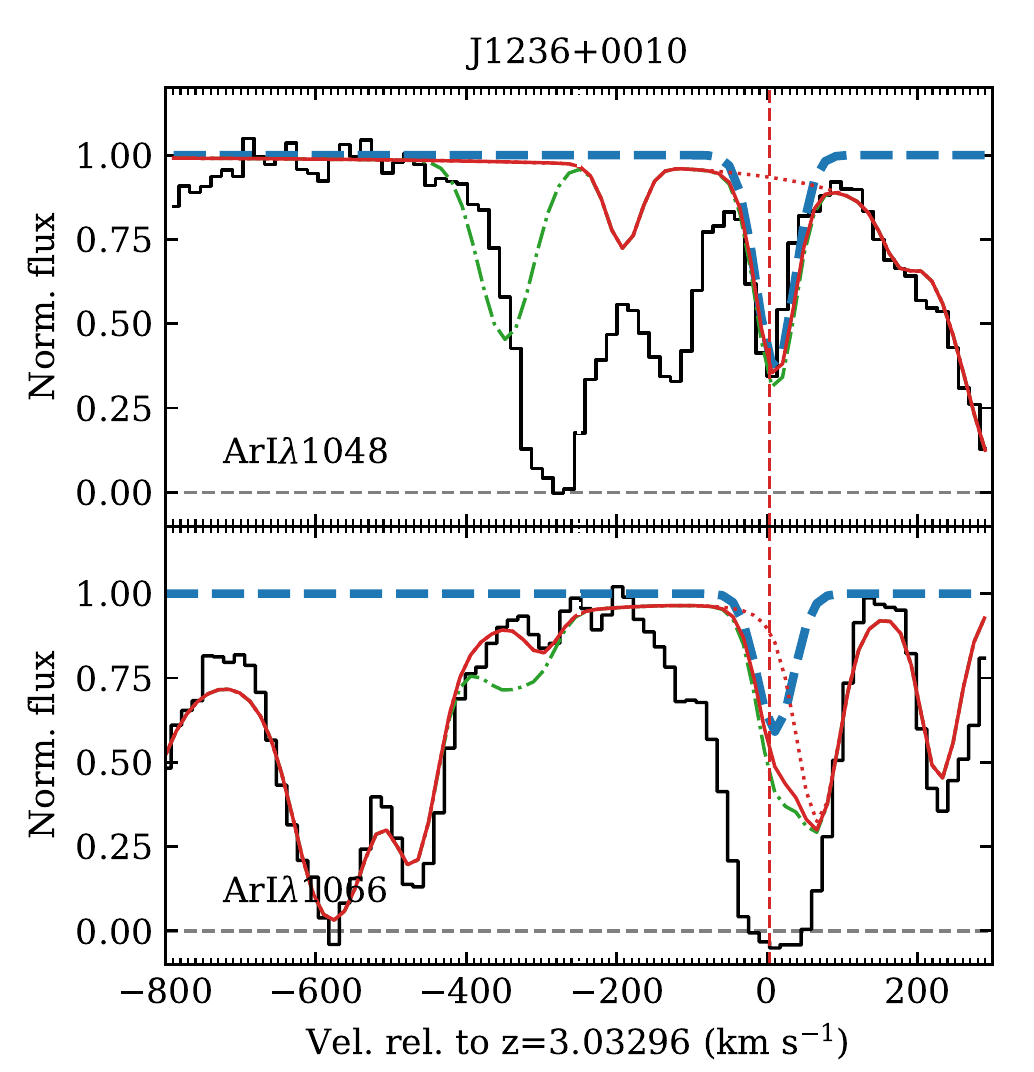}\\
\includegraphics[trim={0.0cm 18.8cm 10.5cm 0.cm}, clip=,width=0.21\hsize]{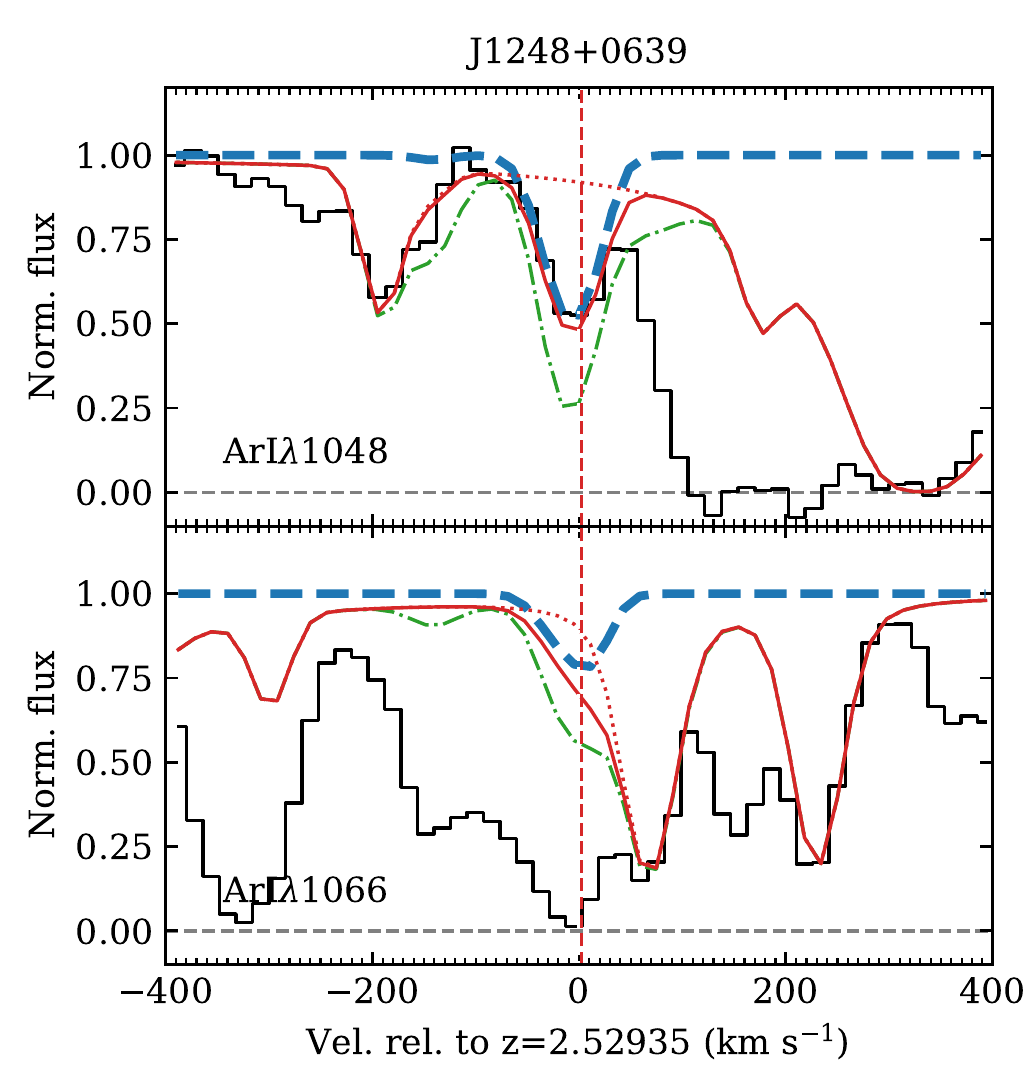}&
\includegraphics[trim={0.0cm 18.8cm 10.5cm 0.cm}, clip=,width=0.21\hsize]{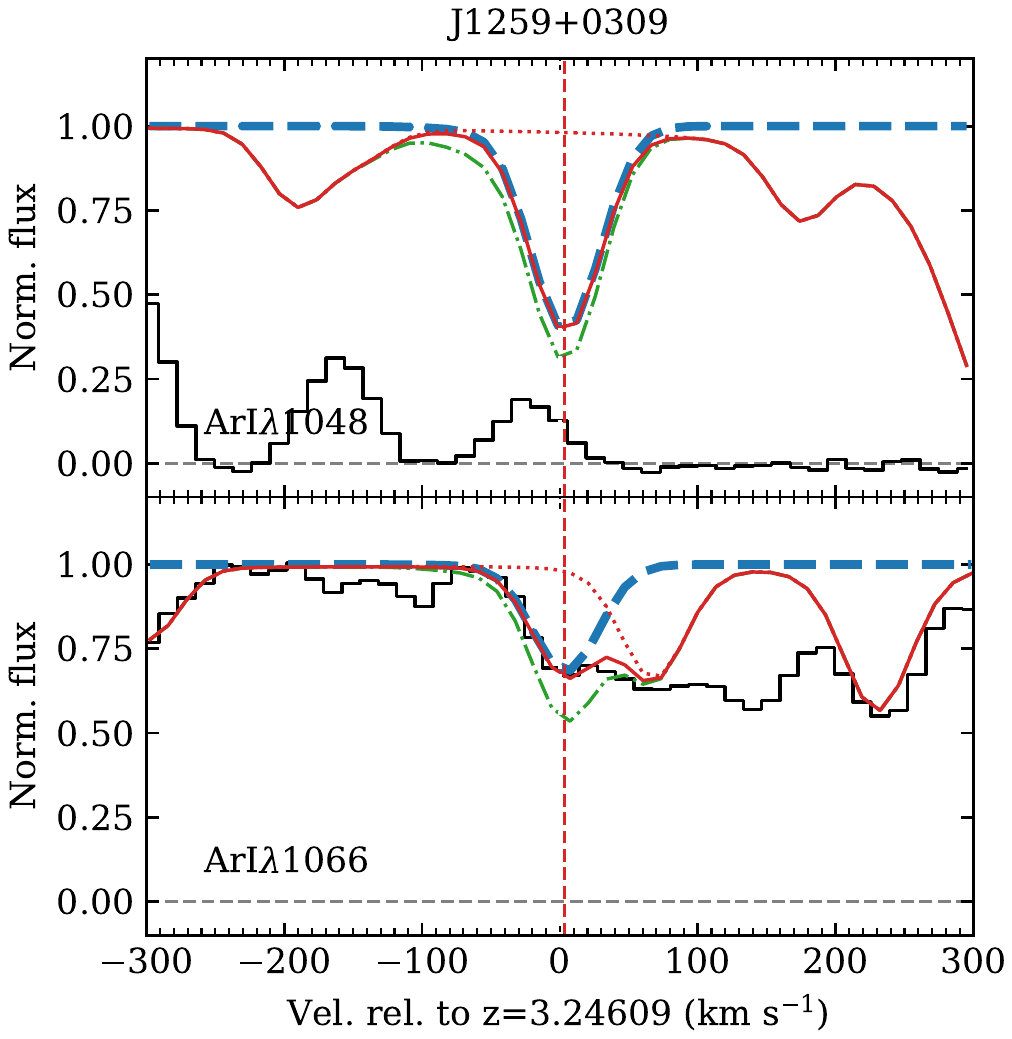}&
\includegraphics[trim={0.0cm 18.8cm 10.5cm 0.cm}, clip=,width=0.21\hsize]{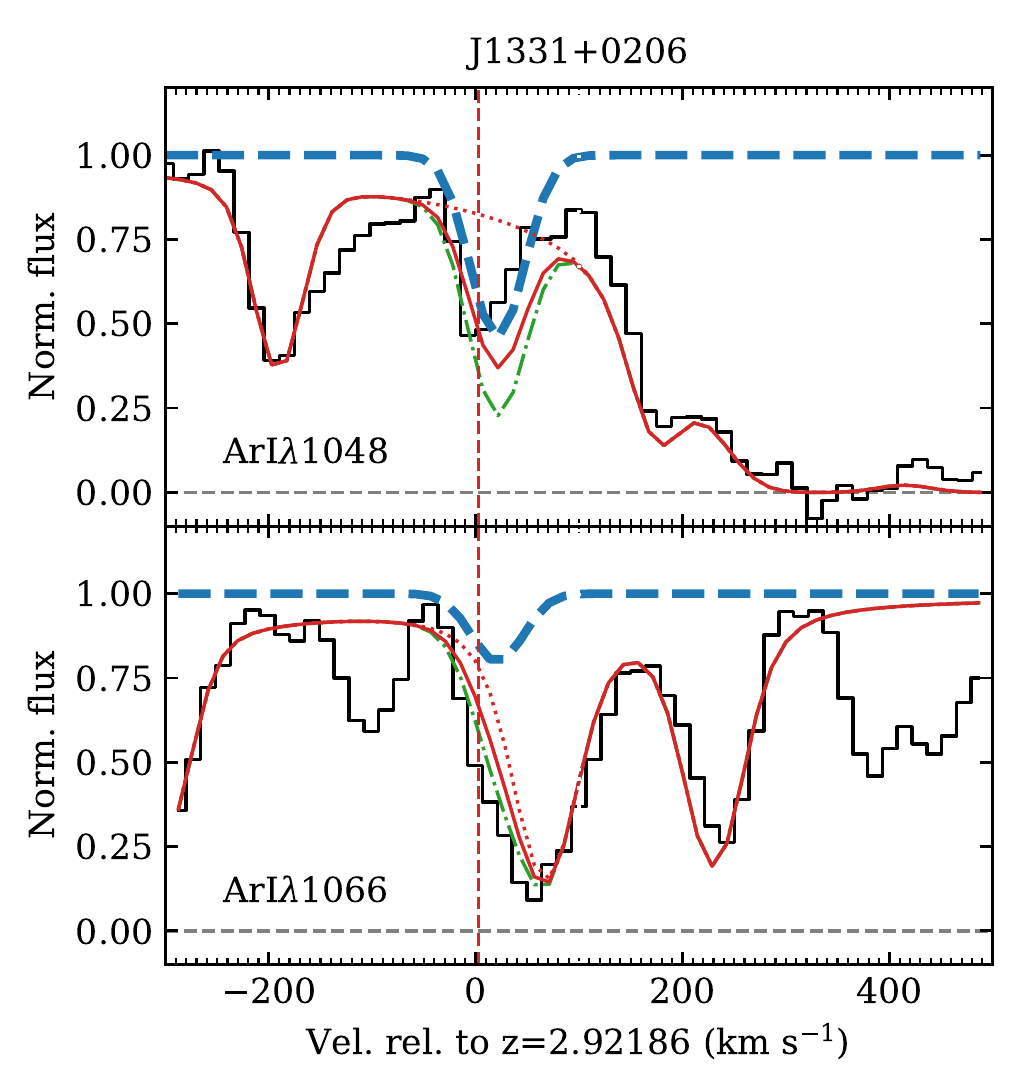}&
\includegraphics[trim={0.0cm 18.8cm 10.5cm 0.cm}, clip=,width=0.21\hsize]{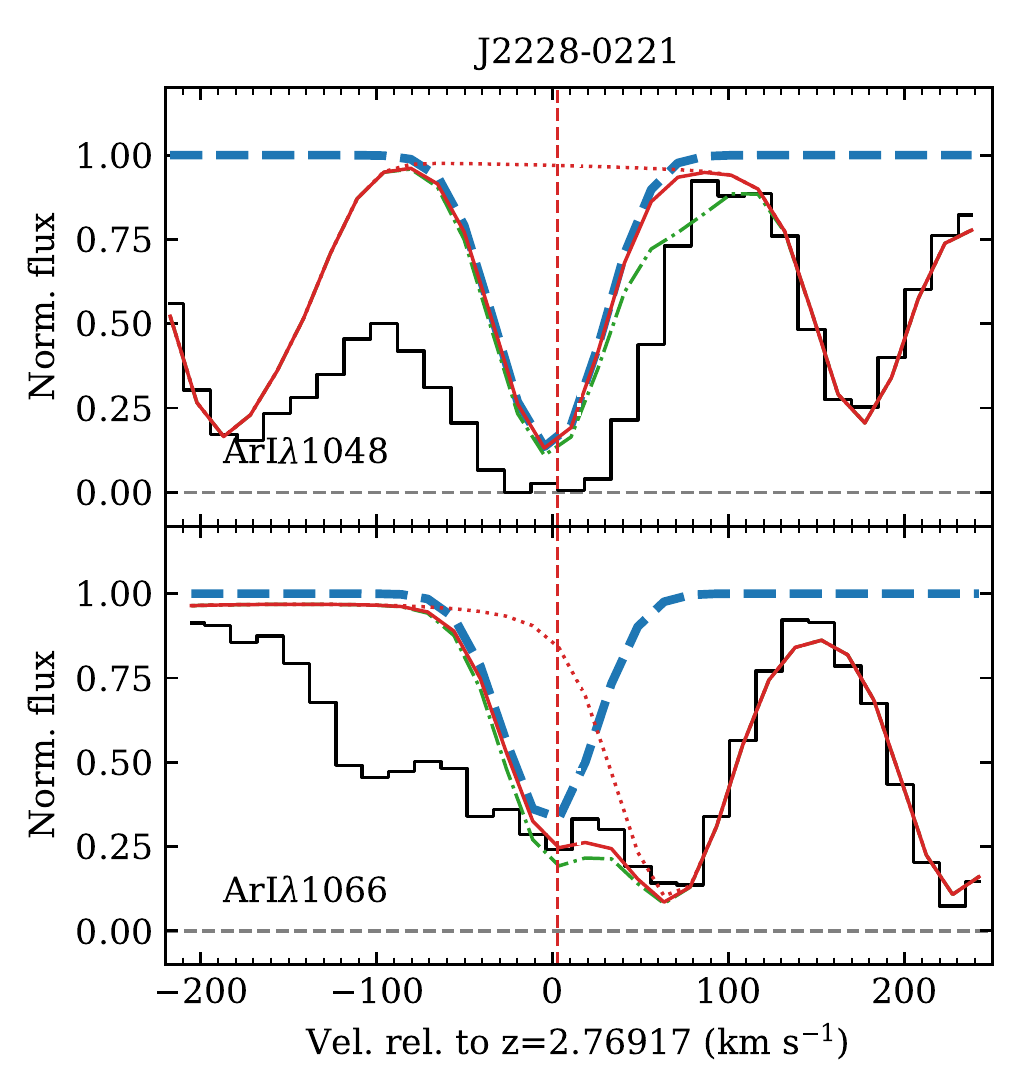}&
\includegraphics[trim={0.0cm 18.8cm 10.5cm 0.cm}, clip=,width=0.21\hsize]{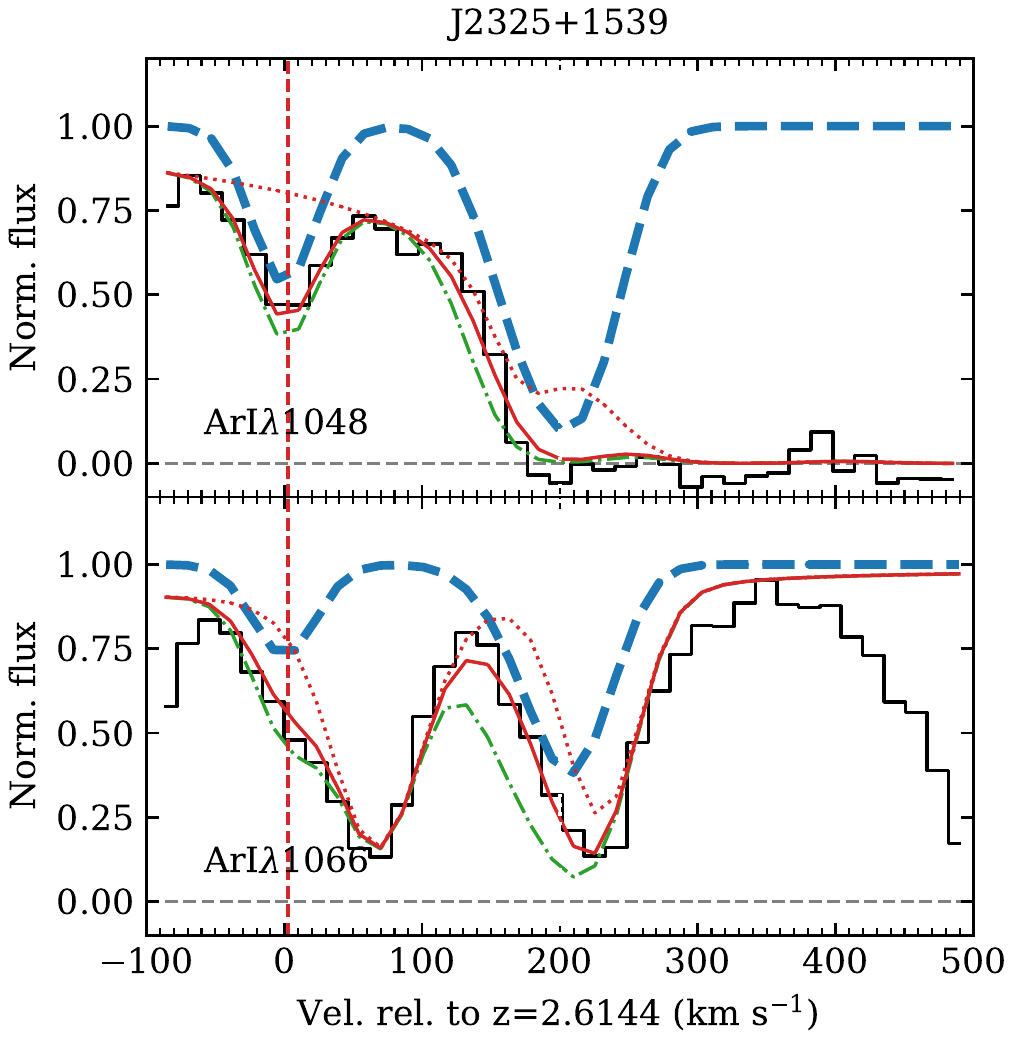}\\
\end{tabular}
    \caption{Neutral argon absorption lines. Dashed blue profile, dotted red and solid red represents the \ArI\, H$_2$ and total absorption, respectively. The green dashed profile represents the expected total absorption obtained by scaling the \SII\ profile assuming all argon in neutral form and Solar [Ar/S] ratio. The vertical dotted line shows the location of the \HH\ component.}
    \label{f:ArIfits}
        \addtolength{\tabcolsep}{+8pt}
\end{figure*}

\clearpage
\section{Measurement of dust extinction \label{a:dust}}

\begin{figure}[!ht]
    \centering
    \includegraphics[width=\hsize]{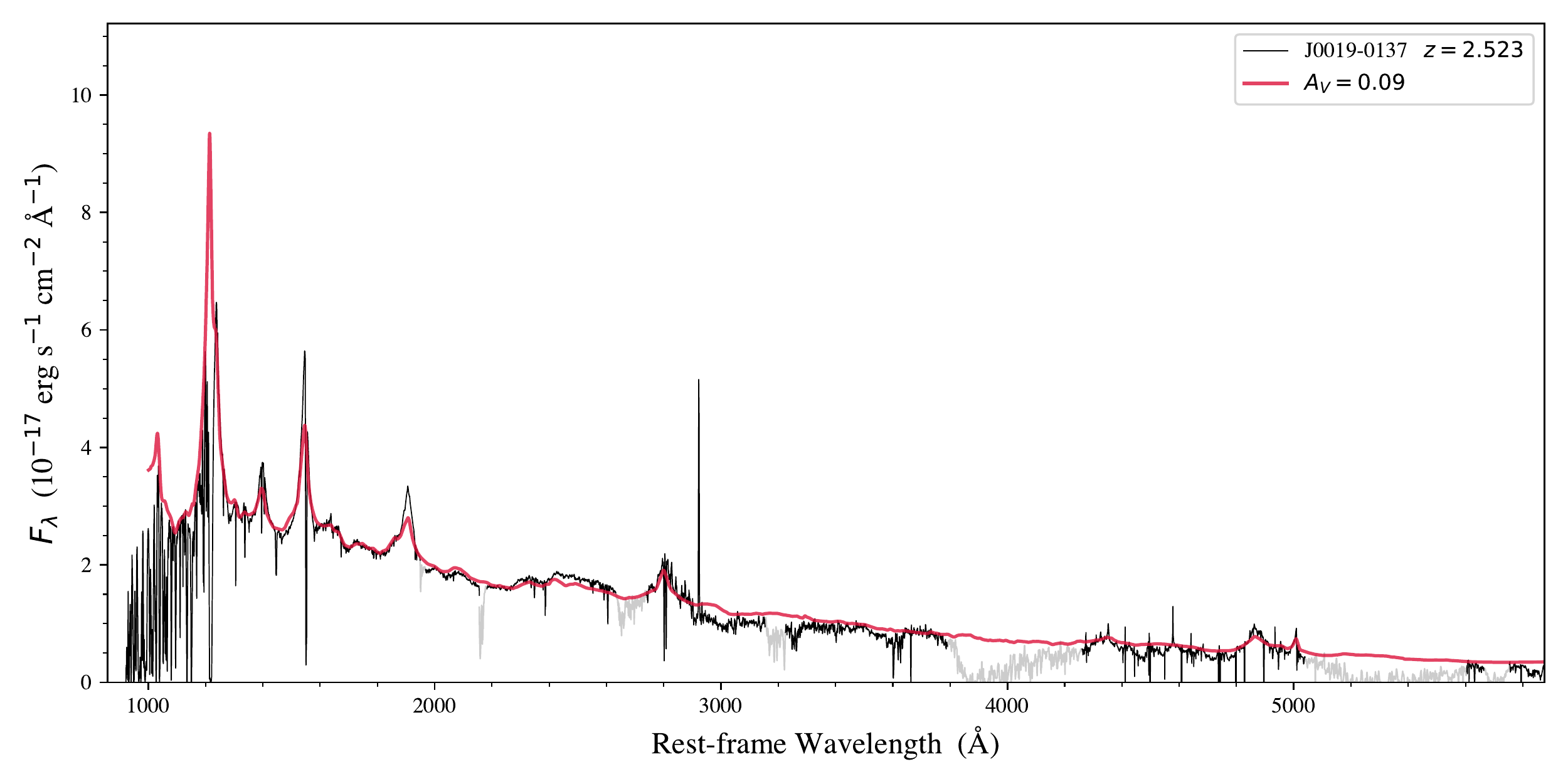}
    \caption{Measurement of the dust extinction through matching the X-shooter spectrum of \Jzzun\ (black) with a reddened quasar composite spectrum from \citet{Selsing2016} (red). Grey regions were masked out during the fit.}
    \label{f:AVJ0019}
\end{figure}

\begin{figure}[!ht]
    \centering
    \includegraphics[width=\hsize]{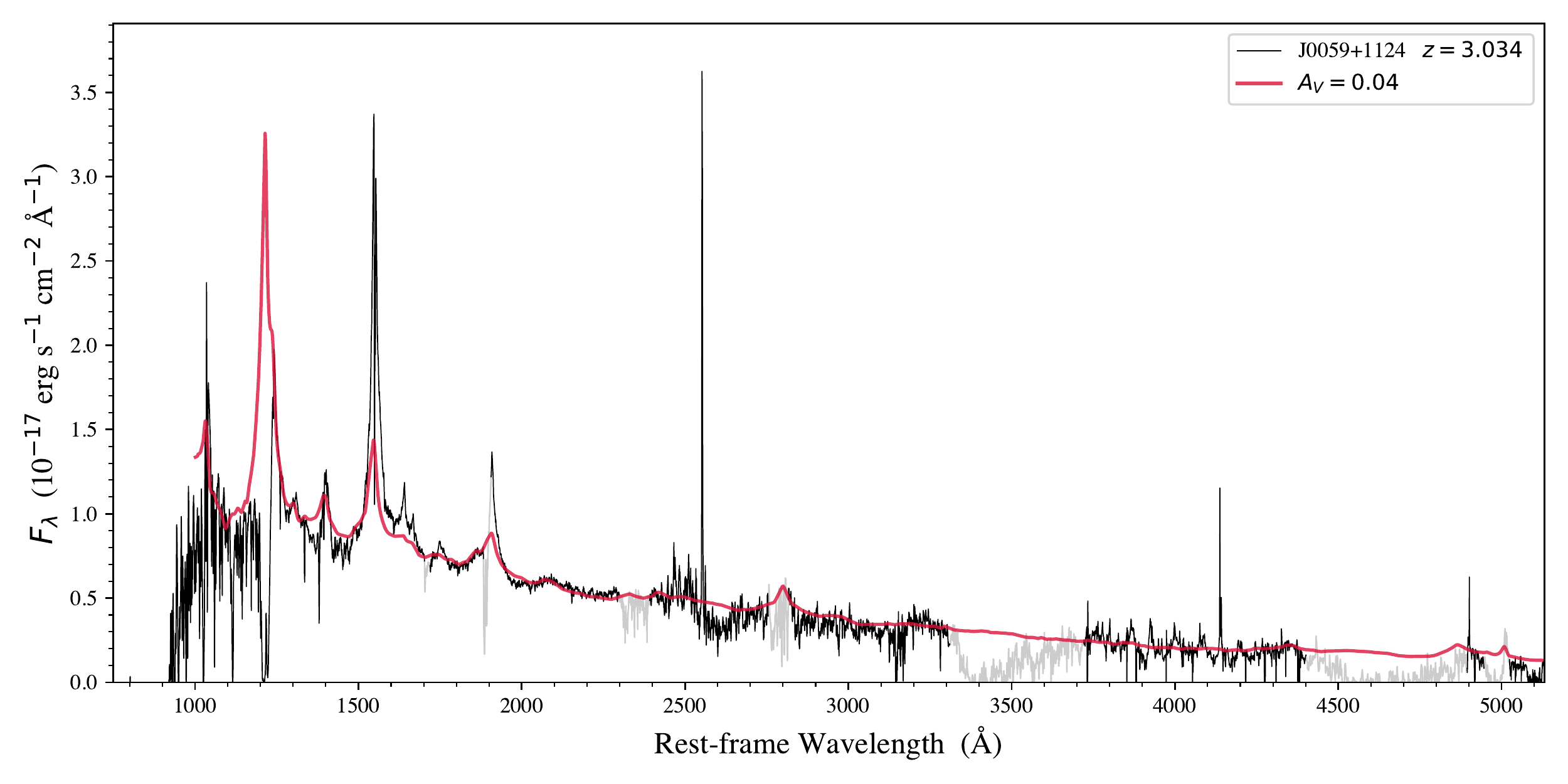}
    \caption{Same as Fig.~\ref{f:AVJ0019} for the quasar \Jzzcn.}
    \label{f:AVJ0059}
\end{figure}

\begin{figure}[!ht]
    \centering
    \includegraphics[width=\hsize]{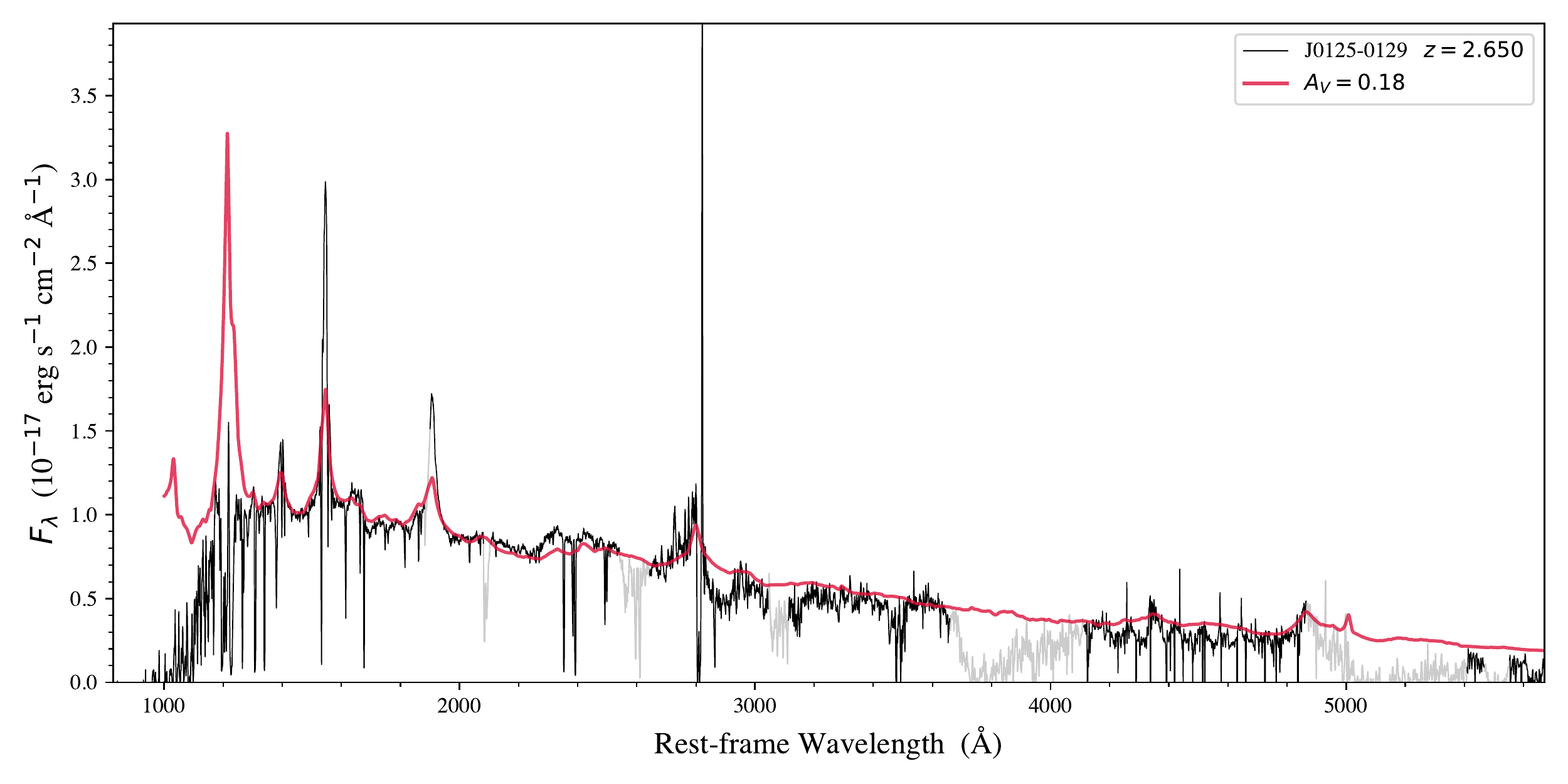}
    \caption{Same as Fig.~\ref{f:AVJ0019} for the quasar \Jzudc.}
    \label{f:AVJ0125}
\end{figure}

\begin{figure}[!ht]
    \centering
    \includegraphics[width=\hsize]{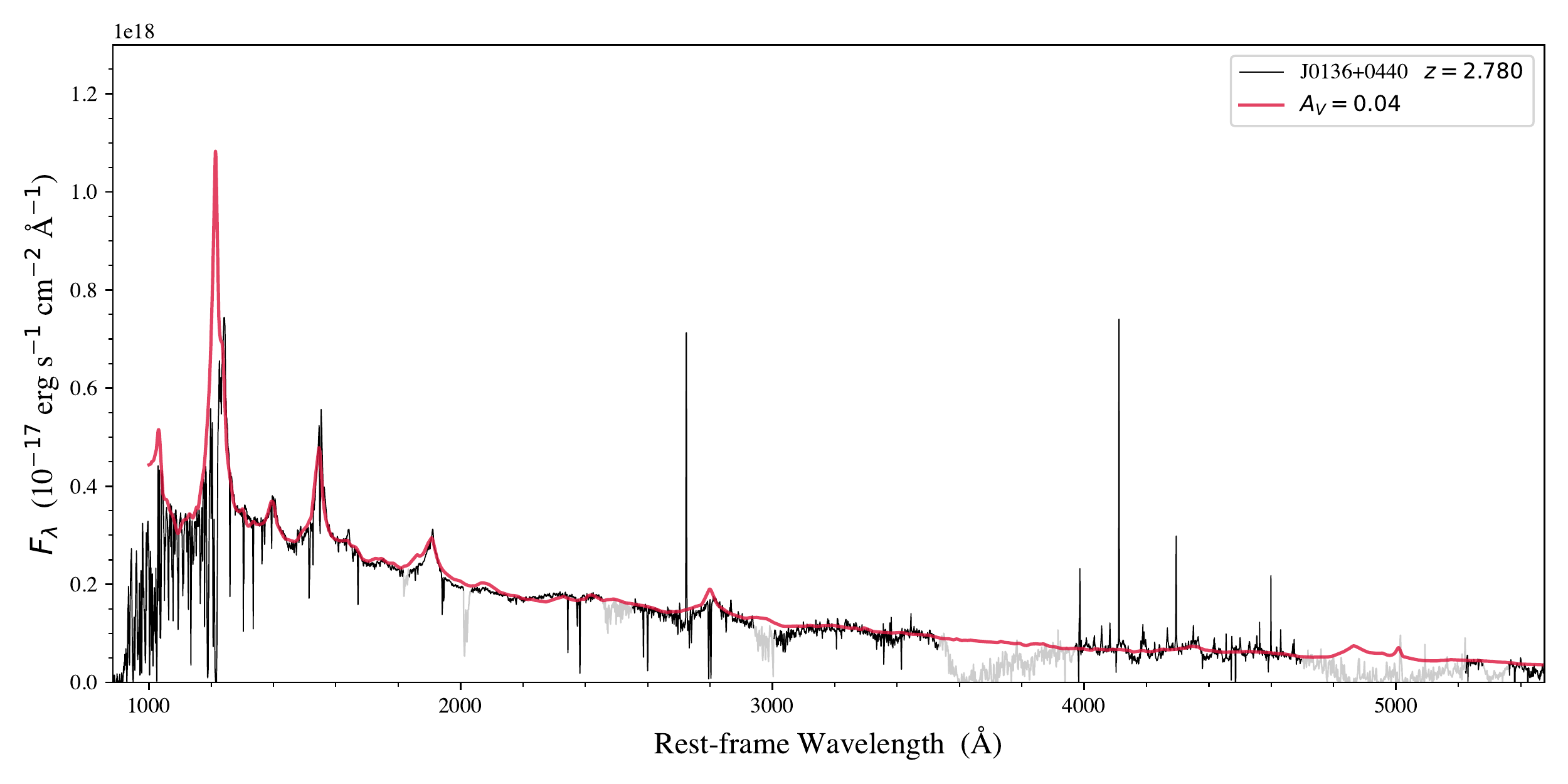}
    \caption{Same as Fig.~\ref{f:AVJ0019} for the quasar \Jzuts.}
    \label{f:AVJ0136}
\end{figure}

\begin{figure}[!ht]
    \centering
    \includegraphics[width=\hsize]{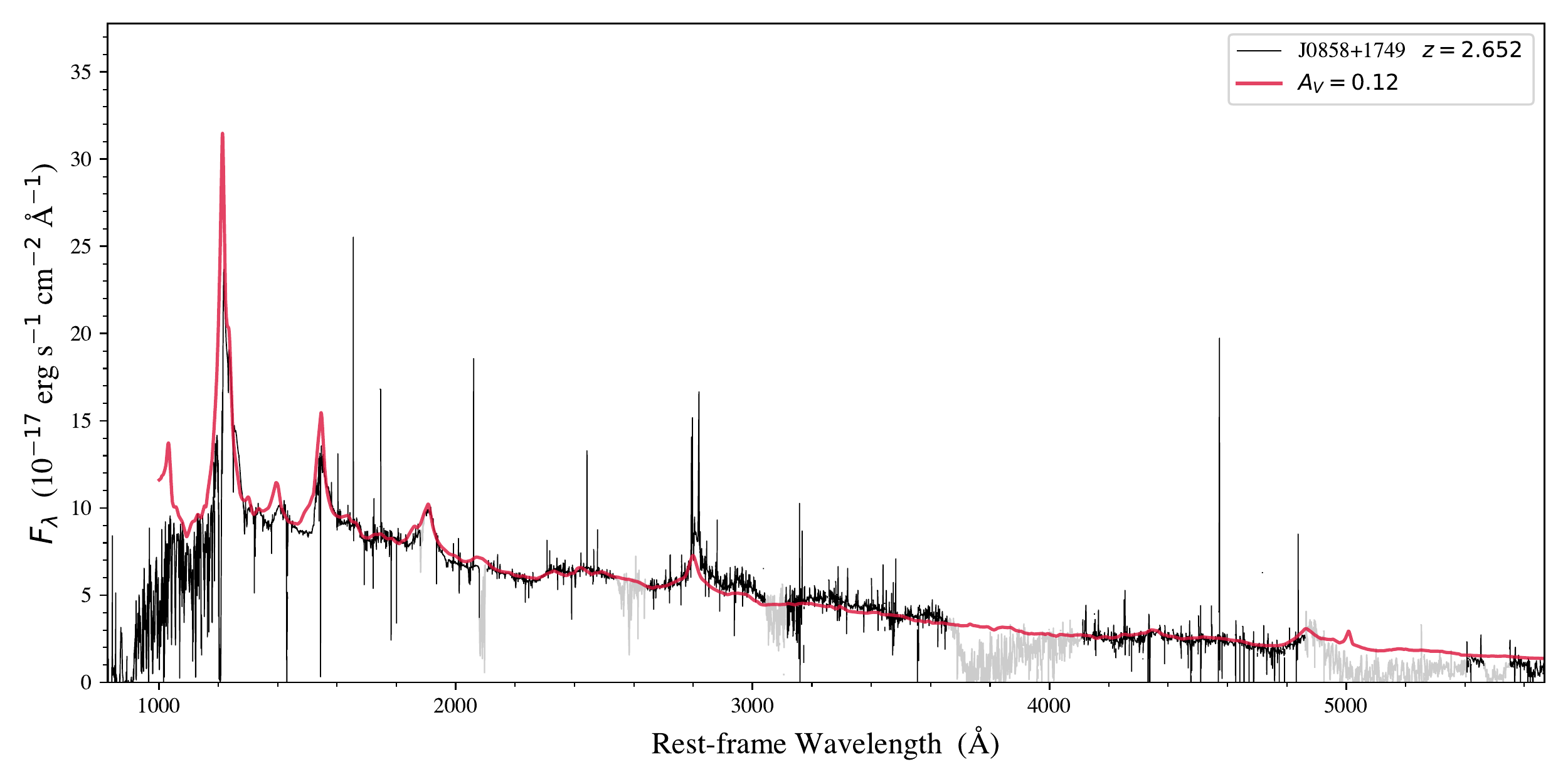}
    \caption{Same as Fig.~\ref{f:AVJ0019} for the quasar \Jzhch.}
    \label{f:AVJ0858}
\end{figure}

\begin{figure}[!ht]
    \centering
    \includegraphics[width=\hsize]{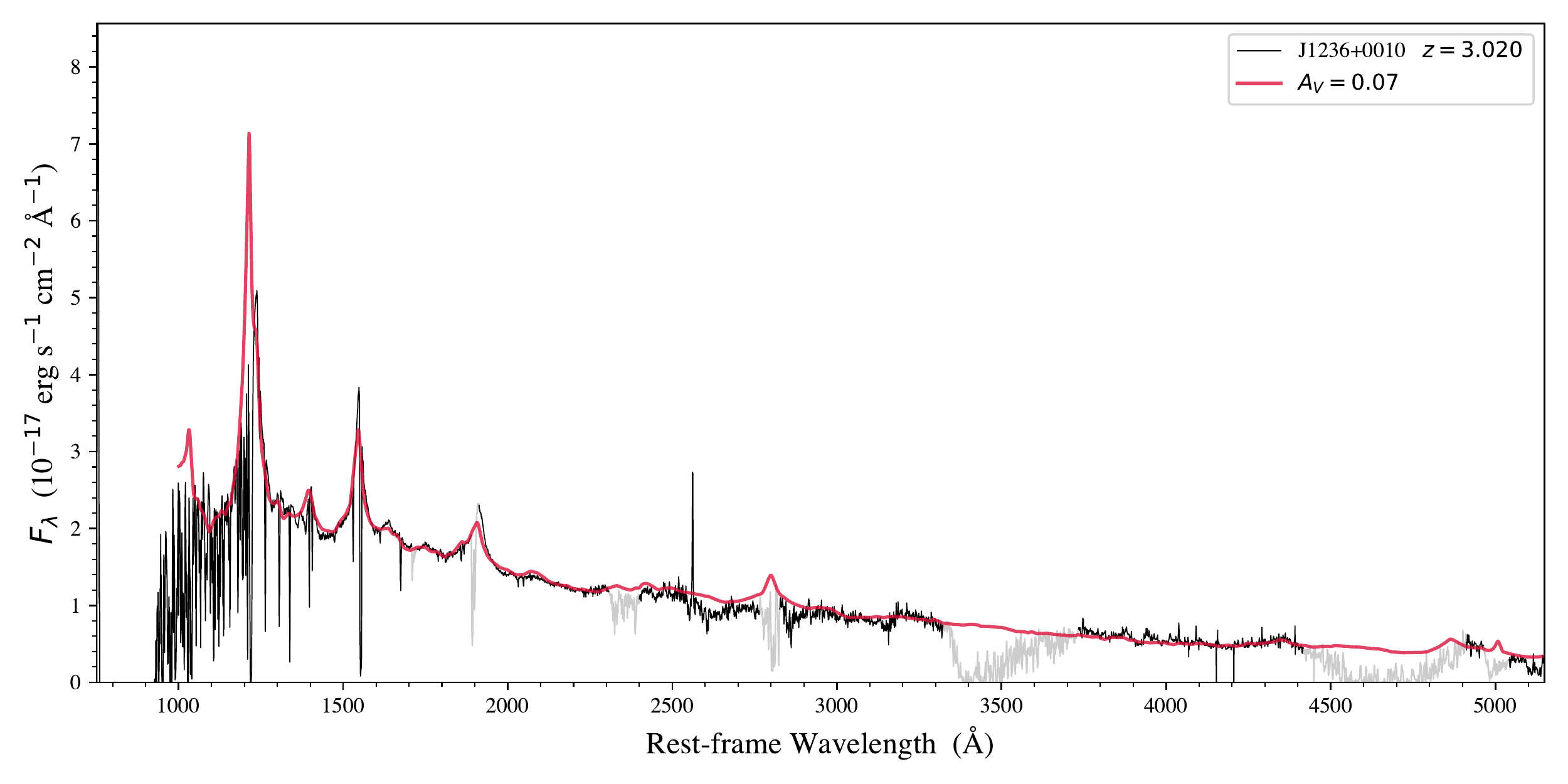}
    \caption{Same as Fig.~\ref{f:AVJ0019} for the quasar \Judts.}
    \label{f:AVJ1236}
\end{figure}

\begin{figure}[!ht]
    \centering
    \includegraphics[width=\hsize]{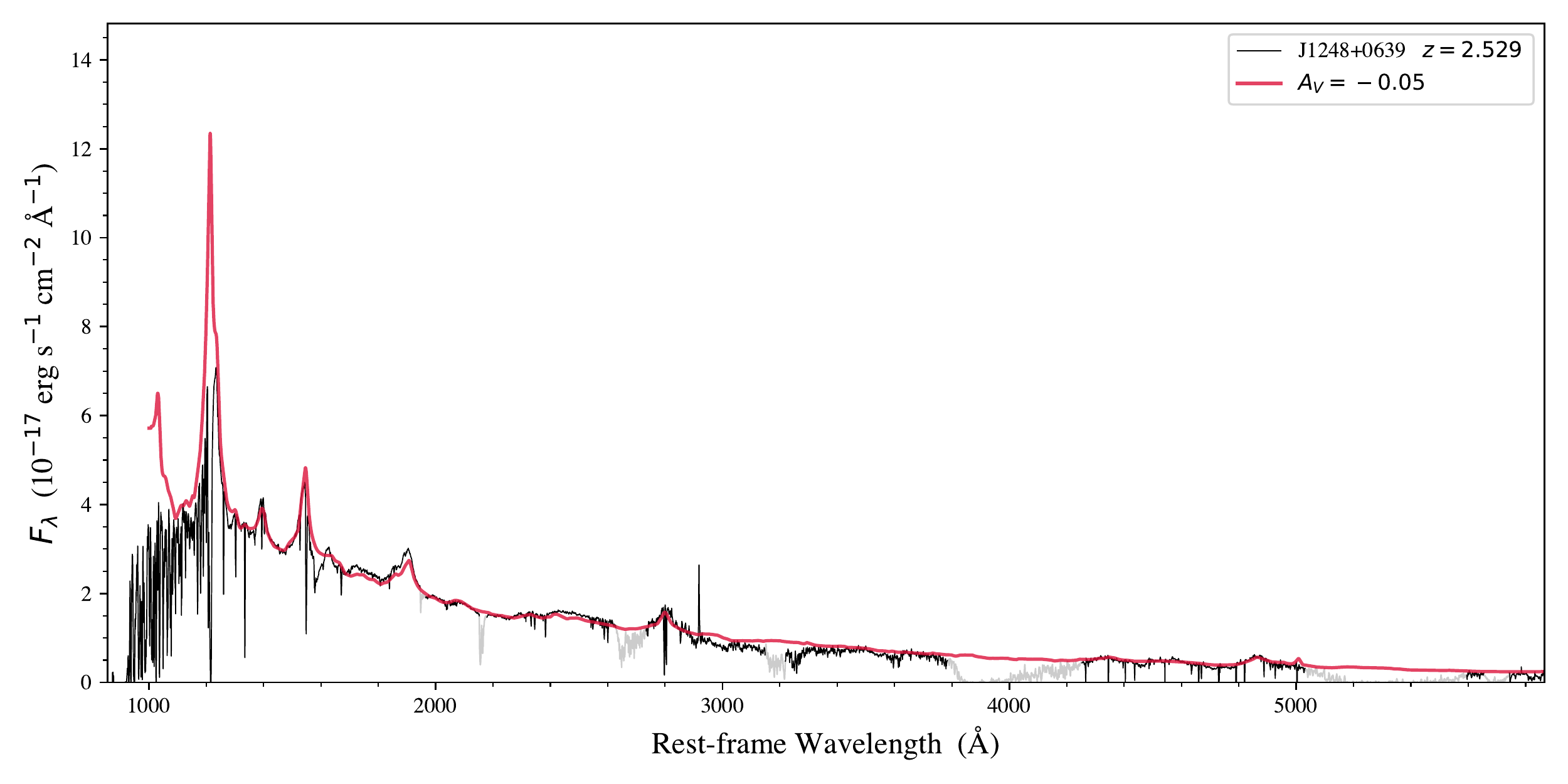}
    \caption{Same as Fig.~\ref{f:AVJ0019} for the quasar \Judqh.}
    \label{f:AVJ1248}
\end{figure}

\begin{figure}[!ht]
    \centering
    \includegraphics[width=\hsize]{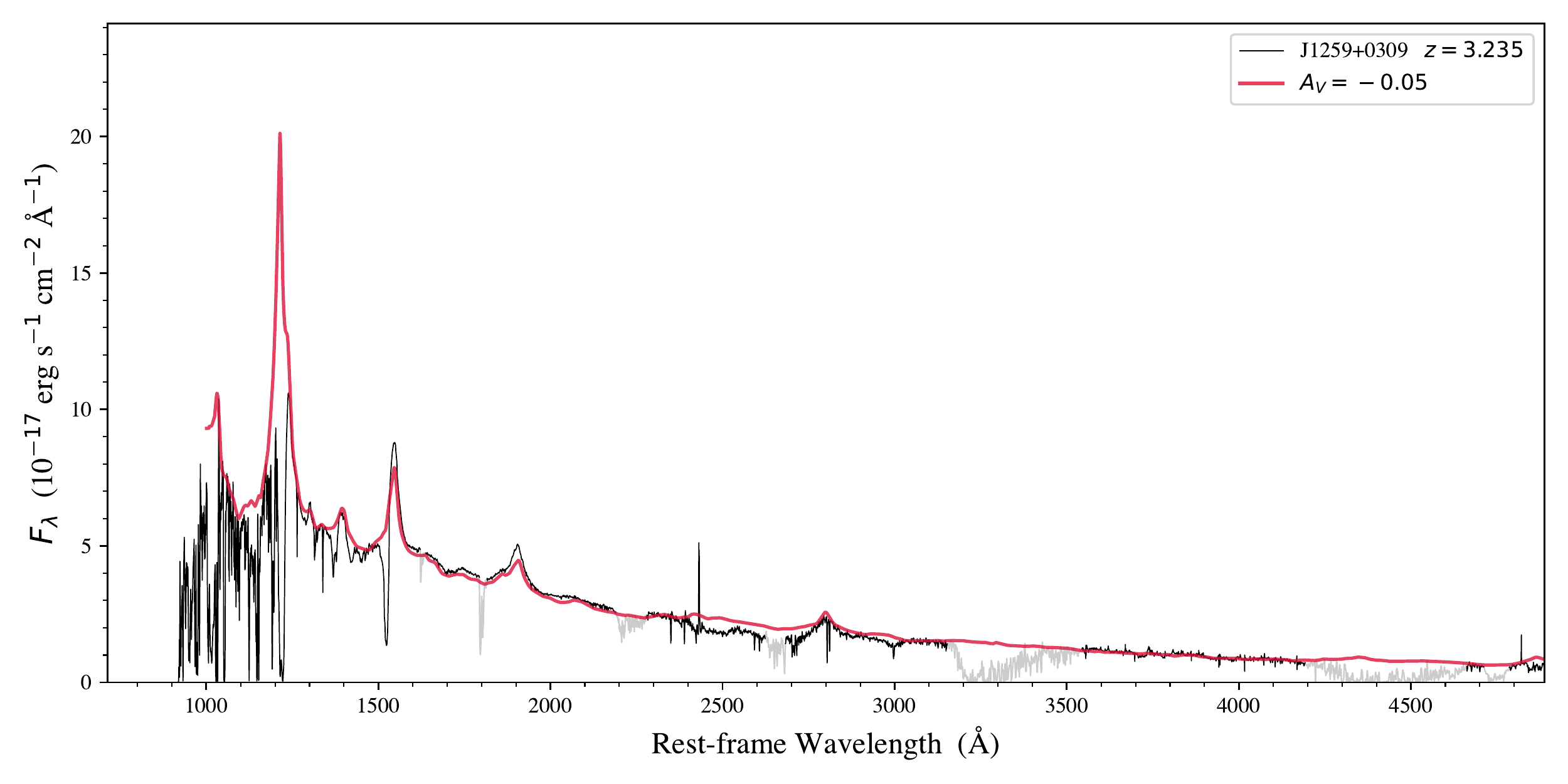}
    \caption{Same as Fig.~\ref{f:AVJ0019} for the quasar \Judcn.}
    \label{f:AVJ1259}
\end{figure}

\begin{figure}[!ht]
    \centering
    \includegraphics[width=\hsize]{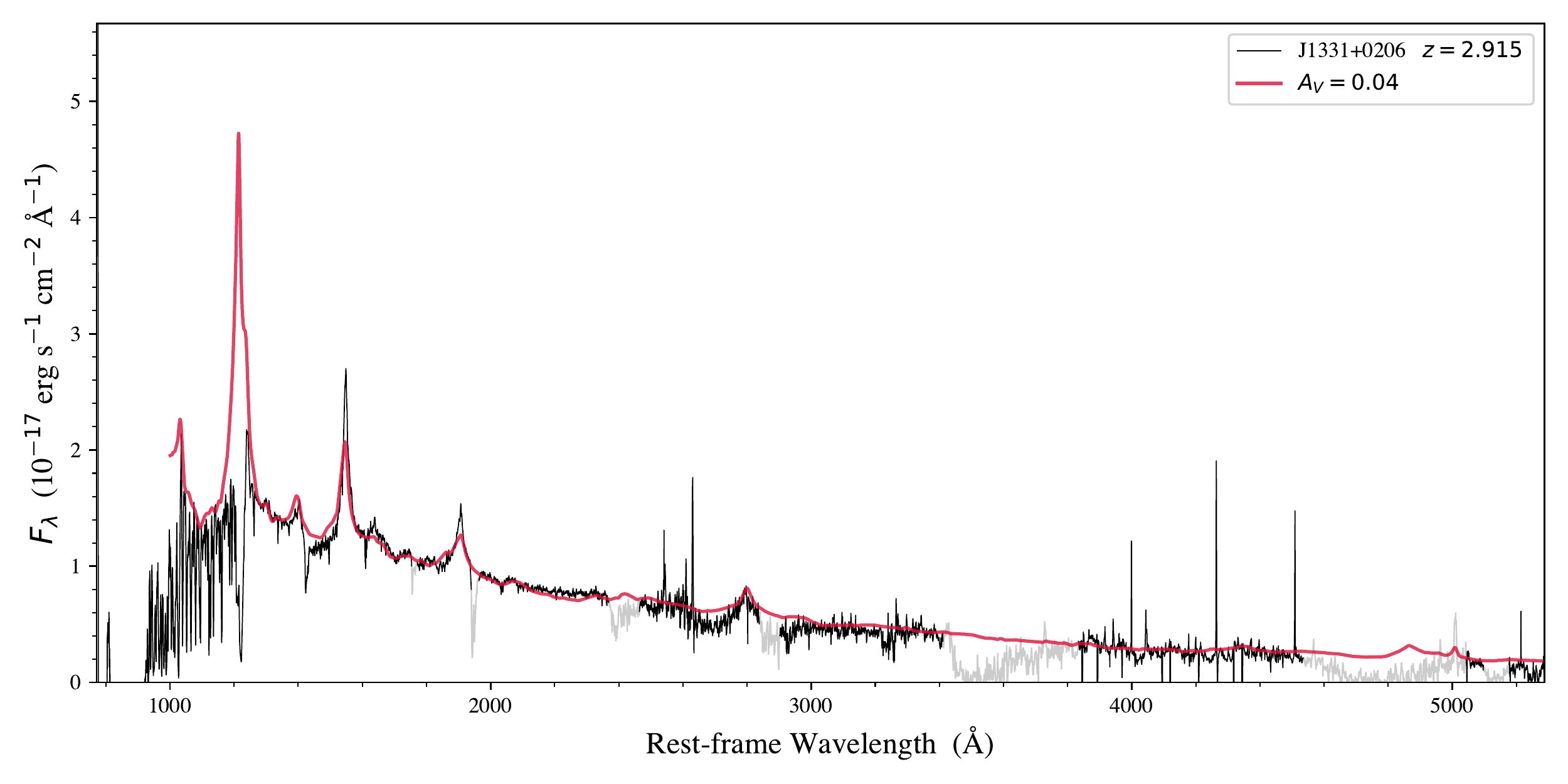}
    \caption{Same as Fig.~\ref{f:AVJ0019} for the quasar \Juttu.}
    \label{f:AVJ1331}
\end{figure}

\begin{figure}[!ht]
    \centering
    \includegraphics[width=\hsize]{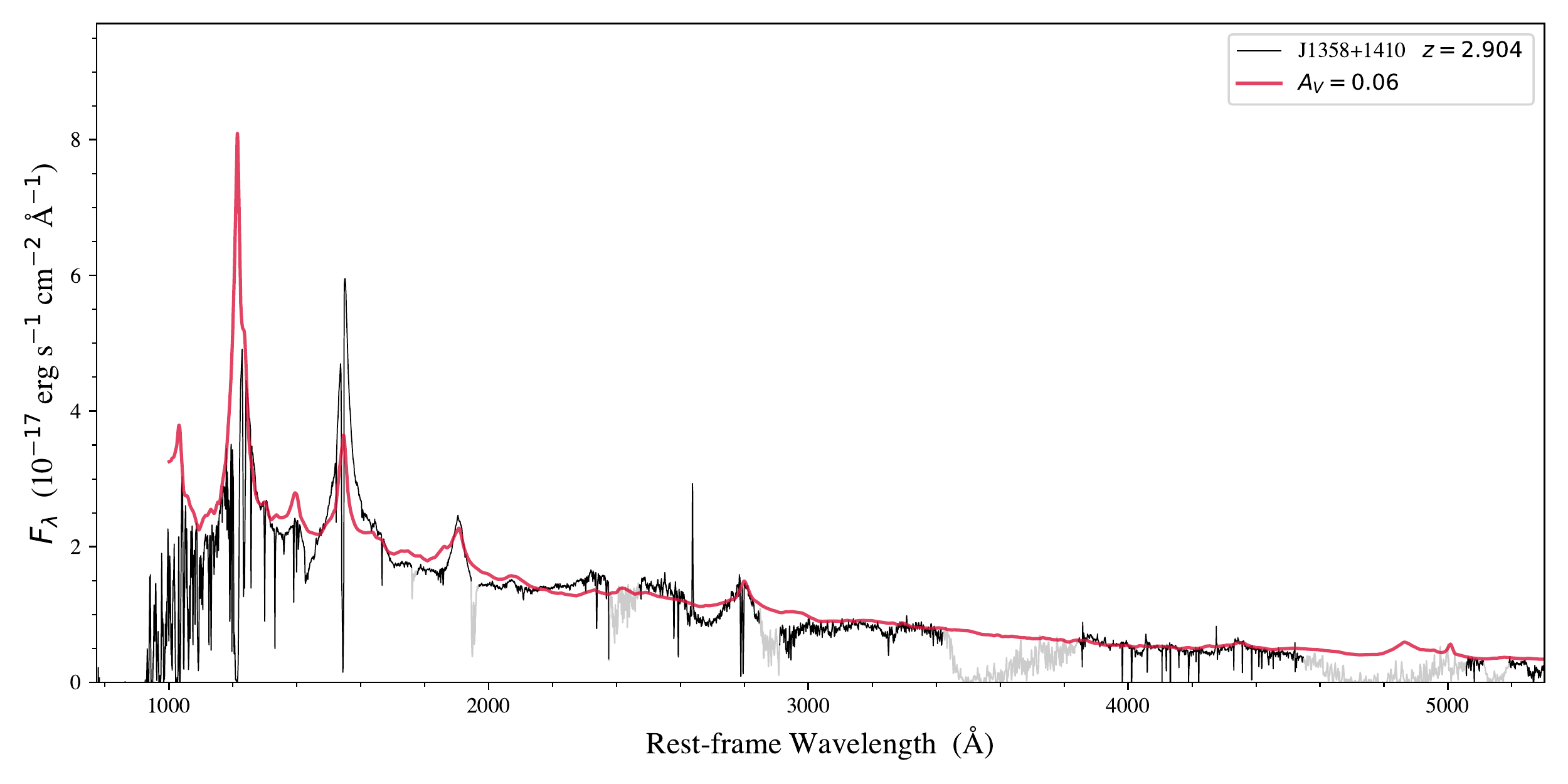}
    \caption{Same as Fig.~\ref{f:AVJ0019} for the quasar \Jutch.}
    \label{f:AVJ1358}
\end{figure}

\begin{figure}[!ht]
    \centering
    \includegraphics[width=\hsize]{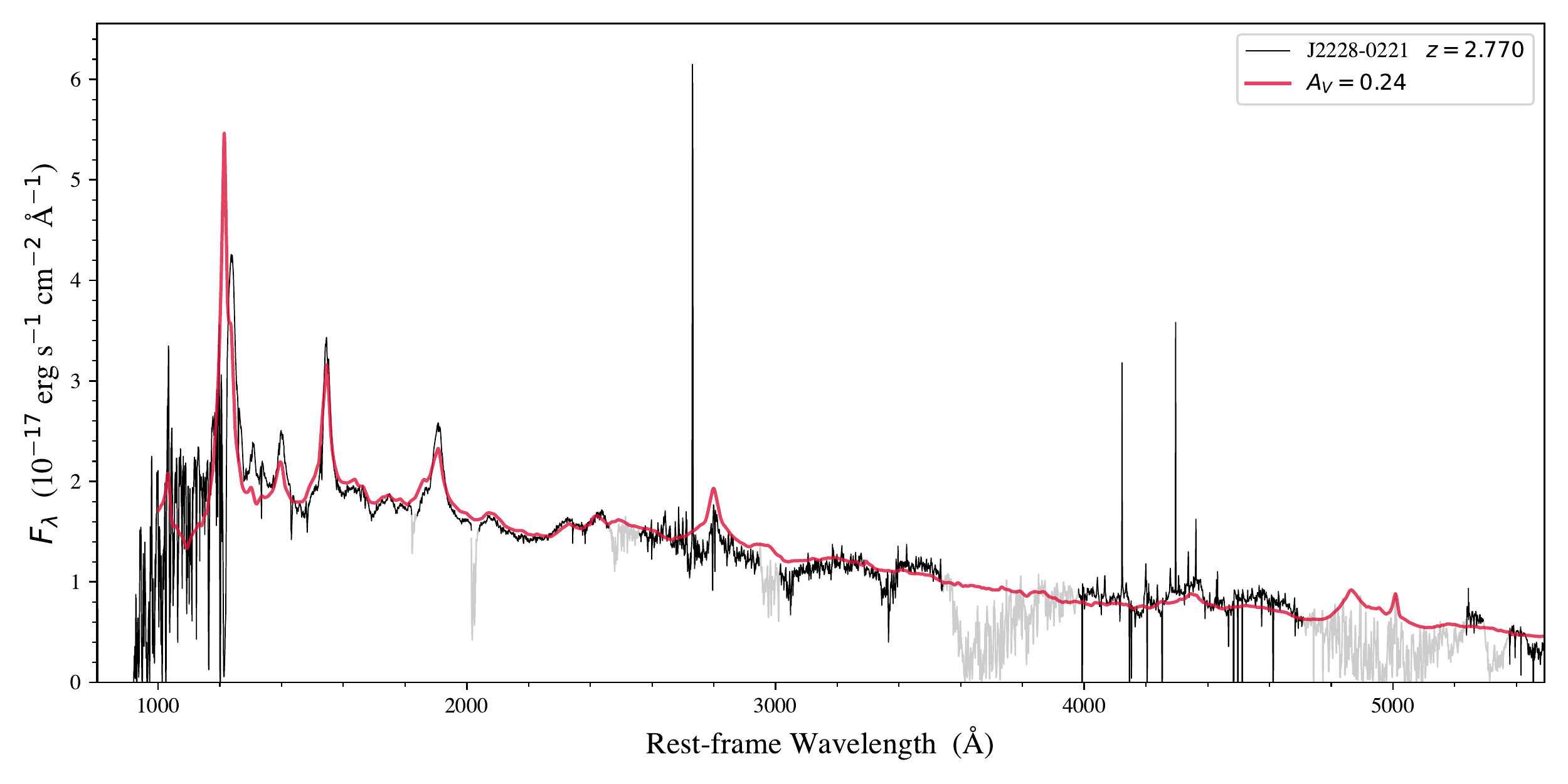}
    \caption{Same as Fig.~\ref{f:AVJ0019} for the quasar \Jdddh.}
    \label{f:AVJ2228}
\end{figure}

\begin{figure}[!ht]
    \centering
    \includegraphics[width=\hsize]{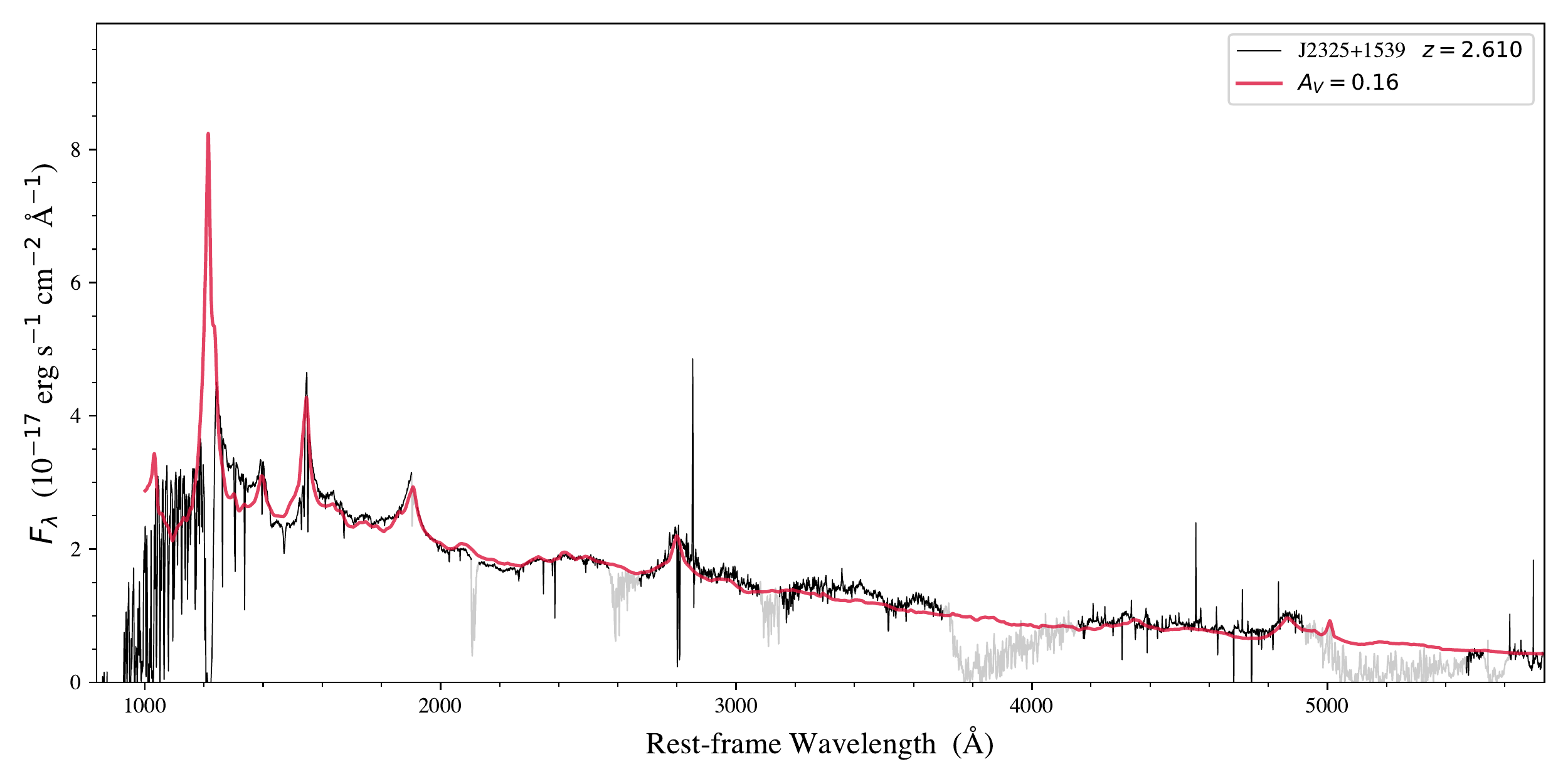}
    \caption{Same as Fig.~\ref{f:AVJ0019} for the quasar \Jdtdc.}
    \label{f:AVJ2325}
\end{figure}

\clearpage
\section{$\Delta v_{90}$ measurements \label{a:dv90}}
%\PN{RC to add figs here}

\addtolength{\tabcolsep}{-3pt}
\begin{figure*}[!t]
    \centering
    \begin{tabular}{ccc}
\includegraphics[trim={0cm 0cm 0.5cm 0cm},width=0.32\hsize]{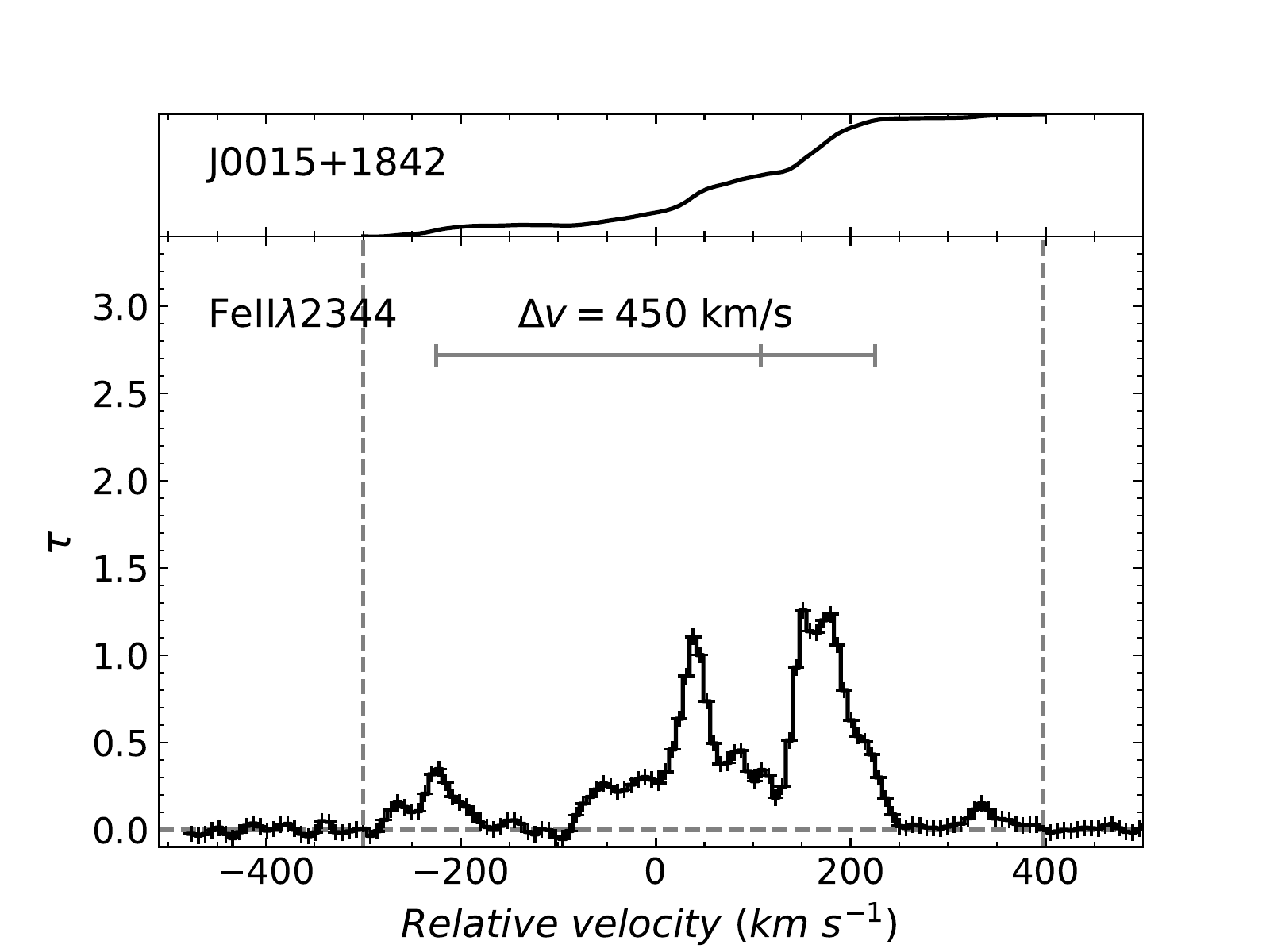} &
\includegraphics[trim={0cm 0cm 0.5cm 0cm},width=0.32\hsize]{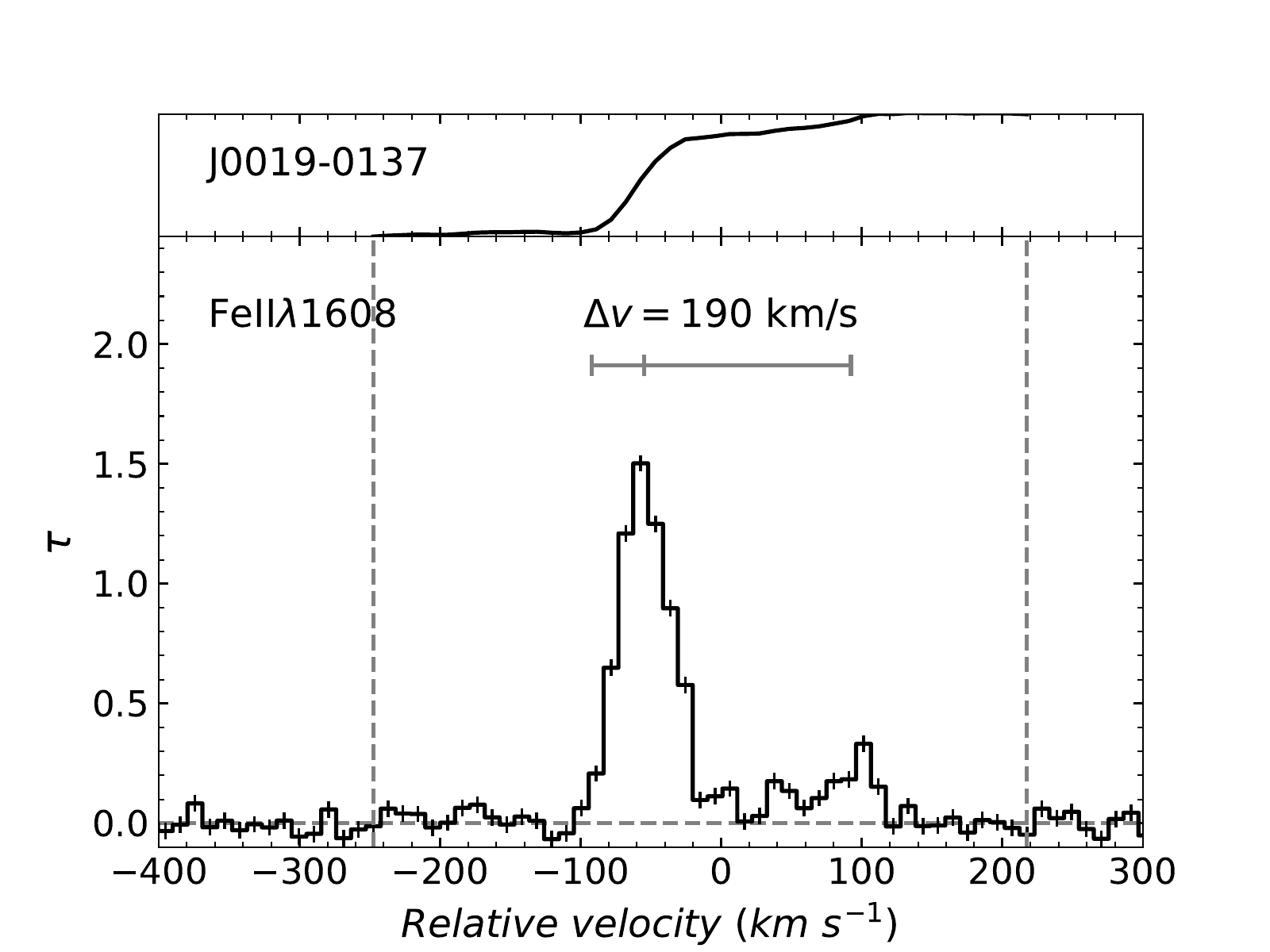} &
\includegraphics[trim={0cm 0cm 0.5cm 0cm},width=0.32\hsize]{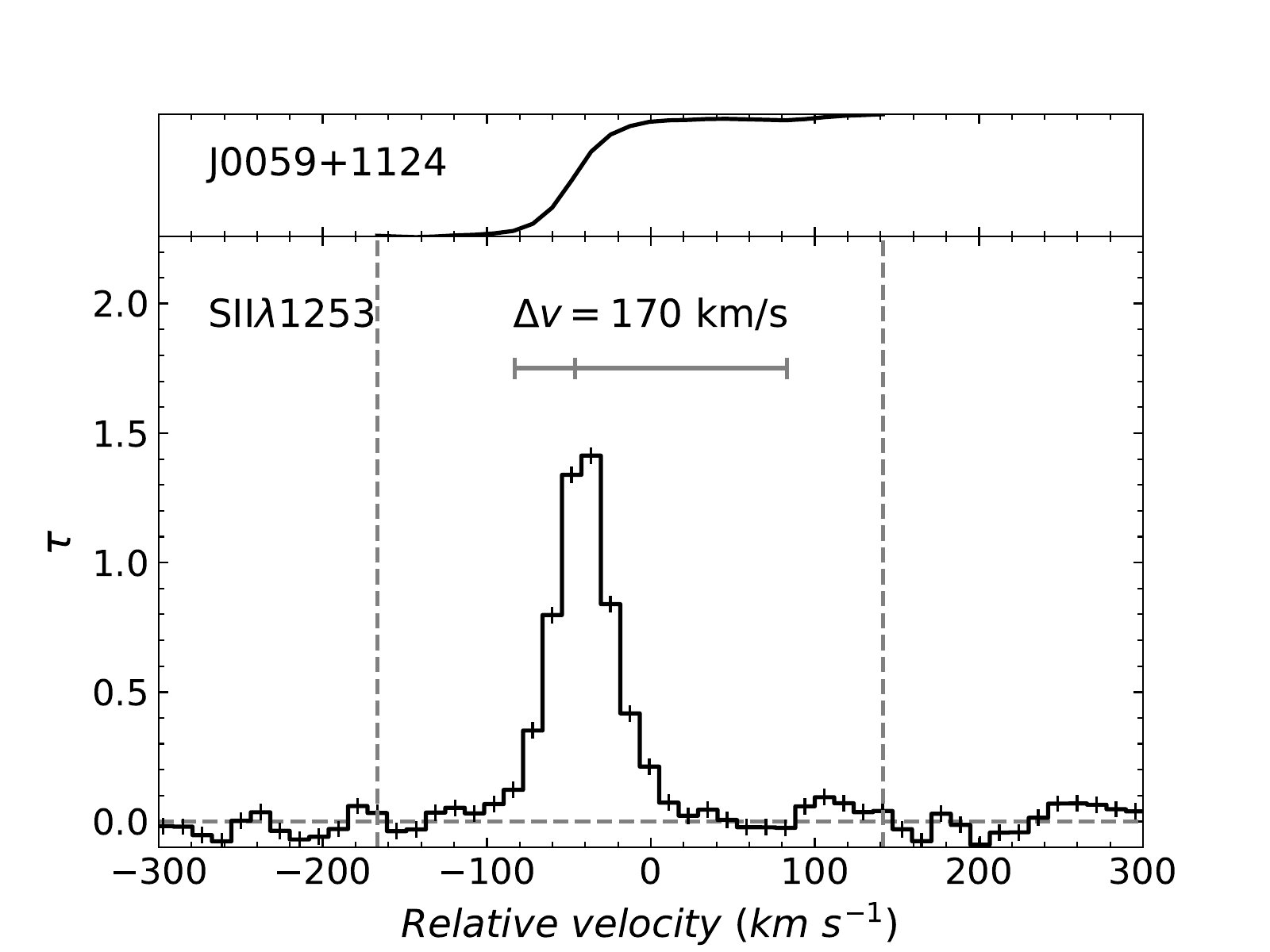}\\
\includegraphics[trim={0cm 0cm 0.5cm 0cm},width=0.32\hsize]{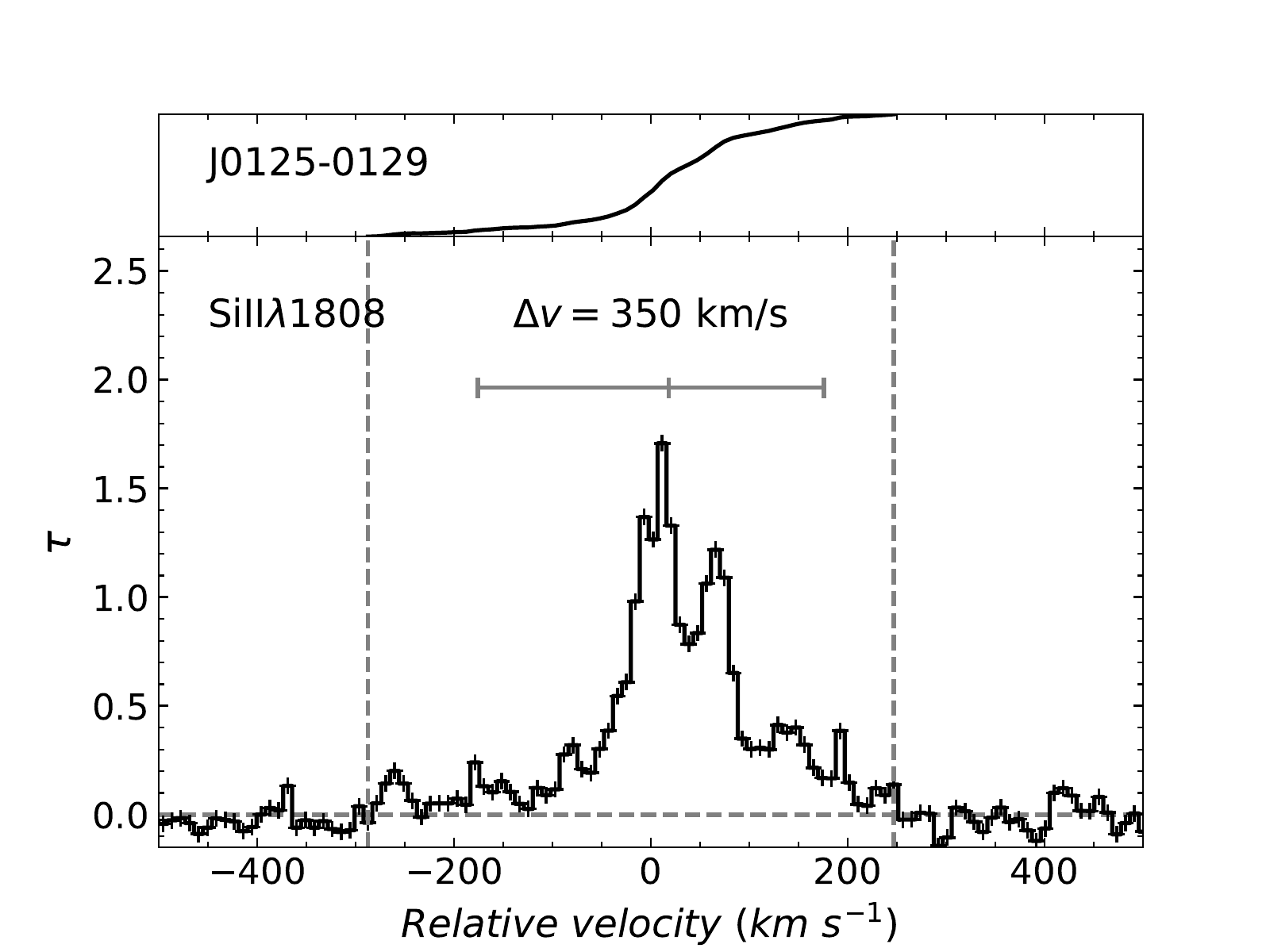} &
\includegraphics[trim={0cm 0cm 0.5cm 0cm},width=0.32\hsize]{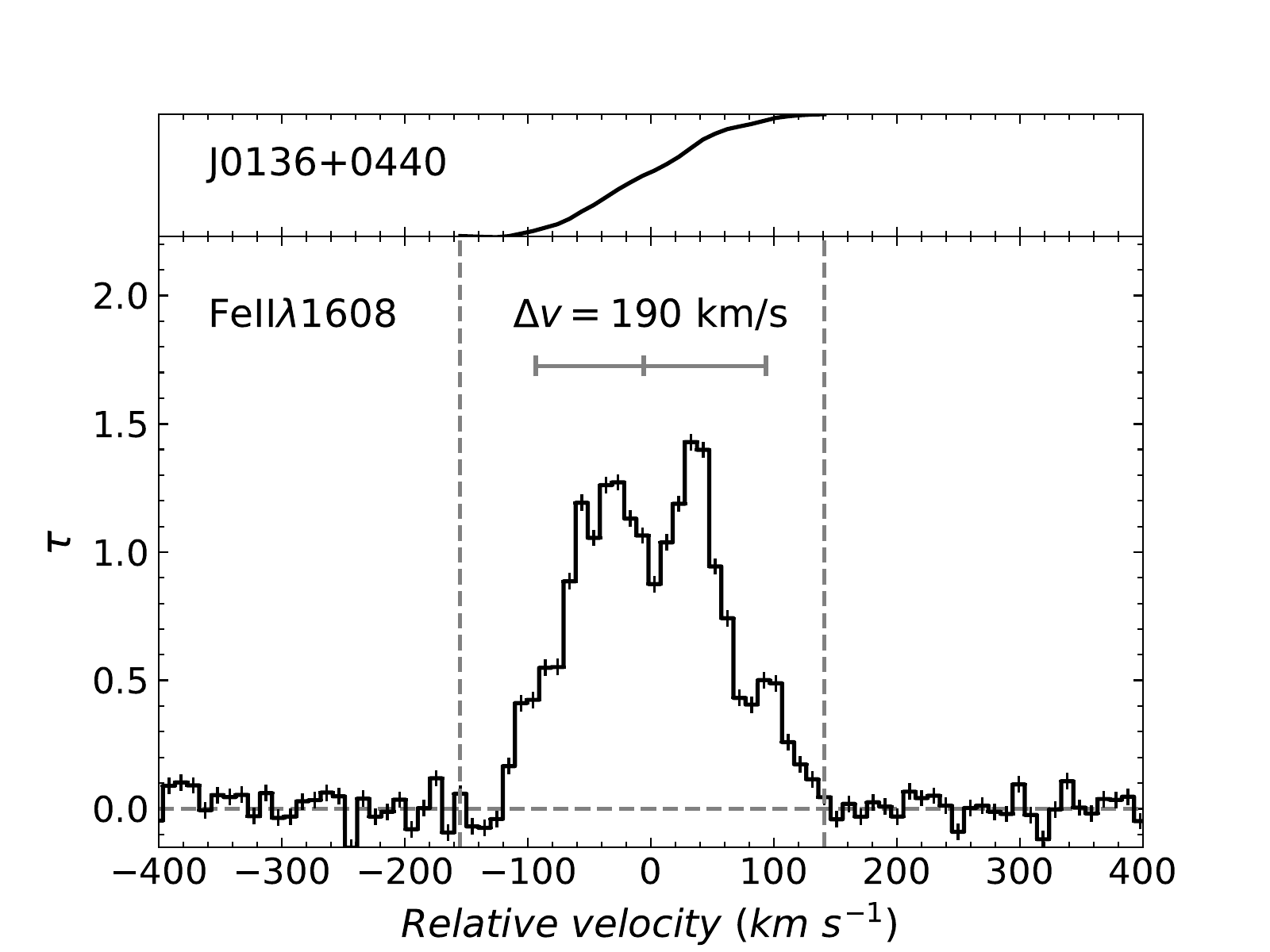} &
\includegraphics[trim={0cm 0cm 0.5cm 0cm},width=0.32\hsize]{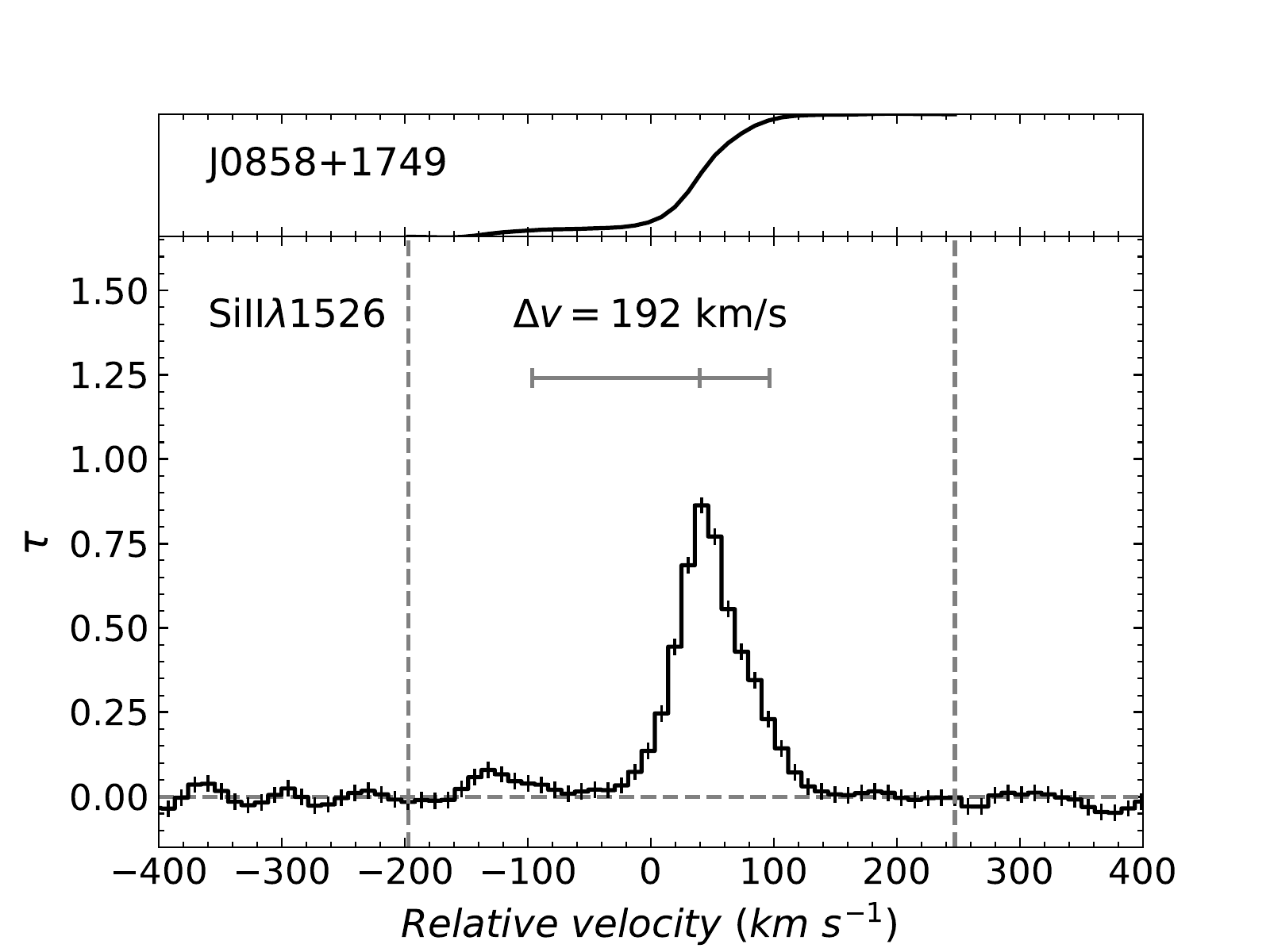}\\
\includegraphics[trim={0cm 0cm 0.5cm 0cm},width=0.32\hsize]{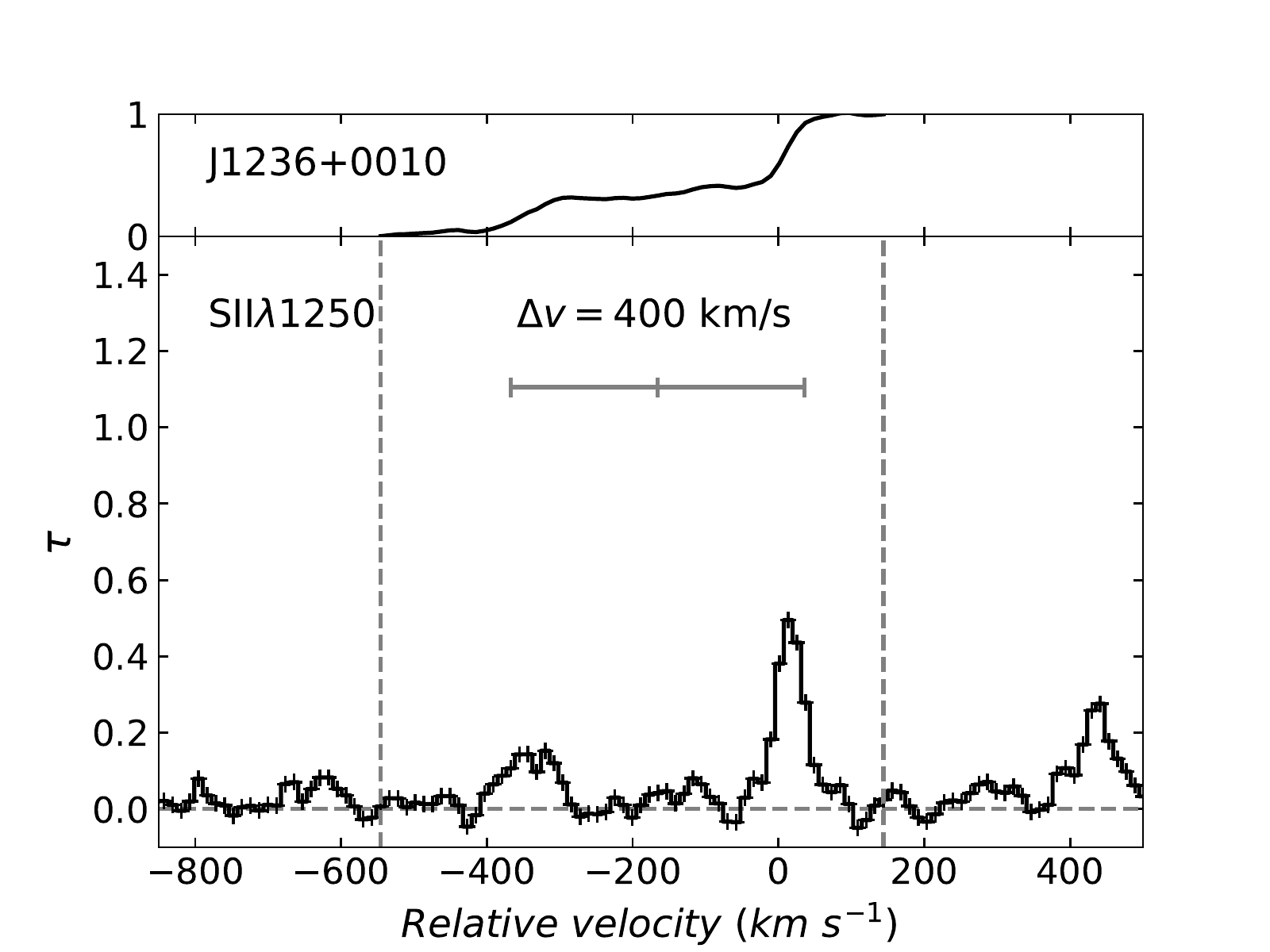} &
\includegraphics[trim={0cm 0cm 0.5cm 0cm},width=0.32\hsize]{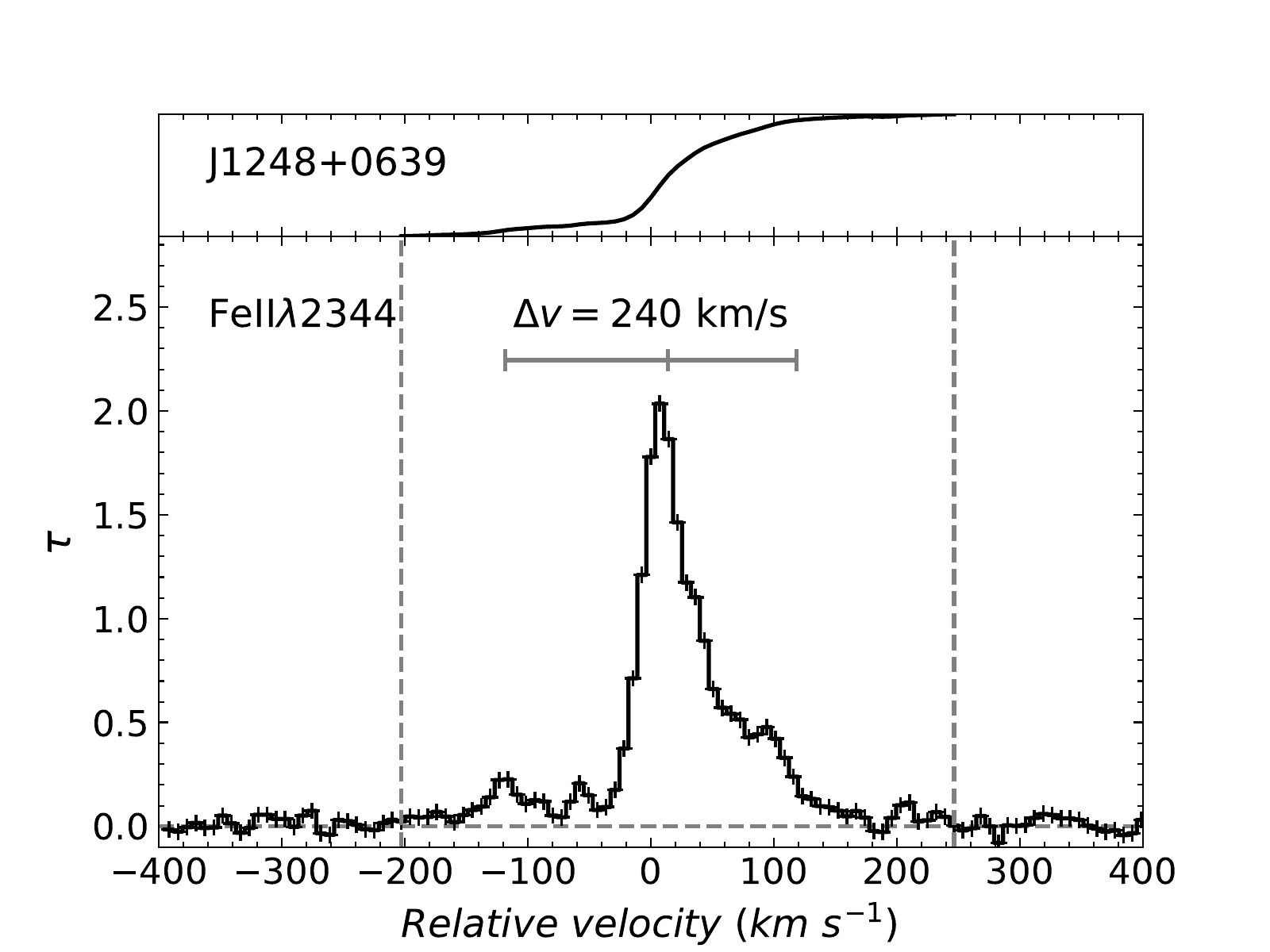} &
\includegraphics[trim={0cm 0cm 0.5cm 0cm},width=0.32\hsize]{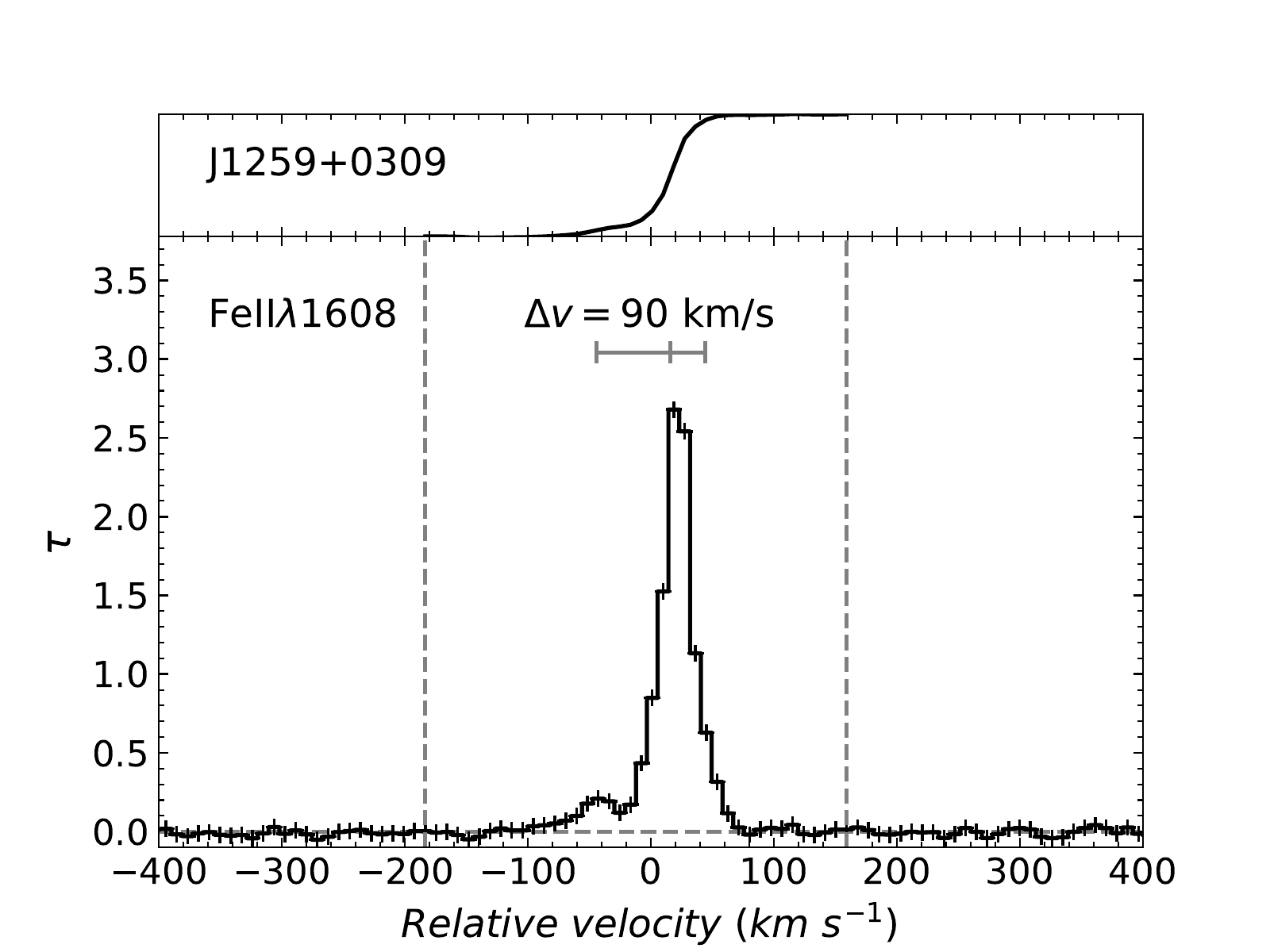} \\
\includegraphics[trim={0cm 0cm 0.5cm 0cm},width=0.32\hsize]{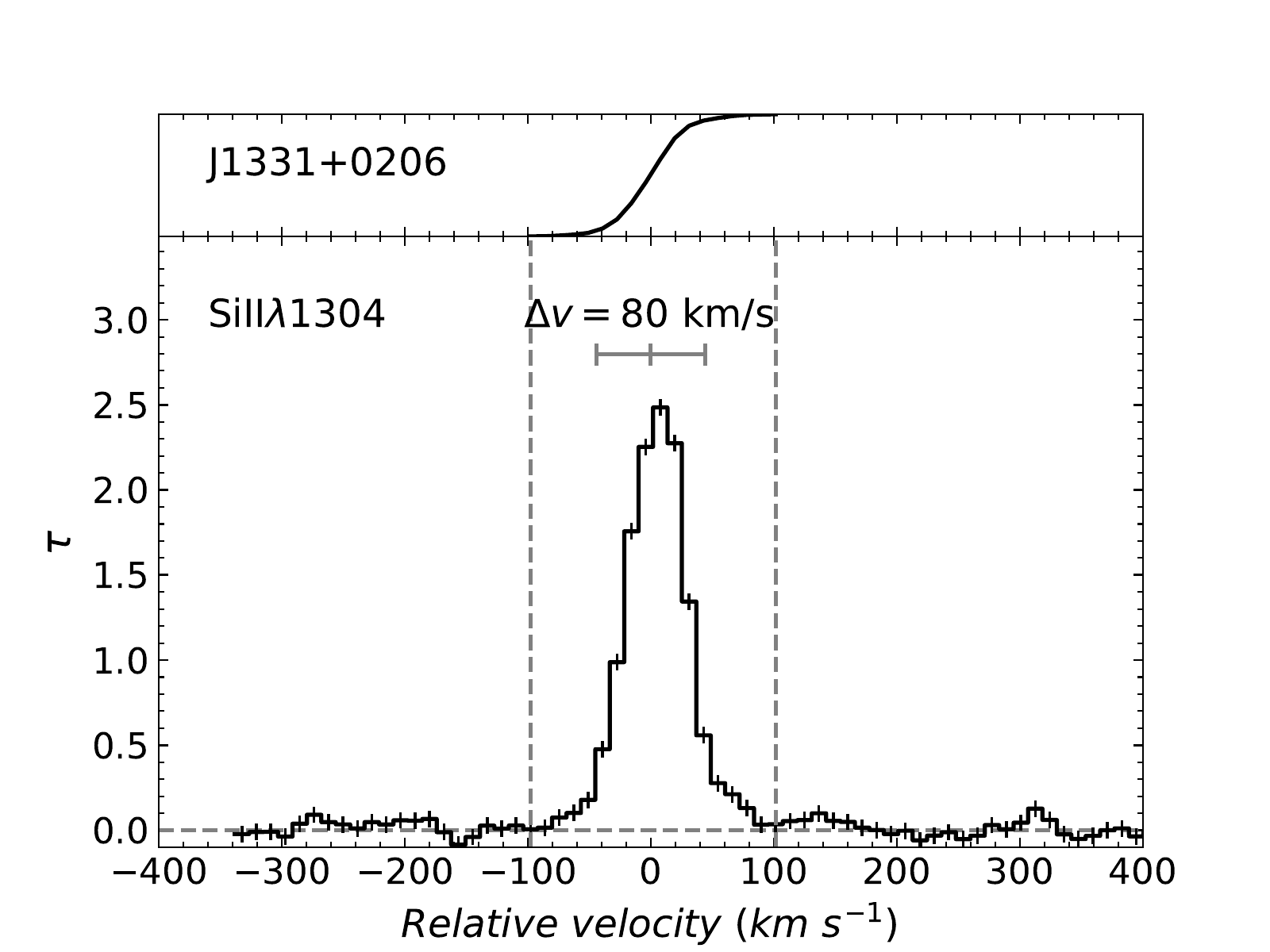} &
\includegraphics[trim={0cm 0cm 0.5cm 0cm},width=0.32\hsize]{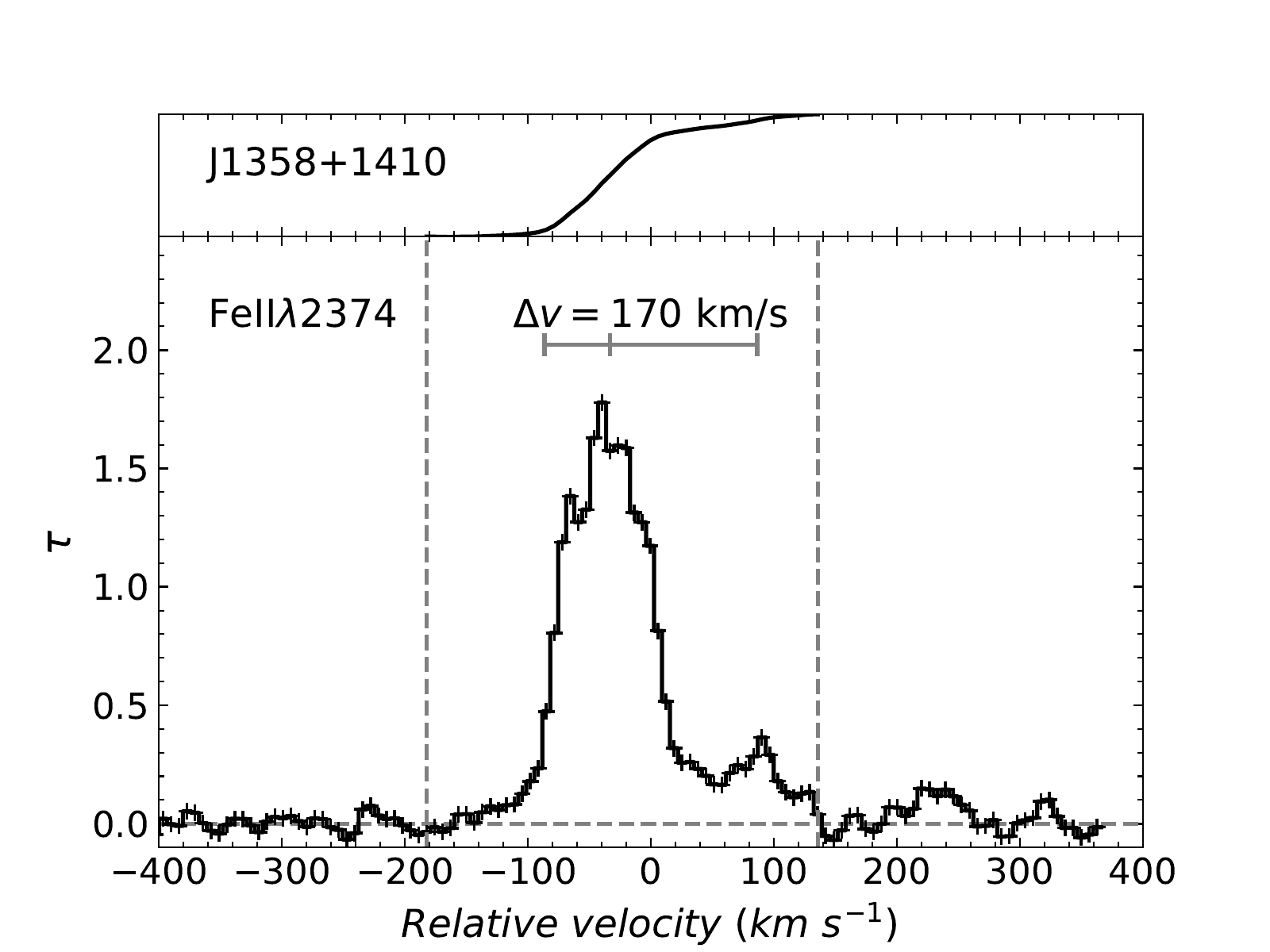} &
\includegraphics[trim={0cm 0cm 0.5cm 0cm},width=0.32\hsize]{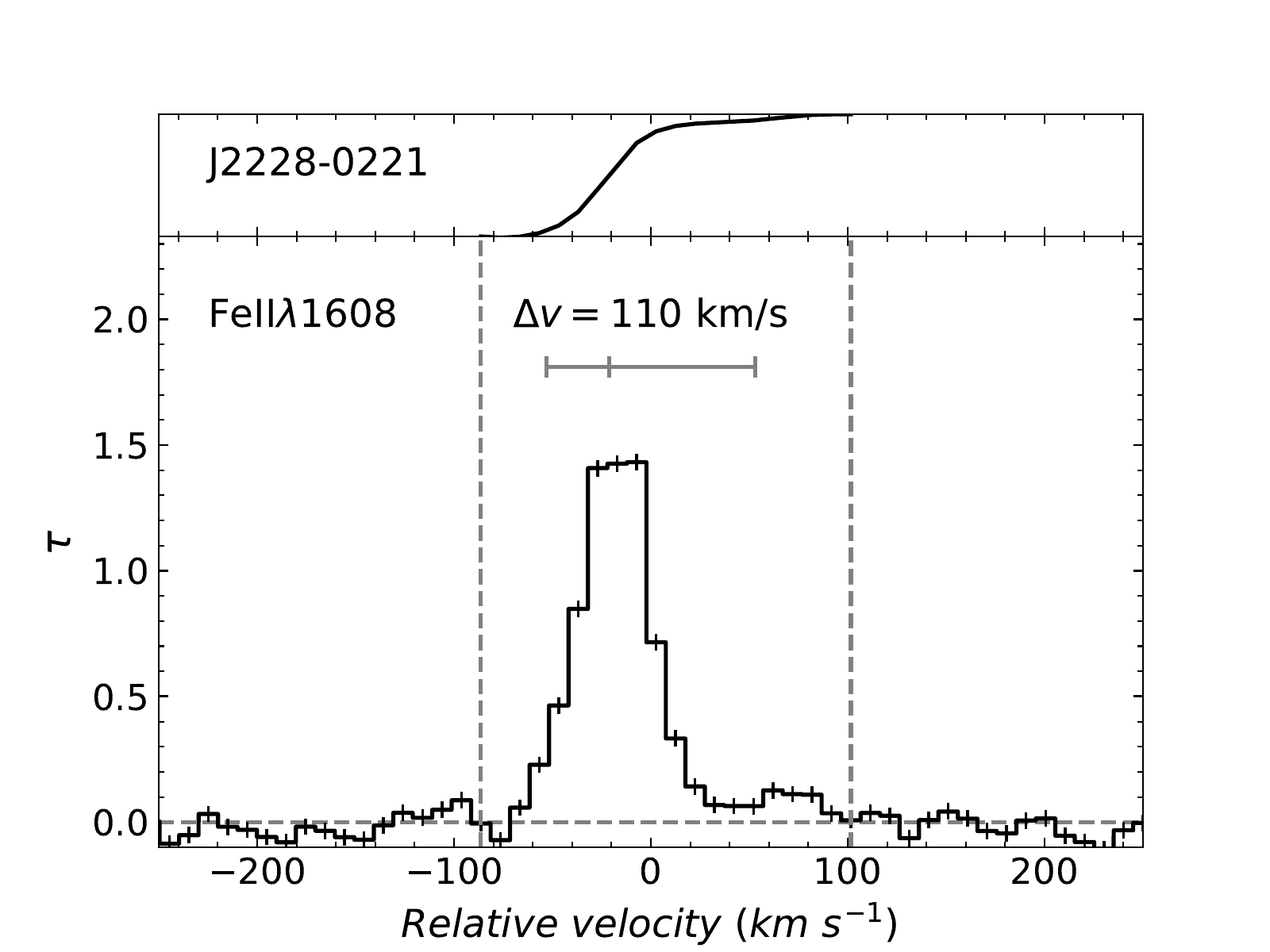} \\
\includegraphics[trim={0cm 0cm 0.5cm 0cm},width=0.32\hsize]{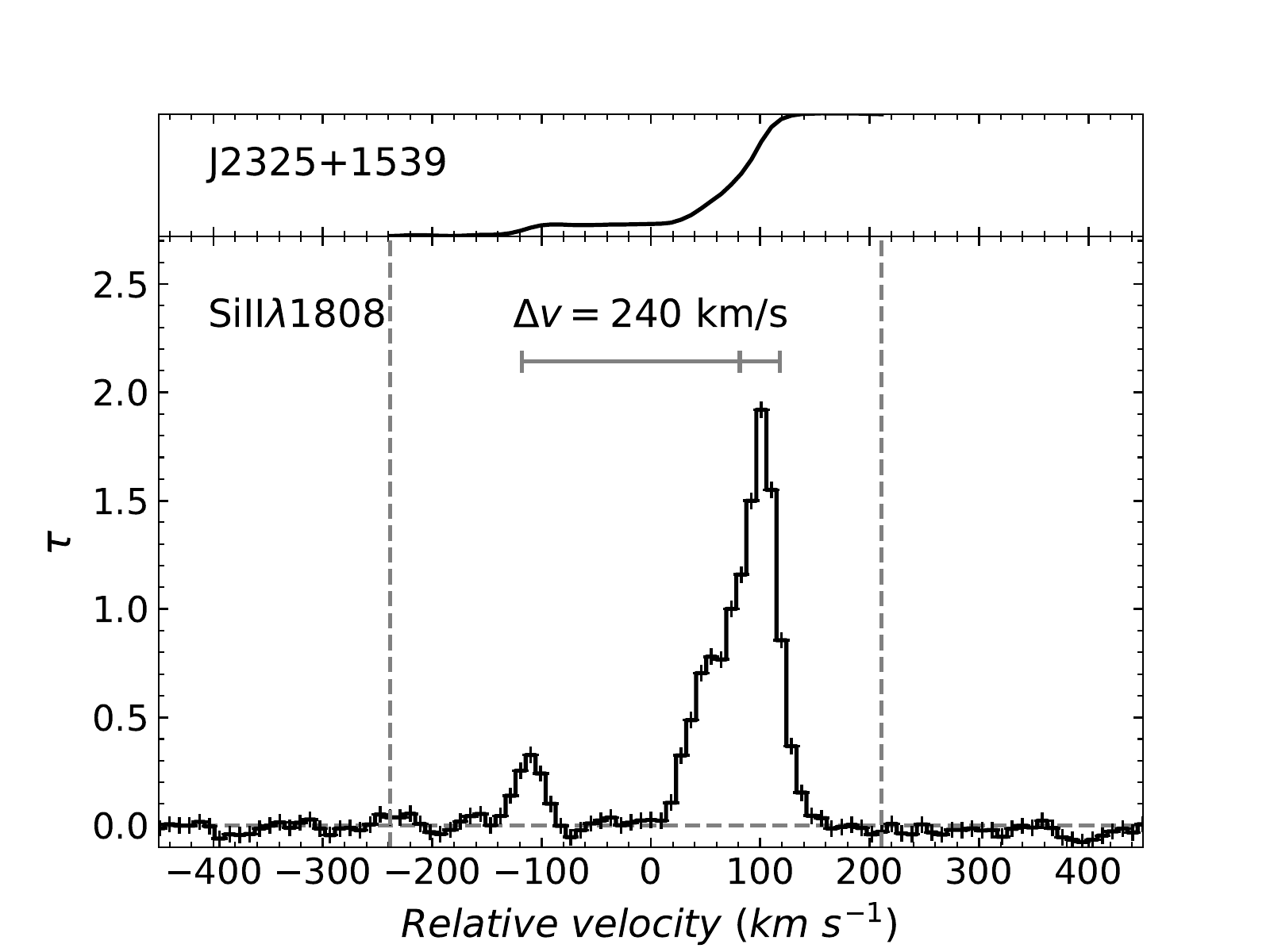} & & \\

\end{tabular}
    \caption{Measurements of $\Delta v_{90}$ for low-ionisation line profiles. The different panels show the apparent optical depth ($\tau$) in the selected transition line for a given system with the cumulative apparent optical depth shown on top. 
    The segments show represent the locations of the 5\%, 50\% and 95\% percentiles.
    }
    \label{f:v90}
\end{figure*}
    \addtolength{\tabcolsep}{+3pt}

\clearpage
\section{Optical/NIR emission lines \label{a:zem}}

\begin{figure}[!h]
    \centering
    \includegraphics[trim={0cm 1cm 0cm 2cm},clip=,width=\hsize]{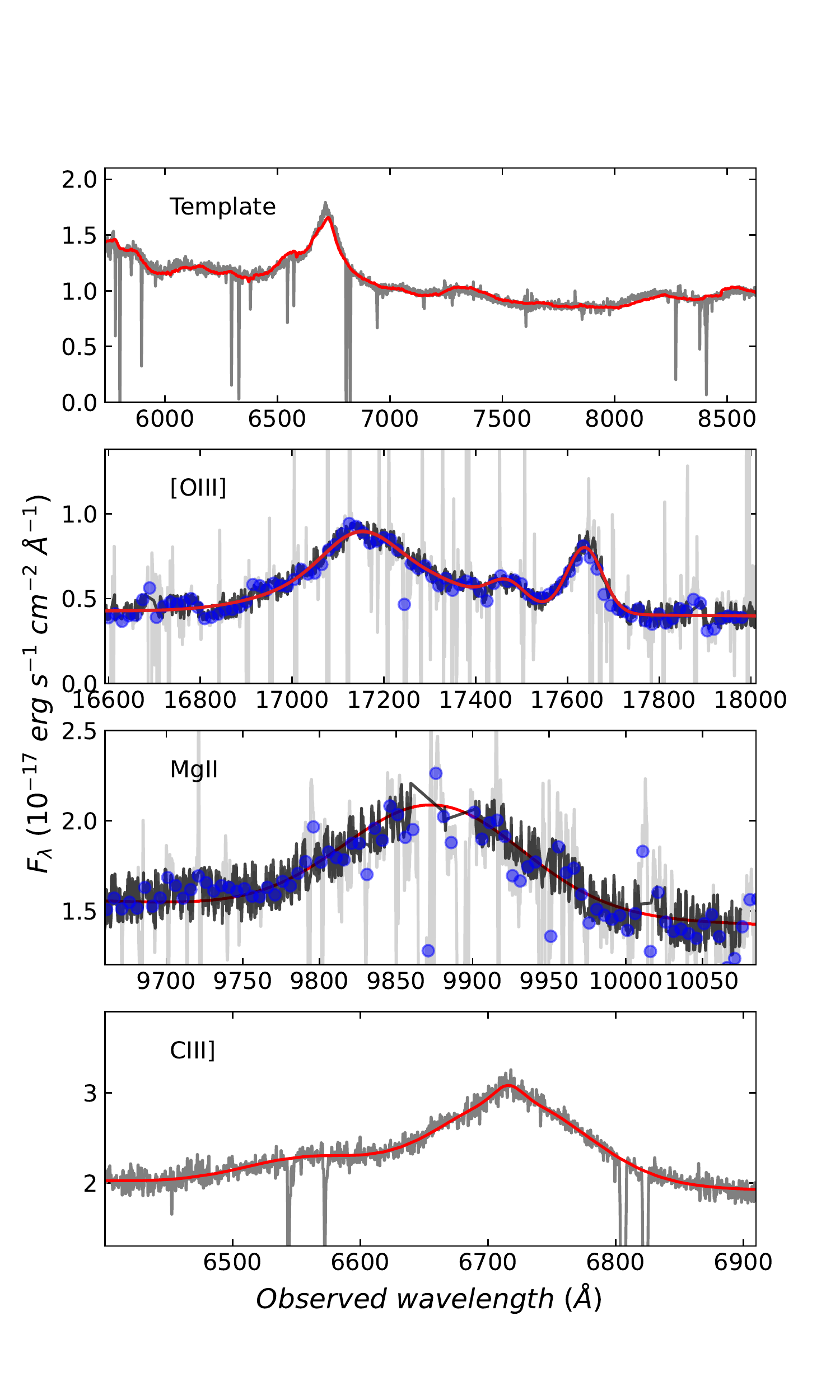}
    \caption{Emission lines from \Jzzun, detected in the visual and near infrared. In the case of NIR, we show in dark gray the spectrum after iteratively rejecting deviant points. Blue dots show median in 20 pixel bins. The best-fit model is shown by the solid red line.  }
    \label{fig:J0019emi}
\end{figure}

\begin{figure}
    \centering
    \includegraphics[trim={0cm 1cm 0cm 2cm },clip=,width=\hsize]{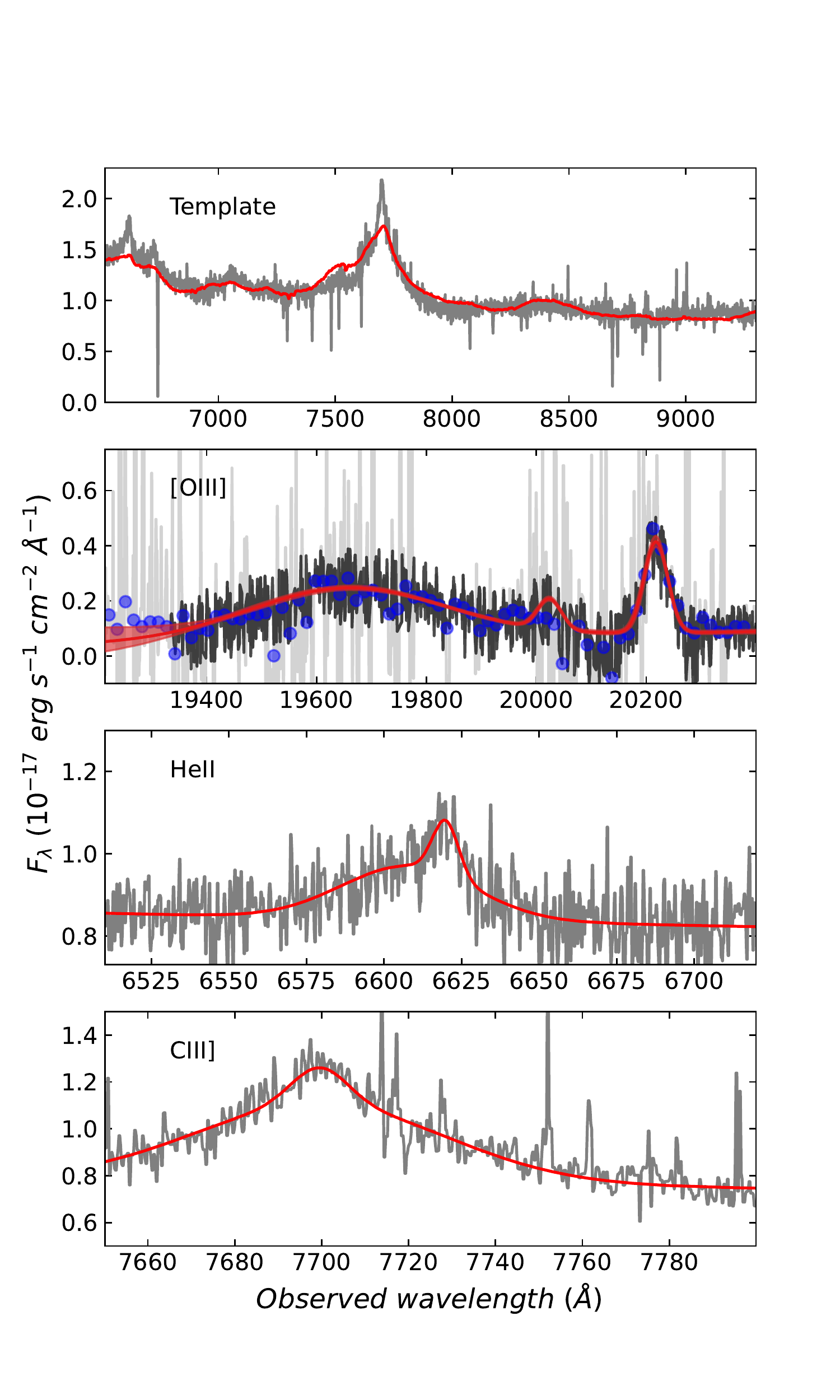}
    \caption{Same as Fig.~\ref{fig:J0019emi} for the quasar \Jzzcn.}
    \label{fig:J0059emi}
\end{figure}

\begin{figure}
    \centering
    \includegraphics[trim={0cm 1cm 0cm 2cm },clip=,width=\hsize]{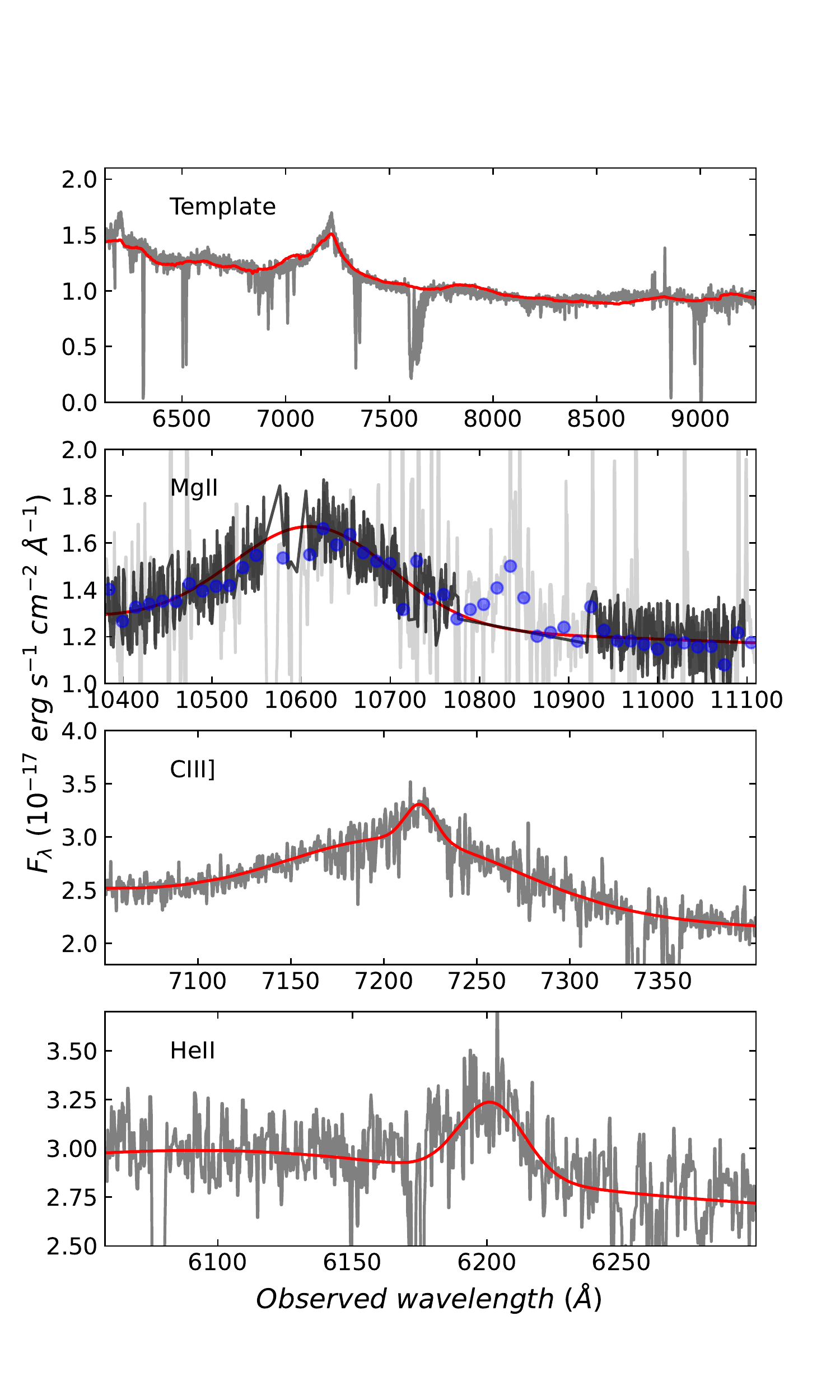}
    \caption{Same as Fig.~\ref{fig:J0019emi} for the quasar \Jzuts.}
    \label{fig:J0136emi}
\end{figure}

\begin{figure}
    \centering
    \includegraphics[trim={0cm 6cm 0cm 2cm },clip=,width=\hsize]{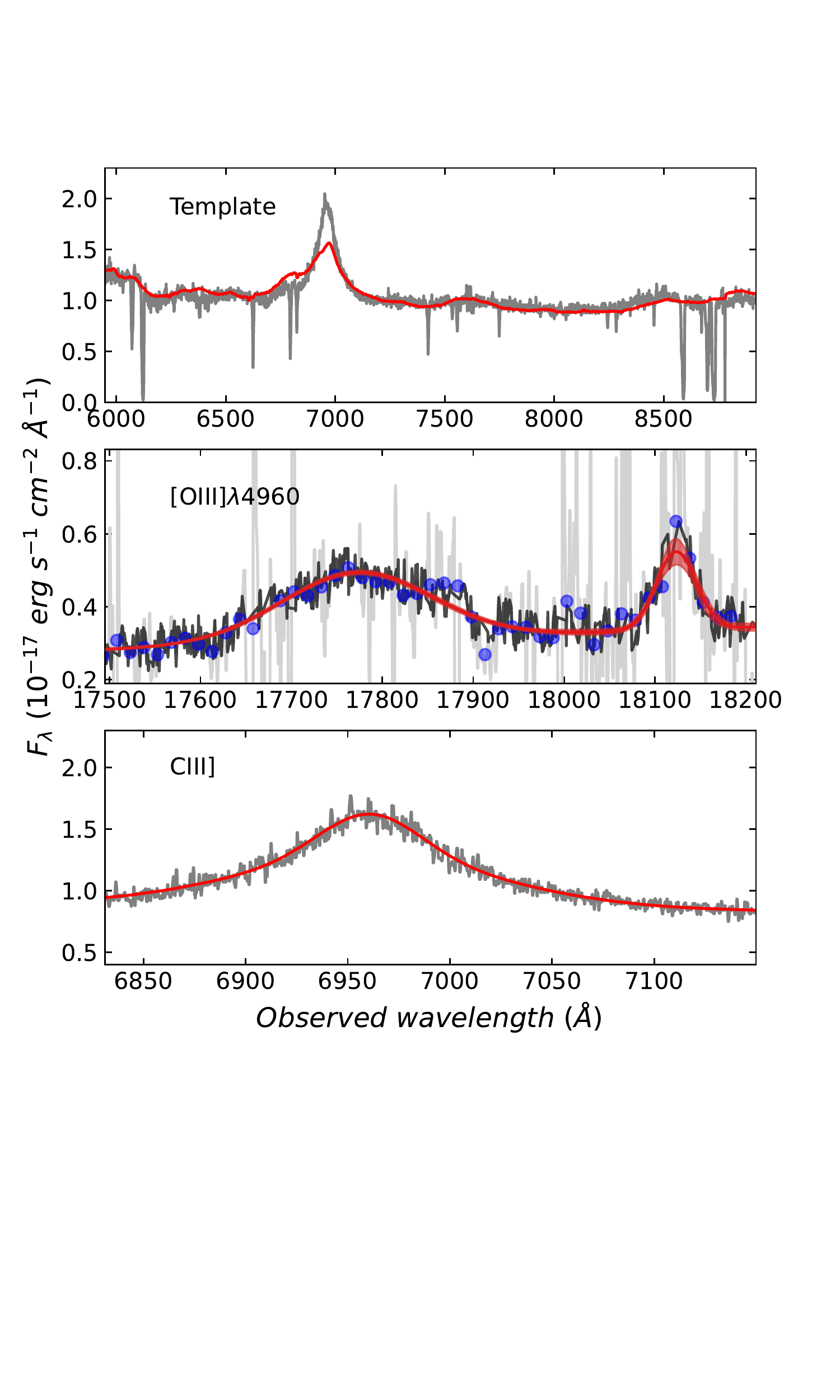}
    \caption{Same as Fig.~\ref{fig:J0019emi} for the quasar \Jzudc.}
    \label{fig:J0125emi}
\end{figure}

\begin{figure}
    \centering
    \includegraphics[trim={0cm 6cm 0cm 2cm },clip=,width=\hsize]{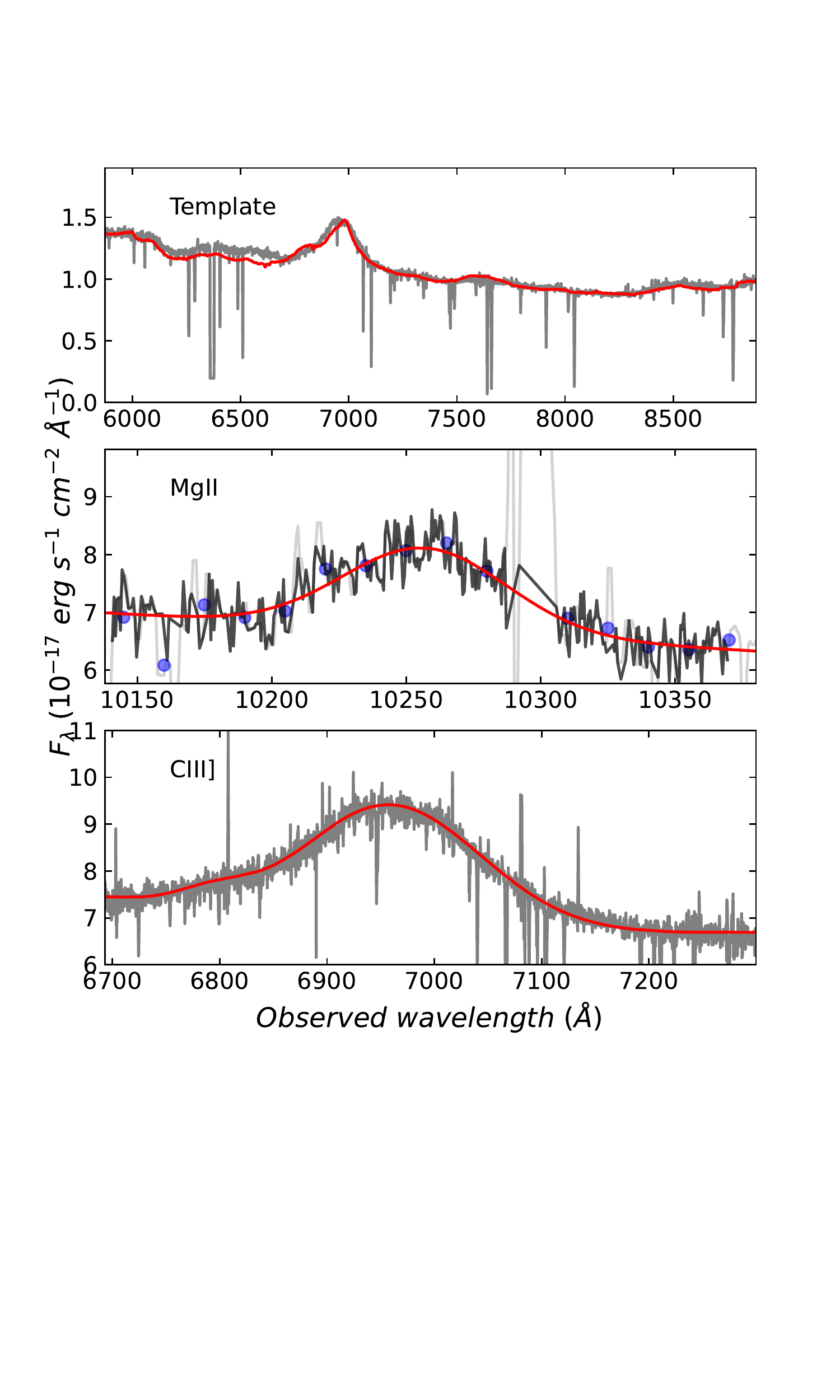}
    \caption{Same as Fig.~\ref{fig:J0858emi} for the quasar \Jzhch.}
    \label{fig:J0858emi}
\end{figure}

\begin{figure}
    \centering
    \includegraphics[trim={0cm 6cm 0cm 2cm },clip=,width=\hsize]{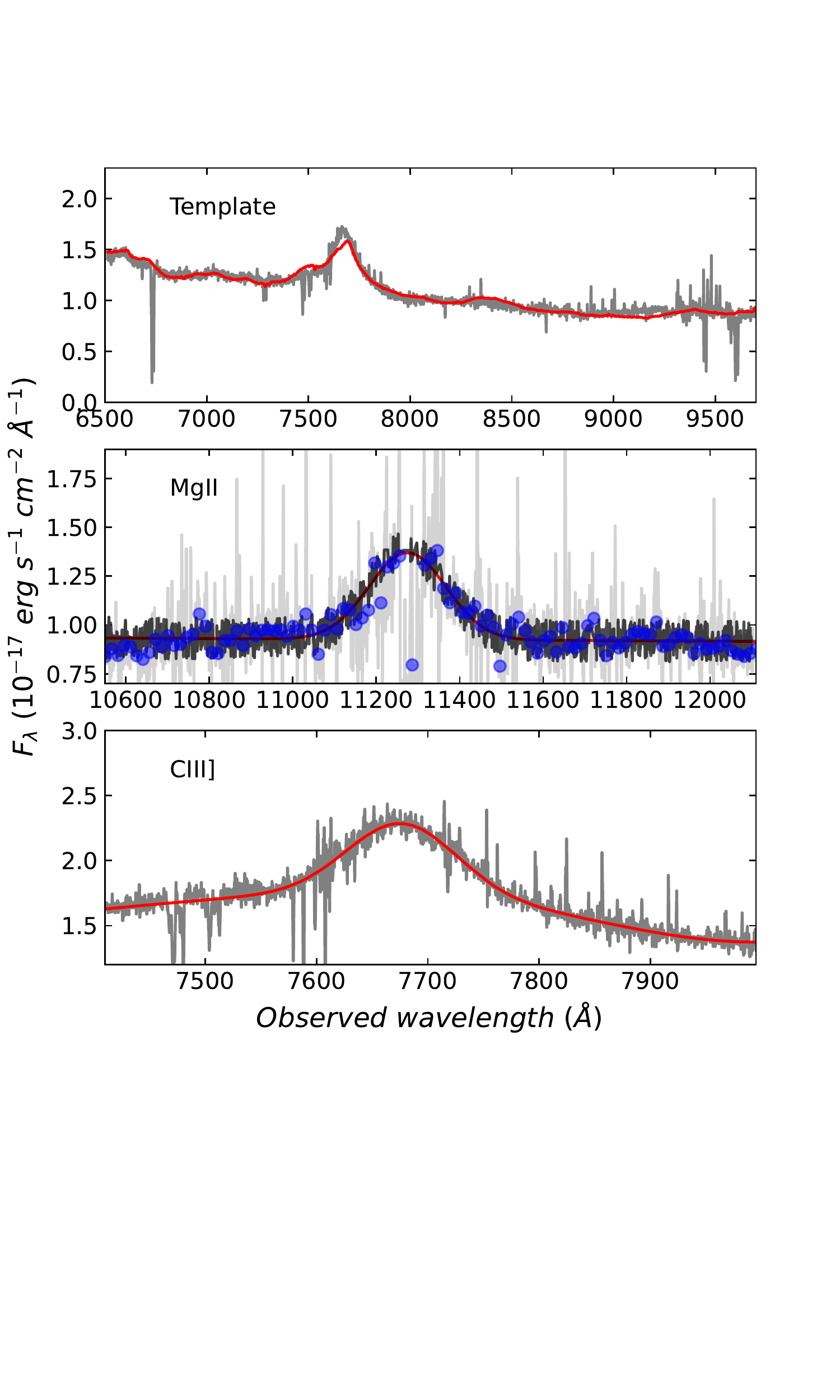}
    \caption{Same as Fig.~\ref{fig:J0019emi} for the quasar \Judts.}
    \label{fig:J1236emi}
\end{figure}

\begin{figure}
    \centering
    \includegraphics[trim={0cm 6cm 0cm 2cm },clip=,width=\hsize]{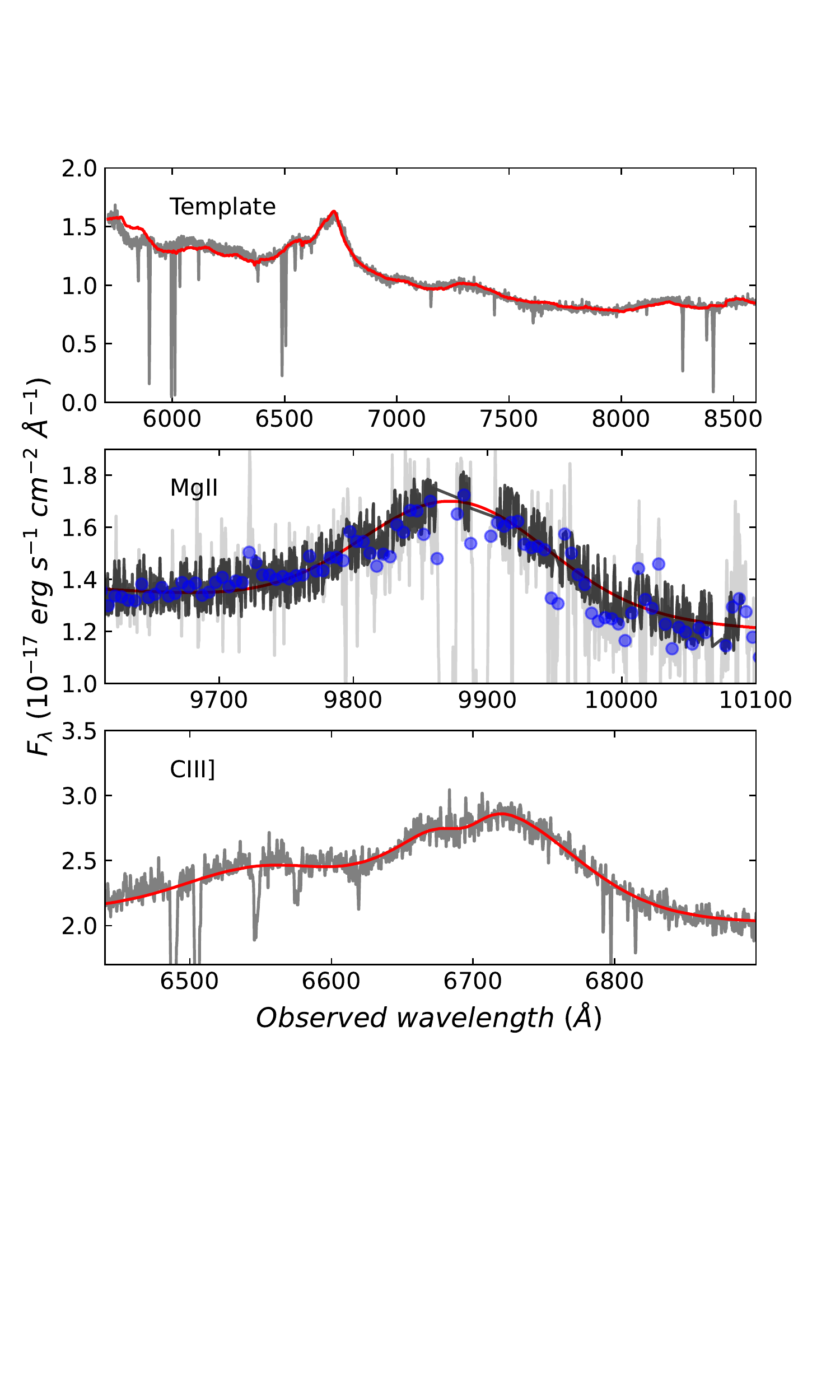}
    \caption{Same as Fig.~\ref{fig:J0019emi} for the quasar \Judqh.}
    \label{fig:J1248emi}
\end{figure}

\begin{figure}
    \centering
    \includegraphics[trim={0cm 6cm 0cm 2cm },clip=,width=\hsize]{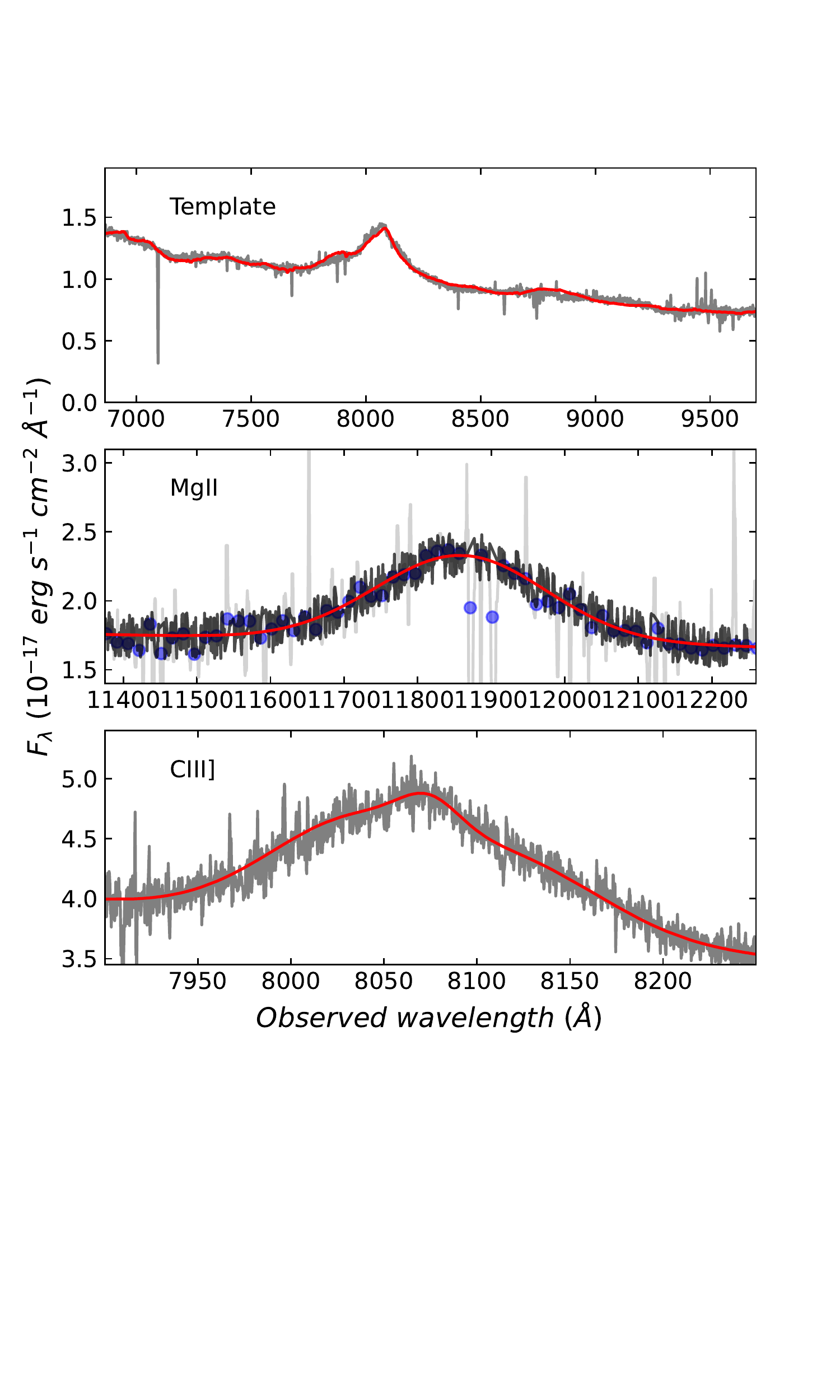}
    \caption{Same as Fig.~\ref{fig:J0019emi} for the quasar \Judcn.}
    \label{fig:J1259emi}
\end{figure}

\begin{figure}
    \centering
    \includegraphics[trim={0cm 1cm 0cm 2cm },clip=,width=\hsize]{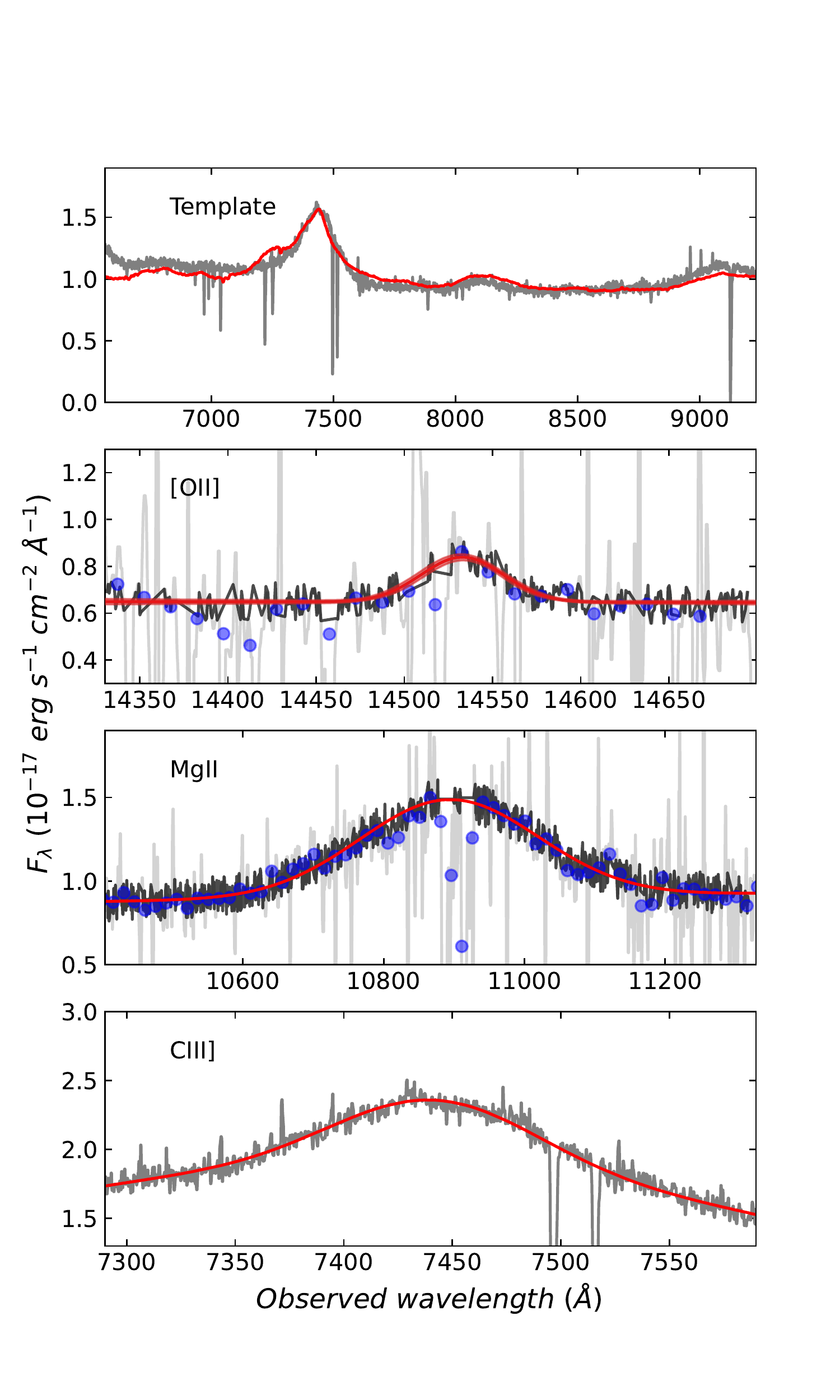}
    \caption{Same as Fig.~\ref{fig:J0019emi} for the quasar \Jutch.}
    \label{fig:J1358emi}
\end{figure}

\begin{figure}
    \centering
    \includegraphics[trim={0cm 11cm 0cm 2cm },clip=,width=\hsize]{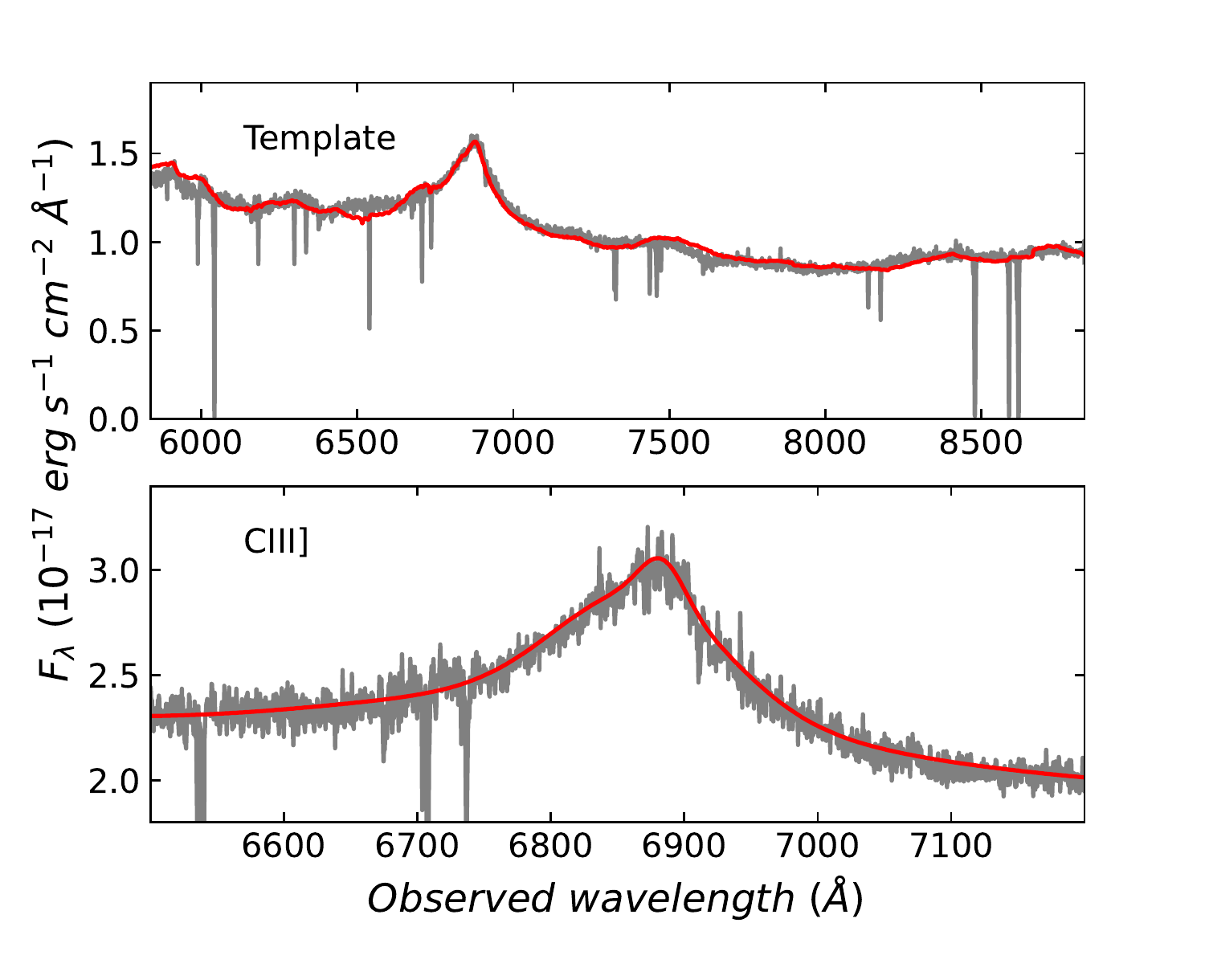}
    \caption{Same as Fig.~\ref{fig:J0019emi} for the quasar \Jdtdc}
    \label{fig:J2325emi}
\end{figure}

\begin{figure}
    \centering
    \includegraphics[trim={0cm 1cm 0cm 2cm },clip=,width=\hsize]{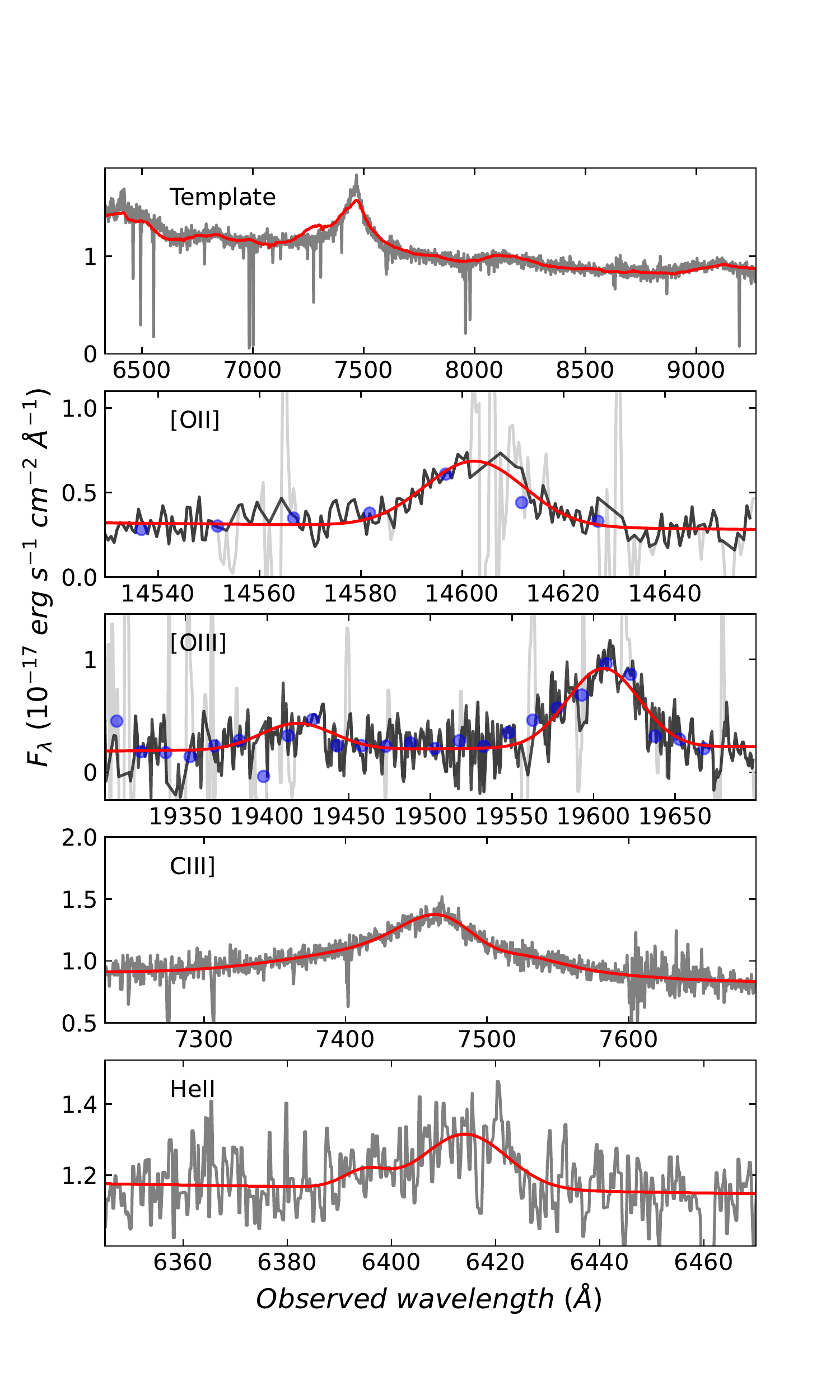}
    \caption{Same as Fig.~\ref{fig:J0019emi} for the quasar \Juttu.}
    \label{fig:J1331emi1}
\end{figure}

\begin{figure}
    \centering
    \includegraphics[trim={0cm 11cm 0cm 2cm },clip=,width=\hsize]{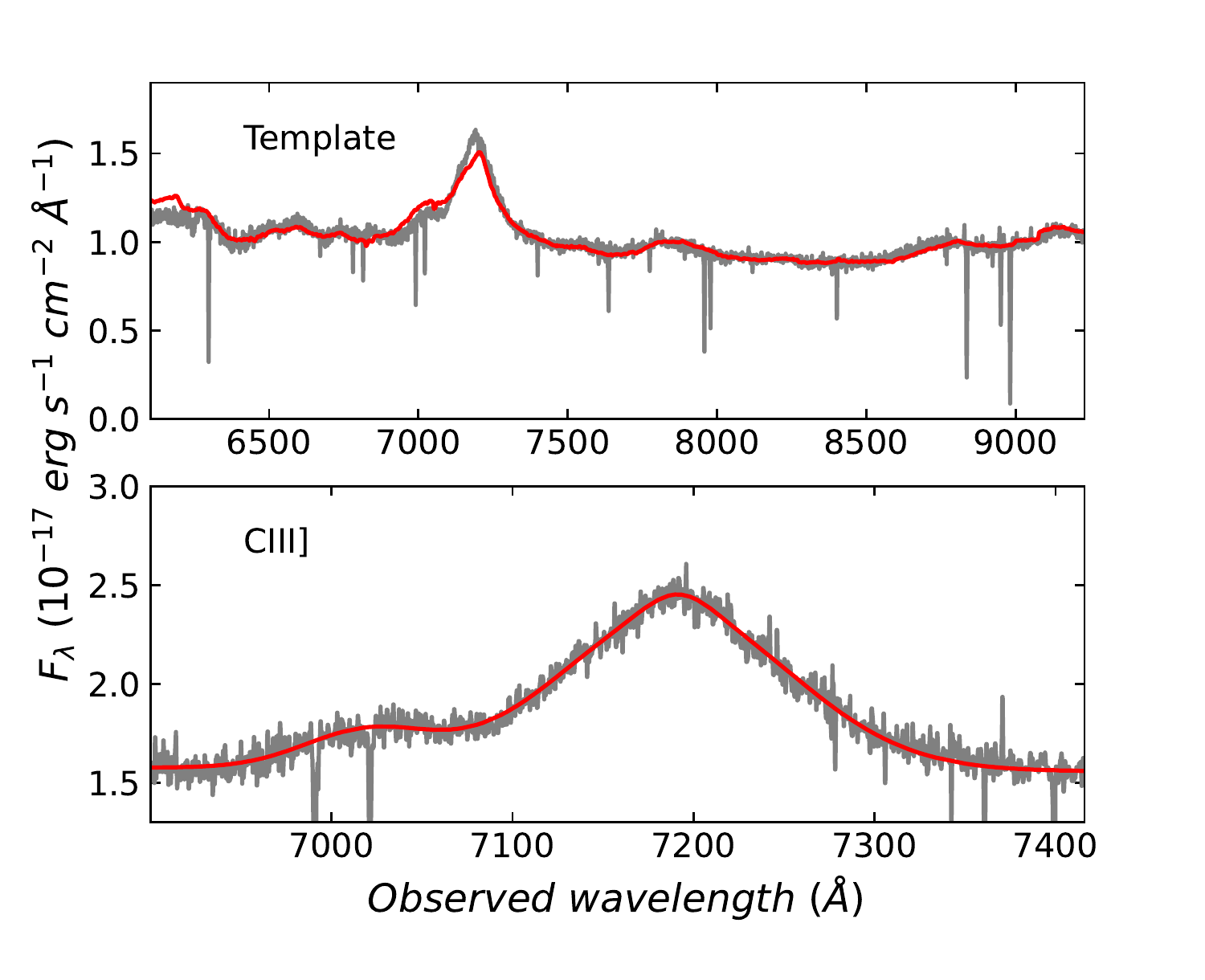}
    \caption{Same as Fig.~\ref{fig:J0019emi} for the quasar \Jdddh}
    \label{fig:J2228emi}
\end{figure}

\clearpage
\section{CO(3-2) emission lines \label{a:zCO}}

\input CO32

\begin{figure*}
    \centering
    \begin{tabular}{cc}
    \includegraphics[width=0.4\hsize]{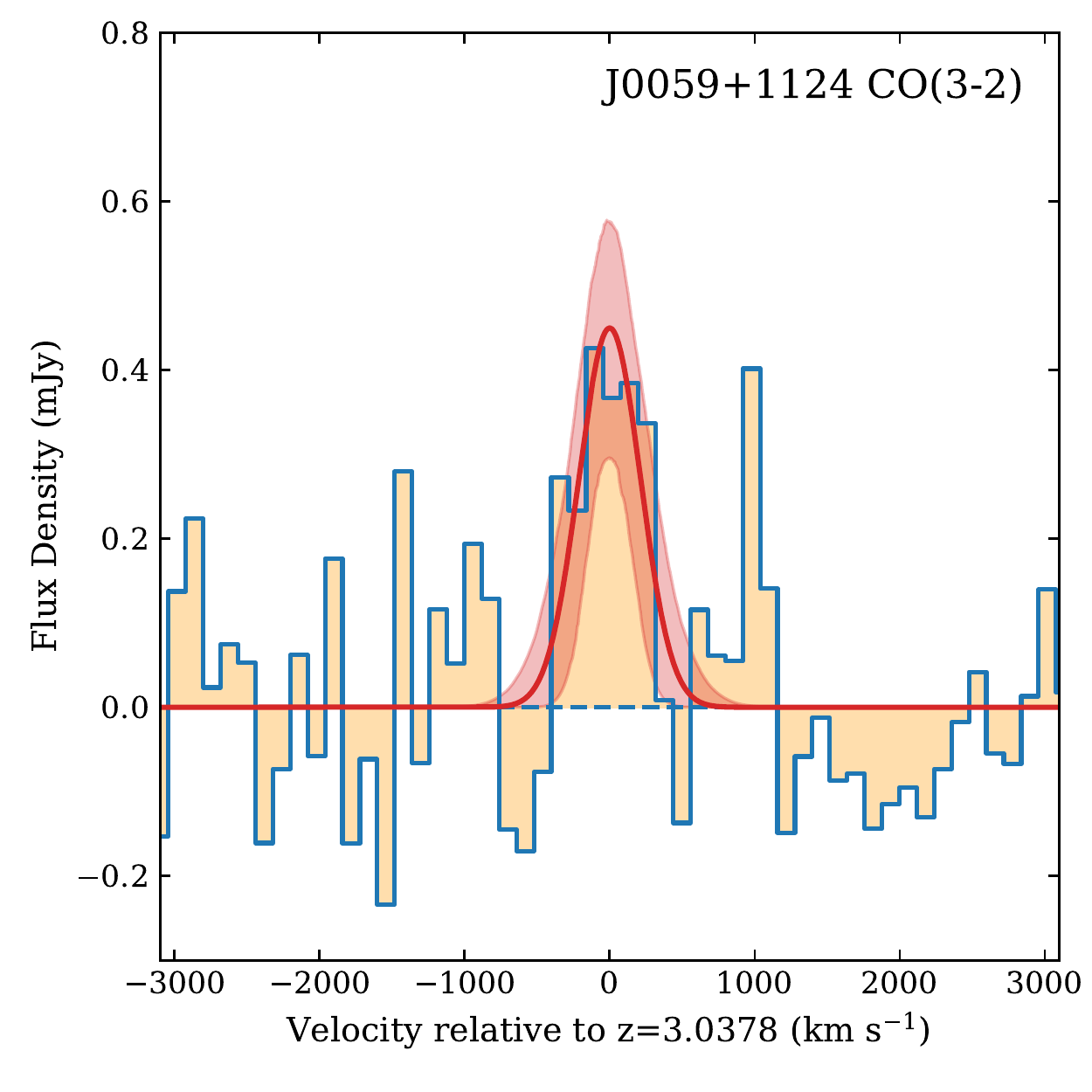}& \\
    \includegraphics[width=0.4\hsize]{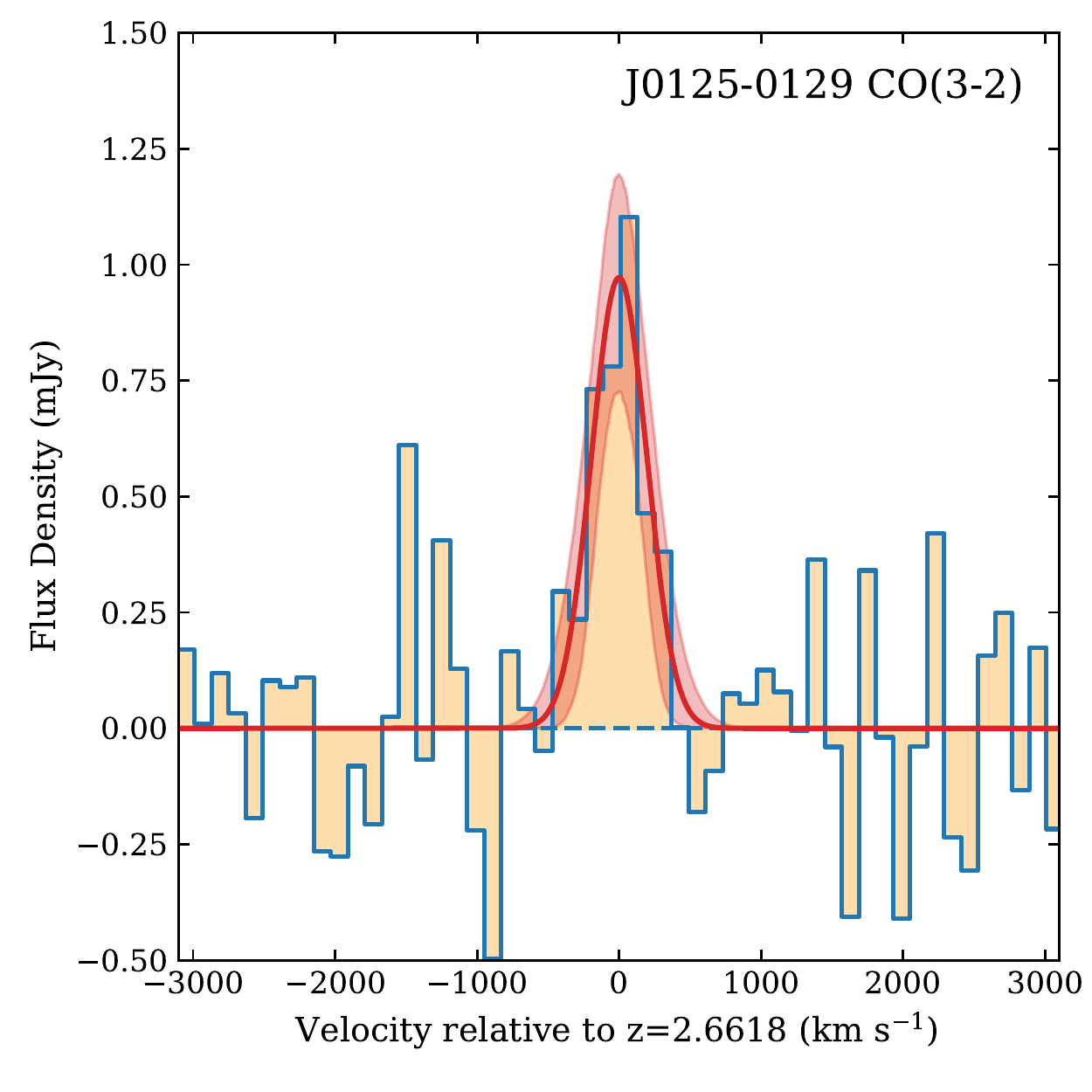}&
    \includegraphics[width=0.4\hsize]{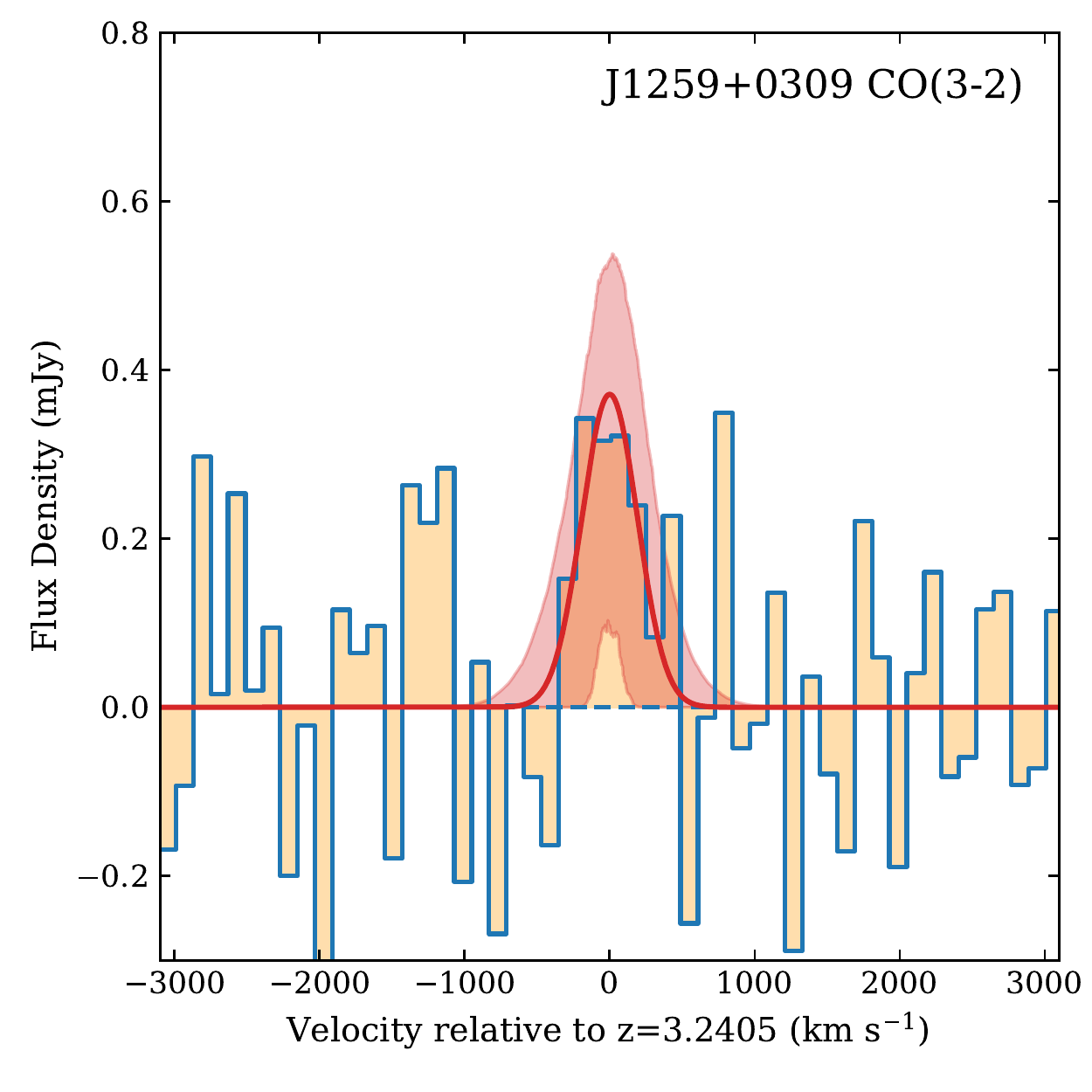}\\
    \includegraphics[width=0.4\hsize]{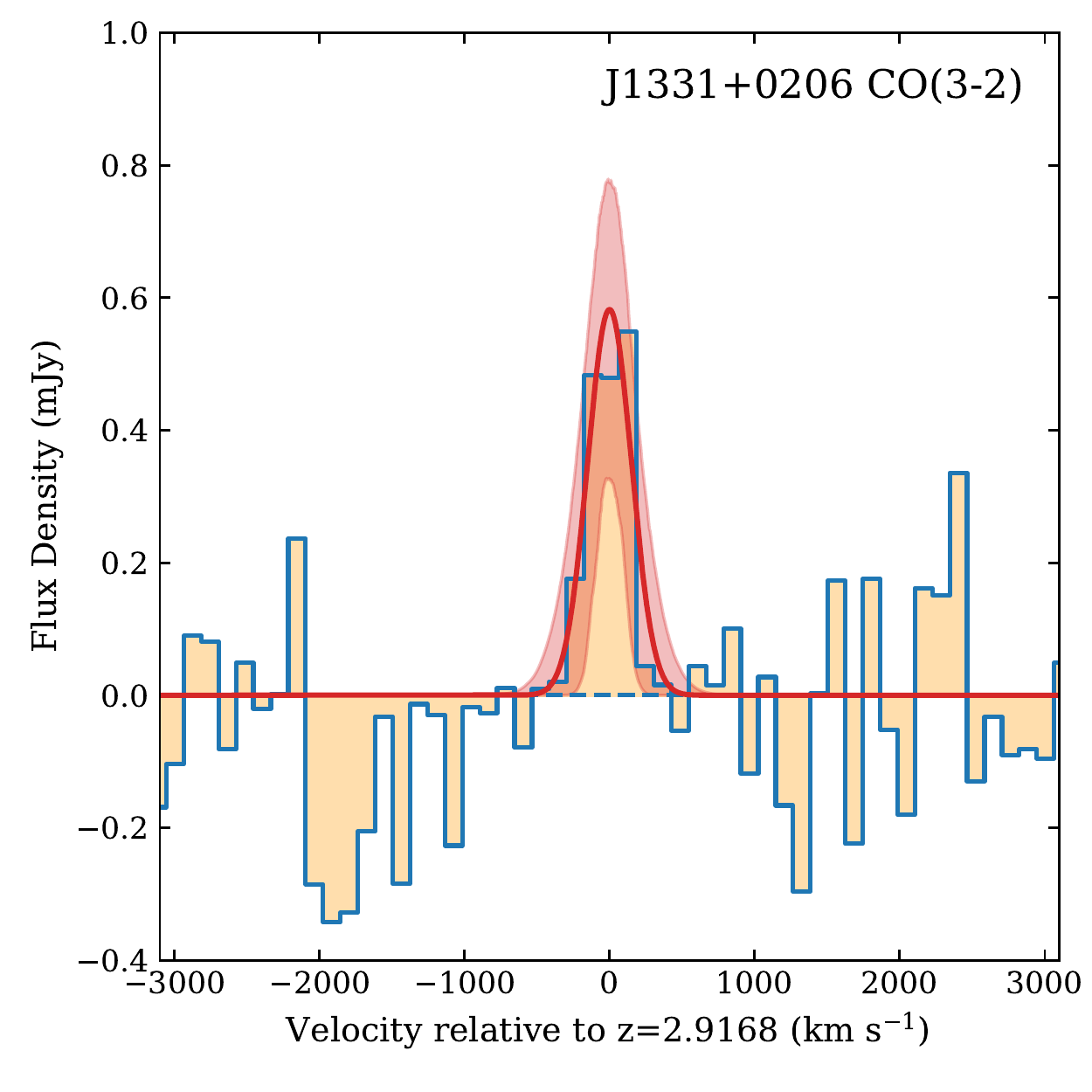}&
    \includegraphics[width=0.4\hsize]{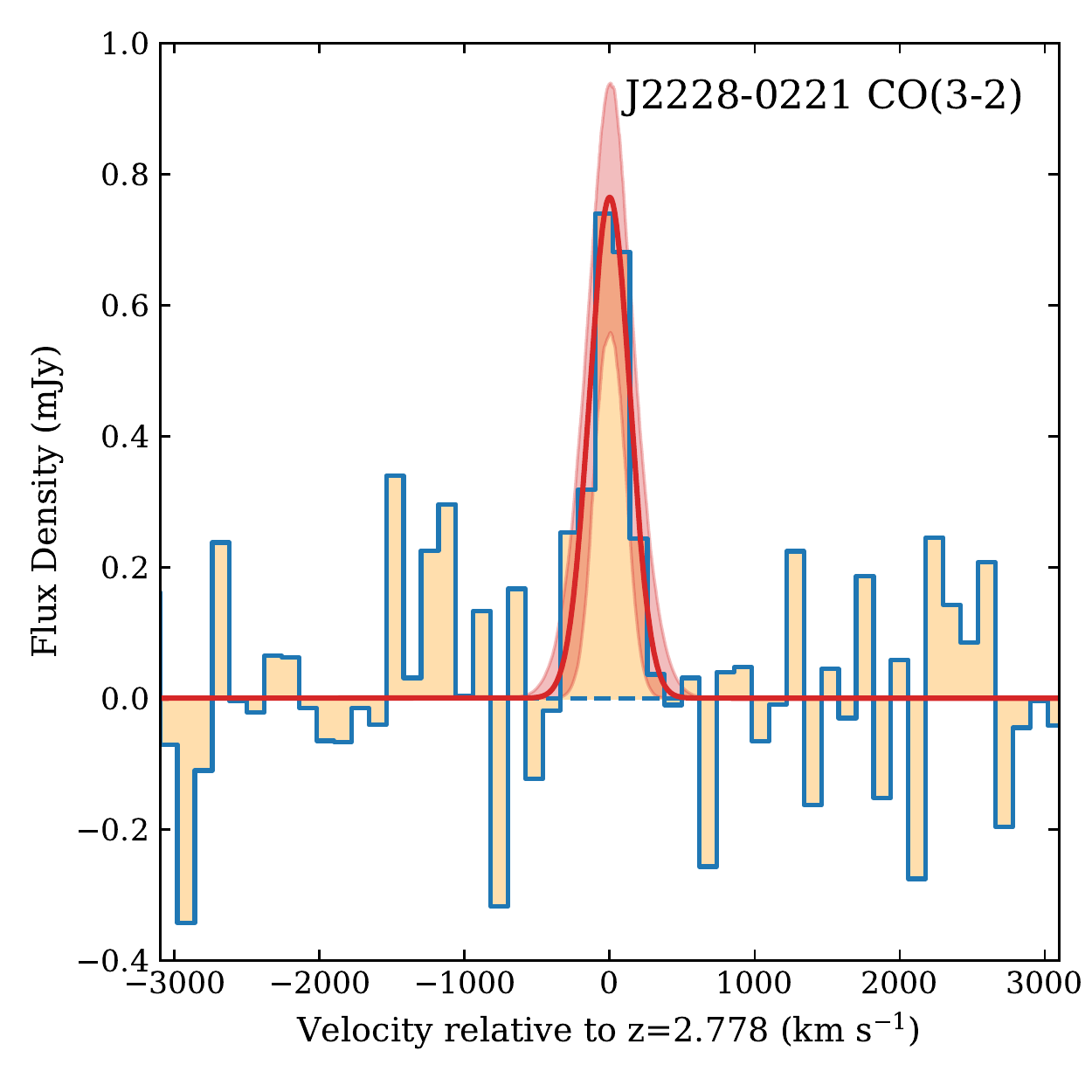}\\
    \end{tabular}
    \caption{Detection of CO(3-2) emission from 3~mm NOEMA spectra extracted in 120~\kms\ bins. A single component Gaussian fit is superimposed in red together with its 90\% confidence interval. A 6$^{\rm th}$ quasar (\Jzzuc) in this sample has also been observed with NOEMA, with the corresponding CO(3-2) detection presented in \citetalias{Noterdaeme2021b}.}
    \label{f:CO32}
\end{figure*}

\end{appendix}
\end{document}

%% file: zem.tex
\begin{table*}[]
    \centering
        \caption{Measurements of the quasar emission lines redshift.    \label{t:zem}}
    \begin{tabular}{c c c c c c c c c c c}
\hline \hline
      Quasar   &  \multicolumn{9}{c}{Redshift}\\
               &  [O\,{\sc ii}] & [\OIII] &  \MgII  & \multicolumn{2}{c}{C\,{\sc iii}]} & \multicolumn{2}{c}{He\,{\sc ii}} & Template & $z_{syst}^i$ & CO(3-2) \\
               &                &         &         & meas.  & corr.            &  meas.   & corr.                         &     &     &   \\
      \hline
      \Jzzuc   &  2.6310        & 2.6309       & --      & --      &  --    & --      & --     &  --     & 2.6310  & 2.6285 \\ 
      \Jzzun   &  --            & 2.5217       & 2.5276  & 2.5185  &   2.5212     & --      & --           & 2.5253  & 2.5217 & --   \\ %
      \Jzzcn   &  --            & 3.0369       & --      & 3.0338  &   3.0369     & 3.0353  & 3.0406      & 3.0395  & 3.0369  & 3.0378 \\ %
      
      \Jzudc   &  --            & 2.6534       & --      & 2.6466  &   2.6494     &  --     & --     & 2.6540  & 2.6534  & 2.6618 \\ %
      \Jzuts   &   --           &  --          & 2.7905  & 2.7820  &   2.7849     & 2.7800  &  2.7814 & 2.7846  & 2.7849   & -- \\ %
      \Jzhch   &  --            & --       & 2.6643      & 2.6450  &   2.6478     & --  & --      & 2.6589  & --  & -- \\ %

      \Judts   &   --           &  --          & 3.0284  & 3.0208  &   3.0239    & --      &  --      & 3.0318  & 3.0284  & --\\ %
      \Judqh   &    --          &  --          & 2.5273  & 2.5206  &   2.5233     & --      & --            & 2.5221  & 2.5273 &  --\\ %
      \Judcn   &  --            &  --          & 3.2365  & 3.2277  &  3.2309      &   --    &  --    & 3.2369  & 3.2365  & 3.2405   \\ %
      \Juttu   &  2.9179        & 2.9147       &    --   & 2.9099  &  2.9129      & 2.9100  & 2.9155 & 2.9153  & 2.9147  & 2.9168          \\ %
      \Jutch   &  2.8992        &  --          &  2.8926 & 2.8971  &   2.9001     & --      & --       & 2.8996  &  2.8926  &  --    \\  %
      \Jdddh   &   --           & --           &         & 2.7676  &   2.7705     &   --    &  --    & 2.7763  &  2.7705 & 2.7780        \\%
      \Jdtdc   &   --           &   --         &  --     & 2.6044  &   2.6072     &  --  & --       & 2.6051  &  2.6072 &  --      \\%
      \hline
    \end{tabular}
\end{table*}

%% file: PH2sum.tex
\begin{table*} 
 \centering 
 \caption{Main properties of the proximate absorption systems \label{t:sum}} 
 \begin{tabular}{ccccccccc}
 \hline \hline
{\large \strut}Quasar & $\log N(\HI)$ & $\log N(\HH)$ & [Zn/H] & [S/H] & [Si/H] & [Fe/H] & A$_{\rm V}$\tablefootmark{a} & $\Delta v_{90}$\tablefootmark{b}  \\
       & (\cmsq)       & (\cmsq)       &        &       &        &        & (mag)       & (\kms)          \\
       \hline
J0015+1842 & 20.71$\pm$0.12 & 19.73 $\pm$ 0.02 & -0.41$\pm$0.13 & -0.54$\pm$0.14 & -0.78$\pm$0.12 & -1.79$\pm$0.12 & 0.40$\pm$0.10 & 450\\
J0019-0137 & 21.02$\pm$0.11 & 20.29 $\pm$ 0.01 & -0.90$\pm$0.12 & -1.00$\pm$0.11 & -0.96$\pm$0.12 & -2.03$\pm$0.11 & 0.08$\pm$0.01 & 190\\
J0059+1124 & 21.77$\pm$0.11 & 19.07 $\pm$ 0.02 & -1.40$\pm$0.15 & -1.37$\pm$0.16 & -1.03$\pm$0.34 & -2.02$\pm$0.13 & 0.04$\pm$0.02 & 170\\
J0125-0129 & 21.90$\pm$0.15 & 20.05 $\pm$ 0.01 & -0.96$\pm$0.15 & -0.96$\pm$0.16 & -0.92$\pm$0.16 & -1.66$\pm$0.15 & 0.19$\pm$0.01 & 430\\
J0136+0440 & 20.73$\pm$0.11 & 18.65 $\pm$ 0.07 & --             & -0.59$\pm$0.11 & -0.80$\pm$0.12 & -1.24$\pm$0.11 & 0.04$\pm$0.01\tablefootmark{c} & 190\\
J0858+1749 & 20.40$\pm$0.11 & 19.72 $\pm$ 0.02 & -0.52$\pm$0.12 & -0.57$\pm$0.12 & -0.96$\pm$0.11 & -2.26$\pm$0.11 & 0.12$\pm$0.01\tablefootmark{c} & 190\\
J1236+0010 & 20.65$\pm$0.11 & 19.75 $\pm$ 0.01 & -0.55$\pm$0.13 & -0.48$\pm$0.11 & -0.71$\pm$0.11 & -1.70$\pm$0.11 & 0.07$\pm$0.02 & 400\\
J1248+0639 & 20.55$\pm$0.11 & 19.76 $\pm$ 0.01 & -0.55$\pm$0.16 & -0.48$\pm$0.19 & -0.89$\pm$0.12 & -1.76$\pm$0.11 & -0.05$\pm$0.02 & 240\\
J1259+0309 & 21.42$\pm$0.11 & 19.10 $\pm$ 0.01 & -1.57$\pm$0.12 & -1.39$\pm$0.11 & -1.33$\pm$0.11 & -1.85$\pm$0.11 & -0.05$\pm$0.01 & 90\\
J1331+0206 & 21.41$\pm$0.11 & 20.11 $\pm$ 0.01 & -1.56$\pm$0.14 & -1.67$\pm$0.12 & -1.61$\pm$0.12 & -1.94$\pm$0.12 & 0.04$\pm$0.02 & 80\\
J1358+1410 & 21.19$\pm$0.11 & 19.85 $\pm$ 0.02 & -0.96$\pm$0.14 & -0.96$\pm$0.11 & -0.87$\pm$0.11 & -1.76$\pm$0.11 & 0.06$\pm$0.03 & 170\\
J2228-0221 & 20.98$\pm$0.11 & 19.44 $\pm$ 0.02 & -0.52$\pm$0.11 & -0.54$\pm$0.11 & -1.00$\pm$0.11 & -1.94$\pm$0.11 & 0.24$\pm$0.03 & 110\\
J2325+1539 & 21.74$\pm$0.11 & 20.25 $\pm$ 0.01 & -1.15$\pm$0.11 & -1.03$\pm$0.11 & -1.14$\pm$0.11 & -1.70$\pm$0.11 & 0.16$\pm$0.03 & 240\\
 \hline 
 \end{tabular} 
 \tablefoot{
 \tablefoottext{a}{Statistical error only, from dispersion between measurements. The systematic error on $A_{\rm V}$ is about 0.07~mag (see text).}
 \tablefoottext{b}{{Measured from the low-ion metal lines (see text).}}
 \tablefoottext{c}{\citet{Balashev2019} quoted A$_V =0.16$ and 0.29 for \Jzuts\ and \Jzhch, respectively. However, these values were obtained with a varying intrinsic power law, as usually done in attempt to separate intrinsic slope from that due to an intervening absorber. Here, as for other quasars in our sample, we consistently re-measured $A_V$ using a fixed power-law to quantify amount of dust in the overall systems.}
 }
 \end{table*}

%% file: J0019met.tex
\begin{table*} 
 \centering 
 \caption{Results of Voigt-profile fitting to singly ionised metal absorption lines proximate to \Jzzun. {The adopted spectral resolutions are 44.3 and 27.0~\kms\ for the UVB and VIS arm, respectively.}%J0019met.tex automatically created from f26.12p3w3381 
 \label{t:J0019met}} 
 \begin{tabular}{ccccccc}
 \hline \hline
{\Large \strut} $z$ & $b$ & \multicolumn{5}{c}{$\log N$ [cm$^{-2}$]} \\
     & (km\,s$^{-1})$ & \SII & \SiII & \FeII & \ZnII & \MnII \\
 \hline
 2.5282 & 19.4$\pm$0.2 & 15.03$\pm$0.02 & 15.50$\pm$0.04 & 14.53$\pm$0.01 & 12.54$\pm$0.05 & 12.52$\pm$0.04 \\
 2.5290 & 20.9$\pm$0.8 & 13.97$\pm$0.10 & 14.01$\pm$0.02 & 13.36$\pm$0.01 & $\ldots$       & $\ldots$       \\
 2.5299 &  5.7$\pm$0.2 & 14.89$\pm$0.06 & 15.29$\pm$0.09 & 13.48$\pm$0.03 & 12.60$\pm$0.05 & 12.06$\pm$0.24 \\
\hline
 Total    &     -        & 15.29$\pm$0.03 & 15.71$\pm$0.04 & 14.59$\pm$0.01 & 12.87$\pm$0.04 & 12.65$\pm$0.07 \\
 \hline 
 \end{tabular} 
 \end{table*}

%% file: J0059met.tex
\begin{table*} 
 \centering 
 \caption{Results of Voigt-profile fitting to singly ionised metal absorption lines proximate to \Jzzcn. {The adopted spectral resolutions are 45.0 and 25.0~\kms\ for the UVB and VIS arm, respectively.}
 %J0059met.tex automatically created from f26.12p3w3730 
 \label{t:J0059met}} 
 \begin{tabular}{cccccccc}
 \hline \hline
{\Large \strut} $z$ & $b$ & \multicolumn{6}{c}{$\log N$ [cm$^{-2}$]} \\
     & (km\,s$^{-1})$ & \SII & \SiII & \FeII & \ZnII & \CrII & \NiII \\
 \hline
 3.0343 & 29.0$\pm$1.1 & 15.22$\pm$0.05 & 15.72$\pm$0.09 & 15.02$\pm$0.06 & 12.38$\pm$0.24 & 12.77$\pm$0.55 & 13.13$\pm$0.70 \\
 3.0345 &  7.8$\pm$2.2 & 15.26$\pm$0.23 & 16.09$\pm$0.45 & 14.77$\pm$0.14 & 12.86$\pm$0.11 & 13.57$\pm$0.10 & 14.10$\pm$0.10 \\
\hline
 Total    &     -        & 15.54$\pm$0.12 & 16.25$\pm$0.32 & 15.21$\pm$0.06 & 12.98$\pm$0.10 & 13.64$\pm$0.11 & 14.14$\pm$0.11 \\
 \hline 
 \end{tabular} 
 \end{table*}

%% file: J1236met.tex
\begin{table*} 
 \centering 
 \caption{Results of Voigt-profile fitting to singly ionised metal absorption lines proximate to \Judts. {The adopted spectral resolutions are 50.0 and 26.5~\kms\ for the UVB and VIS arm, respectively.}
 %J1236met.tex automatically created from f26.12p3w3838 
 \label{t:J1236met}} 
 \begin{tabular}{ccccccc}
 \hline \hline
{\Large \strut} $z$ & $b$ & \multicolumn{5}{c}{$\log N$ [cm$^{-2}$]} \\
     & (km\,s$^{-1})$ & \SII & \SiII & \FeII & \ZnII & \NiII \\
 \hline
 3.0258 &  5.8$\pm$0.5 & $\ldots$       & 13.18$\pm$0.11 & 12.12$\pm$0.15 & $\ldots$       & $\ldots$       \\
 3.0264 & 17.3$\pm$1.8 & $\ldots$       & 13.70$\pm$0.04 & 12.99$\pm$0.03 & $\ldots$       & $\ldots$       \\
 3.0270 &  5.9$\pm$1.0 & $\ldots$       & 13.57$\pm$0.10 & 12.74$\pm$0.07 & $\ldots$       & $\ldots$       \\
 3.0283 & 35.0$\pm$1.0 & 14.71$\pm$0.04 & 14.61$\pm$0.03 & 13.73$\pm$0.02 & 12.02$\pm$0.29 & $\ldots$       \\
 3.0290 & 11.5$\pm$1.0 & $\ldots$       & 14.12$\pm$0.08 & 13.43$\pm$0.05 & 11.73$\pm$0.24 & $\ldots$       \\
 3.0304 & 25.1$\pm$1.0 & $\ldots$       & 13.79$\pm$0.02 & 12.90$\pm$0.04 & $\ldots$       & $\ldots$       \\
 3.0315 & 23.3$\pm$3.4 & $\ldots$       & 13.21$\pm$0.02 & 12.40$\pm$0.11 & $\ldots$       & $\ldots$       \\
 3.0323 & 18.3$\pm$0.3 & $\ldots$       & 13.71$\pm$0.03 & 12.83$\pm$0.05 & $\ldots$       & $\ldots$       \\
 3.0331 & 12.9$\pm$0.4 & 15.30$\pm$0.03 & 15.43$\pm$0.04 & 14.24$\pm$0.02 & 12.69$\pm$0.03 & 13.16$\pm$0.10 \\
 3.0335 & 18.0$\pm$0.9 & $\ldots$       & 13.95$\pm$0.04 & 13.48$\pm$0.04 & $\ldots$       & $\ldots$       \\
\hline
 Total    &     -        & 15.40$\pm$0.03 & 15.55$\pm$0.03 & 14.51$\pm$0.01 & 12.81$\pm$0.06 & 13.16$\pm$0.10 \\
 \hline 
 \end{tabular} 
 \end{table*}

%% file: J1248met.tex
\begin{table*} 
 \centering 
 \caption{Results of Voigt-profile fitting to singly ionised metal absorption lines proximate to \Judqh. {The adopted spectral resolutions are 40.0 and 21.0~\kms\ for the UVB and VIS arm, respectively.}
 %J1248met.tex automatically created from f26.12p3w3327 
 \label{t:J1248met}} 
 \begin{tabular}{ccccccc}
 \hline \hline
{\Large \strut} $z$ & $b$ & \multicolumn{5}{c}{$\log N$ [cm$^{-2}$]} \\
     & (km\,s$^{-1})$ & \SII & \SiII & \FeII & \ZnII & \MnII \\
 \hline
 2.5277 & 19.6$\pm$0.8 & 14.01$\pm$0.08 & 13.62$\pm$0.02 & 13.03$\pm$0.02 & $\ldots$       & $\ldots$       \\
 2.5284 &  8.5$\pm$0.6 & $\ldots$       & 13.35$\pm$0.04 & 12.72$\pm$0.03 & $\ldots$       & $\ldots$       \\
 2.5292 & 18.7$\pm$0.3 & 14.79$\pm$0.03 & 15.08$\pm$0.04 & 14.11$\pm$0.01 & 12.30$\pm$0.07 & 12.27$\pm$0.06 \\
 2.5295 &  2.7$\pm$0.4 & 15.12$\pm$0.26 & 14.45$\pm$0.28 & 13.52$\pm$0.10 & 12.25$\pm$0.14 & 11.53$\pm$0.33 \\
 2.5298 &  3.4$\pm$0.9 & $\ldots$       & 14.25$\pm$0.31 & 12.87$\pm$0.09 & 11.86$\pm$0.33 & 11.53$\pm$0.37 \\
 2.5300 & 44.3$\pm$1.2 & 14.02$\pm$0.18 & 14.37$\pm$0.01 & 13.69$\pm$0.01 & 11.96$\pm$0.53 & $\ldots$       \\
\hline
 Total    &     -        & 15.33$\pm$0.16 & 15.29$\pm$0.05 & 14.37$\pm$0.02 & 12.73$\pm$0.11 & 12.41$\pm$0.08 \\
 \hline 
 \end{tabular} 
 \end{table*}

%% file: J1259met.tex
\begin{table*} 
 \centering 
 \caption{Results of Voigt-profile fitting to singly ionised metal absorption lines proximate to \Judcn. {The adopted spectral resolutions are 50.0 and 28.5~\kms\ for the UVB and VIS arm, respectively.}
 %J1259met.tex automatically created from f26.12p3w3906 
 \label{t:J1259met}} 
 \begin{tabular}{ccccccc}
 \hline \hline
{\Large \strut} $z$ & $b$ & \multicolumn{5}{c}{$\log N$ [cm$^{-2}$]} \\
     & (km\,s$^{-1})$ & \SII & \SiII & \FeII & \ZnII & \CrII \\
 \hline
 3.2453 & 32.2$\pm$0.5 & 13.59$\pm$0.29 & 13.95$\pm$0.02 & 13.53$\pm$0.04 & $\ldots$       & $\ldots$       \\
 3.2461 & 14.5$\pm$0.2 & 15.15$\pm$0.02 & 15.59$\pm$0.02 & 15.01$\pm$0.02 & 12.46$\pm$0.05 & 13.46$\pm$0.03 \\
\hline
 Total    &     -        & 15.17$\pm$0.02 & 15.60$\pm$0.02 & 15.03$\pm$0.02 & 12.46$\pm$0.05 & 13.46$\pm$0.03 \\
 \hline 
 \end{tabular} 
 \end{table*}

%% file: J1331met.tex
\begin{table*} 
 \centering 
 \caption{Results of Voigt-profile fitting to singly ionised metal absorption lines proximate to \Juttu.  {The adopted spectral resolutions are 45.0 and 26.0~\kms\ for the UVB and VIS arm, respectively.}
% J1331_met.tex automatically created from f26.12p3w3807 
\label{t:J1331met}} 
 \begin{tabular}{cccccccc}
 \hline \hline
{\Large \strut} $z$ & $b$ & \multicolumn{6}{c}{$\log N$ [cm$^{-2}$]} \\
     & (km\,s$^{-1})$ & \SII & \SiII & \FeII & \ZnII & \CrII & \NiII \\
 \hline
 2.9221 & 21.5$\pm$0.4 & 14.92$\pm$0.04 & 15.35$\pm$0.05 & 14.97$\pm$0.04 & 12.50$\pm$0.09 & 13.32$\pm$0.07 & 13.78$\pm$0.05 \\
\hline
 Total    &     -        & 14.92$\pm$0.04 & 15.35$\pm$0.05 & 14.97$\pm$0.04 & 12.50$\pm$0.09 & 13.32$\pm$0.07 & 13.78$\pm$0.05 \\
 \hline 
 \end{tabular} 
 \end{table*}

%% file: J1358met.tex
\begin{table*} 
 \centering 
 \caption{Results of Voigt-profile fitting to singly ionised metal absorption lines proximate to \Jutch. {The adopted spectral resolutions are 43.0 and 26.0~\kms\ for the UVB and VIS arm, respectively.}
 %J1358met.tex automatically created from f26.12p3w3767 
 \label{t:J1358met}} 
 \begin{tabular}{cccccccc}
 \hline \hline
{\Large \strut} $z$ & $b$ & \multicolumn{6}{c}{$\log N$ [cm$^{-2}$]} \\
     & (km\,s$^{-1})$ & \SII & \SiII & \FeII & \ZnII & \CrII & \NiII \\
 \hline
 2.8921 & 15.8$\pm$0.6 & 14.79$\pm$0.04 & 15.10$\pm$0.08 & 14.38$\pm$0.03 & 12.12$\pm$0.19 & 12.26$\pm$0.67 & 13.14$\pm$0.16 \\
 2.8926 & 30.6$\pm$1.4 & 15.09$\pm$0.03 & 15.62$\pm$0.04 & 14.66$\pm$0.02 & 12.57$\pm$0.11 & 12.98$\pm$0.18 & 13.51$\pm$0.08 \\
 2.8931 & 48.5$\pm$2.8 & 14.80$\pm$0.05 & 15.28$\pm$0.07 & 13.83$\pm$0.06 & 12.42$\pm$0.16 & 12.66$\pm$0.44 & $\ldots$       \\
 2.8940 & 15.0$\pm$2.3 & $\ldots$       & 13.75$\pm$0.31 & 13.62$\pm$0.04 & $\ldots$       & $\ldots$       & $\ldots$       \\
 2.8943 & 28.4$\pm$1.1 & $\ldots$       & 13.02$\pm$0.28 & 13.46$\pm$0.03 & $\ldots$       & $\ldots$       & $\ldots$       \\
 2.8959 & 23.7$\pm$1.0 & 13.54$\pm$0.57 & 13.66$\pm$0.02 & 13.08$\pm$0.03 & $\ldots$       & $\ldots$       & $\ldots$       \\
\hline
 Total    &     -        & 15.40$\pm$0.02 & 15.87$\pm$0.03 & 14.93$\pm$0.02 & 12.88$\pm$0.08 & 13.20$\pm$0.18 & 13.66$\pm$0.07 \\
 \hline 
 \end{tabular} 
 \end{table*}

%% file: J2228met.tex
\begin{table*} 
 \centering 
 \caption{
 Results of Voigt-profile fitting to singly ionised metal absorption lines proximate to \Jdddh.
 %J2228met.tex automatically created from f26.12p3w3491
  {The adopted spectral resolutions are 45.5 and 28.0~\kms\ for the UVB and VIS arm, respectively.}
 \label{t:J2228met}} 
 \begin{tabular}{cccccccccc}
 \hline \hline
{\Large \strut} $z$ & $b$ & \multicolumn{8}{c}{$\log N$ [cm$^{-2}$]} \\
     & (km\,s$^{-1})$ & \SII & \SiII & \FeII & \ZnII & \CrII & \MnII & \MgII & \TiII \\
 \hline
 2.7691 & 18.5$\pm$0.2 & 15.59$\pm$0.02 & 15.49$\pm$0.03 & 14.50$\pm$0.01 & 13.09$\pm$0.02 & 12.64$\pm$0.13 & 12.64$\pm$0.05 & 15.36$\pm$0.10 & 12.44$\pm$0.17 \\
 2.7701 & 10.3$\pm$0.5 & 13.92$\pm$0.15 & 14.15$\pm$0.06 & 13.28$\pm$0.03 & $\ldots$       & $\ldots$       & $\ldots$       & 14.10$\pm$1.62 & $\ldots$       \\
\hline
 Total    &     -        & 15.60$\pm$0.02 & 15.51$\pm$0.02 & 14.52$\pm$0.01 & 13.09$\pm$0.02 & 12.64$\pm$0.13 & 12.64$\pm$0.05 & 15.38$\pm$0.12 & 12.44$\pm$0.17 \\
 \hline 
 \end{tabular} 
 \end{table*}

%% file: J2325met.tex
\begin{table*} 
 \centering 
 \caption{Results of Voigt-profile fitting to singly ionised metal absorption lines proximate to \Jdtdc. {The adopted spectral resolutions are 50.0 and 26.5~\kms\ for the UVB and VIS arm, respectively.}
 %J2325met.tex automatically created from f26.12p3w3685 
 \label{t:J2325met}} 
 \begin{tabular}{ccccccccccc}
 \hline \hline
{\Large \strut} $z$ & $b$ & \multicolumn{9}{c}{$\log N$ [cm$^{-2}$]} \\
     & (km\,s$^{-1})$ & \SII & \SiII & \FeII & \ZnII & \CrII & \NiII & \MgII & \TiII & \MnII \\
 \hline
 2.6144 &  9.5$\pm$0.3 & 15.19$\pm$0.04 & 15.24$\pm$0.06 & 13.63$\pm$0.02 & 12.80$\pm$0.03 & $\ldots$       & $\ldots$       & 15.14$\pm$0.24 & $\ldots$       & $\ldots$       \\
 2.6163 & 19.4$\pm$0.5 & 15.01$\pm$0.04 & 15.46$\pm$0.03 & 15.05$\pm$0.02 & 12.29$\pm$0.07 & 13.35$\pm$0.03 & 13.80$\pm$0.03 & 15.29$\pm$0.17 & 11.97$\pm$0.78 & 12.73$\pm$0.03 \\
 2.6169 & 23.3$\pm$0.3 & 15.68$\pm$0.02 & 15.96$\pm$0.02 & 15.34$\pm$0.02 & 12.93$\pm$0.04 & 13.68$\pm$0.02 & 14.17$\pm$0.02 & 16.14$\pm$0.03 & 12.90$\pm$0.10 & 13.09$\pm$0.02 \\
\hline
 Total    &     -        & 15.87$\pm$0.02 & 16.14$\pm$0.02 & 15.53$\pm$0.01 & 13.23$\pm$0.02 & 13.84$\pm$0.01 & 14.32$\pm$0.01 & 16.23$\pm$0.04 & 12.95$\pm$0.12 & 13.25$\pm$0.01 \\
 \hline 
 \end{tabular} 
 \end{table*}

%% file: CO32.tex
\begin{table}[!h]
  \caption{Results from Gaussian fits of CO(3-2) emission lines. \label{t:CO32}}
  \begin{tabular}{ccccc}
    \hline \hline
    {\large \strut} Quasar & $z_{CO}$ & FWHM            &  $F_{\rm CO(3-2)}$        & $M(\HH)$     \\
           &         & (\kms)          &  (Jy\,\kms)     & $(10^{10} M_{\odot})$  \\
    \hline
    \Jzzuc & 2.6285  & 1010 $\pm$ 120  & 1.1  $\pm$ 0.2  & 3.4 - 17  \\
    \Jzzcn & 3.0378  & 500  $\pm$ 110  & 0.24 $\pm$ 0.07 & 0.9 - 4.6 \\
    \Jzudc & 2.6618  & 460  $\pm$ 80   & 0.5  $\pm$ 0.1  & 1.5 - 7.4 \\
    \Judcn & 3.2405  & 450  $\pm$ 180  & 0.2  $\pm$ 0.1  & 0.8 - 3.9 \\
    \Juttu & 2.9168  & 350  $\pm$ 100  & 0.22 $\pm$ 0.08 & 0.8 - 4.0 \\
    \Jdddh & 2.7780  & 330  $\pm$ 60   & 0.27 $\pm$ 0.07 & 0.9 - 4.5 \\
    \hline
  \end{tabular}
  \tablefoot{
    Molecular masses are calculated from the CO(3-2) luminosities assuming $r_{31}=0.97$ \citep{Carilli2013}. The range
    corresponds to assumed CO-to-\HH\ conversion factors either typical for quasars $\alpha_{CO}=0.8$\,M$_{\odot} (\kms\,$pc$^2)^{-1}$
    \citep[see e.g.][]{Bolatto2013} or using the standard value $\alpha_{CO}=4$\,M$_{\odot}\,(\kms\,$pc$^2)^{-1}$.
  }
  \end{table}